\pdfoutput=1

\documentclass[%
               pagenumberstop=false,
               BCOR=8mm,
               ngerman,english,
               ]{ktxthss}

\usepackage[%
            linking=true,
            titles=true,
            ]{ktxbbltx}

\usepackage{bm}
\usepackage{adjustbox}
\usepackage{multirow}
\usepackage{diagbox}
\usepackage{xfrac}

\usepackage[%
            colorlinks = true,%
            urlcolor   = Venetian,%
            citecolor  = Lime,%
            linkcolor  = Dodger,%
            linktocpage = true,%
            pdfauthor={Knut Zoch},%
            pdftitle={Cross-section measurements of top-quark pair production in association with a hard photon at 13 TeV with the ATLAS detector},%
            pdfsubject={PhD thesis in experimental particle physics},%
            pdfkeywords={PhD thesis, dissertation, particle physics, high-energy physics, ATLAS experiment, Standard Model, top quark, top-quark properties, top-photon coupling, cross-section measurement, fiducial cross-section, differential cross-section},%
            ]{hyperref}
\usepackage[noabbrev,capitalise]{cleveref}

\bibliography{dissertation-bib}
\usepackage{dissertation-defs}

\newcommand{\HRule}{\rule{\linewidth}{0.5mm}}

\title{%
  \HRule\\[0.4cm]
  \Large
  Cross-section measurements of top-quark\\ pair production in association with a hard photon\\ at 13 TeV with the ATLAS detector
  \HRule\\[1.5cm]
}

\subtitle{%
Dissertation\\[1.0cm]
for the award of the degree\\[0.5ex]
``Doctor of Philosophy'' (Ph.D.)\\[0.5ex]
Division of Mathematics and Natural Sciences\\[0.5ex]
of the Georg-August-Universit\"at G\"ottingen\\[1.0cm]
within the Physics doctoral programme\\[0.5ex]
of the Georg-August University School of Science (GAUSS)\\[2.0cm]
}

\author{%
\large
submitted by\\[1.0cm]
\large
Knut Zoch\\[0.5cm]
\large
from Wittmund\\[2.0cm]
}

\date{}
\publishers{%
\large
G\"ottingen, 2020
}

\uppertitleback{%
\noindent
\underline{Thesis Committee:}\\[0.2cm]
Prof. Dr. Arnulf Quadt\\
{\footnotesize II. Institute of Physics, Georg-August-Universit\"at G\"ottingen}\\[0.2cm]
Prof. Dr. Stan Lai\\
{\footnotesize II. Institute of Physics, Georg-August-Universit\"at G\"ottingen}\\[0.2cm]

\noindent
\underline{Members of the Examination Board:}\\[0.2cm]
Reviewer: \hspace*{5.2em} Prof. Dr. Arnulf Quadt\\
\hspace*{10.2em}{\footnotesize II. Institute of Physics, Georg-August-Universit\"at G\"ottingen}\\[0.2cm]
Second Reviewer: \hspace*{1.8em} Priv.-Doz. Dr. Johannes Erdmann\\
\hspace*{10.2em}{\footnotesize Physics Department, TU Dortmund University}\\[0.2cm]
Additional Reviewer: \hspace*{0.3em} Prof. Dr. Ivor Fleck\\
\hspace*{10.2em}{\footnotesize Physics Department, University of Siegen}\\[0.2cm]

\noindent
\underline{Further members of the Examination Board:}\\[0.2cm]
Prof. Dr. Ariane Frey\\
{\footnotesize II. Institute of Physics, Georg-August-Universit\"at G\"ottingen}\\[0.2cm]
Prof. Dr. Wolfram Kollatschny\\
{\footnotesize Institute for Astrophysics, Georg-August-Universit\"at G\"ottingen}\\[0.2cm]
Prof. Dr. Stan Lai\\
{\footnotesize II. Institute of Physics, Georg-August-Universit\"at G\"ottingen}\\[0.2cm]
Prof. Dr. Steffen Schumann\\
{\footnotesize Institute for Theoretical Physics, Georg-August-Universit\"at G\"ottingen}\\[1.0cm]

Date of the oral examination: 6 July 2020 \\[1.0cm]

\noindent
Reference: II.Physik-UniG\"o-Diss-2020/01  
}

\begin{document}

\frontmatter
\pagestyle{empty}
\maketitle

\pagestyle{plain}


\begin{center}
  \HRule\\[0.4cm]
  \Large\bfseries
  Cross-section measurements of top-quark\\ pair production in association with a hard photon\\ at 13 TeV with the ATLAS detector
  \HRule\\[1.5cm]
\end{center}

\begin{center}
  \bfseries
  \large
  Abstract
\end{center}

\noindent
25 years after the top quark's discovery, the Large Hadron Collider at CERN produces proton-proton collision data on unprecedented scales at unprecedented energies -- and has heralded an era of top-quark precision measurements.
The production of a top-quark pair in association with a photon ($t\bar{t}\gamma$) gives access to the electromagnetic top-photon coupling, one of the fundamental properties of the top quark.
Various extensions of the Standard Model predict modifications of the coupling strength or structure, and deviations from the Standard Model prediction of the $t\bar{t}\gamma$ production cross-section would indicate new physics.
With enough statistics available from the Large Hadron Collider, the electron-muon channel has gained particular interest due to its high signal purity and precise available theory predictions.

This thesis presents results with the full Run~2 dataset collected with the ATLAS detector in proton-proton collisions at the Large Hadron Collider between 2015 and 2018 at 13\,TeV centre-of-mass energy, corresponding to an integrated luminosity of 139\,fb$^{-1}$.
In order to compare the results to fixed-order calculations that include non-doubly-resonant diagrams, a combined measurement of $t\bar{t}\gamma + tW\gamma$ is performed.
The focus is placed on a measurement of the fiducial inclusive cross-section in the electron-muon channel, where exactly one photon, one electron and one muon of opposite charge sign, and at least two jets, one of which must be $b$-tagged, are selected.
Furthermore, the ATLAS data is unfolded to parton level and measurements of differential cross-sections as functions of several observables are presented.
Both fiducial inclusive and differential results are compared to state-of-the-art fixed-order calculations at next-to-leading order in QCD.
An additional focus of the thesis is placed on studies to use machine-learning techniques, in particular deep neural networks, for the identification of prompt photons.

\cleardoublepage




\tableofcontents



\chapter*{Acknowledgements}
\label{sec:preface}

\vspace*{-4pt} 

A doctorate is not something to be achieved without help or guidance from others, most certainly not in experimental particle physics.
The list of people is long and I appreciate everyone I met and worked with over the course of my PhD studies -- \emph{thank you!}

I would like to take the opportunity to thank Arnulf Quadt, my PhD advisor, for his inspiration which had convinced me of particle physics long before even considering a PhD, for the possibility to join his group as a doctoral student and to continue our successful collaboration, for the numerous opportunities of new experiences at conferences, workshops, at \CERN, in Tokyo or in Cambridge, and for the constant support in particle-physics matters and beyond.
I am grateful to Stan Lai, the second member of my thesis committee,
to Johannes Erdmann as the second reviewer of this thesis who had many helpful comments in his convenor role,
and to Ivor Fleck as the additional reviewer of this thesis.
I would like to thank the German Academic Scholarship Foundation (\emph{Studienstiftung des Deutschen Volkes}) for the PhD scholarship that supported me with my research.

I am thankful for the day-to-day supervision by Thomas Peiffer who always had sympathetic ears and helpful advice.
I would like to thank Elizaveta Shabalina for her supervision during my time at \CERN.
Her endless experience in top-quark physics, her positive attitude in spite of conference deadlines, and the discussions with her were most helpful.

An analysis is never a one-person job and I am grateful to everyone involved in the \PPT~project and the \tty \emu analysis.
Specifically, I would like to thank Julien Caudron, Yichen Li, Carmen Diez Pardos, Thomas Peiffer, Elizaveta Shabalina and Joshua Wyatt Smith for their guidance.
As my main contribution to the \tty \emu analysis was to the measurement of the fiducial inclusive \xsec, a special thanks goes to John Meshreki, my \enquote{Siegen counterpart}, who was the lead analyser for the differential measurements.

I am grateful to Osamu Jinnouchi who agreed to the G{\"o}ttingen--Tokyo collaboration and hosted me within his working group at the Tokyo Institute of Technology.
I would like to extend my gratitude to all his group members and everyone else I met in Japan for welcoming me so kindly.
I am equally grateful to everyone I met during my time at \CERN, both at work and beyond (\ie mostly in the mountains).
You made this a truly unique experience.
The same goes to all my colleagues and friends in G{\"o}ttingen.

I would like to thank Boris, Josh, Lisa, Martin and Thomas for their attentive reading of this document and for providing many helpful comments.
Mein letzter Dank geht an meine Eltern f{\"u}r ihre bedingungslose Unterst{\"u}tzung in allen Zeiten.


\chapter*{Contributions by the author}
\label{sec:contributions}

\vspace*{-4pt} 

An experiment as large and complex as the \ATLAS detector needs a collaboration of thousands of physicists to develop and operate the experiment successfully, and to conduct measurements using its recorded collision data.
Since many different steps contribute to \ATLAS publications, these papers are always published in the name of the entire collaboration and are purposely not attributed to individuals.
The author's work documented in this thesis entered two \ATLAS publications: a measurement of the \tty process in the \ljets and dilepton channels using \ATLAS data corresponding to \SI{36}{\ifb} of integrated luminosity~\cite{TOPQ-2017-14}, and a \tty measurement in the \emu channel using \SI{139}{\ifb} of integrated luminosity~\cite{TOPQ-2020-03}.
The following paragraphs are meant to provide a (non-exhaustive) list of the author's contributions to these measurements.

The \SI{36}{\ifb} \tty analysis~\cite{TOPQ-2017-14} uses the \emph{Prompt Photon Tagger} (\PPT) tool in the \ljets channels, a project initiated as a collaborative effort of the G{\"o}ttingen \ATLAS group.
The tool was designed by B.\,V{\"o}lkel and by \JWSmith as part of their MSc thesis and PhD thesis projects, respectively~\cite{Volkel:2017aa,Smith:2018sma}.
After the architecture and the design of the \PPT had been established and finalised, the author took over from B.\,V{\"o}lkel  and contributed to the implementation of the \PPT into the \tty analysis, in particular to the treatment of systematic uncertainties associated with the \PPT (\cf \cref{chap:PPT}).
In addition to the \PPT contributions, the author performed studies related to the implementation of other systematic uncertainties in Ref.~\cite{TOPQ-2017-14}.

The author made central contributions to the \tty \emu measurement~\cite{TOPQ-2020-03} using data corresponding to \SI{139}{\ifb} and was the lead analyser for the fiducial inclusive \xsec measurement:
he wrote the \enquote{ntuple} production software and produced and processed them, studied sample overlap-removal strategies and the categorisation of photons (\cf \cref{sec:simulation-categorisation,sec:simulation-overlap}), conducted event-selection studies (\cf \cref{cha:selection,chap:app-add-controlplots}) and established the treatment of systematic uncertainties and their evaluation (\cf~\cref{chap:systematics,cha:app-red-blue-plots}).
These steps are prerequisites for both the fiducial inclusive and differential \xsec measurements performed in Ref.~\cite{TOPQ-2020-03}.
The latter are only summarised briefly in this thesis (\cf \cref{sec:strategy-differential,sec:results-differential,chap:app-diff-xsec}) as they were not the focus of the author's work.
The lead analyser for these differential measurements was J.\,Meshreki and more details about them are to appear in his PhD thesis.
The author of this thesis conducted the fiducial inclusive \xsec measurement and all related studies and established its strategy (\cf \cref{sec:strategy-fit,sec:results-configuration,sec:results-Asimov,sec:results-data,chap:app-fit-decorrelation}).


\mainmatter
\pagestyle{headings}

\chapter{Introduction}
\label{sec:theory}

\defbibentryset{Schrodinger}{Schrodinger:1926gei,Schrodinger:1926qnk,Schrodinger:1926vbi,Schrodinger:1926xyk}

Elementary particle physics, and in particular high-energy physics, comes with the beauty and the curse of being exceptionally well-described by one single theory: the Standard Model of elementary particles.
It is second to none in both its integrity of its description of phenomena and in the way it changed humankind's perception of the nature of the universe.
The setting stone of today's physics knowledge of elementary particles was laid -- both experimentally and conceptually -- about one hundred twenty years go.

The first elementary particle of the Standard Model, the electron, was discovered by \citeauthor{Thomson:1897cm}~\cite{Thomson:1897cm} in 1897 when experimenting with cathode rays.
Not only did he show the cathode rays to be composed of particles, but he also identified these negatively charged particles to be identical with those radiated from radioactive, heated or illuminated materials.
Conceptually, the works by Planck and Einstein around the turn of the century represented a paradigm shift and were the birth of \emph{modern physics}.
Einstein's \emph{annus mirabilis papers}~\cite{Einstein:01,Einstein:02,Einstein:03,Einstein:04} from 1905 introduced several groundbreaking concepts, including the photoelectric effect, the mass-energy equivalence and special relativity.
On the other hand, Planck's descriptions of black-body radiation spectra, first discussed on conferences in 1900 and then published in 1901~\cite{Planck1901}, imposed the quantisation of energy.
This paved the way for the quantum mechanics of the 1920's:
the wave formulation of \citeauthor{Schrodinger:1926gei}~\cite{Schrodinger} and \citeauthor{Heisenberg:1925aa}'s equivalent formulation of matrix mechanics~\cite{Heisenberg:1925aa,Born:1925aa,Born:1926aa} provided the first conceptually autonomous description of quantum physics.
Dirac combined both quantum mechanics and Einstein's theory of special relativity into the relativistic Dirac wave equation in the late 1920's~\cite{Dirac:01,Dirac:02}, which started a series of developments that culminated in a consistent theory of quantum electrodynamics by 1950~\cite{Tomonaga:1946zz,Schwinger:1948yk,Schwinger:1948iu,Feynman:1949zx,Feynman:1949hz,Feynman:1950ir,Dyson:1949bp,Dyson:1949ha}.
This theory, in conjunction with the electroweak unification and the theory of quantum chromodynamics, is what forms today's Standard Model of elementary particles:
a set of relativistic quantum field theories, all using the same consistent language, to describe the fundamental particles and interactions between them.

The research topic of this thesis is set in the field of top-quark physics, one of the particles where the Standard Model has proven its unprecedented predictive accuracy:
albeit only discovered two decades later in 1995~\cite{Abe:1995hr,D0:1995jca}, the top quark's existence had been postulated as early as in the 1970's.
Among the particles of the Standard Model, it is noteworthy as it is by far the heaviest and, thus, takes on an important role in searches for beyond-Standard-Model physics at higher mass scales.
Physics of the top quark is an open field of research and determining its properties provides valuable insight into the physics of the Standard Model -- and into its possible shortcomings.

More specifically, this thesis focuses on measurements of top-quark pair production in association with a photon, \tty, a process that probes the coupling behaviour of the top quark.
Not only does the measurement of this process test meticulously the coupling behaviour predicted by the Standard Model, but it also comes with sensitivity to beyond-Standard-Model-like physics with anomalous coupling structures.
The thesis presents analysis results of data recorded with the \ATLAS experiment at the Large Hadron Collider at \CERN, the \emph{European Organization for Nuclear Research}.
The examined dataset was taken during the \runii data-taking period in the years 2015 to 2018 from proton-proton collisions at \SI{13}{\TeV} \com energy, and it comprises the largest number of top-quark events recorded to date.
The production \xsec of \tty is measured in a fiducial phase-space volume both inclusively and differentially as a function of several observables in the electron-muon final state.
Preliminary results of the analysis were presented by the author on behalf of the \ATLAS Collaboration at the \enquote{12th International Workshop on Top Quark Physics} (\textsc{top 2019}) in Beijing, China~\cite{Zoch:TOP2019}, and were made public as an \ATLAS conference note~\cite{ATLAS-CONF-2019-042}.
The final results presented here were also submitted as an \ATLAS publication to the peer-reviewed \emph{Journal of High Energy Physics} and have been accepted for publication~\cite{TOPQ-2020-03}.
In this thesis and in the journal publication, the measurement is compared against state-of-the-art theory predictions for the \emu final state~\cite{Bevilacqua:2018woc,Bevilacqua:2018dny}.
An additional focus of this thesis is placed on machine-learning techniques used to identify photons in the \tty \ljets channels in a separate \ATLAS measurement~\cite{TOPQ-2017-14}.

The thesis is organised as follows.
The sections of this chapter first briefly introduce the fundamental concepts and underlying theory of the analysis.
An introduction to the Standard Model of elementary particles is given in \cref{sec:theory-sm}.
Then, \cref{sec:theory-top} discusses the physics of the top quark, before \cref{sec:theory-tty} puts a focus on top quarks in association with photons.
This last section covers both previous experimental results as well as possible interpretations of \tty measurements.
Following that, \cref{sec:exp} introduces the experimental setup of the measurement and discusses the Large Hadron Collider, the \ATLAS experiment and how physics objects are detected and reconstructed with \ATLAS.
\Cref{chap:PPT} presents studies of machine-learning techniques for photon identification and summarises results of an \ATLAS \tty measurement that used these techniques.
The remaining chapters will then focus on the analysis of the \emu final state.
\Cref{chap:simulation} deals with the generation of simulated data using Monte Carlo techniques.
These are pivotal to understand and distinguish contributions in data from various types of processes.
\Cref{cha:selection} summarises how events from the analysed dataset are selected.
\Cref{cha:strategy} discusses the strategy of this analysis for measuring top-quark pair production in association with a photon.
\Cref{chap:systematics} details the systematic uncertainties considered in this measurement.
\Cref{chap:results} presents the results, before a summary and conclusions are given in \cref{chap:conclusions}.
Additional material and studies are summarised in \cref{cha:app-red-blue-plots,chap:app-add-controlplots,chap:app-fit-decorrelation,chap:app-diff-xsec}.

\clearpage

\section{The Standard Model of elementary particles}
\label{sec:theory-sm}

\begin{figure}
  \centering
  \includegraphics[width=0.86\textwidth]{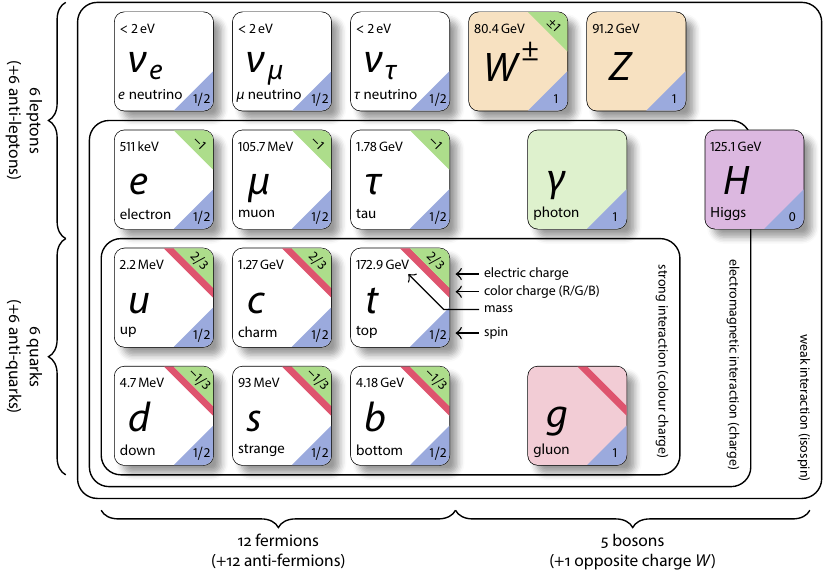}
  \caption[Particles of the Standard Model]{%
    The particles of the Standard Model.
    The twelve fermions are shown on the left, grouped into leptons in the two upper rows and quarks in the two lower rows.
    The coloured boxes on the right represent the gauge bosons and the Higgs boson.
    The quoted mass values are according to Ref.~\cite{PhysRevD.98.030001}.
  }
  \label{fig:sm-overview}
\end{figure}

The Standard Model (\SM) of elementary particles~%
\cite{Glashow:1961tr,Weinberg:1967tq,Salam:1968rm,Glashow:1970gm,Georgi:1972cj,Gross:1973id,Politzer:1973fx,tHooft:1971akt,tHooft:1971qjg,tHooft:1972tcz,tHooft:1972qbu}
is the best and most complete theory of elementary particles and their interactions to date.
The \SM is a quantum field theory based on the $SU(3) \otimes SU(2) \otimes U(1)$ gauge groups and knows two types of elementary particles: twelve \spinhalf fermions that form all visible matter, and \spinone gauge bosons, mediators of the interactions.
Three fundamental interactions are included in the \SM: the strong interaction, mediated by massless gauge bosons called gluons, the weak interaction mediated by the massive $W^{\pm}\mkern-2mu$~bosons and the $Z^0$~boson, and the electromagnetic interaction mediated by massless photons.
The latter two interactions are combined into one consistent theory through electroweak unification~\cite{Glashow:1961tr,Weinberg:1967tq,Salam:1968rm}.
The masses of the elementary particles are generated through \emph{spontaneous symmetry breaking} of the electroweak gauge symmetry~\cite{Anderson:1963pc,Higgs:1964ia,Englert:1964et,Guralnik:1964eu}, which gives rise to the last elementary particle: the Higgs boson.
An overview of all particles of the \SM is given in \cref{fig:sm-overview}, with the twelve fermions on the left, and the gauge bosons and the Higgs boson on the right.

The fermions are separated into quarks and leptons, according to their different coupling behaviour under the three interactions.
Quarks carry colour charge, the quantum number associated with the strong interaction, and therefore interact with gluons; leptons are colour-neutral.
Only those particles with non-zero electromagnetic charge are subject to the electromagnetic interaction, and while this includes all six quarks, only three of the six leptons are charged.
The other three, the uncharged, massless neutrinos of the \SM do not couple to the photon.
In addition to the fermions, the $W^{\pm}\mkern-2mu$~bosons carry electromagnetic charge and thus interact electromagnetically.
The $W^{\pm}\mkern-2mu$~bosons themselves only couple to particles with negative (or left-handed) chirality, that is, to those fermions that transform in a left-handed representation of the Poincar\'e group~\cite{Poincare1906} of special relativity.
To describe this coupling behaviour, the quantum number of the weak isospin is introduced to distinguish left-handed and right-handed fermions, the latter of which carry no isospin.
With this distinction between left-handed and right-handed particles, the charged-current weak interaction violates parity.
The neutral-current weak interaction mediated by the $Z^0$~boson couples to all particles with weak isospin or electromagnetic charge%
\footnote{%
  Henceforth, when referring to $W^{\pm}\mkern-2mu$~bosons and $Z^0$~bosons, the superscript denoting the boson charges is omitted for simplicity.
  Both charge configurations are implied when referring to \Wbosons.
}.
The following paragraphs introduce the strong and electroweak interactions, both essential components to perform top-quark physics at hadron colliders, in more detail.

\paragraph{quantum chromodynamics.}
The $SU(3)$ component of the underlying gauge groups of the \SM describes the strong interaction between elementary particles that carry \emph{colour} charge~\cite{Fritzsch:1973pi,Gross:1974cs,Weinberg:1973un}.
The theory of these interactions is known as quantum chromodynamics (\QCD) and comprises two types of fundamental fields:
the Dirac spinor fields $\psi_{q,a}$ of the colour-charged \spinhalf quarks, where $q = (u,d,c,s,t,b)$ denotes the quark flavour and $a = (1, \dots, n_c)$ refers to the colour degrees of freedom.
In the \SM, $n_c = 3$, and for a visual representation, the three colour states are often denoted as \emph{red}, \emph{green} and \emph{blue} ($r, g, b$).
The other type of fields are the gauge fields $G^C_{\mu}$ of the gluons, corresponding to electrically uncharged, massless particles of spin~1, which mediate the strong interaction of the \QCD and are also subject to self-interaction.
$C = (1, \dots, 8)$ labels the colour index of the gluon fields in this notation, hence, there are \emph{eight} different gluon fields with different colour-anticolour combinations.
With these two types of fields, the \QCD Lagrangian is
\vspace*{0pt minus 2pt}  
\begin{align}
  \label{eq:lagrange_qcd}
  \mathcal{L} = \sum_q
  \conj{\psi}_{q,a} \left( i \gamma^\mu \partial_\mu \delta_{ab} - g_s \gamma^\mu \lambda^C_{ab} G^C_{\mu} - m_q \delta_{ab}  \right) \psi_{q,b}
  - \frac{1}{4} \mathrm{tr} \left( \mathcal{G}_{\mu\nu} \mathcal{G}^{\mu\nu} \right) \, ,
\end{align}
where Einstein notation is used for repeated Latin and Greek indices and sums are implied in the ranges of 1 to 3 and 1 to 4, respectively.
$\gamma^\mu$ are the Dirac gamma matrices.
$\psi_{q,a}$ and $m_q$ are the spinor field and mass of quark $q$, respectively.
The $\lambda^C$ are the generators of the underlying Lie algebra of $SU(3)$, also known as the Gell-Mann $\lambda$-matrices~\cite{GellMann:1962xb}.
The indices $a,b$ denote row and column entries in their $3\times3$ matrix representation.
The combination of $\lambda^C G_\mu^C$ encodes the idea that an interaction with a gluon rotates the colour of a quark within the $SU(3)$ colour space; the quark fields and $\lambda$-matrices are said to be the \emph{fundamental representation} of $SU(3)$.
On the other hand, the gluon fields transform according to the $8\times 8$ adjoint representation of $SU(3)$.
$g_s = \sqrt{4\pi\alpha_s}$ is the coupling constant of the strong interaction.
$\mathcal{G}^A_{\mu\nu}$ is the field strength tensor of the gluon fields $G^A_\mu$, defined as
\vspace*{0pt minus 2pt}  
\begin{align}
  \label{eq:qcd_field_tensor}
  \mathcal{G}^A_{\mu\nu} = \partial_\mu G_\nu^A - \partial_\nu G_\mu^A - g_s f_{ABC} G_\mu^B G_\nu^C \, ,
\end{align}
where again $A,B,C = (1, \dots, 8)$.
$f_{ABC}$ are known as the structure constants of the $SU(3)$ group, defined by the commutators of the $\lambda$-matrices: $\left[ \lambda^A,\lambda^B \right] = 2i f_{ABC}\lambda^C$.

The mathematical structure of the \QCD Lagrangian predicts three different types of interaction vertices in the Feynman calculus: quark-antiquark-gluon vertices ($q\bar{q}g$), and 3-gluon and 4-gluon vertices.
The latter two govern the previously mentioned self-interaction.
\QCD and the strong interaction come with two other peculiarities, the first of which is known as \emph{confinement}:
quarks and gluons are confined into combined, colour-singlet states called \emph{hadrons}, and neither quarks nor gluons are observed as free particles in nature.
Figuratively speaking, colour singlets, or colour-neutral states, can be obtained by either having all three colour types in equal amounts (similar to how a beam of different light colours combines to the colour \emph{white}), or by requiring the total amount of each colour to be zero.
In practice, these two types of colour-neutral hadrons are indeed observed:
one distinguishes \emph{mesons}, a compound state of a quark and an antiquark, and \emph{baryons} as a combination of three quarks or three antiquarks.
More exotic configurations of mesons and baryons, such as tetraquarks ($q\bar{q}q\bar{q}$) and pentaquarks ($qqqq\bar{q}$), have also been observed experimentally.
The second peculiarity is known as \emph{asymptotic freedom}~\cite{Gross:1973id,Politzer:1973fx} and describes the decrease of the strong interaction as the energy-momentum transfer in a process increases.
In a similar fashion, colour-charged particles become asymptotically free as the distance scale decreases.

Divergences arise in \QCD calculations if loop corrections and self-interactions are considered to full extent, but these can be compensated for by applying \emph{renormalisation} techniques~%
\cite{Faddeev:1967fc,tHooft:1971akt,tHooft:1971qjg,tHooft:1972tcz,tHooft:1972qbu,Lee:1972fj,Lee:1974zg,Lee:1972yfa}.
However, the renormalisation introduces an energy-scale dependence to the coupling constant~$\alpha_s$ and restricts its validity to energies close to that scale.
Contrary to what its name suggests, the renormalised coupling constant $\alpha_s$ becomes a running constant $\alpha_s(\mu_R^2)$ and depends on the energy scale $\mu_R$ at which it is evaluated.
$\mu_R$ is called the renormalisation scale and is usually chosen to be at the same order of magnitude as the energy-momentum transfer of the process.
In perturbative \QCD, the running coupling can then be expressed through the \emph{renormalisation group equation} and the perturbative evolution of the \emph{beta function}~%
\cite{Wilson:1969zs,Callan:1970yg,Symanzik:1970rt,Christ:1972ms,Frishman:1973pp}:
\vspace*{0pt plus 3pt}  
\begin{align}
  \label{eq:renorm_group_eq}
  \mu_R^2 \frac{\mathrm{d}\alpha_s}{\mathrm{d}\mu_R^2} = \beta(\alpha_s) = - \left( b_0 \alpha_s^2 + b_1 \alpha_s^3 + b_2 \alpha_s^4 + \dots \right)
  \quad \text{with} ~ b_0 = \frac{11 n_c - 2 n_f}{12\pi} \, .
\end{align}
$n_f$ here denotes the number of quark flavours.
When only considering the leading-order term and as long as $n_f < \frac{11}{2} n_c$, the asymptotic freedom becomes apparent as the solution of \cref{eq:renorm_group_eq} satisfies $\alpha_s(\mu_R^2) \to 0$ in the limit $\mu_R^2 \to \infty$.
In energy regimes, where $m_q \ll \mu_R$, the solution can be expressed in terms of another energy scale~$\LambdaQCD^2$:
\vspace*{0pt plus 3pt}  
\begin{align}
  \label{eq:running_coupling}
  \alpha_s (\mu_R^2, \LambdaQCD^2) = \frac{12\pi}{(11 n_c - 2 n_f) \ln(\mu_R^2/\LambdaQCD^2)} \, .
\end{align}
Here, $\LambdaQCD^2$ depends on the choice of $\mu_R^2$ and represents the scale at which the perturbative approach \enquote{breaks down}.
This point, where the $\alpha_s(\mu_R^2)$ evolution diverges, is known as the Landau pole.
Thus, physics at the scale of $\LambdaQCD^2$ and below would be dominated by non-perturbative effects.
The exact value of $\LambdaQCD^2$ depends on the renormalisation scheme and the order, at which the perturbative series in \cref{eq:renorm_group_eq} is evaluated~\cite{Celmaster:1979km}.
In the commonly used modified minimal-subtraction scheme (denoted $\overline{\mathit{MS}}$), the \textsc{ccfr} Collaboration at the \TEVATRON collider, for example, measured $\LambdaQCD \sim \SI{0.2}{\GeV}$~\cite{Quintas:1992yv}.

\begin{figure}
  \centering
  \includegraphics[width=0.58\textwidth,clip,trim=0 10pt 0 30pt]{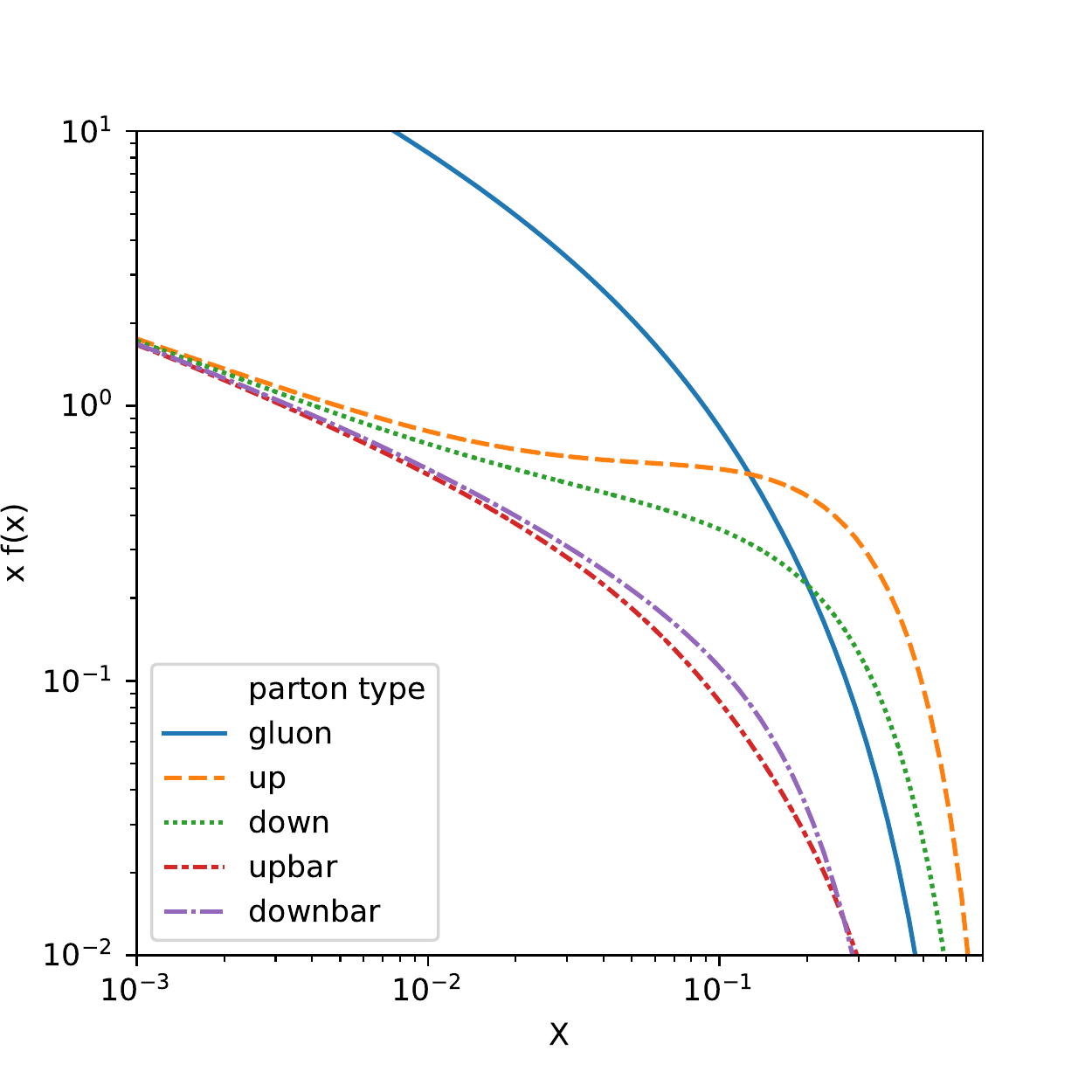}
  \caption[\PDF distributions obtained by the \NNPDF Collaboration]{%
    \PDF distributions at $Q^2 = \SI{3e4}{\GeV \tothe 2}$ obtained by the \NNPDF Collaboration~\cite{Ball:2014uwa} for the lightest two quarks and their antiquarks, as well as for the gluon.
  }
  \label{fig:theory-PDF}
\end{figure}

Physics at hadron colliders such as the \LHC is vastly dominated by \QCD processes, the \xsecs of which can be calculated precisely in perturbative \QCD:
the Feynman calculus gives sets of rules for the calculation of \QCD matrix elements and transition amplitudes, and \xsecs can be obtained from them through Fermi's golden rule.
However, none of these consider the dynamics involved with partons confined in hadronic bound states.
This long-distance regime cannot be described with perturbative \QCD due to occurring soft and collinear singularities, but it can be separated from the short-scale hard interaction according to the \QCD \emph{factorisation theorem}~\cite{Collins:1989gx}.
The factorisation introduces a scale~$\mu_F$, at which the long-distance physics is separated from the hard interaction.
While the latter remains calculable for individual processes via perturbative evolution, the long-distance parton-parton interactions within hadrons in the initial state are parametrised through structure functions $f_i(\xi_i, \mu_F^2)$, known as \emph{parton density functions} (\PDFs)~\cite{Bjorken:1968dy,Feynman:1969ej,Bjorken:1969ja,Bloom:1969kc,Friedman:1972sy}, where $\xi_i$ is the momentum fraction carried by the incoming parton $i$.
The remaining dependency of the \PDFs on the factorisation scale is described by the Dokshitzer-Gribov-Lipatov-Altarelli-Parisi (\DGLAP) equations~\cite{Dokshitzer:1977sg,Gribov:1972ri,Altarelli:1977zs} and is based on \emph{splitting kernels} to describe the splitting probabilities of quarks and gluons.
Similarly to the running coupling constant, the \PDFs are calculated through perturbative evolution around the factorisation scale.
The masses of heavy quarks constitute flavour thresholds in this evolution:
below a heavy quark's mass, this quark is not considered to be part of the hadron content, while it is considered to be part of the hadron content for a scale above its mass.
Then, the quark is included in the \DGLAP equations with massless splitting kernels.
As an example, \PDF distributions at $Q^2 = \SI{3e4}{\GeV \tothe 2} \sim m_t^2$ of the lightest quarks and gluons are shown in \cref{fig:theory-PDF}, as obtained by the \NNPDF Collaboration~\cite{Ball:2014uwa}.

After factorisation, hadron-hadron \xsecs can be calculated by combining \PDFs and partonic \xsecs.
For example, the inclusive \xsec for the production of a top-quark pair (\ttbar pair) at a proton-proton collider can be factorised as~\cite{Collins:1989gx}
\vspace*{0pt plus 2pt}  
\begin{align}
\begin{split}
  \label{eq:hadronic-xsec}
  \sigma(pp \to \ttbar) = \sum_{i,j} \int & \mathrm{d}\xi_i \, \mathrm{d}\xi_j \, f_i(\xi_i, \mu_F^2) \, f_j(\xi_j,\mu_F^2)
  \\ & \times \hat{\sigma}(ij \to \ttbar)(\hat{s}, \xi_i, \xi_j, m_t, \mu_F^2, \mu_R^2) \, ,
\end{split}
\end{align}
where the indices $i, j$ run over gluons, quark flavours and antiquark flavours.
$\hat{\sigma}(ij \to \ttbar)$ denotes the \xsec of the partonic hard interaction, renormalised at scale $\mu_R^2$, factorised at scale $\mu_F^2$, and a function of its effective parton-parton \com energy $\hat{s}$.
In this particular example, the \xsec also depends on the mass $m_t$ of the top quark.

\paragraph{electroweak unification.}
The other two interactions of the \SM, the weak and the electromagnetic interactions, were initially described by two disjoint theories known as quantum flavourdynamics and quantum electrodynamics.
Glashow, Weinberg and Salam introduced electroweak unification~\cite{Glashow:1961tr,Weinberg:1967tq,Salam:1968rm} and showed that the two interactions can be described as a single Yang-Mills field with underlying $SU(2) \times U(1)$ Lie groups.
The two groups introduce the gauge fields $W^i_\mu$ with $i = (1, 2, 3)$ and $B_\mu$ with coupling constants $g$ and $g'$, respectively.
The generators associated with $W^i_\mu$ are known as the weak isospin~$I$, that associated with $B_\mu$ is called the weak hypercharge~$Y$.
The $SU(2)$ part of the theory violates parity and the left-handed fermion fields $\psi_L$ transform as doublets under the group, whereas the right-handed fermion fields transform as $SU(2)$ singlets.
While the weak lepton doublets are identical with the physical mass eigenstates of the leptons, the weak eigenstates of the quarks are admixtures of their mass eigenstates.
The admixture is described through the Cabibbo-Kobayashi-Maskawa (\textsc{ckm}) mixing matrix~\cite{Cabibbo:1963yz,Kobayashi:1973fv}:
\vspace*{0pt plus 2pt}  
\begin{align}
  \label{eq:theory-CKM-matrix}
  \left(\begin{matrix} d' \\ s' \\ b' \\ \end{matrix}\right) =
  \left(\begin{matrix}
      V_{ud} & V_{us} & V_{ub} \\
      V_{cd} & V_{cs} & V_{cb} \\
      V_{td} & V_{ts} & V_{tb} \\
    \end{matrix}\right)
  \left(\begin{matrix} d \\ s \\ b \\ \end{matrix}\right) \, ,
\end{align}
where $d',s',b'$ denote the weak eigenstates, and $d,s,b$ represent the mass eigenstates.
Diagonal matrix elements are close to $\left| V_{ii} \right| \sim 1$.
Off-diagonal elements are suppressed, in particular the admixture described by the corner elements $V_{td}$ and $V_{ub}$.
With the quark admixture, the left-handed weak isospin doublets are
\vspace*{0pt plus 2pt}  
\begin{align}
  \label{eq:theory-weak-eigenstates}
  L_{L1} = \left( \begin{matrix}{\nu_e}_L \\ e_L\end{matrix} \right) \, ,
  \quad
  Q_{L1} = \left( \begin{matrix}u_L \\ d_L'\end{matrix} \right) \, ,
\end{align}
where the index~$1$ refers to the doublets of the first lepton and quark generations.
The doublets for generations 2 and 3 are defined equivalently.
The respective $SU(2)$ singlets are ${\nu_e}_R$, $e_R$, $u_R$ and $d_R'$, and equivalently for generations 2 and 3.

A complex scalar field, known as the Higgs field, is added to the electroweak formalism to generate the masses of the fermion fields and of the weak isospin fields.
This mechanism, proposed independently by multiple groups of people in 1964~\cite{Anderson:1963pc,Higgs:1964ia,Englert:1964et,Guralnik:1964eu}, but often attributed to \citeauthor{Higgs:1964ia,Englert:1964et}, spontaneously breaks the electroweak gauge symmetry and is known as the Brout-Englert-Higgs mechanism.
The excitation of the added field corresponds to a physical scalar particle known as the Higgs boson.
The Higgs field is added as a complex scalar $SU(2)$ doublet $\phi = (\phi^+, \phi^0)$ with a potential $V(\phi)$.
The parameters of the potential can be chosen in such a way that it develops a non-zero vacuum expectation value~$v$ and the electroweak gauge symmetry is broken.
Then, only the physical neutral Higgs scalar $\phi^0 = h$ remains.
\citeauthor{Higgs:1964ia,Englert:1964et} were awarded the 2013 Nobel Prize in Physics after the \ATLAS and \CMS Collaborations at \CERN had announced the discovery of a Higgs-boson-like particle in 2012~\cite{HIGG-2012-27,Chatrchyan:2012xdj}.

With the $SU(2) \times U(1)$ gauge groups and the Higgs mechanism, the Lagrangian of the electroweak sector can then be written as
\vspace*{0pt plus 3pt}  
\begin{subequations}
\begin{flalign}
  \hspace*{1em}
  \mathcal{L}_{\mathrm{EW}} =
  & \sum_i \conj{\psi}_i \left( i \gamma^\mu \partial_\mu - m_i - \frac{m_i h}{v} \right) \psi_i
  \vphantom{\dfrac11 \sum_i}  
  && \text{\llap{\small fermion kinematics + Yukawa}}
  \label{eq:the-lagrangian-ew}
  \\ & - \frac{g}{2\sqrt{2}} \sum_i \conj{\psi}_i \gamma^\mu \left(  1 - \gamma^5 \right) \left( T^+ W_\mu^+ + T^- W_\mu^- \right) \psi_i
  \vphantom{\dfrac11 \sum_i}  
  && \text{\llap{\small weak charged current}}
  \label{eq:the-lagrangian-ew2}
  \\ & - e \sum_i Q_i \conj{\psi}_i \gamma^\mu \psi_i A_\mu
  \vphantom{\dfrac11 \sum_i}  
  && \text{\llap{\small electromagnetic current}}
  \label{eq:the-lagrangian-ew3}
  \\ & - \frac{g}{2 \cos \theta_W} \sum_i \conj{\psi}_i \gamma^\mu \left( g_V^i - g_A^i \gamma^5 \right) \psi_i Z_\mu
  \vphantom{\dfrac11 \sum_i}  
  && \text{\llap{\small weak neutral current}}
  \label{eq:the-lagrangian-ew4}
  \\ & + |D_\mu h|^2 - V(h)
  \vphantom{\dfrac11 \sum_i}  
  && \text{\llap{\small Higgs kinematics + potential}}
  \label{eq:the-lagrangian-ew5}
  \\ & - \frac{1}{4} B_{\mu\nu}B^{\mu\nu} - \frac{1}{4} \mathrm{tr}  \left( W_{\mu\nu} W^{\mu\nu} \right) \, .
  \vphantom{\dfrac11 \sum_i}  
  && \text{\llap{\small gauge field kinematics}}
  \label{eq:the-lagrangian-ew6}
\end{flalign}
\end{subequations}
The sums of the individual terms run over all fermions~$i$.
In \cref{eq:the-lagrangian-ew2}, $g$ is again the coupling constant of the weak isospin fields, and $W^\pm \equiv (W^1 \mp i W^2)/\sqrt{2}$ are the physical representations of the fields, corresponding to the charged \Wbosons.
$T^+$ and $T^-$ are the weak-isospin raising and lowering operators, which generate coupling terms such as $W_\mu^- \overline{e}\gamma^\mu(1-\gamma^5)\nu_e$ and $W_\mu^+ \overline{\nu}_e \gamma^\mu (1-\gamma^5) e$, that is, they connect the weak isospin doublet partners at the interaction vertex.

Instead of the isospin and hypercharge fields $W^3_\mu$ and $B_\mu$, the Lagrangian now considers the physical fields $A_\mu$ of the photon and $Z_\mu$ of the \Zboson .
They are admixtures of the bare fields $B_\mu$ and $W^3_\mu$ according to the electroweak mixing angle $\theta_W$, also known as the Weinberg angle, with
\vspace*{0pt plus 2pt}  
\begin{subequations}
\begin{align}
  \label{eq:theory-gauge-mixture}
  A_\mu &= \phantom{-} B_\mu \cos \theta_W + W^3_\mu \sin \theta_W \, ,\\[0.2ex]
  Z_\mu &= -B_\mu \sin \theta_W + W^3_\mu \cos \theta_W \, ,
\end{align}
\end{subequations}
where $\theta_W$ is defined through the couplings strengths $\theta_W = \tan^{-1} \sfrac{g}{g'}$.
The electromagnetic coupling constant~$e$, corresponding to the elementary charge, is related to them via $e = g \sin \theta_W$.
The masses of the bosons are generated through $M_W = \sfrac{1}{2} \, ev \sin \theta_W$ and $M_Z = \sfrac{1}{2} \, ev \sin \theta_W \cos \theta_W$, where $v$~denotes the vacuum expectation value of the scalar higgs field~$h$.
The photon remains massless after symmetry breaking, $M_\gamma = 0$.
The Higgs boson receives its mass through one of the parameters, $\lambda$, of its potential with $M_h = \lambda v$.
The neutral-current weak coupling in \cref{eq:the-lagrangian-ew4} contains the fermion-specific parameters $g_V^i$ and $g_A^i$ that determine the vectorial and axial-vectorial components.
They relate the weak isospin $I$ or, more precisely, its polarisation along a reference axis $I_3$ with the electromagnetic charge: $g_V^i = I_3^i - 2 Q_i \sin^2 \theta_W$ and $g_A^i = I_3^i$.
Thus, the \Zboson couples to right-handed particles with a purely vectorial coupling, whereas the coupling to left-handed particles has both vectorial and axial-vectorial components.

\begin{figure}
  \centering
  \includegraphics[width=\textwidth]{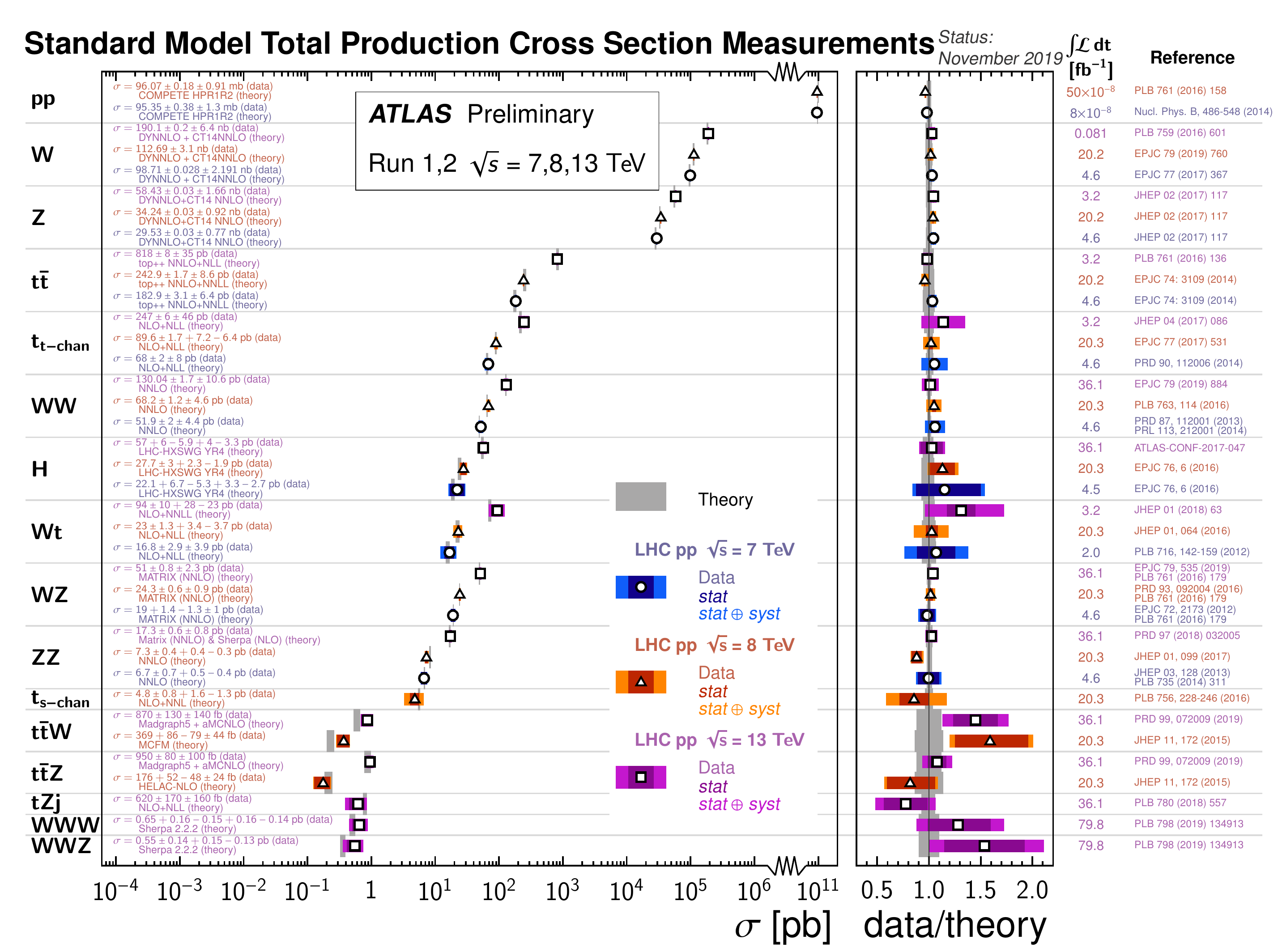}
  \caption[Measurements of \SM \xsecs performed by \ATLAS]{%
    Measurements of \SM \xsecs performed by \ATLAS using proton-proton collision data from \runi and \runii of the \LHC.
    The measurements span over multiple orders of magnitude in their \xsecs and are all in remarkable agreement with the grey boxes of the \SM predictions.
    Figure taken from Ref.~\cite{ATL-PHYS-PUB-2019-024}.
  }
  \label{fig:theory-sm-measurements}
\end{figure}

\paragraph{limitations of the sm.}
The predictions of the \SM have been tested widely and many of the measurements performed, for example with the \ATLAS experiment at the \LHC, are in remarkable agreement with the theory predictions.
\Cref{fig:theory-sm-measurements} gives an overview of \ATLAS measurements undertaken in \runi and \runii, which span multiple orders of magnitude in \xsecs and are all in agreement with the predictions.
However, experimental evidence also suggests that the \SM cannot be a complete theory as it cannot describe some observed phenomena.
Popular examples include the baryogenesis that caused an asymmetry between baryonic matter and antimatter in early stages of the universe~\cite{Kuzmin1985:aa}, or the existence of dark matter in the universe to explain the rotational behaviour of galaxies~\cite{Zwicky:1933gu}.
Several theories and frameworks exist to provide extensions to the \SM model or to embed it into an overarching theory.
Among the most popular are the \emph{Minimal Supersymmetric Standard Model}~\cite{Dimopoulos:1981zb} or the M-theory of superstrings~\cite{Witten:1995ex}, the latter of which attempts to unify the \SM with a theory of quantum gravity by introducing extra dimensions.

\section{Physics of the top quark}
\label{sec:theory-top}

The top quark constitutes the heaviest known elementary particle, and its discovery in 1995 by the \CDF and \dzero Collaborations completed the quark sector of the \SM~\cite{Abe:1995hr,D0:1995jca}.
The top quark is the up-type quark of the third generation and, thus, is the weak-isospin doublet partner of the bottom quark.
Although only discovered twenty years after, there had already been theories about its existence in the 1970's.
In 1973, \citeauthor{Kobayashi:1973fv} proposed to expand the three-flavour quark sector not only by a fourth quark, that was yet to be discovered, but by an entire additional generation, increasing the number of quarks to a total of six~\cite{Kobayashi:1973fv}.
Expanding the quark sector to three generations would solve the fundamental problem of charge-parity violation that had been observed in the 1960's by Cronin and Fitch~\cite{Christenson:1964fg}.
\citeauthor{Kobayashi:1973fv}'s postulation of three left-handed $SU(2)$ quark doublets and six right-handed quark singlets is in agreement with the experimental observations of the last decades, and is what is now known as the quark sector of the \SM.

Their theory picked up pace when the third-generation \taulepton and the bottom quark were discovered in 1975 and 1977, respectively~\cite{Perl:1975bf,Herb:1977ek}.
Not only did these discoveries firmly establish a third generation in both the lepton and the quark sectors, but they also strongly suggested the existence of a second third-generation quark to maintain the quark doublet symmetry.
Several models were proposed without this sixth quark~\cite{TheoryNoT:1,TheoryNoT:2,TheoryNoT:3,TheoryNoT:4} where the bottom quark comes in an $SU(2)$ singlet instead.
However, all of them would spoil the Glashow-Iliopoulos–Maiani mechanism~\cite{Glashow:1970gm} through which flavour-changing neutral currents (\textsc{fcnc}) are suppressed in the \SM.
Without a sixth quark, these currents would manifest in the decay of the bottom quark.
Experimental data, taken for example with the \textsc{cleo} detector in the early 1980's~\cite{Avery:1984ix}, quickly ruled out this possibility.
More indications for a three-generation symmetry came from precision measurements of the \Zboson mass pole.
These allow a determination of the number of neutrinos~\cite{Gaemers:1978fe,Barbiellini:1981zm,Berends:1987zz}.
Various measurements by experiments at the \LEP collider, such as those in Refs.~\cite{Akrawy:1990zy,Adriani:1992zk,Buskulic:1993ke} in the early 1990's, provided evidence for the number of neutrino flavours to be $n_\nu = 3$ as well, thus asking for the sixth quark to be found to maintain equal numbers of quarks and leptons in the \SM.

After its discovery in 1995, the top quark has been under scrutiny at the \TEVATRON and \LHC colliders.
Top quarks are produced in abundance at the \LHC, the operation of which started in 2010.
Combining measurements at the \TEVATRON and the \LHC, the top-quark mass, one of the free parameters of the \SM, has been measured to be~\cite{ATLAS-CONF-2014-008}
\begin{align}
  \label{eq:theory-top-mass}
  m_t = 173.34 \pm 0.27 \stat \pm 0.71 \syst\,\si{\GeV} \, ,
\end{align}
corresponding to relative uncertainties below 0.5\%.
The exceptionally high mass of the top quark translates to a Yukawa coupling with the Higgs field close to unity and assigns a special role to the top quark in the electroweak symmetry breaking of the \SM and in many theories beyond.
In addition, the top quark's mass results in a remarkably short lifetime $\tau_t$.
\citeauthor{Jezabek:1988iv} calculated the top-quark decay width at next-to-leading order (\NLO) in \QCD to be~\cite{Jezabek:1988iv}
\begin{align}
  \label{eq:theory-top-width}
  \Gamma_t = \frac{G_F m_t^3}{8\pi\sqrt 2} \left( 1 - \frac{M_W^2}{m_t^2} \right)^2 \left(1 + 2 \frac{M_W^2}{m_t^2} \right) \left[ 1 - \frac{2\alpha_s}{3\pi} \left( \frac{2\pi^2}{3} - \frac{5}{2} \right) \right] \, ,
\end{align}
resulting in an expected value of $\Gamma_t \approx \SI{1.3}{\GeV}$, when plugging in the top-quark mass from \cref{eq:theory-top-mass} and the \Wboson mass from electroweak fits~\cite{Baak:2014ora,Haller:2018nnx}.
The \NLO prediction is in accordance with the latest direct measurement by \ATLAS which yielded $\Gamma_t = 1.9 \pm 0.5 \,\si{\GeV}$~\cite{ATLAS-CONF-2019-038}.
The width translates to a top-quark mean lifetime at the order of $\tau_t \sim 10^{-24}\,\si{\second}$, which is shorter than the \QCD interaction timescale $\LambdaQCD$ by about one order of magnitude.
As a consequence, the vast majority of top quarks decay before hadronisation occurs -- and thus, studying the properties of the top quark and its decay products provides a unique opportunity to assess quantities of a bare quark.

\begin{figure}
  \centering
  \includegraphics[clip=true, trim=0 2.3cm 0 0]{%
    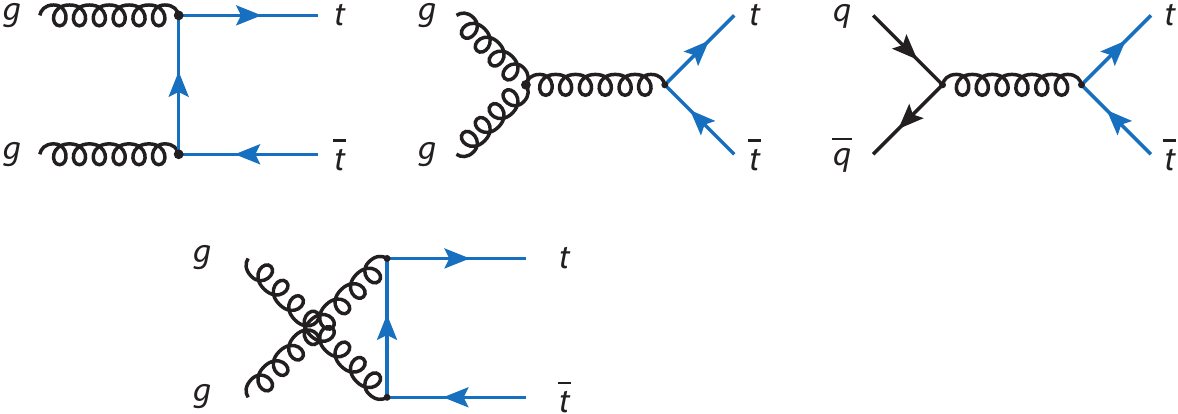}
  \caption[Feynman diagrams for \ttbar production at hadron colliders]{%
    Representative tree-level Feynman diagrams for \ttbar production at hadron colliders.
    The two left-hand diagrams show \tchannel and \schannel diagrams for production via gluon-gluon fusion, the right-hand diagram shows \ttbar production via \qqbar annihilation.
  }
  \label{fig:theory-feynman-ttbar-production}
\end{figure}

\paragraph{pair production.}
The predominant production mode of top quarks at hadron colliders is top-antitop pair production (\ttbar production).
As introduced in \cref{eq:hadronic-xsec}, the hadron-hadron \xsec to create a \ttbar pair can be factorised into the parton distribution functions and a partonic hard interaction.
The \QCD interactions between partons to produce a \ttbar pair include gluon-gluon fusion and quark-antiquark annihilation.
Representative tree-level Feynman diagrams for \schannel and \tchannel gluon-gluon fusion and for \qqbar annihilation are shown in \cref{fig:theory-feynman-ttbar-production}.
While a \com energy of $\sqrt{s} = \SI{1.96}{\TeV}$ during Run~2 of the \TEVATRON required large momentum fractions to be carried by the partons to overcome the \ttbar pair production threshold, the \LHC operates well above the threshold.
As a consequence, whereas production at the \TEVATRON took place primarily through \qqbar annihilation, the Run~2 \com energy of the \LHC at $\sqrt{s} = \SI{13}{\TeV}$ enters a regime of momentum fractions dominated by gluons, and \ttbar pairs are produced through gluon-gluon fusion in about 90\% of the cases.

\begin{figure}
  \centering
  \includegraphics[width=0.9\textwidth]{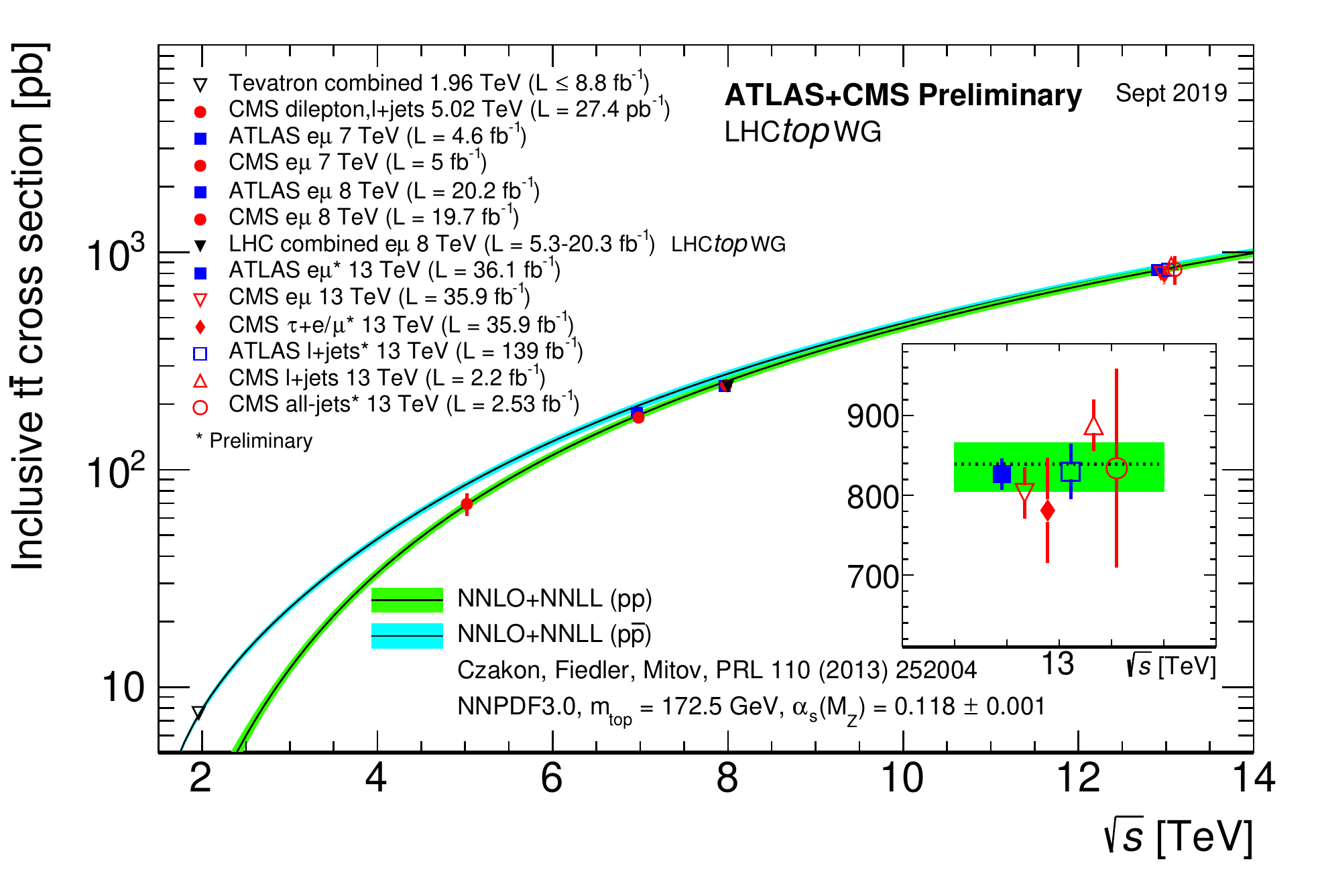}
  \caption[\TEVATRON and \LHC measurements of the \ttbar production \xsec]{
    Various measurements by the \TEVATRON and \LHC Collaborations of the \ttbar production \xsec at different \com energies~%
    \cite{ttbar-xsec:01,ttbar-xsec:02,ttbar-xsec:03,ttbar-xsec:04,ttbar-xsec:05,ttbar-xsec:06,ttbar-xsec:07,ttbar-xsec:08,ttbar-xsec:09,ttbar-xsec:10,ttbar-xsec:11}.
    The plot also shows theory predictions for the \xsecs at $pp$ and $p\bar{p}$ colliders, as calculated with the \Topplusplus programme~\cite{Czakon:2011xx}.
    The theory bands represent uncertainties due to renormalisation and factorisation scale, \PDFs and the strong coupling.
    Measurements at the same \com energy are slightly offset for clarity.
    Figure taken from Ref.~\cite{LHC-TopWG-plots}.
  }
  \label{fig:theory-ttbar-xsec-curve}
\end{figure}

Assuming a top-quark mass of $m_t = \SI{172.5}{\GeV}$, which is the default in all simulations used by \ATLAS, the predicted \LHC \xsec for \ttbar production at a \com energy of $\sqrt{s} = \SI{13}{\TeV}$ is
$\sigma(pp \to t \bar t) = \SI[parse-numbers=false]{832 \,^{+20}_{-29} \, (scales) \pm 35 \, (\PDF+\alpha_S)}{\pb}$,
as calculated with the \Topplusplus programme to next-to-next-to-leading order in \QCD, including soft-gluon resummation to next-to-next-to-leading-log order (see Ref.~\cite{Czakon:2011xx} and references therein).%
\footnote{%
  The first uncertainty comes from the independent variation of the factorisation and renormalisation scales, $\mu_F$ and $\mu_R$, while the second one is associated to variations in the \PDF and $\alpha_S$, following the \textsc{pdf4lhc} prescription with the \textsc{mstw2008} 68\% \textsc{cl} \NNLO, \textsc{ct10} \NNLO and \NNPDF{}\,2.3 five-flavour fixed-flavour-number \PDF sets (see Ref.~\cite{Botje:2011sn} and references therein, and Refs.~\cite{Martin:2009bu,Gao:2013xoa,Ball:2012cx}).
}
Summed in quadrature, the total uncertainty amounts to about 5\%.
Both the \TEVATRON and the \LHC Collaborations have measured the \ttbar \xsec at different \com energies, so far all in agreement with the \SM predictions.
An overview of all measurements in comparison with the theory predictions is shown in \cref{fig:theory-ttbar-xsec-curve}.

\begin{figure}
  \centering
  \includegraphics[]{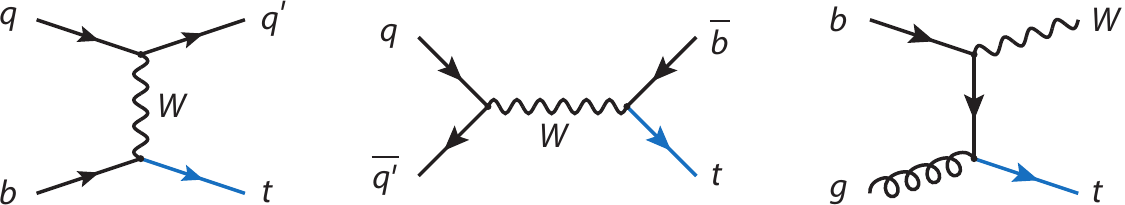}
  \caption[Feynman diagrams of single-top-quark production]{%
    Representative tree-level Feynman diagrams for single-top-quark production at hadron colliders.
    From left to right: \tchannel, \schannel and production with associated \Wboson.
  }
  \label{fig:theory-feynman-stop-production}
\end{figure}

\paragraph{single-top production.}
Hadron colliders also provide opportunities to measure single top quarks through electroweak production diagrams.
At tree level, single top quarks may be produced through \schannel and \tchannel \Wboson exchange, as well as in association with a \Wboson in the final state.
Representative Feynman diagrams of all production modes are shown in \cref{fig:theory-feynman-stop-production}.
Assuming a top-quark mass of $m_t = \SI{172.5}{\GeV}$, \LHC \xsecs for all three modes have been calculated at \NLO in \QCD with \textsc{hathor}~v2.1~\cite{Aliev:2010zk,Kant:2014oha}.
They amount to approximately \SI{10.3}{\pb}, \SI{217}{\pb} and \SI{72}{\pb} for \schannel, \tchannel and associated \tW production, respectively, and are all significantly lower than the \xsec for \ttbar production.
Relative uncertainties range between 4\% and 5\%.%
\footnote{%
  Uncertainties on the \PDF and $\alpha_S$ are calculated using the \textsc{pdf4lhc} prescription~\cite{Botje:2011sn} with the \textsc{mstw2008} 68\% \textsc{cl} \NLO~\cite{Martin:2009iq,Martin:2009bu}, \textsc{ct10} \NLO~\cite{Lai:2010vv} and \NNPDF{}\,2.3~\cite{Ball:2012cx} \PDF sets, added in quadrature to the scale uncertainty.
}
\schannel and \tchannel production favour top quarks over antitop quarks due to the $pp$ initial state at the \LHC, with the ratio predicted to be $R_t = \sigma(tq)/\sigma(\bar{t}q) \sim 1.7$.
Both the \ATLAS and \CMS Collaborations have measured the \xsecs of \tchannel and associated \tW production as well as $R_t$, and the measurements show agreement with the \SM predictions, c.f. for example Refs.~\cite{TOPQ-2017-16,Sirunyan:2018rlu,TOPQ-2015-16}.
Evidence for the \schannel production mode was seen by \ATLAS at a \com energy of \SI{8}{\TeV}~\cite{TOPQ-2015-01}.

Single-top-quark production in association with a \Wboson comes with the peculiarity to show quantum interference effects with \ttbar production at orders higher than leading order in \QCD.
Calculations of \ttbar and \tW production usually use the narrow-width approximation for simulating top-quark decays and, hence, distinguish the doubly and singly resonant Feynman diagrams with two and one top-quark mass resonances.
Modelling interference effects between the two is pivotal and various approaches exist to combine \MC simulations of \ttbar and \tW production, such as the diagram removal (\textsc{dr}) and diagram subtraction (\textsc{ds}) schemes~\cite{Frixione:2008yi,Hollik:2012rc,Demartin:2016axk}.
While the first removes doubly-resonant amplitudes from the \tW calculation, the latter introduces gauge-invariant subtraction terms to cancel the doubly-resonant contributions locally in the \xsec.
Measurements of the \tW production \xsec performed by \ATLAS and \CMS were designed to be insensitive to any interference effects, but a recent \ATLAS measurement targeted a fiducial phase-space region, where these interference effects are significant, to probe the modelling of the interference effects~\cite{TOPQ-2017-05}.
The measurement revealed that only resonance-aware simulations with off-shell top-quark effects at \NLO in \QCD describe the observed spectra in data well, whereas both the \textsc{dr} and the \textsc{ds} schemes diverge in the tails of interference-sensitive observable distributions.

\begin{figure}
  \centering
  \begin{minipage}[b]{0.47\textwidth}
  \centering
    \includegraphics{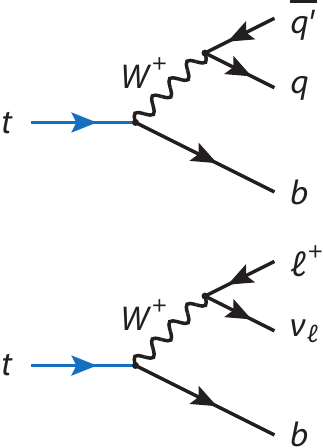}
    \caption[Feynman diagrams of hadronic and leptonic top-quark decays]{%
      Diagrams of hadronic and leptonic decays of the top quark.}
    \label{fig:theory-feynman-top-decays}
  \end{minipage}
  \begin{minipage}[b]{0.50\textwidth}
  \centering
    \includegraphics[width=\textwidth]{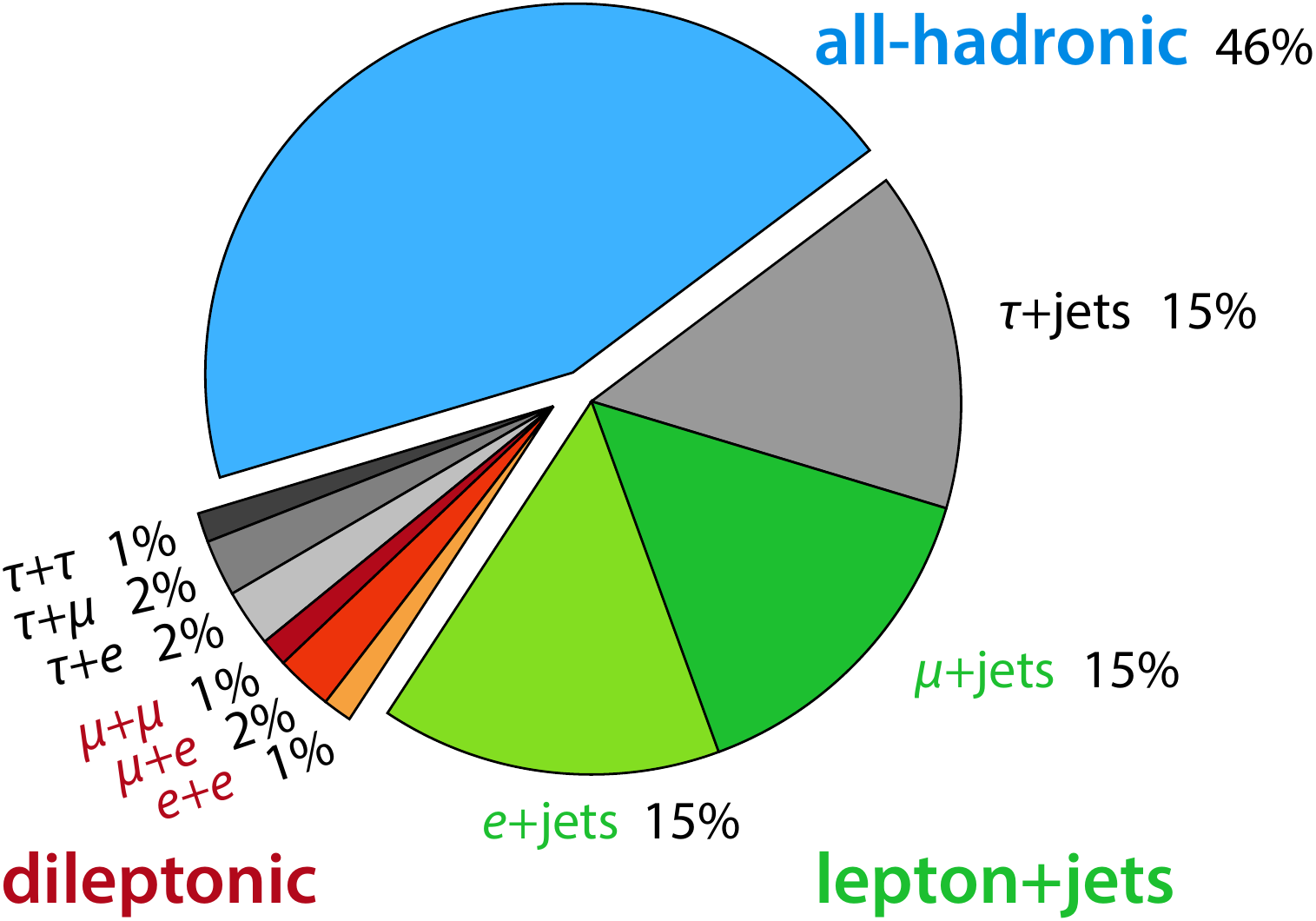}
    \caption[Pie chart of the \ttbar decay channels]{%
      Decay channels of a \ttbar pair and their branching fractions.}
    \label{fig:theory-ttbar-decay-pie}
  \end{minipage}
\end{figure}

\paragraph{top-quark decays.}
Due to its large mass -- and unlike the decay of any other fermion, the decay of the top quark is not suppressed by the massive \Wbosons involved and happens on very short timescales.
In lower-order approximations of the \textsc{ckm} matrix, such as in the Wolfenstein parametrisation~\cite{Wolfenstein:1983yz}, the element $\left| V_{tb} \right| \sim 1$, and the top quark decays almost exclusively into a \Wboson and a bottom quark.
Decays involving strange quarks or down quarks are heavily suppressed due to very weak mixing with the other mass-eigenstate generations.
The \Wboson then decays further into a pair of charged lepton and neutrino or a quark-antiquark pair, with a ratio between the two of approximately $1:2$.
These top-quark decays with subsequent \Wboson decays are called leptonic and hadronic decays, respectively.
Representative Feynman diagrams for both are shown in \cref{fig:theory-feynman-top-decays}.

Similarly, top quarks produced in pairs decay leptonically or hadronically, and one distinguishes three classes of \ttbar decay channels: the dilepton channels, the lepton+jets channels and the all-hadronic channel.
While the first includes those cases, where both top quarks decay leptonically into any combination of charged leptons and neutrinos, cases with one leptonic and one hadronic top-quark decay fall into the definition of the \ljets channels (also known as single-lepton channels).
If both top quarks decay into hadrons, the \ttbar decay is said to be in the all-hadronic decay channel.
An overview of all channels and their branching fractions is given in \cref{fig:theory-ttbar-decay-pie}.
While the all-hadronic channel comes with the largest branching fraction, the final state contains many hadrons, is difficult to resolve experimentally and comes with large background contributions from \QCD activity.
For the experiments, the dilepton channels, and more specifically the \emu channel, provide the cleanest environment to perform precision measurements as there is almost no contamination from background processes in this mixed-lepton-flavour final state.
On the other hand, the two neutrinos of the \emu channel remain undetected and the \ttbar system cannot be reconstructed without ambiguities from the two neutrinos.

\section{Top quarks in association with photons}
\label{sec:theory-tty}

Since the top quark's discovery in the 1990's, many of its properties have been under investigation, ranging from measurements of spin correlations in \ttbar production, to those of \Wboson polarisation in \ttbar decays, and to a measurement of the Yukawa coupling of the top quark in $t\bar{t}H$ final states, c.f. for example Refs.~%
\cite{TOPQ-2015-13,CMS:2018jcg,TOPQ-2016-02,Khachatryan:2016fky,HIGG-2018-13,Sirunyan:2018hoz}.
The last is part of a class of processes where top-quark pairs or single top quarks are produced in association with other elementary particles, which provide access to some of the top quark's most interesting properties: the Yukawa coupling in measurements involving the $tH$ vertex, the weak isospin component $I_3$ in measurements with $tZ$ vertices, and the electromagnetic charge of the top quark in measurements with $t\gamma$ vertices.
Usually denoted as $\ttbar + X$ and $t+X$, these processes are of large interest because many beyond-Standard-Model (\BSM) theories introduce modifications to the top-quark sector and to the couplings of the top quark.
Precise measurements of $\ttbar + X$ and $t+X$ final states provide a unique environment to comb the top-quark sector for any deviations from the \SM and to constrain \BSM theories that introduce modifications to these couplings.

\paragraph{the top-photon vertex.}
The $t\gamma$ vertex and its associated coupling parameter, the top-quark electromagnetic charge, are an open field of research.
While the fermion-photon vertex can be studied directly in $e^+e^- \to q\bar{q}$ production at electron--positron colliders for other quarks, this has not been possible for the top quark due to the required \com energy to produce top-quark pairs.
The top-photon vertex, however, is accessible directly through processes such as \ttbar production with an associated photon in the final state and single-top-quark production in association with a \Wboson and a photon~\cite{Baur:2001si}.
At hadron colliders, the former, henceforth denoted as \tty, is the process with higher \xsec.
Although the top-photon vertex structure is yet to be probed directly, some exotic models suggesting alternative values of the top-quark charge, such as $\left| Q_t \right| = \sfrac{4}{3}$~\cite{Chang:1998pt,Chang:1999zc}, have already been excluded experimentally through measurements of jet charges by the \TEVATRON~\cite{Aaltonen:2013sgl,Abazov:2014lha} and the \LHC~\cite{TOPQ-2011-13,CMS-PAS-TOP-11-031} Collaborations.

Various models exist that predict anomalies in the electric dipole moments of fermions, specifically in the dipole moments of the top quark~%
\cite{Fael:2013ira,AguilarSaavedra:2014vta,Schulze:2016qas,Etesami:2016rwu}.
The electromagnetic vertex factor of the \SM Lagrangian between the photon and fermion $f$ can be extended and generalised through form factors~\cite{Atwood:1991ka,Papavassiliou:1993qe,Baur:2004uw}:
\begin{align}
  \label{eq:theory-EM-generalised}
  \text{$f\gamma$ vertex:} \quad \Gamma^\mu = -i e \left[ \gamma^\mu \left( F_{1V} + \gamma^5 F_{1A} \right)
  + \frac{\sigma^{\mu\nu}q_\nu}{2m_f} \left(i F_{2V} + \gamma^5 F_{2A} \right) \right] \, ,
\end{align}
where $\sigma^{\mu\nu} = \frac{i}{2}\left[ \gamma^\mu,\gamma^\nu \right]$, and the form factors $F_i$ are functions of $s = q^2$.
$F_{1V}$ and $F_{1A}$ are the vectorial and axial-vectorial form factors, respectively, and in the \SM they are $F_{1V} = Q_f$ and $F_{1A} = 0$ at leading order.
$F_{2V}$ is the form factor of the magnetic dipole moment (\MDM) of fermion $f$, and $F_{2A}$ is the form factor of the electric dipole moment (\EDM), both of which represent tensor-like contributions to the coupling vertex.
Neither the \MDM nor the \EDM contribute to the coupling at leading order in the \SM.
The dipole moments only become non-zero, when higher-order quantum loop corrections are considered, but remain small: the \SM predicts $F_{2V} = 0.02 \, Q_t$ for the top-photon vertex from one-loop quantum corrections.
$F_{2A}$ receives a non-zero value only from three-loop corrections~\cite{Bernreuther:2005gq}.
Enhancements in the \MDM and \EDM form factors could hint towards \BSM physics.
\Cref{eq:theory-EM-generalised} can be reshaped to an effective \tty Lagrangian:
assuming the \SM-like coupling behaviour with additional \MDM and \EDM contributions, it may take the form~\cite{Bouzas:2012av,AguilarSaavedra:2008zc}:
\vspace*{0pt plus 2pt}  
\begin{align}
  \label{eq:theory-tty-effective}
  \mathcal{L}_{\tty} = - e \conj{\psi}_t \left[
  Q_t \gamma^\mu + \frac{i \sigma^{\mu\nu} q_\nu}{m_t} \left(d_V + i \gamma^5 d_A \right)
  \right] \psi_t A_\mu \, ,
\end{align}
where $\psi_t$ denotes the spinor of the \spinhalf top-quark fields and $A_\mu$ is the photon gauge field from \cref{eq:the-lagrangian-ew3}.
$d_V$ and $d_A$ can be related to $F_{2V}$ and $F_{2A}$, respectively, and only differ from them in some constants.
Anomalies in the \MDM and \EDM of the top quark introduced by \BSM physics would manifest in a modified coupling in the above Lagrangian.

\vspace*{0pt plus 2pt}  

\paragraph{interpretations in the context of eft.}
Modifications of the coupling can also be studied in the context of effective field theory~(\EFT) in a model-independent way~\cite{Buchmuller:1985jz}.
The \EFT approach assumes that additional heavy fields of scale $\Lambda$ are beyond the probed energy range and that the \SM describes the physics below this scale well as an \emph{effective} field theory.
The heavy fields would be suppressed with $1/\Lambda$ and they would only manifest in modified coupling behaviours in the probed low-energy regime.
Their implications could then be described by an expansion of the dimension-four \SM Lagrangian:
\begin{align}
  \label{eq:theory-EFT-Lagrangian}
  \mathcal{L}_{\mathrm{eff}} = \mathcal{L}_{\mathrm{SM}}^{(4)} + \frac{1}{\Lambda^2} \sum_i C_i O_i^{(6)} + \dots
\end{align}
It can be shown that there are no effective dimension-five operators $O^{(5)}$ with only fermion fields and gauge-boson fields, which maintain gauge invariance and affect the top-quark sector, thus, the lowest relevant order are the effective dimension-six operators $O_i^{(6)}$.
Operators of higher orders would be suppressed by higher powers of $\Lambda$ and, hence, would only contribute weakly to the low-energy regime.
The dimension-six operators $O_i^{(6)}$ are invariant under the \SM gauge groups $SU(3) \times SU(2) \times U(1)$, assuming that the breaking of the $SU(2) \times U(1)$ of the \SM is indeed a phenomenon connected to the Fermi scale, not the \EFT scale $\Lambda$.
$C_i$ are known as Wilson coefficients~\cite{Wilson:1969zs} and describe the strength of the modifications introduced by the dimension-six operators $O_i^{(6)}$.
Lists of all possible operators for effective expansions of the \SM have been compiled~\cite{Buchmuller:1985jz,Grzadkowski:2010es}, but the list of those relevant for electroweak couplings of the top quark can be reduced to only eight effective operators.
Ref.~\cite{AguilarSaavedra:2008zc} showed that only two of them contribute to $d_V$ and $d_A$ in \cref{eq:theory-tty-effective}:
\begin{subequations}
\begin{align}
  \label{eq:theory-EFT-tty-moments}
  \delta d_V &= \frac{\sqrt{2}}{e} \operatorname{Re} \left[ \cos \theta_W C_{uB\phi}^{33} + \sin \theta_W C_{uW}^{33} \right] \frac{v m_t}{\Lambda^2} \, , \\
  \delta d_A &= \frac{\sqrt{2}}{e} \operatorname{Im} \left[ \cos \theta_W C_{uB\phi}^{33} + \sin \theta_W C_{uW}^{33} \right] \frac{v m_t}{\Lambda^2} \, .
\end{align}
\end{subequations}
Here, $v$ is the vacuum expectation value of the Higgs field.
The Wilson coefficients $C_{uB\phi}^{33}$ and $C_{uW}^{33}$ describe interactions between the gauge fields of the $SU(2)_L$ and the $U(1)_Y$ and third-generation quarks.
In addition, $C_{uG\phi}^{33}$ introduces possible modifications to the interaction between gluons and third-generation quarks and is, thus, relevant for \tty production.
The corresponding dimension-six operators are defined as
\begin{subequations}
\begin{align}
  \label{eq:teory-EFT-operators}
  O_{uB\phi}^{33} &= (\conj{Q}_L \sigma^{\mu\nu} t_R) \tilde{\phi} B_{\mu\nu} \, , \\[0.6ex]
  O_{uW}^{33} &= (\conj{Q}_L \sigma^{\mu\nu} \tau^i t_R) \tilde{\phi} W_{\mu\nu}^i \, , \\[0.6ex]
  O_{uG\phi}^{33} &= (\conj{Q}_L \sigma^{\mu\nu} \lambda^a t_R) \tilde{\phi}G_{\mu\nu}^a \, .
\end{align}
\end{subequations}
$B_{\mu\nu}$, $W_{\mu\nu}^i$ and $G_{\mu\nu}^a$ are the field strength tensors of the $U(1)_Y$, $SU(2)_L$ and $SU(3)_C$ gauge fields, respectively.
$\tau^i$ and $\lambda^a$ are the generators of the $SU(2)$ and $SU(3)$ with indices as introduced before.
$Q_L$ is the left-handed doublet of the third quark generation, $t_R$ is the right-handed top-quark singlet.
$\phi$ is the Higgs doublet of the \SM, with $\tilde{\phi} = i\tau^2 \phi^*$.

\begin{figure}
  \centering
  \includegraphics[scale=0.9]{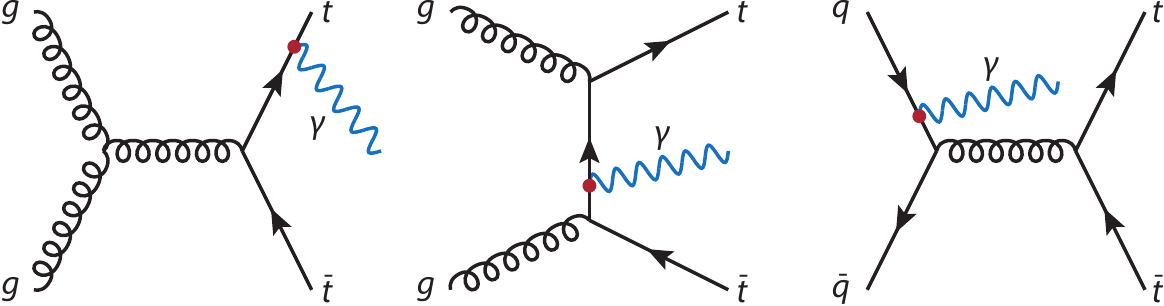}
  \caption[Feynman diagrams for radiative top-quark production]{%
    Representative Feynman diagrams for photon radiation through radiative top-quark production.
    From left to right, the photon is radiated by (1)~a top quark, which goes on-shell, (2)~a \tchannel top-quark exchange, (3)~an initial-state quark.
  }
  \label{fig:theory-feynman-tty-rad-prod}
\end{figure}

\begin{figure}
  \centering
  \includegraphics[scale=0.9]{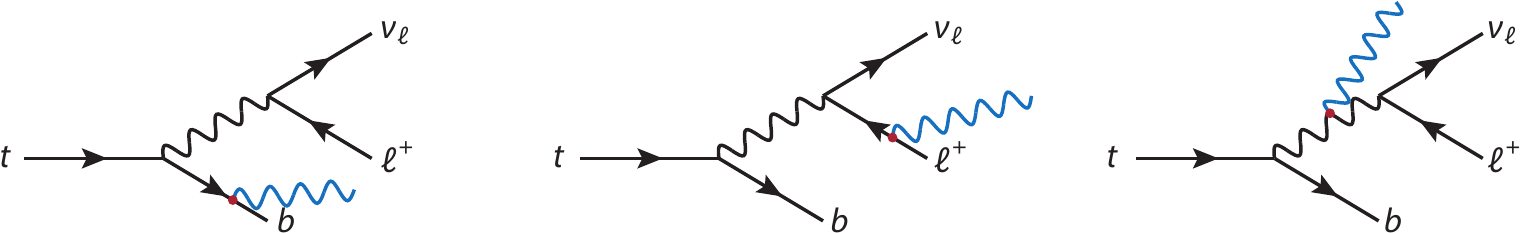}
  \caption[Feynman diagrams for radiative top-quark decay]{%
    Representative Feynman diagrams for photon radiation through radiative top-quark decay.
    From left to right, the photon is radiated by (1)~a \bquark, (2)~a decay product of the \Wboson, (3)~the \Wboson itself.
  }
  \label{fig:theory-feynman-tty-rad-decay}
\end{figure}

\paragraph{theory computations.}
The production of a top-quark pair in association with a photon, \tty, is always an \emph{inclusive} process and does not necessarily include a $t\gamma$ vertex.
In addition to the na{\"i}ve picture of a photon radiated by one of the top quarks in a \ttbar final state, photons may be radiated by any of the charged particles involved in the process.
Assuming stable top quarks for an initial simplification, a photon may be radiated by one of the top quarks before it goes on-shell, by a \tchannel top-quark exchange, or by an initial-state quark if the \ttbar production takes place via \qqbar annihilation.
All of these are collectively known as \emph{radiative top-quark production}, representative Feynman diagrams for which are shown in \cref{fig:theory-feynman-tty-rad-prod}.
In addition, when considering the instability of the top quarks and including their immediate decay products, photons may be radiated by the \bquarks, by the \Wbosons, or by any of the charged decay products of the \Wbosons.%
\footnote{%
  Naturally, the fraction of photons radiated by the charged decay products of the \Wbosons is enhanced in leptonic \Wboson decays as the fermion-photon coupling goes with $Q_f^2$.
}
On-shell top quarks may also radiate a photon and go off-shell before their decay.
These are collectively known as \emph{radiative top-quark decay}, for which representative Feynman diagrams are shown in \cref{fig:theory-feynman-tty-rad-decay}.
Together, radiative production and radiative decay are observed as a single process when searching for final-state signatures associated with \tty production, and they cannot be disentangled.
For example, the sought final states could be
\begin{subequations}
\begin{align}
  pp &\to \ttbar(\gamma) \to bq\overline{q}'b \mu \nu_\mu \gamma \, ,\\
  pp &\to \ttbar(\gamma) \to b e \nu_e b \mu \nu_\mu \gamma \, ,
\end{align}
\end{subequations}
which would correspond to the $\mu$+jets and \emu decay channels, respectively.
When looking for these final-state signatures, the irreducible background contributions are similar to those seen for \ttbar production, albeit with smaller \xsecs and with an additional photon.
Again, dilepton channels, in particular the \emu channel, provide the cleanest environment for precision measurements with almost no background contributions.

\begin{figure}
  \centering
  \includegraphics{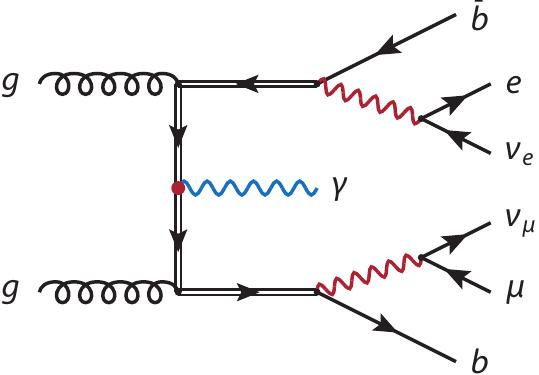}
  \hspace*{1cm}
  \includegraphics{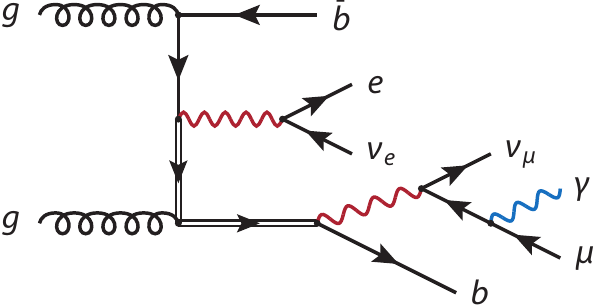}
  \caption[Resonant Feynman diagrams contributing to \WbWby production]{%
    Two representative Feynman diagrams contributing to \WbWby production in the \emu final state at \LO in \QCD.
    The top-quark mass resonances are marked with double-lined arrows.
    The left-hand side shows an example of a doubly-resonant diagram, the right-hand side one of a singly-resonant diagram.
    The \Wboson is marked in red, the final-state photon in blue.
  }
  \label{fig:theory-tty-resonances}
\end{figure}

\begin{figure}
  \centering
  \includegraphics{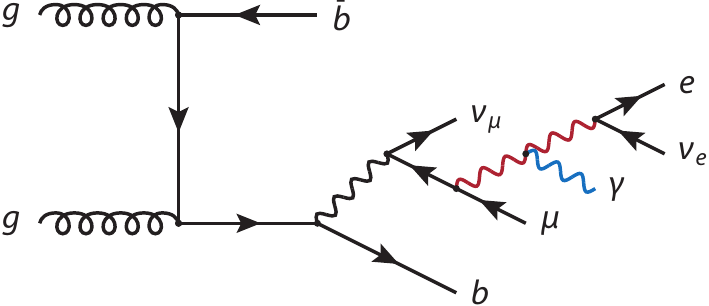}
  \caption[Non-resonant Feynman diagram contributing to \WbWby production]{%
    Representative non-resonant Feynman diagram contributing to \WbWby production in the \emu final state at \LO in \QCD.
    Although no top-quark mass resonances are present, diagrams such as this one contribute to the same final state as those shown in \cref{fig:theory-tty-resonances}.
    The \Wbosons are marked in red, the final-state photon in blue.
  }
  \label{fig:theory-tty-resonances2}
\end{figure}

As for associated \tW and \ttbar production, single-top-quark production in association with a \Wboson and a photon, \tWy, interferes with \tty production when the top-quark decays are simulated and when the computation includes higher orders in \QCD:
 doubly-resonant and singly-resonant Feynman diagrams contribute to the same final state, and they cannot be considered separately.
Representative diagrams in the \emu final state for both cases are shown in \cref{fig:theory-tty-resonances}.
In addition, even non-resonant diagrams contribute, an example of which is depicted in \cref{fig:theory-tty-resonances2}.
In practice, first calculations of \tty production at \NLO in \QCD were done in \citeyear{Duan:2009ph}, which assumed stable top quarks where the problem of interference does not occur, c.f. Refs.~\cite{Duan:2009ph,Duan:2011qg,Maltoni:2015ena}.
Electroweak corrections at \NLO were added to the results in Ref.~\cite{Duan:2016qlc}, albeit still assuming stable top quarks.
They proved to become sizeable for large photon transverse momenta and large invariant masses $M_{\ttbar}$, but to remain small for the overall total \xsecs.
First realistic theory predictions with unstable top quarks, including photon radiation by decay products of the top quark, were presented in Ref.~\cite{Melnikov:2011ta} in \citeyear{Melnikov:2011ta}.
These new calculations revealed a significant \tty \xsec increase due to the added radiative top-quark decay.
With the kinematic cuts used for that computation, the fraction of photons radiated in top-quark decays was found to be approximately \SI{50}{\percent}, with an increase towards smaller photon transverse momenta.

\begin{figure}
  \centering
  \includegraphics[width=0.48\textwidth]{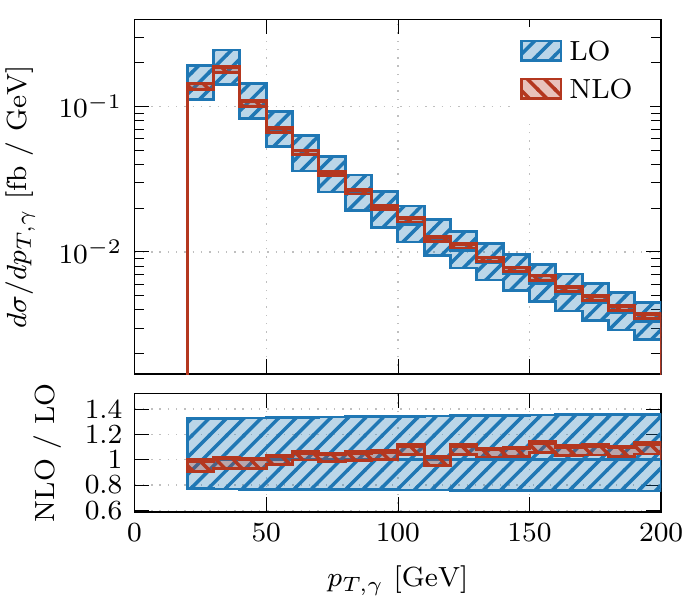}
  \includegraphics[width=0.48\textwidth]{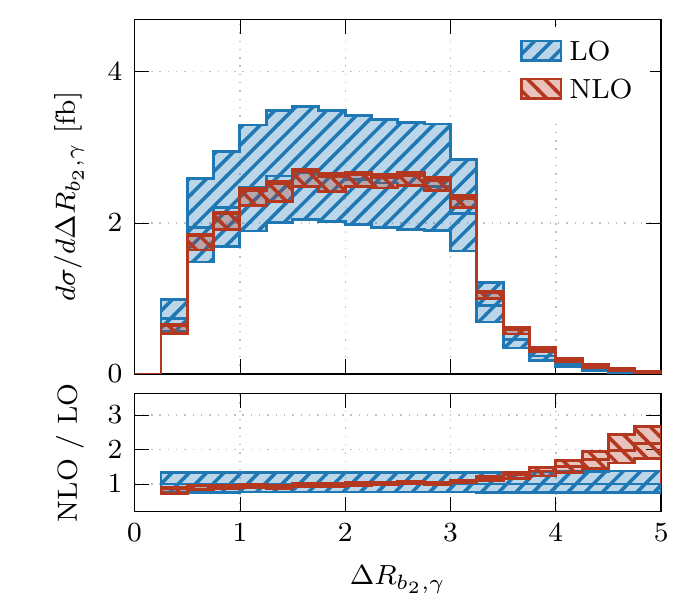}
  \caption[Differential distributions for \tty \LO and \NLO predictions]{%
    Two examples of differential distributions presented in Ref.~\cite{Bevilacqua:2018woc}:
    the left-hand side shows the photon transverse momentum, one of the observables insensitive to \NLO corrections in \QCD;
    the right-hand side shows the separation $\Delta R = \sqrt{\Delta\phi^2+\Delta\eta^2}$ of the photon and the softer of the two \bjets in the plane of azimuthal angles $\phi$ and pseudorapidities $\eta$, which is highly sensitive to \NLO corrections in its shape.
    For both distributions, the scales were chosen dynamically, \cf \cref{sec:strategy-prediction} for details of the computation.
  }
  \label{fig:theory-worek-examples}
\end{figure}

Although constituting a significant progress towards a more realistic picture of \tty production, Ref.~\cite{Melnikov:2011ta} still used the narrow-width approximation for the decays of the top quarks and, hence, only included doubly-resonant Feynman diagrams in the calculation.
The first full description of \tty at \NLO in \QCD, including all resonant and non-resonant diagrams, interference terms and off-shell effects of the top quarks and \Wbosons was given by \citeauthor{Bevilacqua:2018woc} in Ref.~\cite{Bevilacqua:2018woc} in \citeyear{Bevilacqua:2018woc}.
The publication presents \tty \xsec calculations in the \emu final state, or, more specifically, for matrix elements with $e^+\nu_e\mu^- \overline{\nu}_\mu b \bar{b} \gamma$ final states at a \com energy of $\sqrt{s} = \SI{13}{\TeV}$ at the \LHC.
This includes all doubly-resonant, singly-resonant and non-resonant diagrams, including the examples shown in \cref{fig:theory-tty-resonances,fig:theory-tty-resonances2}.
The calculations reveal that, although \NLO corrections in \QCD have little impact on the total \xsec of the examined fiducial phase space%
\footnote{%
  When evaluated at fixed renormalisation and factorisation scales, the total \xsec of the fiducial phase space differs considerably, but this difference becomes negligible with a dynamical choice of the scales of $\mu_R = \mu_F = \ST/4$~\cite{Bevilacqua:2018woc}.
  Here, \ST is defined as the scalar sum of the transverse momenta of all final-state particles, including the missing transverse momentum from the neutrinos.
  More details on the choice of scales are detailed in \cref{sec:strategy-prediction}.
}%
, shape distortions between \LO and \NLO of more than \SI{100}{\percent} are observed for some differential distributions of dimensionless observables.
In particular, separations $\Delta R = \sqrt{\Delta\phi^2+\Delta\eta^2}$ in azimuthal angles $\phi$ and pseudorapidities $\eta$, such as those of the hard photon and the softer of the two \bjets, show large discrepancies between \LO and \NLO, as depicted in \cref{fig:theory-worek-examples}.
Both shown distributions are sensitive to the top-quark charge and \EDM or \MDM contributions to the top-photon coupling~\cite{Baur:2001si,Baur:2004uw}.
Hence, precise theory calculations for \SM-like \tty production are needed to reach precision levels, at which the coupling can be probed for possible modifications.

Comparisons done by the same authors show that the included off-shell effects of the top quark only play a small role in the calculation of the inclusive \xsec~\cite{Bevilacqua:2019quz}.
Differential distributions of dimensionless observables seem to be insensitive to these effects as well, and, hence, are described well by the narrow-width approximation for top quarks.
However, full \NLO computations in \QCD become necessary for dimensionful observables, such as transverse momenta, where differences of up to \SI{60}{\percent} are observed in differential distributions.
The theory predictions of Ref.~\cite{Bevilacqua:2018woc} are used as a reference for the measurement presented in this thesis.
They are summarised in \cref{sec:strategy-prediction} in more detail.

\paragraph{experimental status.}
The window to measurements of \tty production was first opened at the \TEVATRON where the \xsec for \ttbar production had already been measured with high precision, e.g. in Ref.~\cite{Aaltonen:2010ic}.
The \CDF Collaboration reported first evidence of \tty production in \ppbar collisions at a \com energy of $\sqrt{s} = \SI{1.96}{\TeV}$~\cite{Aaltonen:2011sp}, with data corresponding to an integrated luminosity of \SI{6.0}{\ifb}.
The measured value, extrapolated from the \xsec of the probed \ljets channels to the total \xsec, is in good agreement with the theory prediction for the examined fiducial phase space.
\num{30} \tty candidate events are observed, which would correspond to a fluctuation of the background-only hypothesis by \num{3.0} standard deviations.
Additionally, the ratio between \tty and \ttbar production \xsecs, where many systematic uncertainties cancel, was measured to be $\mathcal{R} = 0.024 \pm 0.009$, to be compared with the \SM prediction of $\mathcal{R} = 0.024 \pm 0.005$ obtained from theoretical prediction of the \tty and \ttbar \xsecs~\cite{Aaltonen:2011sp}.

\tty production was observed experimentally by the \ATLAS Collaboration in $pp$ collisions at the \LHC at a \com energy of $\sqrt{s} = \SI{7}{\TeV}$ in \citeyear{TOPQ-2012-07}~\cite{TOPQ-2012-07}.
The examined data corresponds to \SI{4.59}{\ifb}, and the measurement was again performed in the \ljets channels.
Good agreement was found with the theory prediction at \NLO in \QCD for the examined phase space.
The \num{140} and \num{222} candidate events in the \ejets and \mujets channels, respectively, correspond to a fluctuation of \num{5.3} standard deviations of the background-only hypothesis, hence the claim of observation.
Both the \ATLAS and \CMS Collaborations remeasured the \tty production \xsec in the \ljets channels at \SI{8}{\TeV} at the \LHC~\cite{CMS-TOP-14-008,TOPQ-2015-21} and found good agreement with the \SM prediction.
For the first time, the \ATLAS measurement included differential distributions of the transverse momentum and the pseudorapidity of the photon.
At $\sqrt{s} = \SI{13}{\TeV}$, \ATLAS performed first inclusive and differential measurements with partial \runii data~\cite{TOPQ-2017-14}, corresponding to an integrated luminosity of \SI{36.1}{\ifb}.
The analysis used machine-learning algorithms to identify prompt photons in events, the tools for which are discussed in \cref{chap:PPT} of this thesis.
In addition, due to the increased statistics at higher \com energies, the measurements are also performed in the dilepton channels for the first time.
In addition to the inclusive \xsecs in the decay channels, various differential distributions are measured.
All of them show good agreement with the \SM predictions.

\begin{figure}
  \centering
  \includegraphics[width=\textwidth,clip=true,trim=0 8pt 0 8pt]{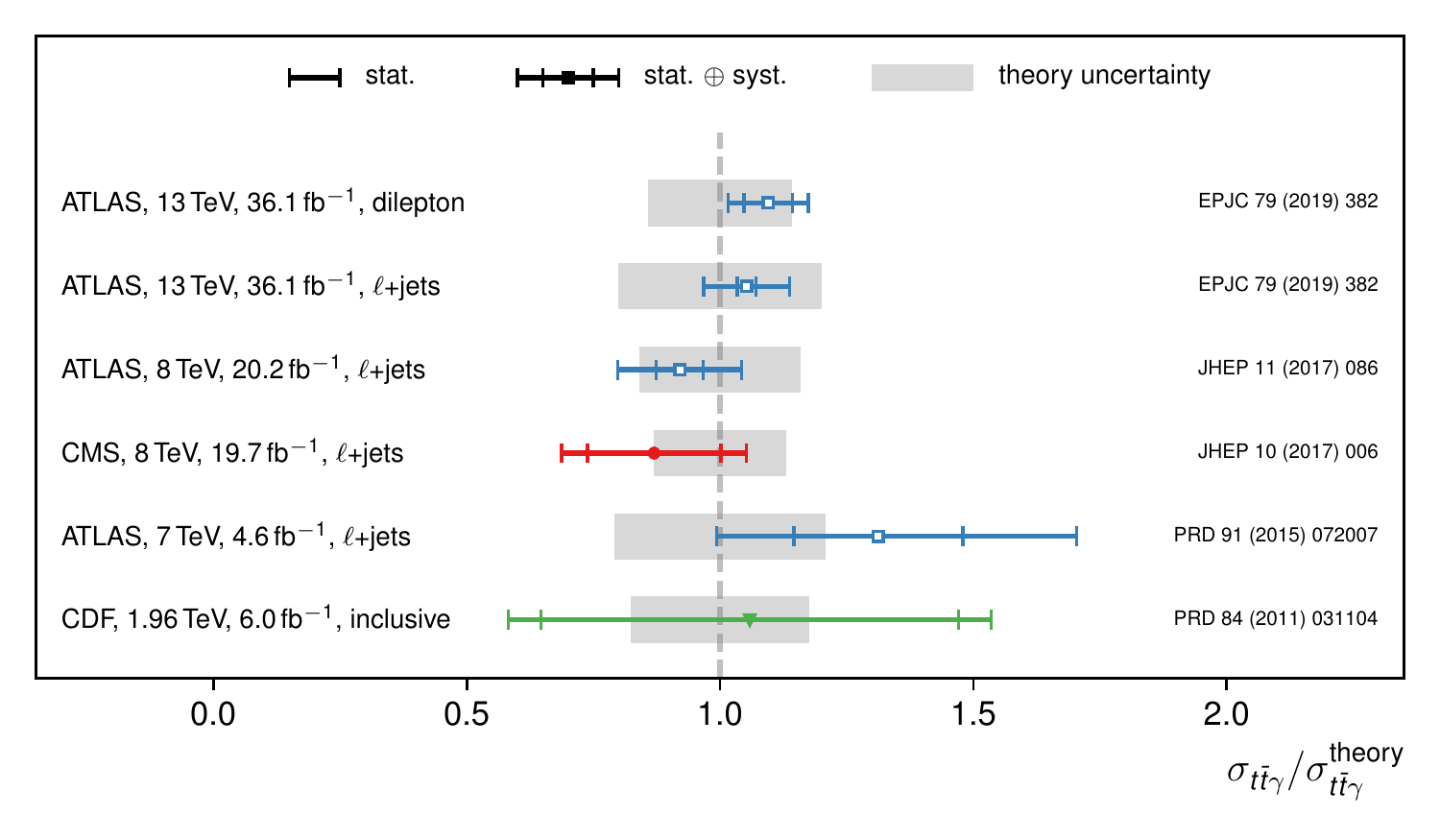}
  \caption[Previous measurements of \tty production at hadron colliders]{%
    Previous measurements of \tty production at hadron colliders~%
    \cite{Aaltonen:2011sp,TOPQ-2012-07,CMS-TOP-14-008,TOPQ-2015-21,TOPQ-2017-14}.
    As a figure of merit, the measured \xsec over \SM prediction is given.
    The inner error bars are statistical uncertainties only, the outer bars are combined statistical and systematic uncertainties.
    The colours of the bars and marker styles indicate different collaborations.
    The grey blocks represent theory uncertainties.
  }
  \label{fig:theory-tty-measurements}
\end{figure}

A summary of all measurements of \tty production is given in \cref{fig:theory-tty-measurements}.
As all measurements are done in different fiducial phase spaces, the given figure of merit is the measured \xsec over the respective \SM prediction.
The plot demonstrates the decrease of statistical uncertainties over time, and the increasing precision of the experimental measurements.
The grey bars in the background show the uncertainties on the theory predictions.

\chapter{Experimental setup of the measurement}
\label{sec:exp}

As with every measurement of a fundamental quantity in physics, the measurement of a \xsec in particle physics needs an experimental setup.
The one used for this analysis, however, is one of a kind:
the data is provided by means of proton-proton collisions by the Large Hadron Collider (\LHC), the largest particle-physics accelerator ever built, and it is recorded with the \ATLAS experiment, the (by volume) biggest particle-physics detector built to date.
Both are hosted by the \emph{European Organization for Nuclear Research}, known as \CERN (French: \emph{Conseil Europ{\'e}en pour la Recherche Nucl{\'e}aire}), a European organisation and particle-physics research laboratory, located at the Franco-Swiss border near Geneva.
As of early 2020, \CERN holds 23~member states.
Its main role is to provide particle accelerators and particle-physics research infrastructure to its member states.
The individual experiments are run by international collaborations formed by national research organisations and institutes of the \CERN member states and other associated nations.
For example, the \ATLAS Collaboration operates the \ATLAS detector.

The following sections introduce the experimental setup used for this \xsec measurement.
\Cref{sec:exp_lhc} describes the \CERN accelerator infrastructure and, in particular, the \LHC, the collisions of which are analysed in this thesis.
In \cref{sec:exp_atlas}, the \ATLAS experiment is introduced briefly.
\ATLAS consists of many individual components and each of them returns highly complex electric signals of the collisions recorded.
To \enquote{make sense} of these, to accumulate and combine the signals and to reconstruct the physics objects that triggered them in the detector, is a non-trivial process that requires elaborate data processing.
The reconstruction and identification of particles are summarised in \cref{sec:exp_objects}.

\section{The Large Hadron Collider (LHC)}
\label{sec:exp_lhc}

The Large Hadron Colliders (\LHC)~\cite{Evans:2008zzb} is the largest and most powerful particle accelerator ever constructed.
It is built underneath the \CERN research laboratory in a concrete-lined tunnel at a depth of approximately \SI{100}{\m} under ground%
\footnote{%
  The actual depth varies with the terrain between 50~and \SI{175}{\m}.
}
due to its sheer dimensions.
The \LHC is a circular collider with a total circumference of \SI{26.7}{\km} and covers an area between the centre of Geneva and the French Jura Mountains, crossing the Franco-Swiss border at a total of four points.
By design, the \LHC accelerates protons in two beam pipes up to energies of \SI{7}{\TeV}.
The two counter-rotating beams intersect at four interaction points and are brought to collision.
The acceleration of the protons is done in sixteen radio-frequency caverns, and the beams are bent on a quasi-circular path by some \num{1200} superconducting dipole magnets.
Additional quadrupole magnets keep the beams focused, and magnets of higher multi-pole orders are used to further correct imperfections in the beam geometry.
All magnets are made of Nb-Ti alloy and operate below their superconducting transition temperatures, at approximately \SI{1.9}{\K} (below \SI{-271}{\celsius}).
This requires extensive cooling of the \LHC with superfluid helium.

\begin{figure}
  \centering
  \includegraphics[width=0.95\textwidth, clip=true, trim=0 100pt 0 500pt]{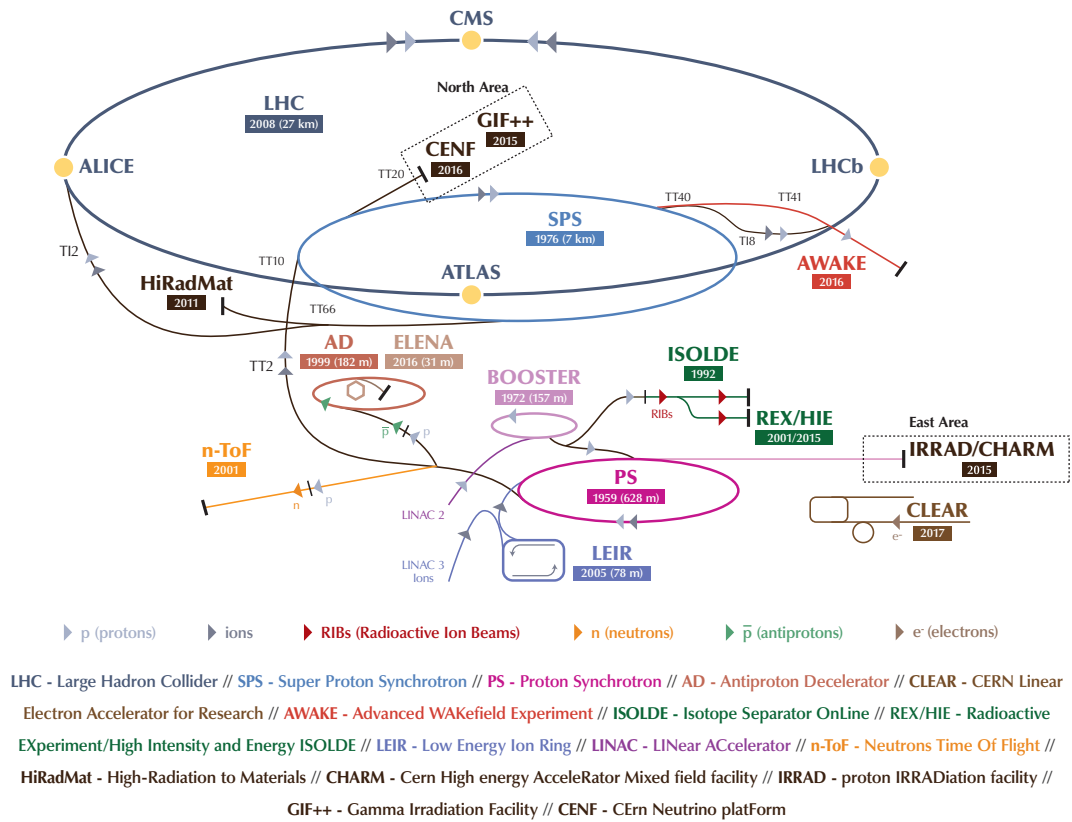}
  \caption[\CERN accelerator complex including the \LHC]{%
    The \CERN accelerator complex with all pre-accelerator stages of the \LHC.
    The chain for proton-proton collisions starts with a proton beam in the linear accelerator \textsc{linac2}, from where the protons are injected into the \textsc{booster} ring.
    The \textsc{booster} sends them into the \textsc{ps}, from where they continue into the \textsc{sps}.
    Only then they are injected in both directions into the \LHC ring.
    In the \LHC ring, the proton beams reach their final energies before they are brought to collision at four intersection points.
    \textcopyright{}\CERN}
  \label{fig:exp-CERN-accelerators}
\end{figure}

Before injected into the \LHC ring, the protons undergo several pre-stages of acceleration.
The entire \CERN accelerator complex is depicted in \cref{fig:exp-CERN-accelerators}.
First, the electrons are stripped off of hydrogen atoms, which are then inserted into a linear-accelerator facility called \textsc{linac2}.
Here, the protons reach energies of \SI{50}{\MeV}.
They are then injected into the \textsc{booster} ring and accelerated to energies of \SI{1.4}{\GeV}.
The \textsc{booster} sends them into the \textsc{ps}, where they reach \SI{25}{\GeV}.
The following stage is the \textsc{sps}, where they are accelerated up to \SI{450}{\GeV}, before entering the \LHC ring.
The transition from the \textsc{sps} into the \LHC ring is where the protons are split into two separate beams and are inserted at two injection points in opposite directions.
Although physicists refer to particle \emph{beams} in the \LHC, the protons are actually not arranged as continuous streams, but as bunches of approximately \num{e11} protons.
Depending on the filling scheme of the \LHC, there may be up to \num{2808} bunches in each of the two beams, resulting in bunch collision rates of up to \SI{40}{\mega\hertz}.
This corresponds to intervals between collisions of only \SI{25}{\nano\s}.
By design, the \LHC is capable of a maximum beam energy of \SI{7}{\TeV}, and a resulting \com energy of $\sqrt{s} = \SI{14}{\TeV}$.
The instantaneous luminosity is designed to reach $\mathcal{L} = \SI{e34}{\cm\tothe{-2}\s\tothe{-1}}$~\cite{Evans:2008zzb}.

\begin{figure}
  \centering
  \includegraphics[width=0.56\textwidth]{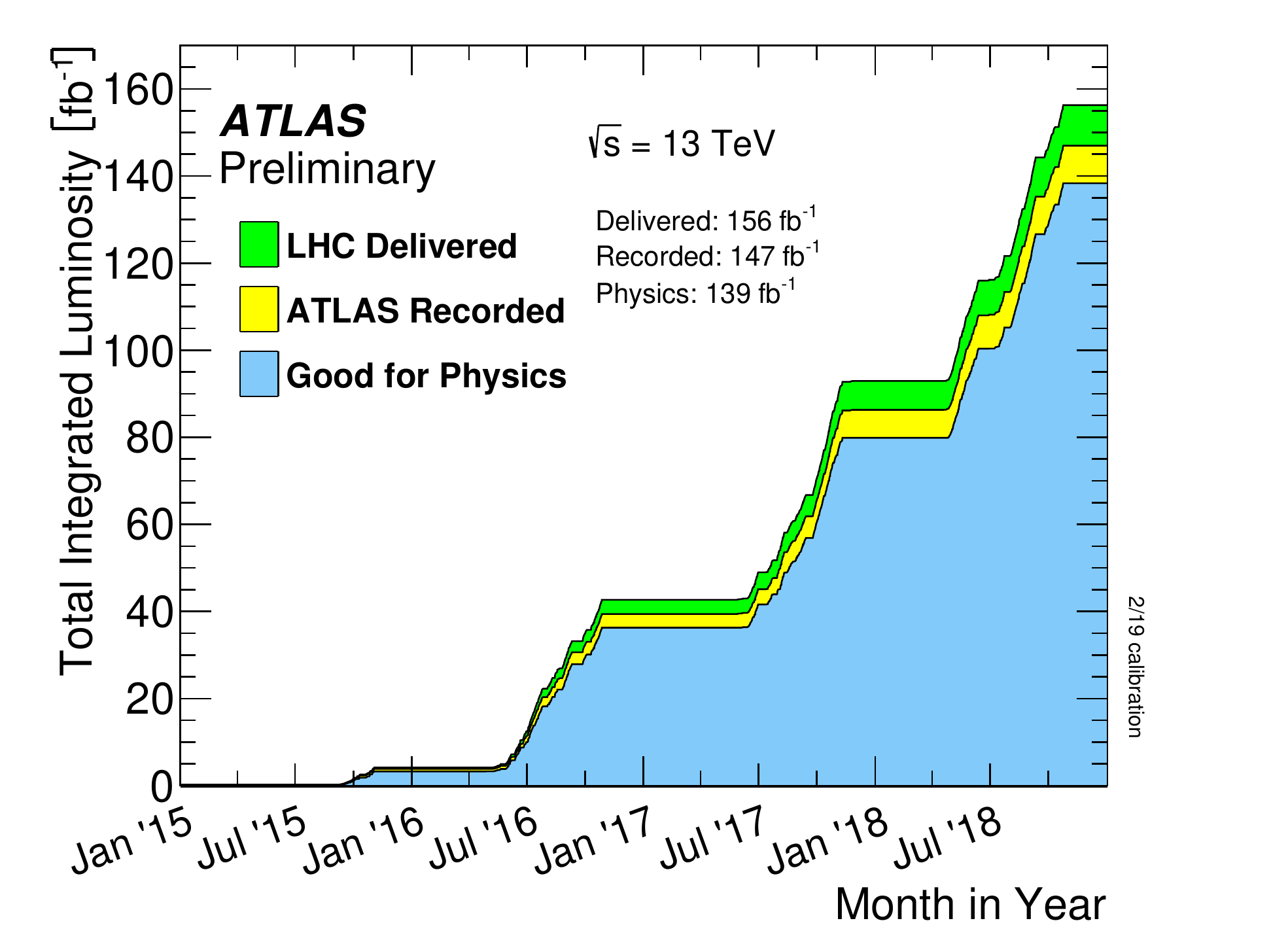}
  \caption[Total integrated luminosity during \runii]{%
    Total integrated luminosity of proton-proton collisions at the \LHC during \runii (2015--2018).
    The plot shows the luminosity values delivered by the \LHC, the luminosity fraction recorded with \ATLAS and the luminosity fraction, where \ATLAS was fully operational and stable data-taking conditions were maintained (\enquote{Good for Physics}).
    The size of the dataset taken under these conditions corresponds to an integrated luminosity of \SI{139}{\ifb}.
    Figure taken from Ref.~\cite{ATLAS-lumi}.
  }
  \label{fig:exp-lumi}
\end{figure}

Operation of the \LHC was commenced in September 2008, but the physics programme was delayed until 2010 due to a magnet quench incident in the first days of initial testing that caused extensive damage to the machine.
The physics programme of the \runi started in March 2010, which was when the \LHC first collided beams at a \com energy of \SI{7}{\TeV}.
Proton-proton collisions at that energy were continued until the end of 2011, with a total integrated luminosity provided by the \LHC of about \SI{5.5}{\ifb}.
The beam energies were increased to \SI{4}{\TeV} for 2012, resulting in a successful year of operation:
data corresponding to almost \SI{23}{\ifb} of integrated luminosity was taken that year.
\runi was followed by a two-year shutdown period to allow repairs and upgrades of the accelerator infrastructure and the \LHC experiments.
\runii commenced in March 2015 with increased beam energies of \SI{6.5}{\TeV}, resulting in proton-proton collisions at a \com energy of \SI{13}{\TeV}.
Only interrupted by short year-end shutdown periods and a few heavy-ion collision runs, proton-proton data was taken continuously until the end of 2018, marking the end of \runii.
The luminosity delivered between 2015 and 2018 amounts to \SI{156}{\ifb}, as shown in \cref{fig:exp-lumi}, where the accumulated luminosity is displayed.

\begin{figure}
  \centering
  \includegraphics[width=0.48\textwidth]{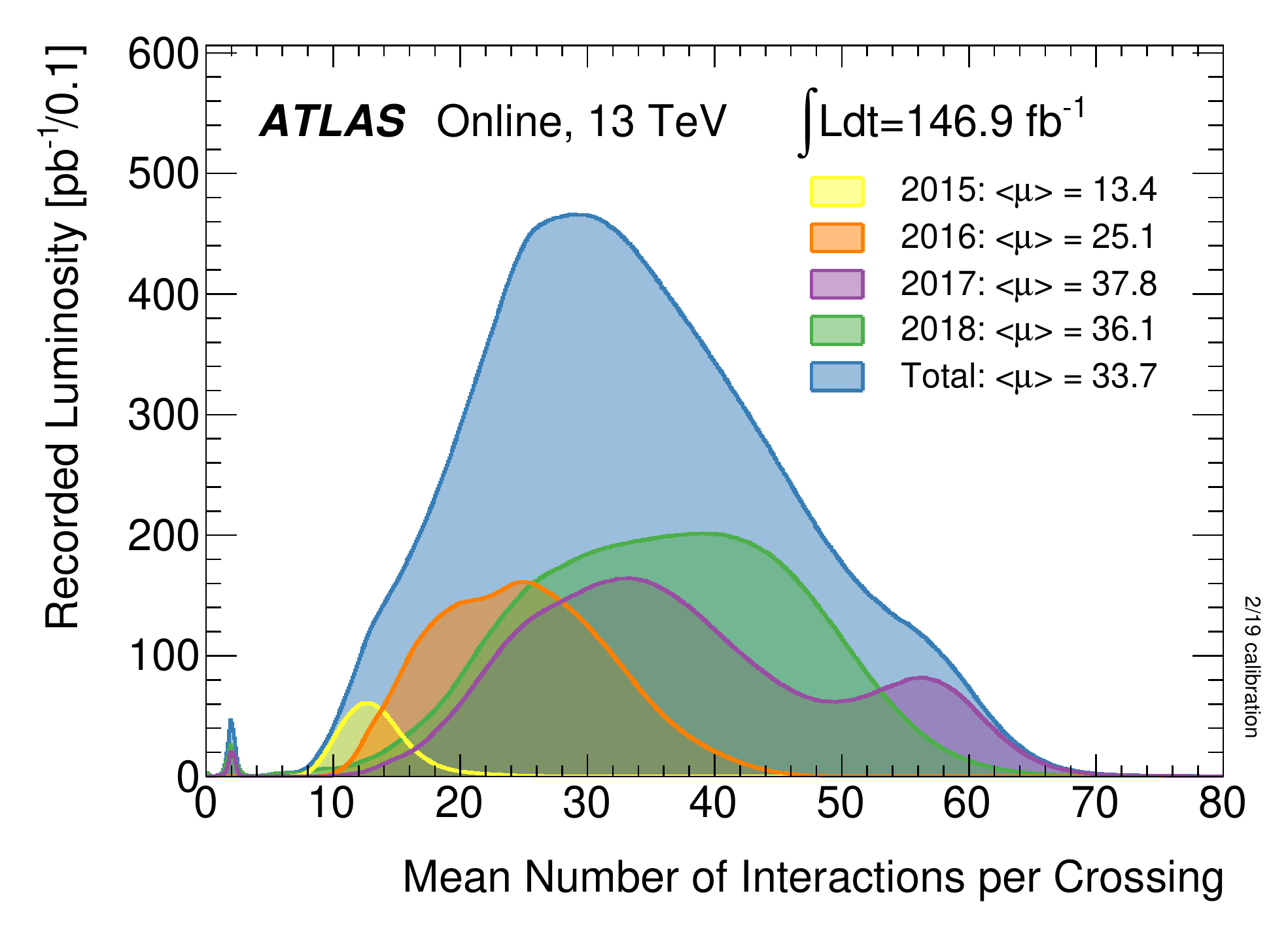}
  \caption[Mean number of interactions per bunch crossing during \runii]{%
    The mean number of interactions per bunch crossing (pile-up) during \runii, as recorded with \ATLAS.
    The plot shows the profiles of pile-up conditions in the four years of operation.
    A characteristic double-peak structure is observed for 2017 as the filling scheme of the \LHC was adjusted mid-year.
    Figure taken from Ref.~\cite{ATLAS-lumi}.
  }
  \label{fig:exp-pileup}
\end{figure}

Proton-proton collisions with large beam energies and under high beam intensity provide a very busy environment to study elementary particles.
The four intersection points of the \LHC beams are used by four experiments to shed light on complementary aspects of high-energy physics:
the \ATLAS and \CMS experiments~\cite{PERF-2007-01,CMS-TDR}, located at the \LHC tunnel entry points 1 and 5, respectively, are large, general-purpose detectors that perform precision measurements of \SM properties and explore the highest possible energy regimes.
The \LHCb experiment~\cite{LHCb-TDR}, located at point 8, is a forward detector to study the physics of $B$~mesons, and \ALICE~\cite{ALICE-TDR}, located at point 2, is focused on heavy-ion collisions.
The detectors are challenged by the proton-proton collisions not only due to high collision rates, also because the \LHC beam intensity is high enough to create multiple interactions per proton bunch crossing.
\Cref{fig:exp-pileup} shows the profile of the mean number of interactions per bunch crossing, as recorded by the \ATLAS detector during \runii.
The profile underlines the busy environment, with up to 70 collisions taking place in some of the \LHC fills in 2017 due to special fill conditions.
Overall, on average \ATLAS recorded $\left<\mu\right> = 33.7$ interactions per bunch crossing during \runii.
The \ATLAS experiment, by which the data analysed in this thesis was taken, is introduced in more detail in the following section.

\section{The ATLAS experiment}
\label{sec:exp_atlas}

\begin{figure}
  \centering
  \includegraphics[width=0.92\textwidth]{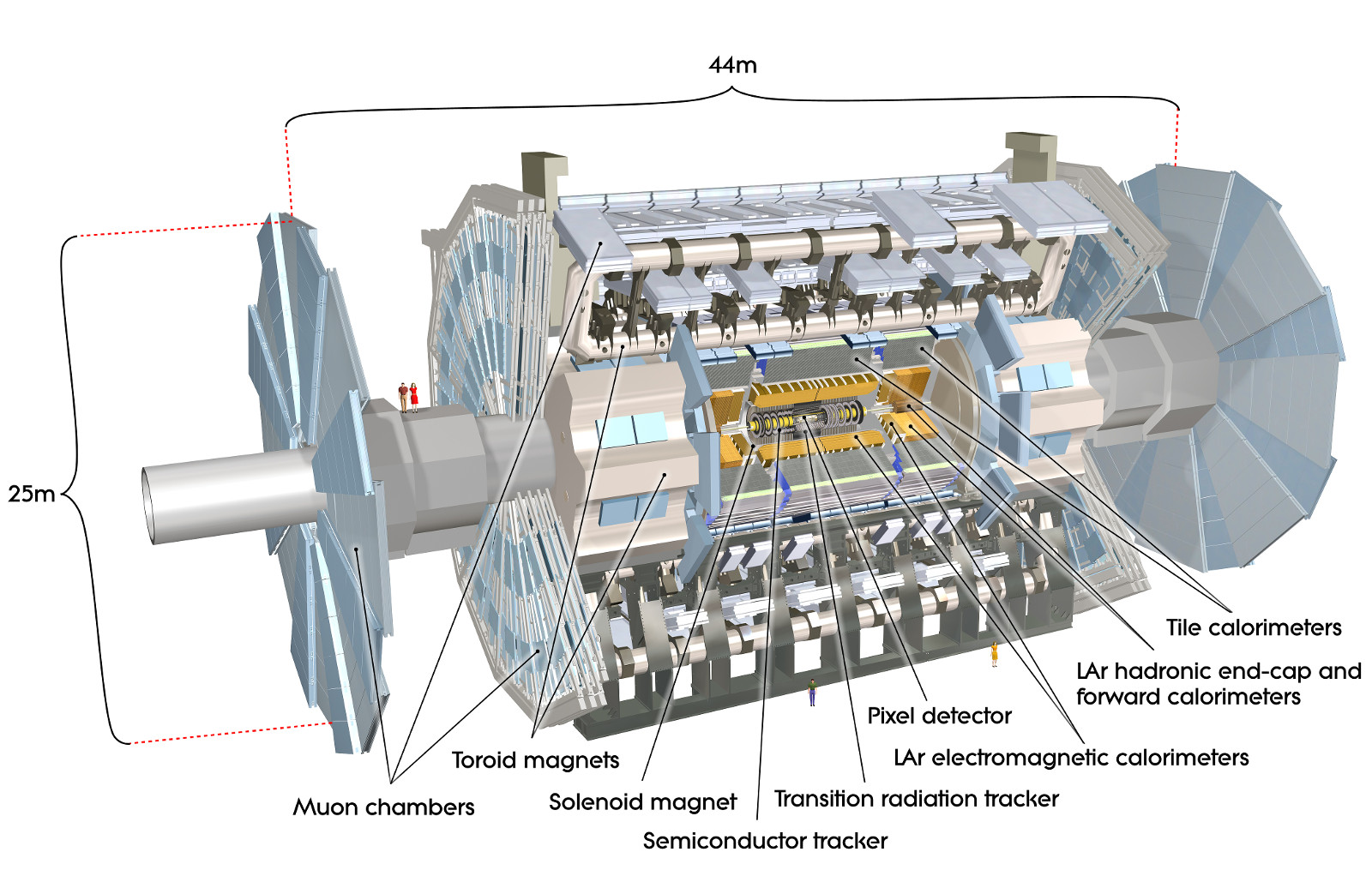}
  \caption[Components of the \ATLAS detector]{%
    Overview of the components of the \ATLAS detector~\cite{PERF-2007-01}.
    To visualise the sheer size of the \SI{44}{\m} long and \SI{25}{\m} tall apparatus, humans are shown for scale.
    \textcopyright{}\CERN
  }
  \label{fig:exp-ATLAS-overview}
\end{figure}

The \ATLAS experiment, where \ATLAS is a contraction of \emph{A Toroidal LHC ApparatuS}, is a general-purpose experiment designed to explore the high-energy, high-luminosity physics regime at the \LHC.
It is located at one of the four interaction points, where the two \LHC beams intersect, and faces its partner experiment, \CMS, on the opposite side of the \LHC ring.
\ATLAS is by volume the largest of the four \LHC experiments and measures \SI{44}{\m} in length and \SI{25}{\m} in height, with a weight of approximately \SI{7000}{\tonne}.
The profile of the detector and its onion-shell-like components are depicted in \cref{fig:exp-ATLAS-overview}.
It is designed to cover almost the full $4\pi$ solid angle around the interaction point and consists of multiple components, introduced in the following in more detail:
the Inner Detector to measure tracks of particles and to identify interaction vertices, the calorimeters to quantify energies of particles, and the Muon Spectrometer to measure transverse momenta of muons.
The detection systems are supplemented by the detector magnets:
a solenoid magnet built around the Inner Detector to bend the tracks of charged particles, and a system of toroid magnets built outside the calorimeters to supply the Muon Spectrometer with a magnetic field for bent muon tracks.

\ATLAS uses its own coordinate system with its origin placed at the (theoretical) interaction point at the centre of the detector.
The $z$-axis is defined to run along the beam axis, with the $x$--$y$ plane transverse to the beam axis.
The $x$-axis and $y$-axis point towards the centre of the \LHC ring and towards the earth surface, respectively.
Event kinematics within \ATLAS usually do not use the $x$--$y$ coordinates, however, but are described through transverse components -- for example, the transverse momentum \pT of a particle or the missing transverse momentum \MET.
As a second coordinate, the azimuthal angle $\phi$ around the beam axis is used.
The third component of the polar coordinate system, the polar angle $\theta$, is replaced by the rapidity
\begin{align}
  \label{eq:exp-rapidity}
  y \equiv \frac{1}{2} \ln \left( \frac{\left| \vec{p} \right| - p_L}{\left| \vec{p} \right| + p_L} \right) \, ,
\end{align}
where $p_L$ is the longitudinal component of the particle's momentum.
$y$ is preferred over $\theta$ as differences in the rapidity are invariant under Lorentz boosts along the beam axis.
For massless particles, or if the mass is negligibly small compared to the momentum ($\left| \vec{p} \right| \gg m$), the rapidity is equivalent to the pseudorapidity $\eta = - \ln \tan \sfrac{\theta}{2}$, which is generally used within \ATLAS.
Differences in solid angle between two objects can then be expressed as
\begin{align}
  \label{eq:exp-dR-distance}
  \Delta R = \sqrt{ \Delta \phi^2 + \Delta \eta^2} \, ,
\end{align}
where $\Delta \phi$ and $\Delta \eta$ are the differences in azimuthal angles and pseudorapidities between the two objects, respectively.
This notation will be used in the following.

\begin{figure}
  \centering
  \includegraphics[width=0.73\textwidth]{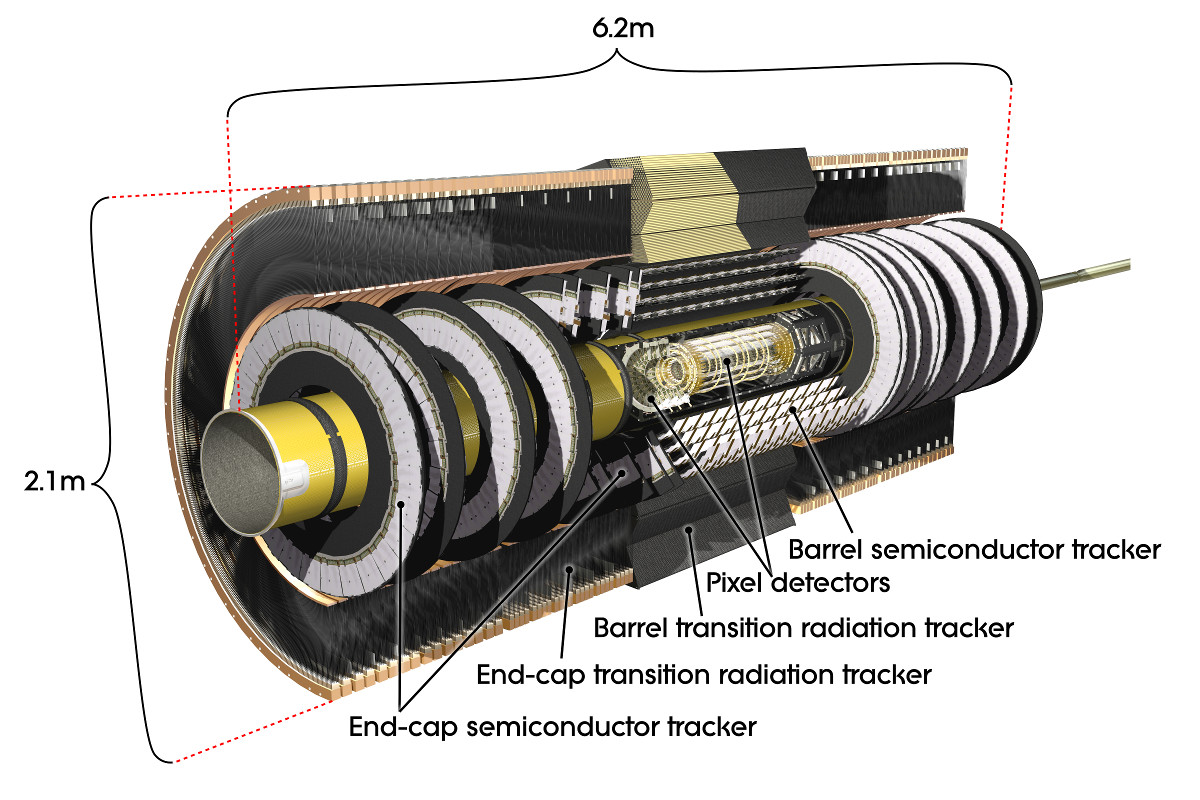}
  \caption[The \ATLAS Inner Detector]{%
    Overview of the \ATLAS \ID system~\cite{PERF-2007-01}.
    The \ID is the innermost component of the \ATLAS detector and is built directly around the beam pipe and the interaction point.
    It consists of barrel-shaped structures in the central part and disk-like structures, called end-caps, in the forward regions.
    \textcopyright{}\CERN
  }
  \label{fig:exp-ATLAS-ID}
\end{figure}

\paragraph{inner detector.}
Tracking in \ATLAS is based on particle hits in the Inner Detector (\ID)~\cite{ATLAS-TDR-4,ATLAS-TDR-5}, a system of three sub-components that is located closest to the beam pipe of the \LHC.
In conjunction with the solenoid magnet placed around it, the \ID system provides accurate determination of particles' momenta and vertex identification.
The solenoid~\cite{ATLAS-TDR-9} has a total length of \SI{5.3}{\m} with a diameter of \SI{2.5}{\m} and provides a magnetic field of approximately \SI{2}{\tesla}.
By order of distance from the beam pipe, the components of the \ID, all immersed in the solenoid field, are the \PIXEL detector, the Semiconductor Tracker (\SCT) and the Transition Radiation Tracker (\TRT).
All three components consist of barrel-shaped structures in the central region around the interaction point, the extents of which vary among the components (the \TRT barrel only covers $\abseta \lesssim 0.8$, whereas the \PIXEL barrel extends up to $\abseta \lesssim 1.8$).
Additionally, end-cap structures are placed in the forward regions of larger \abseta.
The layout of the \ID system is depicted in \cref{fig:exp-ATLAS-ID}.

Not shown is the innermost barrel layer of the \PIXEL detector, the \emph{Insertable B-Layer} (\IBL)~\cite{ATLAS-TDR-2010-19}, which was added for \runii to improve the tracking performance.
Apart from \IBL, the \PIXEL system consists of three other barrel layers, placed at distances of \SIlist{33.2;50.2;88.5;122.5}{\milli\m} from the beam pipe, respectively.
The barrel layers are supplemented by three disks each in the two forward regions.
The \PIXEL detector is based on silicon semiconductor technology, provides more than \num{100} million readout channels and is designed to come with high spatial resolution to record hits of particles and to reconstruct particle trajectories.
The high resolution in the \etaphi plane and along the beam axis is vital to identify primary interaction vertices, especially with large mean numbers of interactions during \runii.
Information from the \PIXEL system is also used to reconstruct secondary vertices of particle decays that occur within the detector volume.

Additional tracking information is provided by the \SCT, a semiconductor strip detector, which consists of eight layers of strips in its barrel component and two end-caps.
Half of the barrel strips are aligned with the beam axis and provide high-granularity $\phi$ coordinates, while the other half is tilted by a stereo angle of \SI{40}{\milli\radian} to also provide hit coordinates in $z$~direction.
The third sub-component, the \TRT, is a straw tube detector, with the tubes placed in parallel to the beam pipe in the barrel component and radially in the end-caps.
Apart from additional tracking information, the signal amplitude of the \TRT is sensitive to the Lorentz factor~$\gamma$ of traversing particles.
Thus, amplitude differences can be used to identify light, ultra-relativistic electrons, and to distinguish them from pions and other hadrons.

\begin{figure}
  \centering
  \includegraphics[width=0.78\textwidth]{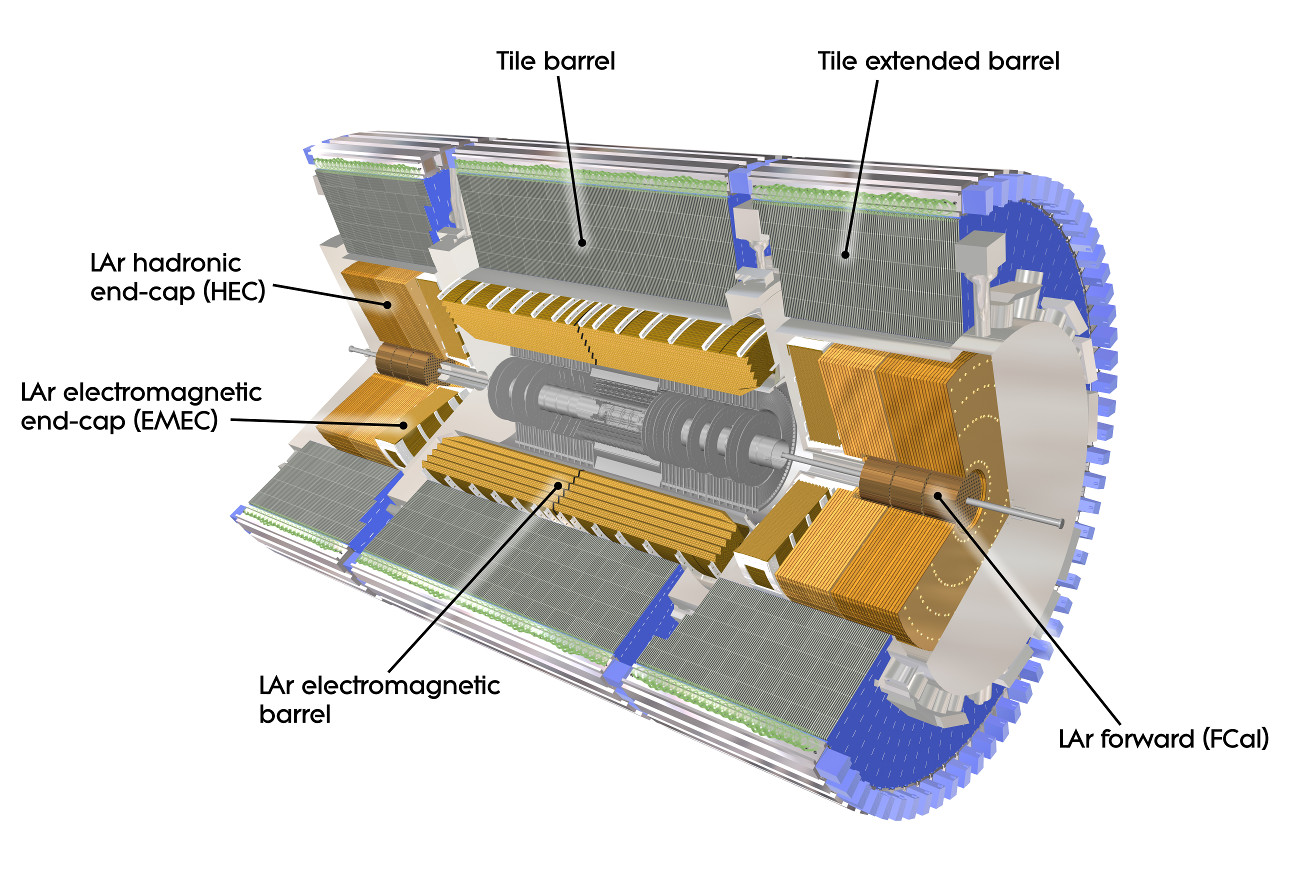}
  \caption[The \ATLAS calorimeters]{%
    Overview of the \ATLAS \ECAL and \HCAL calorimeters~\cite{PERF-2007-01}.
    As with the \ID system depicted in \cref{fig:exp-ATLAS-ID}, the calorimeters consist of barrel components in the central part and end-caps in the forward regions.
    \textcopyright{}\CERN
  }
  \label{fig:exp-ATLAS-calorimeters}
\end{figure}

\paragraph{calorimeters.}
The energies of hadrons, electrons and photons in \ATLAS are determined with two different types of calorimeters~\cite{ATLAS-TDR-2,ATLAS-TDR-3}:
the Electromagnetic Calorimeter (\ECAL) and the Hadronic Calorimeter (\HCAL).
The combined system is designed to cover pseudorapidities up to $\abseta < 4.9$.
With $\abseta < 3.2$, the \ECAL coverage includes that of the \ID system, and the \ECAL is designed to have a fine granularity for precision measurements of electron and photon energies.
On the other hand, the coarser granularity of the \HCAL satisfies the physics requirements for jet reconstruction and \MET measurements.
The overall thicknesses of the calorimeters amount to more than twenty radiation lengths for the \ECAL and approximately ten interaction lengths for the \HCAL, both in the barrel and end-cap regions, which provides good containment of both electromagnetic and hadronic showers and minimises punch-through effects into the Muon Spectrometer system.
The layout of the calorimeter system is depicted in \cref{fig:exp-ATLAS-calorimeters}.

The \ECAL is divided into a barrel structure (at $\abseta < 1.475$) and two end-cap components (at $1.375 < \abseta < 3.2$).
All parts are constructed as a sampling calorimeter, with lead absorber plates and active liquid argon (LAr) layers:
the atoms of the lead plates are ionised, and the produced particles trigger electric signals in the LAr components.
Those components covering $\abseta < 2.5$ are segmented in three calorimeter sections in depth, while the others have two sections in depth and a coarser granularity.
For $\abseta < 1.8$, an additional presampler corrects for electron and photon energy losses upstream of the calorimeter.
This presampler consists of thin layers of active LAr material with thicknesses of \SIlist{1.1;0.5}{\centi\m} in the barrel and end-cap regions, respectively.

The core of the \HCAL are sampling calorimeters with steel absorber plates and scintillating tiles as active material:
the tile barrel at $\abseta < 1.0$ and the tile extended barrels in the range $0.8 < \abseta < 1.7$.
All are segmented in depth in three layers, and extend from an inner radius of \SI{2.28}{\m} to an outer radius of \SI{4.25}{\m}.
The tile calorimeters are supplemented by LAr hadronic end-cap systems in the range $1.5 < \abseta < 3.2$.
The end-caps consist of two wheels on each side of the detector, each with two segments in depth.
An additional LAr forward calorimeter provides even higher coverage in \abseta and consists of three modules:
the first is made of copper and measures predominantly electromagnetic showers, whereas the other two are made of tungsten and focus on hadronic interactions.

\begin{figure}
  \centering
  \includegraphics[width=0.64\textwidth]{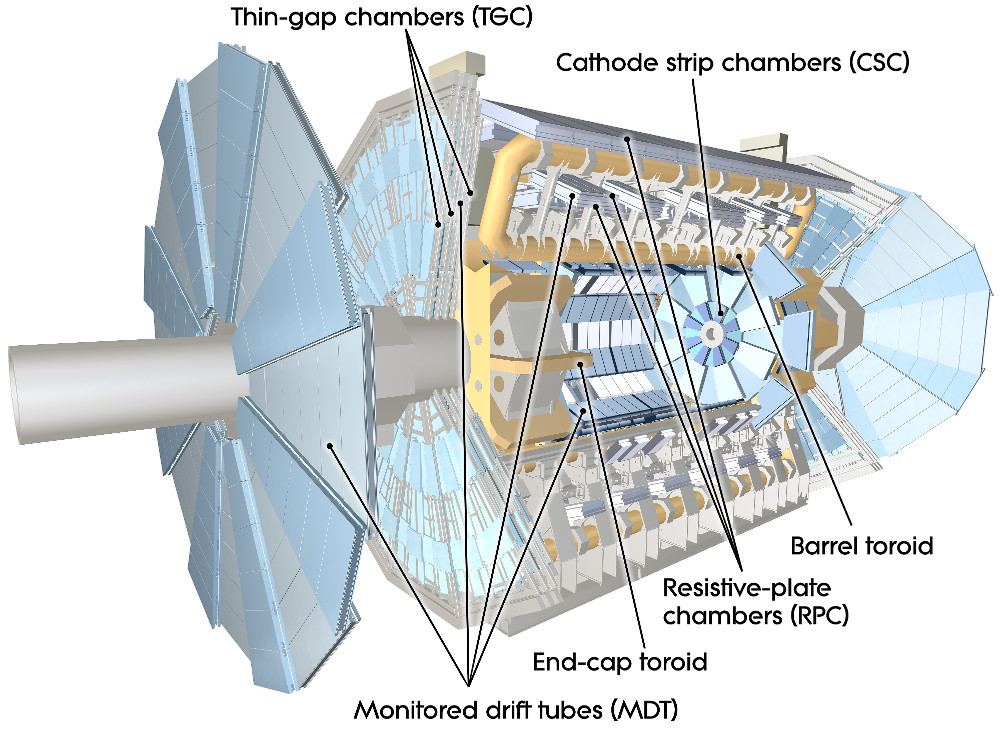}
  \caption[The \ATLAS Muon Spectrometer]{%
    Overview of the \ATLAS \MS system and the toroid magnets~\cite{PERF-2007-01}.
    High-precision tracking is provided by the monitored drift tubes and additional cathode strip chambers in the innermost layer in the forward regions.
    The muon trigger system consists of resistive plate chambers in the barrel region and cathode strip chambers in the forward regions.
    \textcopyright{}\CERN
  }
  \vspace*{3pt}  
  \label{fig:exp-ATLAS-MS}
\end{figure}

\paragraph{muon spectrometer.}
Muons only deposit a small fraction of their energies in the \ATLAS calorimeters and an additional system is needed to detect them with high precision:
the Muon Spectrometer (\MS)~\cite{ATLAS-TDR-10} is the outermost of \ATLAS's \enquote{onion shells}.
Consisting of several systems of chambers, muon momenta are measured in the \MS based on the magnetic deflection of muon tracks in the air-core toroid magnets~\cite{ATLAS-TDR-7,ATLAS-TDR-8} that are incorporated in the barrel and end-cap structures of the \MS system.
In the range $\abseta < 1.4$, a magnetic field, approximately orthogonal to the muon tracks, is provided by the barrel toroid.
It consists of eight radially assembled coils, with total lengths of more than \SI{25}{\m}.
The generated magnetic field has a strength of approximately \SI{4}{\tesla}.
For $1.6 < \abseta < 2.7$, muon trajectories are bent by the two smaller end-cap toroids, which are inserted into the ends of the barrel toroid and align with the inner solenoid magnet.
They each consist of eight racetrack-like coils.
In the transition region, in the range $1.4 < \abseta < 1.6$, magnetic deflection is provided by both the barrel and the end-cap toroids.
An overview of all components of the \MS system and the toroid magnets is given in \cref{fig:exp-ATLAS-MS}.

High-precision tracking is provided by two separate muon chamber systems:
monitored drift tubes, arranged in three layers, cover most of the \abseta range.
The layers are oriented cylindrically around the beam pipe in the barrel region, and perpendicular to the beam in the transition and end-cap regions.
They are supplemented by cathode strip chambers with higher granularity in the innermost layer for pseudorapidity ranges $2.0 < \abseta < 2.7$.
An additional, separate chamber system acts in the range $\abseta < 2.4$ and consists of resistive plate chambers in the barrel region and thin gap chambers in the forward regions.
Apart from providing trigger information on well-defined muon \pT thresholds, information from this system is also used to identify bunch-crossings and to supplement the tracking chambers with orthogonal measurements of muon coordinates.

\paragraph{trigger and data acquisition.}
The Trigger and Data Acquisition (\TDAQ) system~\cite{TRIG-2016-01} of the \ATLAS detector is the central point to acquire event data from the individual subsystems, process and filter the data, and forward it to permanent storage.
With high bunch-crossing rates of up to \SI{40}{\mega\hertz} during \runii and high levels of pile-up, only a small fraction of events can be read out and put into storage.
To minimise the dead time of individual components, decisions about events and whether to store them require a fast-response system.
The \TDAQ system in \runii comprises two trigger levels: a hardware based low-level trigger (\Lone) and a software based high-level trigger (\HLT).

The \Lone trigger consists of a central trigger decision unit that receives information from the low-granularity components of the calorimeters and the muon-trigger system.
This unit also sets preventive dead times to avoid overlapping readout windows and overflowing buffers in the readout components.
After an event is accepted by the \Lone trigger system, it is stored temporarily in a dedicated buffer system off-detector, called the \emph{Read-Out System}.
In addition, the \Lone system defines \emph{regions of interest} in the \etaphi plane to highlight candidate objects, such as muons, electromagnetic clusters or large total transverse momentum.
Once the event is \Lone-accepted and put into the Read-Out System, these regions of interest are then forwarded to the \HLT.
There, they are reconstructed regionally using a dedicated computing farm.
At that stage, the \HLT accesses data from all \ATLAS components and performs a more sophisticated, software-based trigger decision on candidate objects.
Events accepted by the \HLT are then written to disk and stored permanently.
The \Lone trigger reduces the initial \SI{40}{\mega\hertz} of collisions to an approximate trigger rate of \SI{100}{\kilo\hertz}, which are further reduced by the \HLT to the order of \SI{1}{\kilo\hertz}.

\section{Physics object reconstruction in ATLAS}
\label{sec:exp_objects}

The proton-proton collisions of the \LHC create a plethora of particles with every bunch crossing at rates of up to \SI{40}{\mega\hertz}.
Many of the proton-proton collisions result in elastic scatterings that are of little interest to the high-energy physics programme of the \ATLAS experiment.
However, even if only hard-scattering interactions are considered, the instantaneous luminosity of the \LHC beam is large enough to produce dozens of these with every bunch crossing.
The vast majority of particles produced in hard-scattering events is scattered with large transverse momentum, but decays on very short timescales.
Only a small fraction of particles from the primary interaction exists long enough to be detected directly with the \ATLAS detector.
The others disintegrate into lighter, more stable particles, which in turn may be detected -- or continue decaying in a chain until stable particles are created.
Out of all types of elementary and composite particles, only fourteen have mean free paths longer than \SI{500}{\micro\meter}, thus, have a chance to interact with the detector materials and enable detection.
These are (including antiparticles for fermions): muons, electrons, photons, pions, kaons, protons and neutrons.
The latter four of these are not elementary, but compound particles, made from colour-charged elementary particles and created following the colour confinement of \QCD.
Colour-charged particles produce entire sprays of colour-neutral compound particles, known as \emph{jets}.
Jets involving bottom quarks have unique properties and can often be identified as \bjets.
These and the other objects reconstructed with \ATLAS are described briefly in the following paragraphs.

\paragraph{muons.}
Muons only interact weakly with the detector material, and therefore do not leave significant energy deposits in the calorimeters.
Instead, muon reconstruction algorithms rely on track information from the \ID and the \MS systems of \ATLAS, and are only supplemented by calorimeter information.
Generally, the \ATLAS reconstruction of muons is based on four classes of candidates~\cite{PERF-2015-10}:
(1) segment-tagged muons, where an \ID track is reconstructed, but the muon candidate only crossed one layer of the \MS chambers.
This may occur if the muon carries little transverse momentum or traverses a \MS region with lower acceptance.
(2) calorimeter-tagged muons, where a calorimeter signature compatible with a minimum ionising particle is associated to an \ID track.
This type of muon candidate recovers regions of low \MS acceptance, for example, where the \MS chambers are not fully instrumented to allow cabling of the more central detector components.
(3) extrapolated muons, which are only based on \MS trajectories and mostly used in regions of $\abseta > 2.5$ beyond the coverage of the \ID system.
(4) combined (CB) muons, based on independent tracks in the \MS and the \ID systems that are matched to a combined track through a global refit.

This analysis only uses CB muons as they provide the highest reconstruction efficiencies and cover the relevant detector regions within $\abseta < 2.5$.
The majority of muons uses an outside-in recognition, where muon candidates are first identified in the \MS chambers and then extrapolated to the detector core to match them with an \ID track.
About 0.5\% of muons are reconstructed with the complementary inside-out approach.
Identified CB muon candidates need to fulfil a set of \emph{medium} quality criteria~\cite{PERF-2015-10} to suppress background candidates, mainly originating from hadron decays.
The quality criteria exploit characteristic kinks of muon tracks coming from in-flight hadron decays, which degrade the fit quality of the combined \ID and \MS track.
As a figure of merit, a parameter called $q/p$ significance is used, a quantity to describe differences in the charge/momentum ratio between the \ID and the \MS tracks, with uncertainties on both ratios taken into account.
The $q/p$ significance is required to be below a fixed-cut value.

Muon candidates are required to be isolated using cuts on track-based and calorimeter-based isolation variables~\cite{PERF-2015-10}.
For low-\pT muons, the track-based isolation variable, \ptvarcone{30}, is defined as the scalar sum of all track transverse momenta with $\pT > \SI{1}{\GeV}$ in a variable-radius cone of $\Delta R$ around the muon candidate with transverse momentum $\pT^\mu$.
The variable radius of $\Delta R = \text{min}\left( \SI{10}{\GeV} / \pT^\mu , 0.3 \right)$ is chosen to optimise background suppression.
For high-\pT muons, a fixed-radius cone of $\Delta R = 0.2$ is used and cuts are placed on the corresponding track-based isolation variable \ptcone{20}.
Simultaneously, the variable $\ettopocone$, defined as the sum of the energy of topological clusters~\cite{PERF-2011-03} around the muon, after subtracting the muon energy itself, is used to require calorimeter-based isolation.
In the applied \emph{FCTight\_FixedRad} isolation menu (tight fixed-cut isolation with fixed-radius requirements in the high-\pT regime), the muons are required to simultaneously fulfil
\begin{subequations}
\begin{align}
  \label{eq:object-muon-iso}
  \text{for $\pT < \SI{50}{GeV}$:}&& \quad
  \ptvarcone{30} < 0.04 \, \pT^\mu \quad \text{and} & \quad
  \ettopocone < 0.15 \, \pT^\mu\, , \\[0pt plus 3pt]  
  \text{for $\pT > \SI{50}{GeV}$:}&& \quad
  \ptcone{20} < 0.04 \, \pT^\mu \quad \text{and} & \quad
  \ettopocone < 0.15 \, \pT^\mu\, .
\end{align}
\end{subequations}
The reconstruction efficiencies of muons are measured in both data and simulation, and scale factors based on comparisons between the two are extracted to correct the reconstruction efficiencies in \MC simulation.
The list of muon candidates is shortened further by requiring calibrated transverse momenta of $\pT > \SI{25}{\GeV}$ and pseudorapidities of $|\eta| < 2.5$.
\ATLAS-recommended requirements on the association of the muon candidate to the primary vertex are also applied:
firstly, the difference in the $z$-axis between the track origin and the primary vertex, when expressed at the beam line, must be $|\Delta z_0 \sin(\theta)| <\SI{0.5}{\mm}$.
Secondly, the transverse impact parameter of the muon track $d_0$, defined as the point of closest approach of the track in the transverse plane to the primary vertex, is required to fulfil $|d_0|/\sigma(d_0) < 3$.

\paragraph{electrons.}
As electrons carry electromagnetic charge, they leave tracks in the \ID system of \ATLAS before they hit the calorimeters.
When entering the \ECAL and interacting with its material, electrons lose a significant amount of their energy through Bremsstrahlung.
Bremsstrahlung photons then convert into electron--positron pairs, which in turn interact with the \ECAL material.
The electrons, positrons and photons are all collimated and are reconstructed typically as a single cluster in the \ECAL system.
However, Bremsstrahlung interactions can also occur in the volume of the \ID, before entering the calorimeters.
Therefore, multiple tracks in the \ID matched to topological clusters (\emph{topo-clusters}) in the calorimeters must also be considered as electron candidates.

Electron candidates in \ATLAS are built in four steps~\cite{PERF-2017-01,EGAM-2018-01}: the formation of topo-clusters in the calorimeter cells, the reconstruction of tracks in the \ID, the matching of topological clusters with one or multiple \ID tracks, and the building of superclusters from these matched candidates.
The topo-cluster reconstruction uses a 4--2--0 algorithm, which first looks for calorimeter cells with a significant energy deposit (four times larger than a pre-defined noise threshold $\sigma^{\text{EM}}_{\text{noise,cell}}$).
Neighbouring cells with $E^{\text{EM}}_{\text{cell}} \geq 2 \sigma^{\text{EM}}_{\text{noise,cell}}$ are then added to the proto-cluster.
Afterwards, a crown of nearest-neighbour cells is added to the cluster independent of their energy.
The proto-cluster is split if multiple local energy maxima exist.
Tracks are then formed from hits in the \PIXEL and \SCT systems according to methods detailed in Ref.~\cite{PERF-2017-01} using the \ATLAS Global $\chi^2$ Track Fitter~\cite{Cornelissen:2008zza} and Gaussian-sum filters~\cite{Fruehwirth:1987fm,Fruehwirth:2003131,ATLAS-CONF-2012-047} to account for possible Bremsstrahlung losses.
The tracks are matched to the topo-clusters by imposing cuts on $\left| \eta_{\text{track}} - \eta_{\text{clus}} \right|$ and $(\phi_{\text{track}} - \phi_{\text{clus}})$.
In a final step, track-matched topo-clusters are used as seeds to form superclusters that incorporate possible satellite clusters into the topo-clusters, with more details on the procedure given in Ref.~\cite{EGAM-2018-01}.
The formed superclusters are then paired with tracks with the identical matching procedure as the one used for topo-clusters.
The energies of electron candidates are calibrated using methods detailed in Refs.~\cite{PERF-2017-03,EGAM-2018-01}.

Further identification criteria are imposed to improve the purity of selected electrons.
Various parameters of the electron candidate are used to discriminate prompt electrons and hadronic energy deposits faking an electron signature:
properties of the primary electron track, the lateral and longitudinal development of the shower in the \ECAL, and the spatial compatibility of the primary electron track with the supercluster.
The used \emph{TightLH} identification working point~\cite{PERF-2017-01,EGAM-2018-01} uses a likelihood function, defined as the product of probability density functions for signal-like electron candidates, $P_{S,i}(x_i)$, evaluated at value $x_i$ for parameter $i$.
A second likelihood function $L_B$ for background-like candidates is constructed and their ratio is used as a discriminant $d_L$:
\vspace*{0pt plus 4pt}  
\begin{align}
  \label{eq:exp_el_likelihood}
  d_L = \text{ln} \left( \frac{L_S(\vec{x})}{L_B(\vec{x})} \right)
  \quad \text{with}~~
  L_{S/B}(\vec{x}) = \prod_{i=1}^n P_{S/B,i}(x_i) \, .
\end{align}
The probability density functions are extracted from $Z \to ee$ and $J/\Psi \to ee$ events with a tag-and-probe method~\cite{EGAM-2018-01}.
The \emph{TightLH} identification working point imposes a cut on the likelihood discriminant and requires $E/p < 10$ and $\pT > \SI{10}{GeV}$ for the primary electron track, targeting identification efficiency values of approximately 80\%~\cite{EGAM-2018-01}.

Electron candidates are required to be isolated using a \emph{Gradient} isolation working point with a target efficiency of $\epsilon = 0.1143 \cdot \pT + 92.14\%$, corresponding to 90\% at $\pT = \SI{25}{\GeV}$ and 99\% at $\pT = \SI{60}{\GeV}$~\cite{PERF-2017-01,EGAM-2018-01}.
The isolation is imposed through cut maps on the calorimeter isolation variable \etcone{20} and the track isolation variable \ptvarcone{20}, derived from $Z \to ee$ and $J/\Psi \to ee$ events.
\etcone{20} is the topo-cluster energy in a cone of fixed size $\Delta R = 0.2$ around the barycentre of the electron candidate, corrected for leakage and pile-up and with the core energy of the electron candidate subtracted.
\ptvarcone{20} is defined the same way as \ptvarcone{30} for muon isolation, with a variable-radius cone that maxes out at $\Delta R = 0.2$.

The reconstruction, identification and isolation efficiencies of electrons are measured in both data and simulation, and scale factors based on comparisons between the two are extracted to correct the efficiencies in \MC simulation.
This analysis only considers electrons with calibrated $\pT > \SI{25}{\GeV}$ and $|\eta_{\mathrm{clus}}| < 2.47$, excluding the crack region of the \ECAL in the area $1.37 < |\eta_{\mathrm{clus}}| < 1.52$ due to lower acceptances.%
\footnote{$\eta_{\mathrm{clus}}$ denotes the pseudorapidity of the supercluster associated to the electron candidate.}
As done for muons, requirements on the association of the candidate to the primary vertex are applied:
firstly, the track must be in close proximity to the primary vertex along the $z$-axis with $|\Delta z_0 \sin(\theta)| < \SI{0.5}{\mm}$.
Secondly, the transverse impact parameter must fulfil $|d_0|/\sigma(d_0) < 5$.

\paragraph{photons.}
Similarly to electrons, photon undergo energy losses in the \ECAL system due to conversion into electron--positron pairs, which in turn interact with the \ECAL material through Bremsstrahlung processes.
However, the conversion into an electron--positron pair might already occur before the photon enters the calorimeters.
Therefore, the reconstruction procedure of photon candidates in \ATLAS is optimised separately for photons that convert before reaching the \ECAL system (\emph{converted photons}), and for those photons not associated with a conversion (\emph{unconverted photons}).
Since they share much of their signature with electron candidates, the reconstruction of photons and electrons is done in parallel~\cite{PERF-2017-02,EGAM-2018-01}: converted photons are identified as superclusters associated with a conversion vertex instead of an electron track, and unconverted photons as superclusters matched to neither an electron track nor a conversion vertex.

The formation of topo-clusters and the reconstruction of tracks are identical to those of electron candidates.
Then, in an additional step, conversion vertex reconstruction is performed with tracks loosely matched to a topo-cluster.
Two-track conversion vertices are formed from two opposite-charge tracks, the vertex of which is compatible with a massless particle.
Single-track vertices are those without hits in the innermost, most sensitive components of the \ID system.
To increase the purity of converted photons, the tracks associated with conversion vertices must have a high probability to be electron tracks as determined by the \TRT.
If multiple vertices are matched to a topo-cluster, two-track vertices including hits in \PIXEL and \SCT are preferred over two-track vertices with \TRT hits only, which are in turn preferred over single-track vertices.
Topo-clusters are then used as supercluster seeds for photons, regardless of any matching to tracks or conversion vertices.
The methods to build superclusters are largely identical to those used for electrons, with more details on the procedure given in Ref.~\cite{EGAM-2018-01}.
In addition, if a photon supercluster seed is matched to a conversion vertex, a satellite cluster is added to that supercluster if its conversion vertex is identical to that of the supercluster seed, or if its best-matched track is compatible with the conversion vertex of the supercluster seed.
Because electron and photon superclusters are built independently, the same seed cluster can produce both electron and photon candidates.
With methods detailed in Ref.~\cite{EGAM-2018-01}, trivial ambiguities between them are detected and resolved.
Remaining ambiguous candidates are kept, but marked as such.
The energies of photon candidates are then calibrated using methods detailed in Refs.~\cite{PERF-2017-03,EGAM-2018-01}.

\begin{figure}
  \centering
  \includegraphics[width=0.9\textwidth]{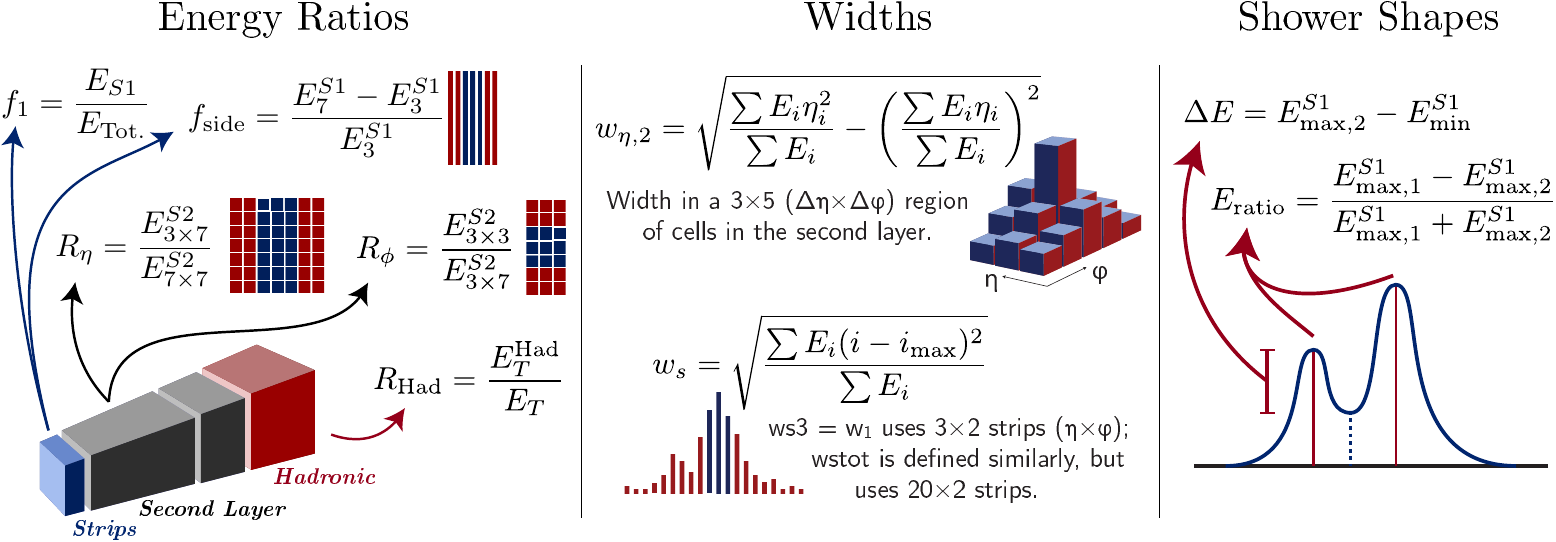}
  \caption[Shower-shape parameters considered in photon identification]{%
    Shower-shape parameters of photons considered for the \emph{Tight} identification working point, as presented in Ref.~\cite{Saxon:1746004}.}
  \label{fig:exp-photon-shower-shape}
\end{figure}

Refs.~\cite{PERF-2017-02,EGAM-2018-01} report observed discrepancies between data and simulation in the peak positions of the shower-shape parameters of photons, pointing towards a mismodelling of the lateral shower development in \MC simulation.
The discrepancies are mitigated by applying data-driven shifts to the variables for simulated photons, commonly referred to as \emph{fudge factors}.
After mitigation, further identification criteria are imposed to improve the purity of prompt, isolated photons, and to reject secondary photons from hadronic decays or hadronic activity faking photon signatures.
The used \emph{Tight} identification working point~\cite{PERF-2017-02,EGAM-2018-01} uses one-dimensional cuts on parameters of the lateral and longitudinal shower evolution of the photon candidate.
In particular, parameters using the first layer of the \ECAL play an important role in rejecting $\pi^0$ decays into highly collimated photon pairs.
An overview of all shower-shape parameters used in the \emph{Tight} working point is given in \cref{fig:exp-photon-shower-shape}.
The cut-based selection is optimised separately in bins of \abseta and \ET to accommodate varying shower shapes due to the geometry of the detector.
For low-\ET photons, $Z \to \ell\ell\gamma$ and \Zjets events are used as a source of signal-like and background-like photons, respectively, whereas for $\ET > \SI{25}{\GeV}$ the cuts are derived from \plusjets{\gamma} and dijet samples~\cite{EGAM-2018-01}.

Photon candidates must be isolated, imposed through fixed cuts on calorimeter and track isolation variables.
The used \emph{FixedCutTight} working point~\cite{PERF-2017-02,EGAM-2018-01} requires $\etcone{40} < 0.022 \cdot E_T(\gamma) + \SI{2.45}{\GeV}$ for calorimeter isolation, and $\ptcone{20} < 0.05 \cdot E_T(\gamma)$ for track isolation if the photon candidate has a matched conversion vertex.
\etcone{40} is the topo-cluster energy in a cone of fixed size $\Delta R = 0.4$ around the barycentre of the photon candidate, with the core energy of the photon candidate subtracted.
\ptcone{20} is the scalar sum of all track transverse momenta with $\pT > \SI{1}{\GeV}$ in a cone of fixed size $\Delta R = 0.2$ around the photon candidate.

The reconstruction, identification and isolation efficiencies of photons are measured in both data and simulation, and scale factors based on comparisons between the two are extracted to correct the efficiencies in \MC simulation.
The analysis then only considers photons with calibrated $\ET > \SI{20}{\GeV}$ and $|\eta_{\mathrm{clus}}| < 2.37$, again excluding the \ECAL crack region $1.37 < |\eta_{\mathrm{clus}}| < 1.52$ due to lower acceptance values.

\paragraph{jets.}
As they carry colour charge, quarks and gluons -- created in the primary interaction or as secondary particles in decays -- cannot be observed as free particles.
Due to the colour confinement of \QCD, they form hadronic bound states and produce a spray of highly collimated, colour-neutral particles in the detector.
To describe the initial colour-charged particle, these sprays are reconstructed experimentally as objects called \emph{jets}, cone-like structures propagating through the detector.
Those particles in a jet with electromagnetic charge leave bent tracks in the \ID system, and they all deposit their energy in the \ECAL and \HCAL systems.
In \ATLAS, jets are reconstructed from topological clusters in the calorimeters.

In a first step, topological clusters are reconstructed in the \ECAL and \HCAL systems.
These are then combined using the \antikt jet algorithm~\cite{Cacciari:2008gp} in the \textsc{fastjet} implementation~\cite{Cacciari:2011ma}.
As opposed to simple cone-finding algorithms that identify coarse regions of energy flow in an event, the \antikt algorithm fulfils two crucial jet-algorithm properties: it is both collinear and infrared safe.
That is, the jet properties are not affected by collinear splittings or by infinitely soft, infrared emissions.
Like the \kt algorithm, \antikt recombines topological energy clusters sequentially, based on a distance measure $d_{ij}$ between the individual entities and the distance $d_{iB}$ between entity~$i$ and the beam, defined as:
\begin{subequations}
\begin{align}
  \label{eq:exp-jet-antikt}
  d_{ij} &= \text{min}\left( \pT^{2a} (i), \pT^{2a} (j) \right) \frac{\Delta R_{ij}^2}{R^2} \, ,\\[0pt plus 3pt]  
  d_{iB} &= \pT^{2a} (i) \, ,
\end{align}
\end{subequations}
where $\pT(i)$ is the transverse momentum of entity~$i$, and $\Delta R_{ij}$ is the distance between entities~$i$ and~$j$ in the $\eta$--$\phi$ plane.
Going sequentially through pairs $i,j$ of entities, the smaller of the distances $d_{ij}$ and $d_{iB}$ is identified, and if it is $d_{ij}$, the entities~$i$ and~$j$ are recombined.
If it is $d_{iB}$, the entity~$i$ is called a jet and removed from the list of entities.
Afterwards, the distances are recalculated and the procedure continues sequentially until no entities are left.

For the \kt algorithm, the exponent is $a=1$, whereas for \antikt, it is chosen to be $a=-1$.
This gives precedence to harder entities -- and softer entities tend to get recombined with harder entities much before they get clustered amongst themselves.
This is a key difference to the \kt algorithm, where softer entities are clustered first and then combined with harder entities only towards the end of the sequential algorithm.
The \antikt algorithm results in more cone-like jets than clusters combined with the \kt algorithm~\cite{Salam:2009jx}.
This analysis uses \antikt jets with a distance parameter of $R=0.4$, and jets are only considered if they fulfil $\abseta < 2.5$ and $\pT > \SI{25}{\GeV}$.

As jets are very complex objects reconstructed from a spray of dozens of particles, their energies need to be calibrated through reference objects and through evaluating the detector response in simulation.
The calibration is performed in a total of six steps, with details on the procedures of each given in Ref.~\cite{PERF-2016-04}:
(1), the jet origin is corrected to point to the primary vertex instead of to the centre of the detector, which improves the resolution in $\eta$ significantly.
The procedure is identical to that described in Ref.~\cite{PERF-2012-01}.
(2), pile-up contributions originating from the same bunch crossing (in-time pile-up) and from neighbouring bunch crossings (out-of-time pile-up) are removed from the jet energy.
This correction is based on the area the jet covers in the $\eta$--$\phi$ plane and on the jet's transverse momentum density in that area.
(3), a second pile-up correction removes residual jet-energy dependencies on the number of reconstructed primary vertices and on the number of bunch crossings~$\mu$ per event, based on \MC-truth information.
(4), the \emph{absolute} jet energy calibration corrects the reconstructed jet four-momenta to the particle-level energy scales.
This step also removes biases in the $\eta$ calibration of the jet energies.
(5), the \emph{Global Sequential Calibration}, first explored at $\sqrt{s} = \SI{7}{\TeV}$ in Ref.~\cite{PERF-2011-03}, corrects the jet energy scales further using a combination of observables from the calorimeters and the \ID and \MS systems.
Residual dependencies on the flavour composition of the jets and the energy distribution within the jet are removed, both of which vary significantly between quark-initiated and gluon-initiated jets.
(6), final \emph{in-situ} corrections are applied to account for jet-energy mismatches between data and simulation.
They are based on well-measured reference objects, such as photons and \Zbosons, and on the multi-jet balance of events.
The included $\eta$-intercalibration corrects the jet energy scales of forward jets to that of central jets using dijet events.

Imbalances in these dijet events are also used to calibrate the resolution of the jet energies, determined in measurements similar to those detailed in Ref.~\cite{PERF-2011-04}.
The method assumes an approximate scalar balance between the two jets' transverse momenta, and asymmetries observed between these momenta are used to determine the energy resolution.
The results are combined with a second in-situ technique, known as the \emph{bi-sector method}.

Pile-up corrections are applied to the jet energy scales, but in-time and out-of-time pile-up activity can also mimic jet signatures in the \ATLAS detector.
A multivariate tool, known as the jet vertex tagger (\JVT)~\cite{ATLAS-CONF-2014-018}, tests the compatibility of the tracks associated with a jet with the primary vertex, and it provides a discriminant output.
The \JVT output is constructed from a two-dimensional likelihood, using the scalar sum of the track transverse momenta associated with a jet, and a constructed parameter called \emph{corrected jet vertex fraction}.
The response of the \JVT discriminant is measured in both data and simulation, and scale factors based on comparisons between the two are extracted to correct the tagging efficiencies in \MC simulation.
Jets with $\pT < \SI{60}{\GeV}$ in this analysis are required to pass a cut on the discriminant of $\JVT > 0.59$ to reduce jets from pile-up.

\paragraph{$b$-jets.}
Compared to jets originating from other hadrons, jets from hadrons with \bquarks have unique properties that can be used to identify them.
\ATLAS uses algorithms for \bjet identification that exploit the long lifetime, the high mass and the high decay multiplicity of hadrons with \bquarks~\cite{FTAG-2018-01}.
With a mean lifetime at the order of \SI{1.5}{\pico\s}, they have a significant mean flight length in the \ATLAS detector.
Decays of \bhadrons can thus be identified through a vertex displaced from the primary vertex of the hard interaction.
\ATLAS uses a two-level approach, where the first stage are three low-level algorithms that reconstruct the characteristic features of \bhadron decays:
(1), the algorithms \IPtwoD and \IPthreeD~\cite{ATL-PHYS-PUB-2017-013} exploit the large impact parameters of tracks from \bhadron decays.
(2), \SVone~\cite{ATL-PHYS-PUB-2017-011} performs inclusive reconstruction of secondary vertices that could come from \bhadron decays.
(3), \JetFitter~\cite{ATL-PHYS-PUB-2018-025} attempts to reconstruct the full \bhadron to \chadron decay chain.

In the second stage, these low-level algorithms are combined in high-level tools using multivariate techniques.
This analysis uses the \MVtwo~\cite{ATL-PHYS-PUB-2017-013} algorithm based on a boosted-decision-tree discriminant.
The discriminant uses gradient boosting on a total of \num{1000} trees with their depth parameter set to \num{30}.
As training input, a hybrid $t\bar{t} + Z'$ \MC simulation is used.
Apart from the low-level algorithms, the transverse momenta \pT and pseudorapidities \abseta of the jets are included in the training to exploit correlations.
For the training, \pT and \abseta spectra of \bjets and \cjets are reweighted to match those of light jets to avoid possible training biases.
In addition, to achieve better \cjet rejection, their fractions are enhanced with respect to light jets in the training data.
Operating points of fixed \btag efficiency are defined at \SIlist{60;70;77;85}{\percent}, the last of which is used for this analysis.
Evaluated on a baseline \ttbar \MC simulation, the \cjet and light-jet rejection values amount to approximately \numlist{2.7;25}, respectively.
In addition to the fixed-efficiency operating points, the \MVtwo discriminant distribution is divided into five pseudo-continuous bins, defined by the selection cuts used to define the fixed-efficiency operating points.
The number of fixed-efficiency operating points passed by a \bjet candidate can then be evaluated as a \btag \emph{score}.
The \btag efficiencies and rejections are calibrated using methods detailed in Ref.~\cite{FTAG-2018-01}, and correction factors are applied to the simulated samples to compensate for differences in the efficiencies between data and simulation.

\paragraph{missing transverse momentum.}
The \com energy in hard-scattering interactions between protons is not constant, but determined in a probabilistic way through the parton density functions of the protons.
Thus, the \com system of the partonic hard interaction might be boosted along the beam axis.
However, the total transverse momentum in the initial state, or its magnitude, colloquially referred to as the \enquote{total transverse energy} and denoted \ET, is expected to be zero due to energy-momentum conservation.
This is exploited to estimate the invisible, missing component of momentum in the transverse plane of the final state, \MET, that remains undetected by the \ATLAS detector.
On the one hand, the detector does not cover the full solid angle and has regions of lower acceptance, where final-state particles remain undetected.
On the other hand, the detector may be insensitive to some particles, and they might not trigger any detection signals at all.
In the \SM, neutrinos are the only type of particles that pass through the detector undetected.
But also many \BSM theories predict weakly interacting particles that could trigger an excess of missing transverse energy in the detector.

The calculation of \MET relies on the reconstruction and calibration of all other object candidates~\cite{PERF-2016-07}.
This includes muons, electrons, photons, jets and hadronically-decaying \tauleptons that originate from the primary interaction.
To avoid double counting of energy deposits in the calorimeters, a dedicated overlap-removal procedure is performed among the object candidates.
The calculation of \MET includes an additional \emph{soft} signal, which comprises all well-identified tracks in the \ID system that are not associated to any physics object candidate.
\MET is then constructed from the transverse momentum vectors of all hard objects and from those of the soft signals.
The $x$/$y$-components of the missing transverse momentum, $E^{\text{miss}}_{x(y)}$, are calculated as
\begin{align}
  \label{eq:exp-met-calc}
  E^{\text{miss}}_{x(y)} = ~~ - \sum_{i \in \{\text{hard objects}\}} p_{x(y),i} ~~ - \sum_{j \in \{\text{soft signals}\}} p_{x(y),j} \, ,
\end{align}
from where the overall magnitude is calculated as
\begin{align}
  \label{eq:exp-met-calc2}
  \MET = \sqrt{ \left( E^{\text{miss}}_x \right)^2 + \left( E^{\text{miss}}_y \right)^2 } \, .
\end{align}
For the hard objects, identification and selection criteria are imposed, as detailed in Ref.~\cite{PERF-2016-07}.
Tracks considered for the soft signal must fulfil $\pT > \SI{400}{\MeV}$ and they need to be associated with the primary vertex: $d_0 < \SI{0.5}{\milli\metre}$ and $z_0 \sin(\theta) < \SI{1.5}{\milli\metre}$.
Additional $\Delta R$ distance criteria to other objects are imposed.
The scale and resolution of \MET are calibrated using $Z \to \mu\mu$ events without genuine missing transverse energy, where any detected \MET is due to the limited acceptance of the detector, or due to limited resolution in the detection of other objects.
$W \to e\nu$ and $W \to \mu\nu$ final states are used to calibrate \MET with genuine contributions from undetected neutrinos.

\paragraph{overlap removal.}
To avoid the same calorimeter energy deposits or the same tracks to be associated to multiple objects, an overlap-removal procedure is applied among the object candidates.
After compiling lists of candidates, the following procedures are applied sequentially:
if a muon candidate is reconstructed including calorimeter information and if that muon shares a track with an electron candidate, the muon candidate is removed.
Afterwards, electron candidates that share tracks with muons are removed.
To avoid ambiguities between jets and electrons, jet candidates within a cone of $\Delta R = 0.2$ in the \etaphi plane around electrons are removed.
Any electron candidates that subsequently remain within $\Delta R = 0.4$ of a jet are removed.
If a jet candidate is within a cone of $\Delta R = 0.2$ to a muon and has only two or fewer associated tracks, the jet is removed.
On the other hand, muon candidates are removed if they are closer than $\Delta R = 0.4$ to a jet and if that jet has more than two associated tracks.
Finally, photon candidates in the vicinity of electrons and muons are removed by imposing $\Delta R (\gamma, \ell) \geq 0.4$, which also reduces the fraction of photons radiated by charged leptons.
Jet candidates within a cone of $\Delta R = 0.4$ around the remaining photons are removed.



\chapter{Photon identification with neural networks}
\label{chap:PPT}

\vspace*{0pt plus 5pt}  

The identification of prompt photons, \ie of those that come from the hard interaction, is pivotal for every hadron-collider analysis with photons in the final state.
Photon identification in \ATLAS, introduced in the previous chapter, is based on working points that impose one-dimensional cuts on parameters of the lateral and longitudinal shower evolution of the photon candidate.
These are also the criteria used for identifying photons in the later chapters of this thesis and in Ref.~\cite{TOPQ-2020-03}.
However, over the course of this thesis, studies were performed to improve photon identification with the use of machine-learning techniques, in particular, of neural networks.
The idea is simple: while identification criteria with one-dimensional cuts only take advantage of each observable's power to discriminate prompt photons from other detector activity, neural networks are multivariate analysis tools that exploit correlations and non-linear relations between these observables.

In a collaborative effort of the G{\"o}ttingen \ATLAS group, the \emph{Prompt Photon Tagger} (\PPT) was developed, a generic tool to distinguish prompt photons from hadronic activity that fakes photon signatures in the detector (called \emph{hadron-fake} photons).
The tool was applied to \tty events in the \ljets channels and results using the \PPT were published in Ref.~\cite{TOPQ-2017-14}.
Large contributions to the tool were made by B.\,V{\"o}lkel in his MSc thesis project~\cite{Volkel:2017aa}, in particular to the development of the architecture and training of the neural network.
The application to \tty events and the estimation of \tty-specific uncertainties were done in close collaboration with \JWSmith who has shown studies and results using the \PPT in his PhD thesis~\cite{Smith:2018sma}.
The following paragraphs introduce briefly some of the necessary concepts of machine learning for constructing a neural network.
Then, the architecture and training procedure of the \PPT are summarised.
Afterwards, the application of the \PPT to \tty events is detailed and some of the results obtained in Ref.~\cite{TOPQ-2017-14} with the \PPT are shown.

\vspace*{0pt plus 3pt}  

\paragraph{neural networks.}
In the broadest sense of the word, machine learning is the \enquote{field of study that gives computers the ability to learn without being explicitly programmed} (A.\,Samuel, 1959)%
\footnote{%
  This quote is often attributed to A.\,Samuel including a citation of his 1959 paper~\cite{Samuel1959:aa}.
  However, the reference does not actually contain this quote.
  Nonetheless, it still holds an appropriate definition of the term machine learning and may be understood as a gist of Samuel's paper.
}.
From an engineering perspective, a machine given a certain task would learn from experiences by maximising a provided measure of its performance.
In the example of a machine-learning tool charged with the task to identify prompt photons, this tool would use \emph{labelled} data in its training process and would try to maximise its classification performance into true positives and true negatives, keeping type-I and type-II errors as little as possible.
Labelled data in this case refers to data points which carry information about the true origin of the photon to enable evaluation of the predicted classification labels.
Often associated with neural networks, machine learning comprises all sorts of models, ranging from support-vector machines to decision-tree forests.

\begin{figure}
  \centering
  \includegraphics[scale=1.2]{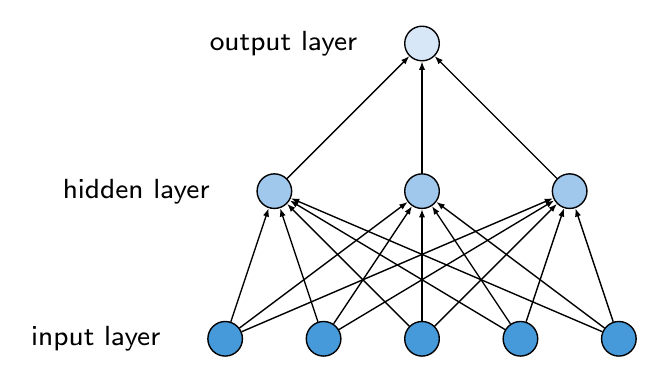}
  \caption[Representative network structure of a multi-layer perceptron]{%
    Representative network structure of a multi-layer perceptron.
    The information of such a feed-forward network flows in one direction only: from the input to the output neurons.
    The hidden layer is densely connected with input and output.
  }
  \label{fig:PPT-perceptron}
\end{figure}

Historically, it seemed logical to seek inspiration from the structure of biological brains to build \emph{artificial} neural networks and to create \enquote{intelligent} machines.
The idea of artificial neural networks is not new at all, but has been around since as early as 1943 when the first landmark paper was published~\cite{Culloch1943:aa}.
The most straight-forward design for a neural network is a composition of multiple layers of \emph{neurons}, and the neurons of each layer are connected with those of the next.
The first of such models, the \emph{perceptron}, was proposed by \citeauthor{Rosenblat1958:aa} in 1957~\cite{Rosenblat1958:aa}.
An example of a perceptron structure with one hidden layer between input and output is depicted in \cref{fig:PPT-perceptron}.
With no interconnections between neurons of the same layer, the design is a \emph{feed-forward} neural network as it does not contain any recurrent connections, such as cycles or loops.
The learning process is then a strengthening or weakening of the neurons' connections following the \emph{Debian} learning strategy:
when a neuron triggers another neuron, the connection between the two is strengthened~\cite{Hebb1949:aa}.

Developments in the field of neural networks came in three waves~\cite{Goodfellow2016:aa}:
after the initial and secondary waves in the 1940s to 1960s, and 1980s and 1990s (known as connectionism), respectively, most researchers had abandoned the idea of \enquote{deep} neural networks%
\footnote{%
  The attribute \enquote{deep} for neural networks is -- in many circumstances -- just a matter of preference, but has been used widely throughout the community.
  In most cases, \enquote{deep} refers to any neural network with more than one hidden layer between input and output.
}
because they were considered untrainable.
The third wave hit the machine-learning community in 2006 when \citeauthor{Hinton2006:aa}~\cite{Hinton2006:aa} successfully trained a deep neural network to perform handwritten digit classification.
They identified some of the problems with training deep networks.
This and many other soon-following developments, \eg Refs.~\cite{Bengio2006:aa,Ranzato2006:aa}, revealed the potential of \emph{deep learning}, a term coined in this third wave.
Since then, especially with alleviating the \emph{problem of vanishing gradients}~\cite{Glorot2010:aa}, deep learning has developed at an incredible pace and has conquered many areas of industry and research, including particle physics.
Comprehensive overviews of machine learning in particle physics are available \eg in Ref.~\cite{Guest:2018yhq}.

Going back to a tool given a task to identify prompt photons, a modern deep neural network would consist of multiple densely-connected, hidden layers with a multitude of nodes (\ie neurons) each.
Charged with classifying data instances into classes $A$ and $B$, such networks consist of one single output node, the value of which reflects a confidence/probability of a data instance to be class $A$.
Each of the node connections is attributed with a weight, which are the trainable parameters of the model, \ie the parameters the model \enquote{learns} from data.
As a performance measure, the model is assigned a \emph{loss} function.
Most neural networks for binary classification use the \emph{binary cross-entropy}
\begin{align}
  \label{eq:ppt-binary-crossent}
  \operatorname{Loss}(\vec{\theta}) = - \frac{1}{m} \sum_{i=1}^m \left[ y_i \log \estimate{p_i}(\vec{\theta}) + ( 1- y_i) \log ( 1 - \estimate{p_i}(\vec{\theta})) \right] \, ,
\end{align}
where $m$ is the number of instances used for a single training step (the \emph{batch size}), and $y_i$ is the \emph{target value} of instance $i$, that is, which class the instance belongs to:
$y_i = 1$ if the instance is (truly) class $A$, 0 otherwise.
$\estimate{p_i}(\theta)$ is the estimated probability for instance $i$ to be class $A$, and it is a function of the model parameters $\vec{\theta}$.
The higher the estimated probabilities $\estimate{p_i}$ for instances with $y_i = 1$, and the lower for instances with $y_i = 0$, the smaller the cross-entropy, thus, improving performance of the binary classification.

After each training cycle, the value of the loss function is calculated.
Then, the connection weights are updated through a process known as back-propagation~\cite{Rumelhart:1986aa,LeCun1987:aa}:
going back layer by layer from output to input, the gradient of the network's loss value is calculated with respect to all its trainable parameters $\vec{\theta}$.
Depending on the size of the gradient at each connection, the weights are adjusted to reduce the loss function value.
The gradient calculation relies on monotonic and, at best, continuously differentiable outputs of the nodes to maximise the efficiency of the training process.
Therefore, the raw output scores of the nodes are modified through \emph{activation functions} before propagated to the next layer.
Using non-linear activation functions enables the network to develop non-trivial relations between the nodes.
Popular non-linear activation functions are
\begin{subequations}
\begin{align}
  \label{eq:ppt-activation-functions}
  \text{ReLU (rectified linear unit):} \quad g(z)_i &= \max(0, z_i) \\
  \text{Sigmoid:} \quad g(z)_i &= \frac{1}{1 + \exp(-z_i)} \\
  \text{Softmax:} \quad g(z)_i &=  \frac{\exp(z_i)}{\sum_j \exp(z_j)}
\end{align}
\end{subequations}
where $z_i$ is the raw output score of node $i$.
Note that ReLU is not differentiable at $z=0$, but is nonetheless popular due to its simplicity.
The softmax function is different than the others as it considers the scores of all output nodes of the same layer.
The softmax outputs fulfil $\sum_i g(z)_i = 1$ and, hence, enable a probability interpretation even with multiple output nodes.
It is used frequently for the output of a multiclass classifier to give class probabilities.
For binary classification, where one single output node is sufficient, other saturating activation functions, such as sigmoid, allow probability interpretations as well.

\vspace{0pt plus 3pt}  

\begin{figure}
  \centering
  \includegraphics[width=0.48\textwidth,clip,trim=0 20pt 5pt 15pt]{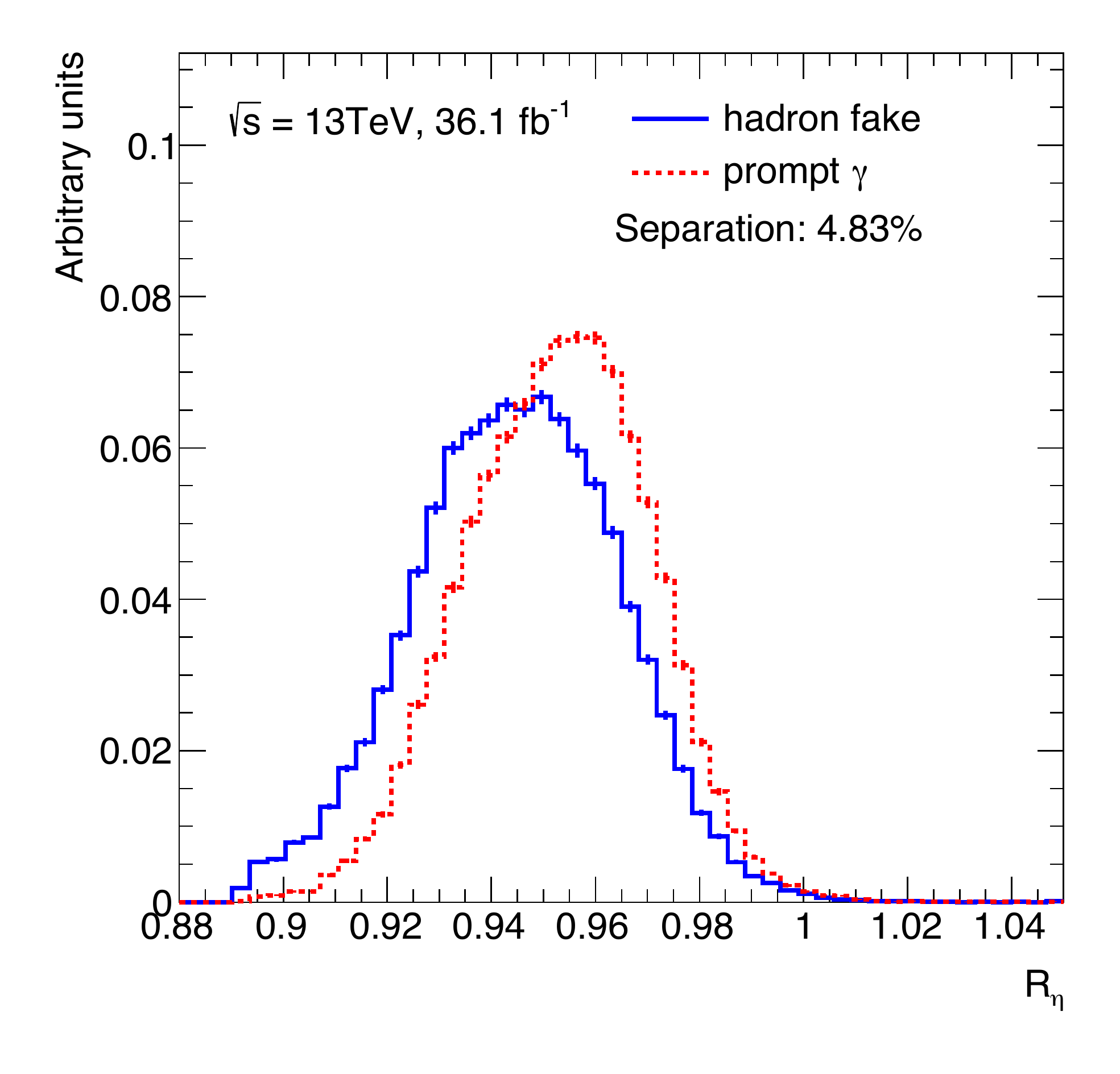}
  \includegraphics[width=0.48\textwidth,clip,trim=0 20pt 5pt 15pt]{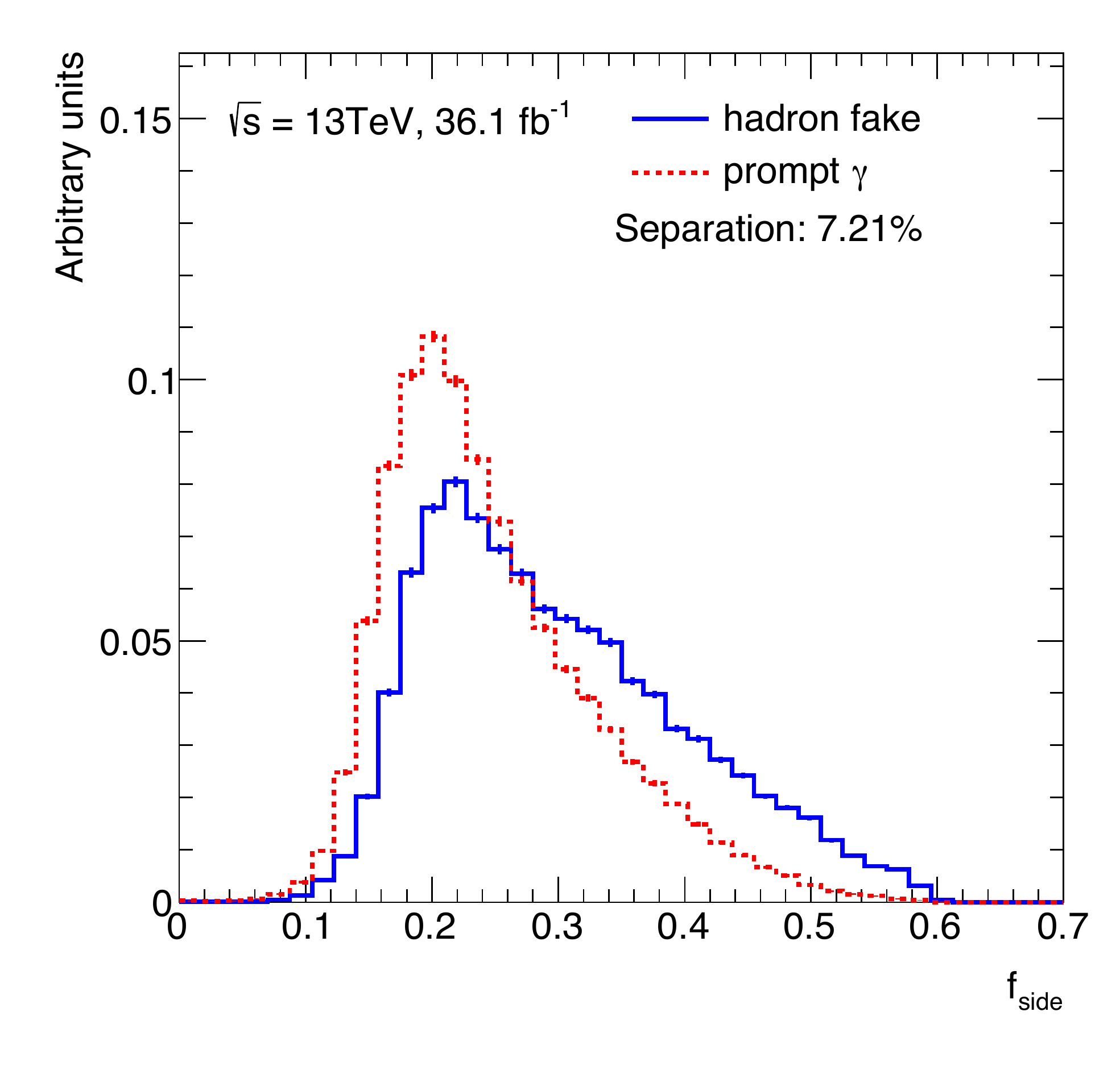}
  \caption[Distributions of two representative shower-shape variables]{%
    Distributions of two representative shower-shape variables: the energy ratio of $3\times7$ to $7\times7$ cells in $\eta\times\phi$ coordinates in the second \ECAL layer, $R_\eta$, and the energy outside the 3 central strips, but within 7 strips in the first \ECAL layer, $f_{\text{side}}$.
    The plots compare the predicted distributions for prompt photons and hadron-fake photons.
  }
  \label{fig:ppt-shower-shapes}
\end{figure}

\paragraph{ppt architecture.}
The \PPT is designed as a feed-forward neural network to perform binary classification of prompt photons and hadron-fake photons.
The input variables are chosen to be a subset of the shower-shape variables also used for the photon identification working points in \ATLAS, as described in \cref{sec:exp_objects}.
Detailed definitions of all considered shower-shape variables are given in \cref{tab:ppt-showershapes}.
Many of them are either based on energy ratios or shower widths in the \ECAL system, but others also consider hadronic leakage of photon candidates into the \HCAL.
Example distributions of two shower-shape variables are shown in \cref{fig:ppt-shower-shapes}.
The left-hand plot shows $R_\eta$, the ratio of $3\times7$ to $7\times7$ cells in $\eta\times\phi$ coordinates in the second layer of the \ECAL.
The right-hand plot shows $f_{\text{side}}$, the energy outside the inner 3, but within 7 strips in the first \ECAL layer.

\begin{table}
	\centering
	\caption[Definition of photon shower-shape variables in \ATLAS]{%
    Definition of the shower-shape variables used for the cut-based photon identification in \ATLAS and considered as an input for the \PPT, \cf also \cref{fig:exp-photon-shower-shape}.
  }
	\label{tab:ppt-showershapes}
  \renewcommand{\arraystretch}{1.2}
  \newcommand{\centeredparbox}[1]{\parbox[c]{\hsize}{\vspace*{.3em}#1\vspace*{.3em}}}
	\begin{tabular}{c p{0.78\textwidth}}
    \toprule
		\multicolumn{2}{l}{Hadronic leakage}\\
    \midrule
		$\begin{matrix} R_{\text{had}} \\ \text{or}~R_{\text{had1}} \end{matrix}$ & \centeredparbox{Transverse energy leakage in the \HCAL normalised to $\ET(\gamma)$ in the \ECAL. In the region $0.8\leq |\eta|\leq 1.37$, the entire energy of the photon candidate in the \HCAL is used ($R_{\text{had}}$), while in the region $|\eta| < 0.8$ and $|\eta| > 1.37$ the energy of the first layer of the \HCAL is used ($R_{\text{had1}}$).}\\
    \midrule
		\multicolumn{2}{l}{Energy ratios and width in the second layer of \ECAL}\\
    \midrule
		$R_\eta$ & Energy ratio of $3\times 7$ to $7\times 7$ cells in $\eta\times\phi$ coordinates.\\
		$R_\phi$ & Energy ratio of $3\times 3$ to $3\times 7$ cells in $\eta\times\phi$ coordinates.\\
		$w_{\eta 2}$ & Lateral width of the shower, using a window of $3\times 5$ cells. \\
    \midrule
		\multicolumn{2}{l}{Energy ratios and widths in the first (strip) layer of \ECAL}\\
    \midrule
		$w_{s\,3}$ & Shower width along $\eta$, using $3 \times 2$ strips around the largest energy deposit.\\
		$w_{\text{s\,tot}}$ & Shower width along $\eta$, using $20 \times 2$ strips around the largest energy deposit.\\
		$f_{\text{side}}$ & \centeredparbox{Energy outside the 3 central strips but within 7 strips, normalised to the energy within the 3 central strips.}\\
		$E_{\text{ratio}}$ & \centeredparbox{Ratio between difference of the first and second energy maximum divided by their sum ($E_{\text{ratio}}=1$ if there is no second maximum).}\\
		$\Delta E$ & \centeredparbox{Difference between the second energy maximum and the minimum found between first and second maximum ($\Delta E = 0$ if there is no second maximum).}\\
    \bottomrule
	\end{tabular}
\end{table}

The two plots show the predicted distributions for prompt photons in dashed red, and those for hadron-fake photons in solid blue.
The two were created with two dedicated \MC simulations that are also used within \ATLAS to calibrate the cut-based photon identification working points.
The first \MC sample simulates \QCD Compton processes and contains prompt photons only.
The other is a dijet sample with hadron-fake photons.
For the histograms in \cref{fig:ppt-shower-shapes} and for the training of the \PPT, the photon candidates are required to pass the same kinematic requirements as those listed in \cref{sec:exp_objects}, but no identification or isolation criteria are imposed.
Based on comparisons between prompt photons and hadron-fake photons such as those shown in \cref{fig:ppt-shower-shapes}, the input variables for the \PPT were chosen.
As a figure of merit, the \emph{separation power} of each of these shower-shape variables was calculated based on the distributions obtained from the two \MC samples.
The separation is defined as
\begin{align}
  \label{eq:ppt-separation-power}
  S = \frac{1}{2} \sum_{i \in \text{bins}} \frac{ (s_i - b_i)^2 }{ s_i + b_i}
\end{align}
where $s_i$ and $b_i$ are the numbers of signal-like and background-like photons in bin~$i$, respectively.
Six shower-shape variables were chosen as input for the \PPT, listed in \cref{tab:ppt-separation} with their separation powers.

The neural network of the \PPT is constructed using the \textsc{keras} library~\cite{keras}, with the implementation back-end provided by \textsc{tensorflow}~\cite{tensorflow}.
It is injected into the analysis software using the \textsc{lwtnn} library~\cite{dguest2017:lwtnn}.
To determine the best architecture of the network, given the six input variables, grid searches were performed in the hyper-parameter space.
The final architecture of the \PPT uses an input layer with six nodes for the input observables, three densely-connected hidden layers with 64, 40 and 52 nodes, respectively, and a single-node output layer.
The first of the three hidden layers uses ReLU activation functions, while the other two use softmax.
The output layer uses a sigmoid activation function.
Between the first and second hidden layer, and between the second and third hidden layer, batch normalisation~\cite{Ioffe2015:batch} is implemented to enable higher learning rates and to overcome problems with vanishing gradients in the nodes.
The \emph{Adam} optimiser~\cite{kingma2014adam} is used to speed up the overall training process.
The training uses the two dedicated \MC simulations described above and applies no further event selection apart from the photon kinematic requirements.
The datasets, about one million events with prompt photons and two hundred thousand events with hadron-fake photons, are split randomly into train and test sets of \SI{80}{\percent} and \SI{20}{\percent} size, respectively.
The class of events with hadron-fake photons is reweighted to match the number of training instances in the prompt-photon class.
The training then evaluates mini-batches of \num{10000} events for each training step.
It is stopped after \num{300} epochs, \ie when the entire set of instances has been seen by the model \num{300} times.

\begin{table}
	\centering
	\caption[Separation powers of the \PPT input variables]{%
    Separation powers of those shower-shape variables included into the \PPT.
    The six variables with the highest separation values were chosen from \cref{tab:ppt-showershapes}.
    }
	\label{tab:ppt-separation}
	\begin{tabular}{l S}
  \toprule
	variable & {Separation value} \\
  \midrule
	$R_{\text{had}}$  & \SI{3.33}{\percent} \\
	$R_\eta$          & \SI{4.83}{\percent} \\
	$R_\phi$          & \SI{7.01}{\percent} \\
	$w_{\eta 2}$      & \SI{2.01}{\percent} \\
	$w_{\eta 1}$      & \SI{4.14}{\percent} \\
	$f_{\text{side}}$ & \SI{7.21}{\percent} \\
  \bottomrule
	\end{tabular}
\end{table}

\begin{figure}
  \centering
  \includegraphics[width=0.46\textwidth,clip,trim=0 0 30pt 30pt]{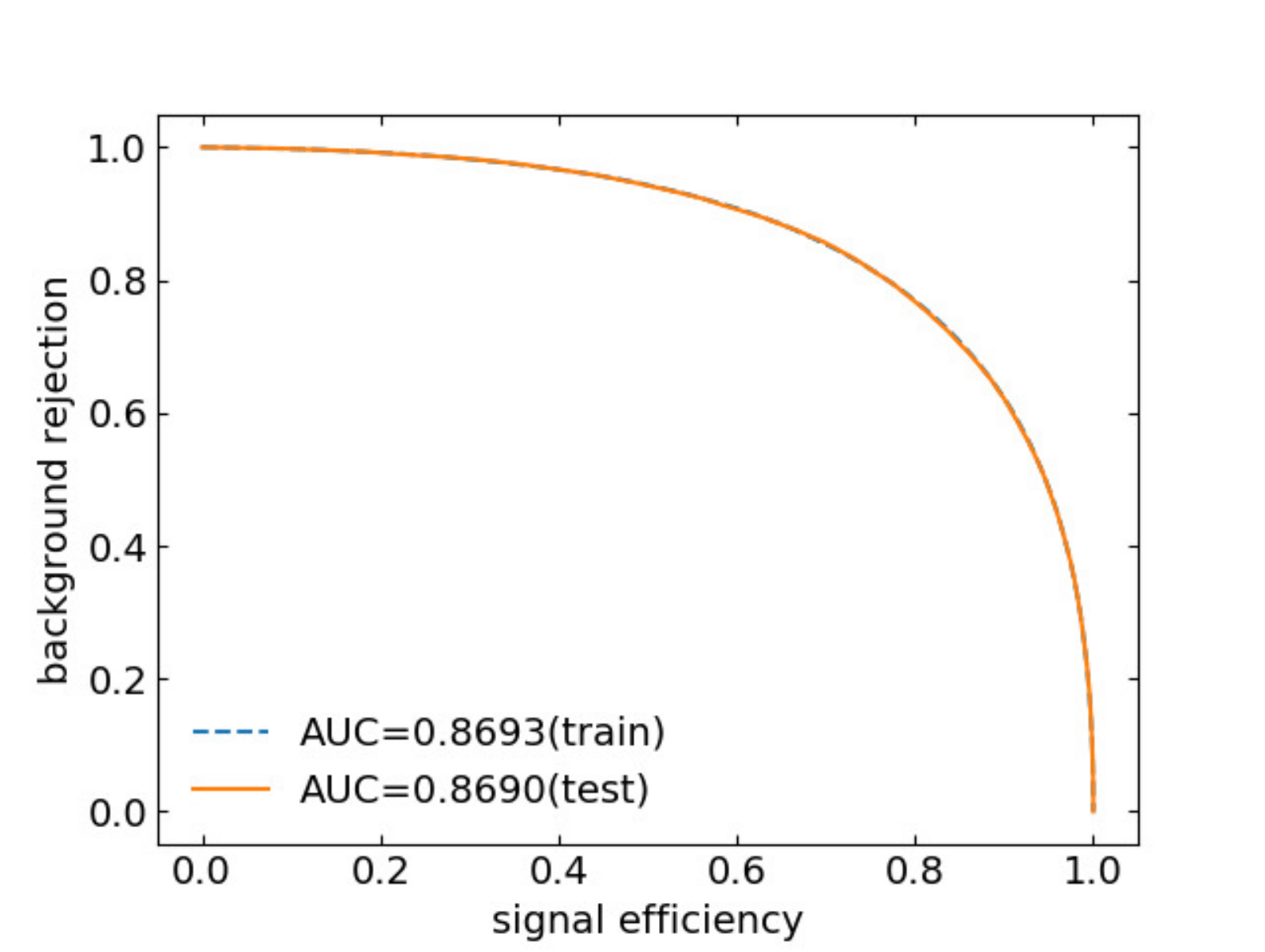}
  \includegraphics[width=0.46\textwidth,clip,trim=0 0 30pt 30pt]{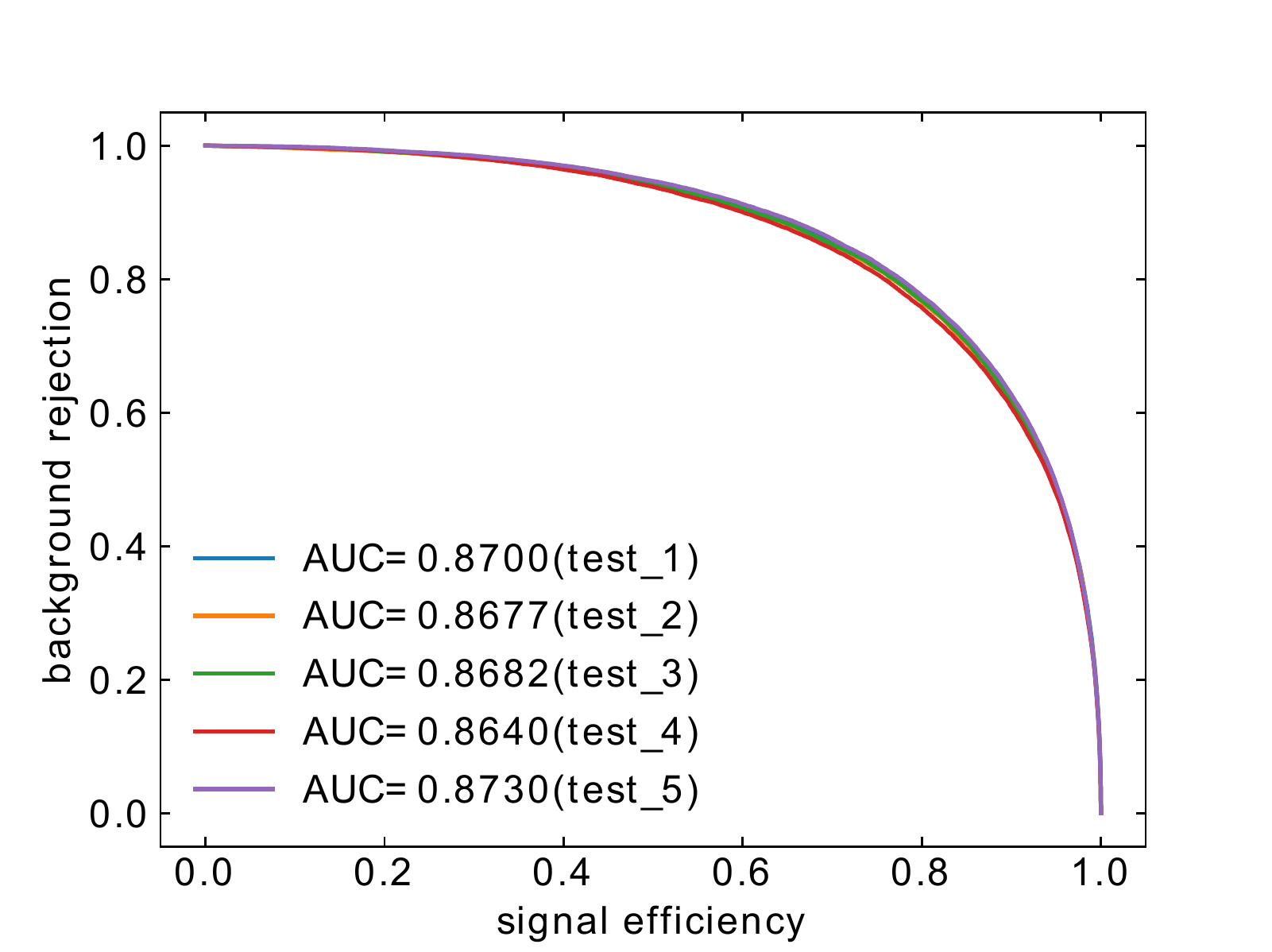}
  \caption[\ROC curves and 5-fold cross validation results of the \PPT]{%
    \ROC curves for train and test sets on the left, which show little difference.
    \ROC curves of the five test sets of the 5-fold cross validation on the right.}
  \label{fig:ppt-ROC-curves}
\end{figure}

The performance of the model is evaluated using the receiver-operator-characteristic (\ROC) curves:
in this two-dimensional graph, the signal efficiency is plotted against the background rejection for all output values of the binary classifier.
The obtained \ROC curves for train and test sets are shown on the left-hand side in \cref{fig:ppt-ROC-curves}.
The larger the area under the \ROC curve, the higher the signal efficiency and background rejection.
The trained network reaches area-under-curve values of \num{0.8693} and \num{0.8690} for train and test sets, respectively.
With little difference between the two, the model shows no sign of overtraining.
As an additional test of the stability of the network, a 5-fold cross validation is performed:
the input data is split into five equal sets, and for each possible combination among the five, four sets are used for training, whereas the fifth serves as a test set.
The resulting five models show almost no performance differences -- as seen in the \ROC curves of their test sets in \cref{fig:ppt-ROC-curves} on the right-hand side.

\begin{figure}
  \centering
  \includegraphics[width=0.48\textwidth,clip,trim=0 0 0 20pt]{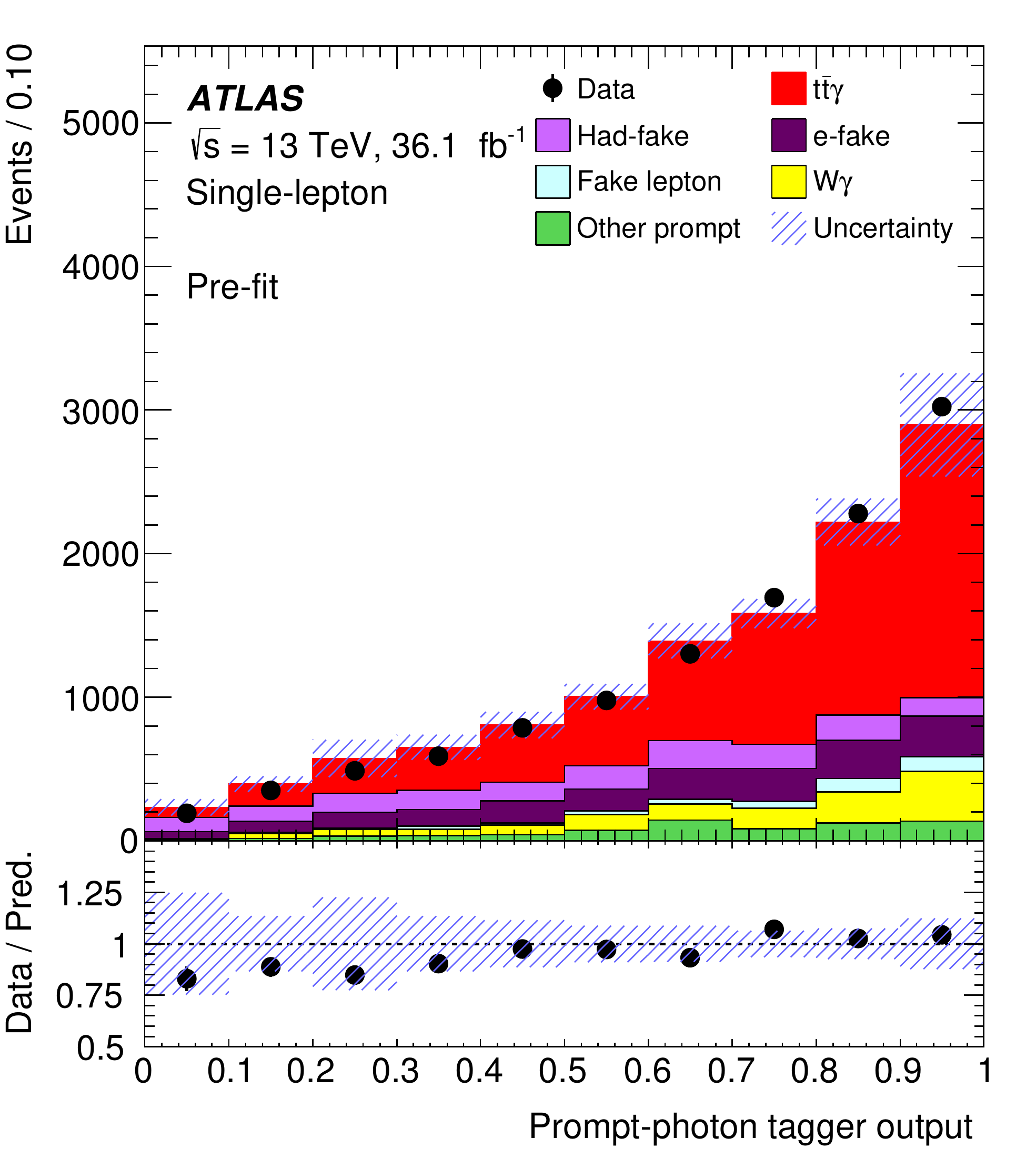}
  \caption[Control plot of the \PPT output in signal region]{%
    Control plot of the \PPT output in the combined \ejets and \mujets signal region.
    The classification into prompt and fake photons is done similarly to what is described in \cref{sec:simulation-categorisation}.
    With \PPT-specific scale factors applied to the predictions, good agreement between data and \MC simulation is observed.
    The hatched uncertainty bands include combined statistical and systematic uncertainties including systematics assigned to the \PPT.
    Figure taken from Ref.~\cite{TOPQ-2017-14}.
  }
  \label{fig:ppt-controlplot}
\end{figure}

\paragraph{application to analysis.}
With input variables purely related to the shower shapes of photon candidates, the \PPT is a generic tool to perform binary classification of those candidates into prompt and hadron-fake photons.
\tty measurements, in particular in the \ljets final states, where a significant background contribution from processes containing hadron-fake photons is expected, benefit from such a classification tool.
The \ATLAS \tty analysis using \SI{36}{\ifb} of \runii data~\cite{TOPQ-2017-14} used the \PPT to increase separation between prompt photon candidates and hadron-fake photons.
In the combined \ejets and \mujets channels, the analysis predicts about \SI{12}{\percent} contribution from hadron-fake photons.
\Cref{fig:ppt-controlplot} shows the distribution of the \PPT output in the combined signal region.
The \tty signal category and those background categories with prompt photons, such as \Wy, show a strong slope towards the right-hand side of the distribution, whereas the \emph{Had-fake} category, the category with hadron-fake photons, is distributed equally over the shown spectrum.
The categorisation into events with prompt or fake photons uses \MC-truth information and is done similarly to what is described in this thesis in \cref{sec:simulation-categorisation}.

\begin{figure}
  \centering
  \includegraphics[width=0.48\textwidth,clip,trim=0 10pt 0 10pt]{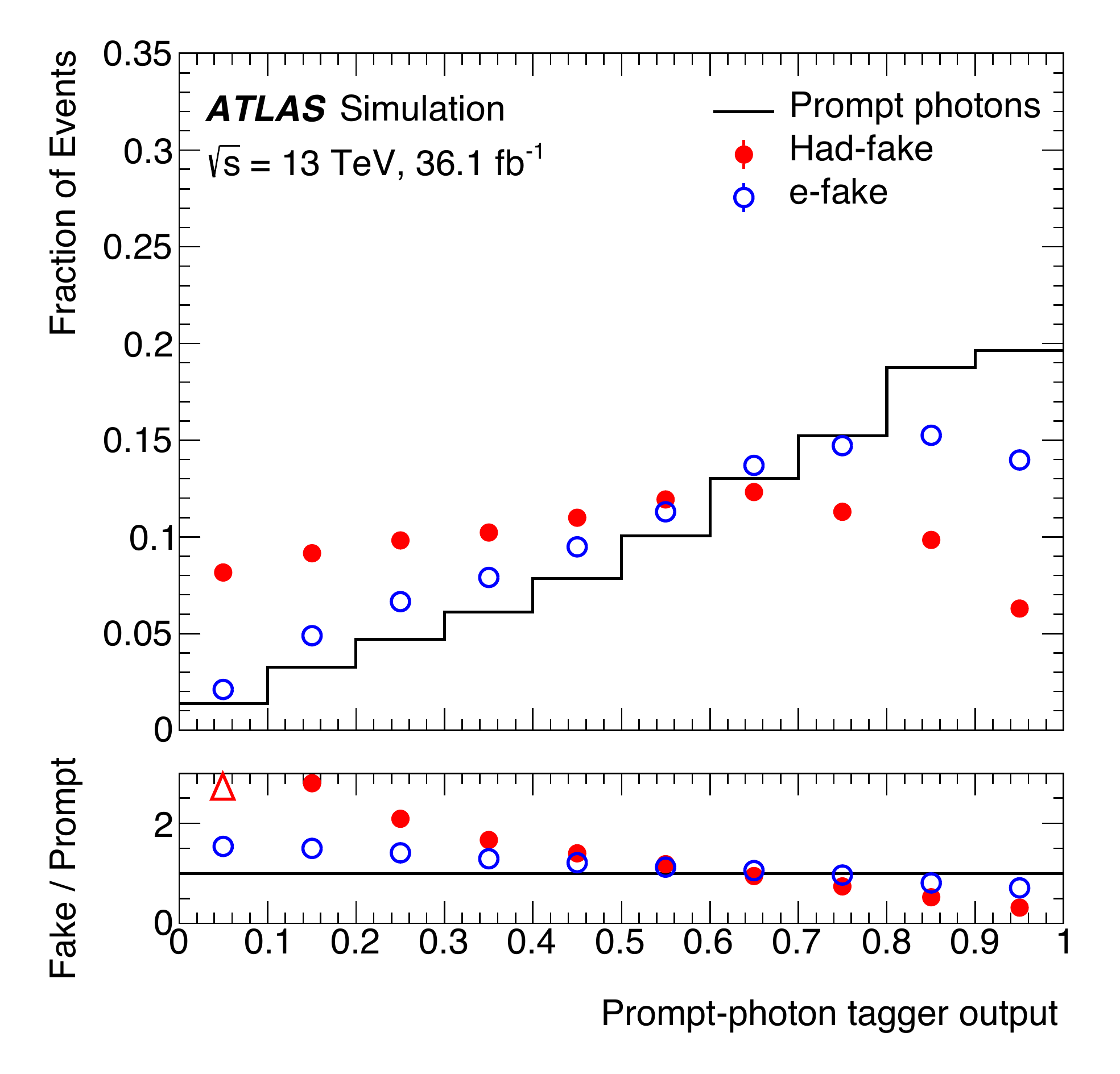}
  \caption[Response of the \PPT to different photon types]{%
    Output of the \PPT for three different event categories used in the \ljets signal regions: prompt photons, hadron-fake and electron-fake photons.
    The categorisation is done similarly to what is described in \cref{sec:simulation-categorisation}.
    Figure taken from Ref.~\cite{TOPQ-2017-14}.
  }
  \vspace*{3pt}  
  \label{fig:ppt-response}
\end{figure}

For a better comparison of the \PPT response to prompt photons and fake activity, \cref{fig:ppt-response} shows the \PPT response for prompt photons, hadron-fake photons and electron-fake photons.
The latter are electrons that fake photon signatures in the \ATLAS calorimeters.
To allow a direct comparison of the shapes, each distribution is normalised to unity.
The ratio at the bottom of the figure confirms separation power of the \PPT output for prompt and hadron-fake photons.
For electron-fake photons, that were not considered in the training of the tool, little separation power is observed to the prompt category as the neural network did not \enquote{learn} to pick up any differences between these two photon types.

Although named \emph{tagger}, which usually implies that the classification tool makes a class prediction based on a decision boundary in its output distribution, the \PPT was not used as such in the \SI{36}{\ifb} analysis.
Instead, the continuous distribution, as shown in \cref{fig:ppt-controlplot}, was used as an input variable into another neural network that was trained at event level.
This event-level binary classifier was then used to obtain maximum separation between the \tty signal and all background contributions and to extract a fiducial \xsec of the \tty process.
Among the input variables to the event-level classifier, the \PPT showed the highest separation power between \tty events and events with hadron-fake photons.

\begin{figure}
  \centering
  \includegraphics[width=\textwidth,clip,trim=0 12pt 0 6pt]{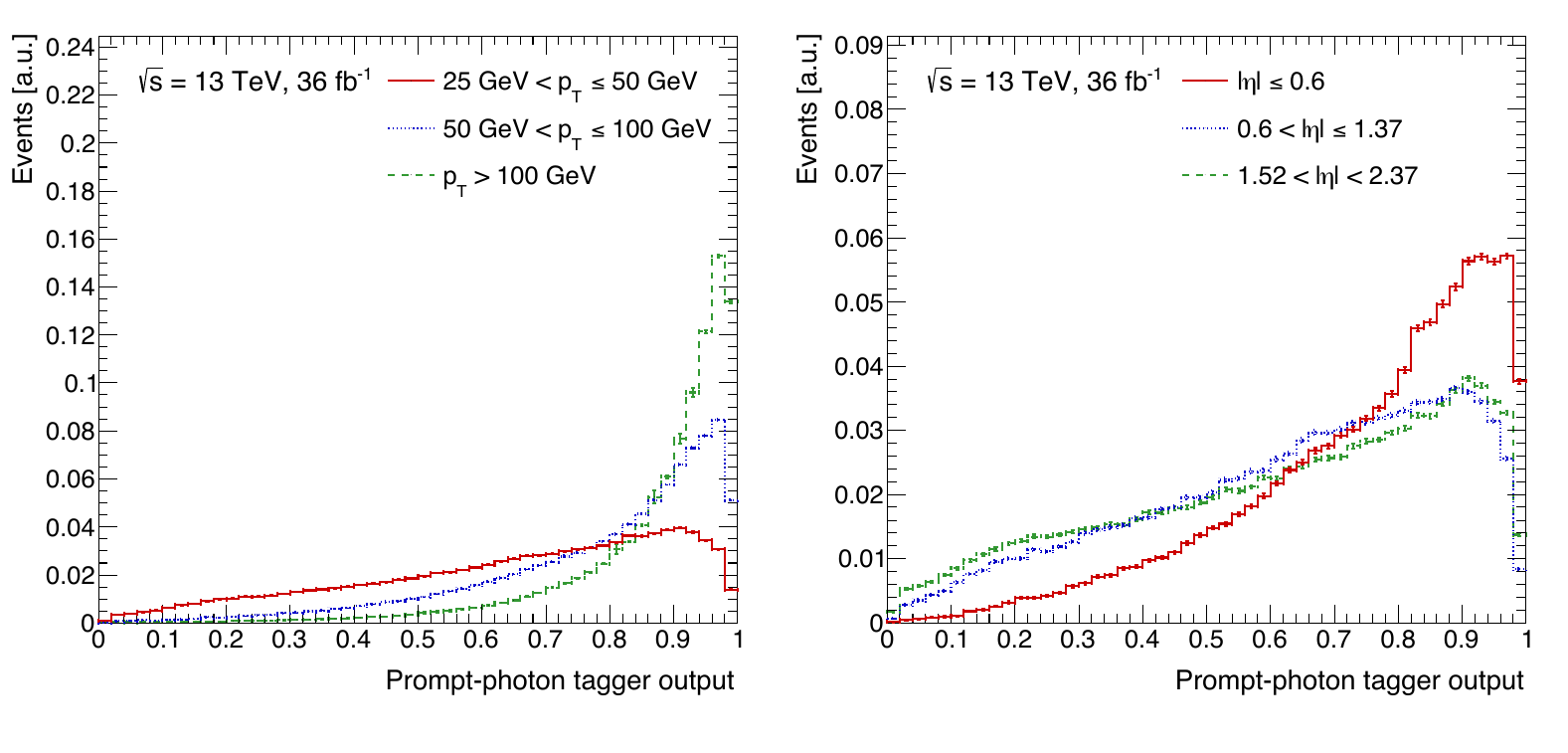}
  \caption[Dependence of the \PPT on photon kinematics]{%
    Dependence of the \PPT output on the transverse momentum and pseudorapidity of the photons used in the training.
    The plots show the \PPT response to prompt photons from the training dataset in three different bins of \pT and \abseta.
  }
  \label{fig:PPT-pT-eta}
  \vspace*{3pt}  
\end{figure}

As the \PPT was purely trained on \MC simulation, it relies heavily on the correct simulation of the lateral and longitudinal photon shower development to also perform well on \ATLAS data.
This is of particular importance given the observed discrepancies between data and simulation in the lateral shower-shape variables that are mitigated by fudge factors, as detailed in \cref{sec:exp_objects}.
In addition, although identical in their shower development for given kinematics, the kinematic distributions of photons in a \tty analysis are different from those used in the training of the \PPT.
As \cref{fig:PPT-pT-eta} shows, the \PPT responds differently to high-\pT photons than to those with low transverse momentum, and the \PPT spectrum of central photons is different from that observed for photons with high \abseta.
To account for all these effects, simulation-to-data scale factors are derived in three control regions of the \tty analysis to correct the \PPT distributions in \MC simulation to those observed in data.
One region is dedicated to estimating mismodelling of prompt photons, the other two account for discrepancies between simulation and data for hadron-fake photons.

A dedicated prompt-photon control region enriched in $Z \to \ell\ell\gamma$ events is defined by requiring two charged leptons of the same flavour, but opposite charge, and their invariant mass must be between \SIlist{60;100}{\GeV}.
No requirements on the number of jets are imposed.
The photon candidates must pass both the identification and isolation criteria of the signal region.
The event selections of the two hadron-fake control regions are similar to those of the \ljets signal region:
exactly one photon, at least four jets, one of which must be \btagged with the \SI{77}{\percent} operating point of the \MVtwo tagger, and either one muon or one electron.
To make an event selection that is orthogonal to that of the \ljets signal region, photons in this control region must \emph{fail} the isolation requirements.
In addition, $\ptcone{20}(\gamma) > \SI{3}{\GeV}$ is required to further remove contamination from prompt photons.
As only shower-shape variables enter the \PPT that are also used for the photon identification working points in \ATLAS, and as those shower-shape variables are largely uncorrelated with photon isolation criteria~\cite{PERF-2017-02,EGAM-2018-01}, little absolute correlation between the failed isolation criteria and the \PPT is expected.
However, as the isolation-fail photons and photons in the signal region might have different kinematic distributions, a second hadron-fake control region is derived from the selection criteria of the \ljets signal region, where the photons must pass the isolation criteria, but \emph{fail} the identification requirements.

The simulation-to-data scale factors from the control regions are then derived as follows:
the regions are split in slices of photon \abseta and transverse momentum to account for differences in the \PPT response to different kinematics.
In each slice, the total predicted events in \MC simulation are scaled to match the event yields in data.
Then, the observed slopes between data and simulation are extracted.
Due to a remaining \tty signal contamination of approximately \SI{7}{\percent}, the isolation-fail hadron-fake control region varies the \tty signal by \SI{\pm 50}{\percent}, and the scenario with the largest discrepancy between simulation and data is chosen.
The extracted slopes from the prompt-photon control region and the isolation-fail hadron-fake control region are applied as correction factors to those categories with prompt photons and hadron-fake photons in the signal region, respectively.
That is, the scale factors for hadron-fake photons are applied to the \emph{Had-fake} category of \cref{fig:ppt-controlplot}, and the prompt-photon scale factors to the signal category and to all background categories with prompt photons.
To estimate uncertainties conservatively on these two sets of simulation-to-data scale factors, they are switched on and off separately and the resulting effects are symmetrised and used as uncertainties on the scale factors.
In addition, the extracted slope from the identification-fail hadron-fake control region is applied as an uncertainty on the hadron-fake simulation-to-data scale factors.

\begin{figure}
  \centering
  \includegraphics[width=0.7\textwidth]{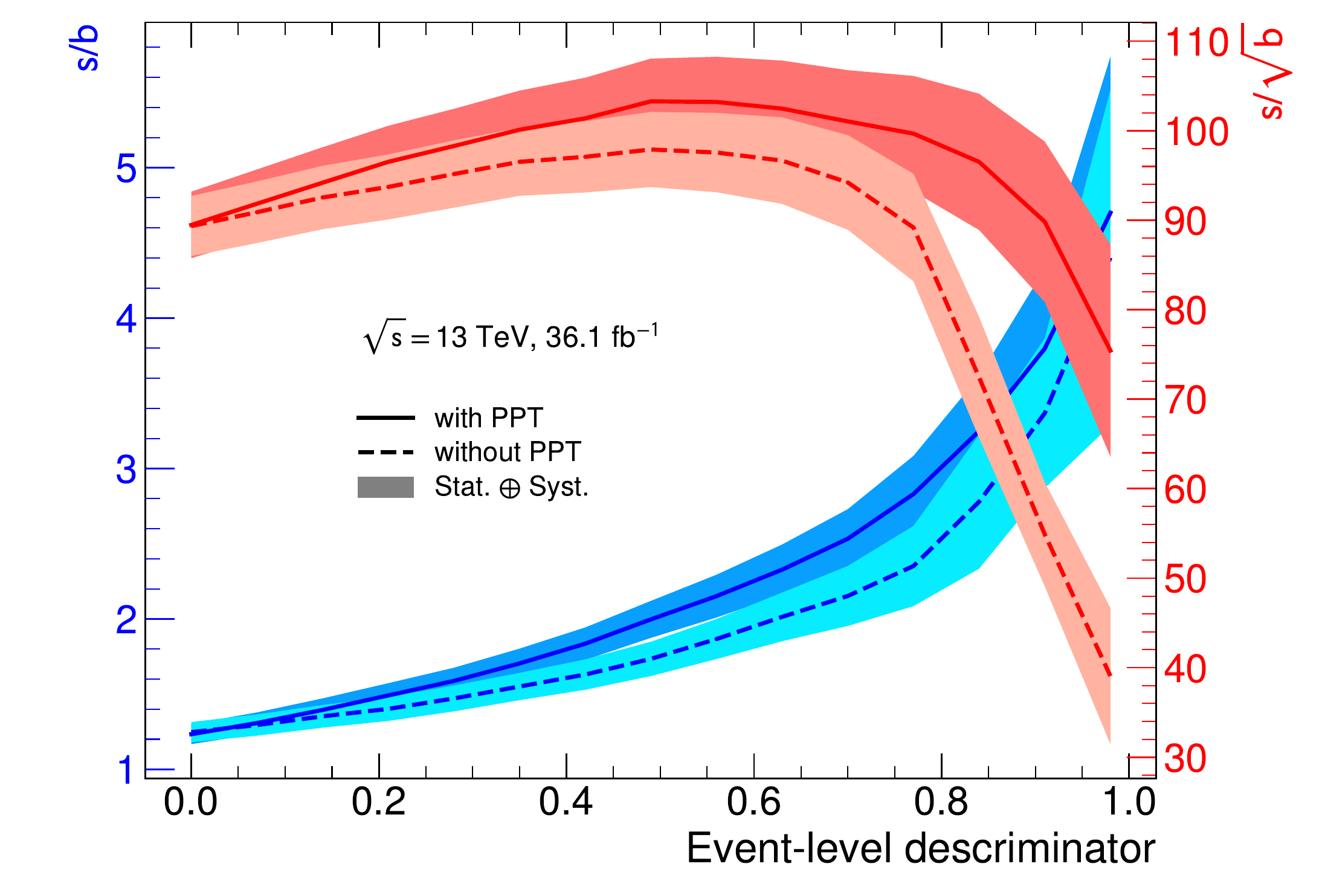}
  \caption[Impact of the \PPT on the analysis results]{%
    Impact of the \PPT on the results of the \SI{36}{\ifb} analysis.
    The plot shows the resulting both signal-over-background ratio and the significance $S/\sqrt{B}$ for different cuts on the event-level classifier that uses the \PPT as an input.
    Figure taken from Ref.~\cite{Smith:2018sma}.
  }
  \label{fig:PPT-impact}
\end{figure}

\Cref{fig:PPT-impact}, taken from Ref.~\cite{Smith:2018sma}, presents a study of the impact of the \PPT on the \tty \xsec measurement performed with \SI{36}{\ifb}.
As figures of merit, the plot shows both the signal-over-background ratio and the significance $S/\sqrt{B}$.
They are plotted as a function of the event-level classifier output, that was once trained with the \PPT and once without the \PPT.
The shown signal-over-background ratios and significances are calculated as if a cut were placed on the event-level classifier at the point of evaluation.
The significance shows strong improvements towards the right-hand side of the distribution when the \PPT is included into the training.
For example, for a cut value of \num{0.9} on the classifier output, the scenario with the \PPT gives about $S/\sqrt{B} \sim 90$, whereas the scenario with no \PPT yields $S/\sqrt{B} < 60$.
On the other hand, including the \PPT introduces additional uncertainties, given as shaded uncertainty bands.
They become more apparent as the cut on the event-level classifier output is tightened.

While the \ljets channels of \tty production deal with significant contributions from processes with hadron-fake photons, the dilepton channels provide a much cleaner environment for measurements.
The \xsec measurement performed with \SI{36}{\ifb} also used an event-level classifier in these channels, although with a focus on separating the \tty signal from prompt-photon backgrounds, the most dominant source of background in dilepton final states.
With predicted contributions of only \SI{5}{\percent} from hadron-fake processes, the \PPT was not included into that classifier.
Consequently, it was also not considered for the analysis of the very clean \emu channel presented in the remaining chapters of this thesis -- but it would, however, be of use again for future measurements in the \ljets channels.


\chapter{Simulation of signal and background}
\label{chap:simulation}

Although quality criteria and dedicated selection cuts can be applied to \ATLAS data to increase the fraction of a certain process in the set of selected events, there is no inherent way to know the exact composition of the resulting set of events.
Simulated events from Monte Carlo (\MC) generators are compared with \ATLAS data and are used to estimate this composition and the strength of the signal in \ATLAS data.
This analysis considers \MC simulations of the signal process and of all background processes with significant expected contributions in a \tty event selection in the \emu channel.
After event generation, the response of the \ATLAS detector to these \MC-generated events is simulated with \Geant~\cite{Agostinelli:2003aa}, as available within the \ATLAS simulation infrastructure~\cite{SOFT-2010-01}.
Because simulating the full detector response is computationally expensive, the fast-simulation package \textsc{atlfast-ii} (\atlfast) is used for some of the \MC-generated events, which parameterises hadronic showers in the calorimeters to speed up simulation.

To account for additional proton-proton interactions from the same and neighbouring bunch crossings, known as in-time and out-of-time pile-up, respectively, the hard-scattering events are superimposed with minimum-bias interactions generated with \Pythia~\cite{Sjostrand:2006za,Sjostrand:2014zea}.
These interactions use an \ATLAS set of tuned generator parameters called \emph{A3}~\cite{ATL-PHYS-PUB-2016-017} and the \NNPDFLO \PDF set~\cite{Ball:2012cx}.
They are then reweighted to match the pile-up conditions observed in data.
In practice, three independent sets of events are generated for each \MC simulation and are reweighted to reflect the data-taking conditions in 2015--16, 2017 and 2018, respectively.
The three different classes of \MC simulation are referred to as \MC production \emph{sub-campaigns} and are named \emph{mc16a}, \emph{mc16d} and \emph{mc16e} within the \ATLAS simulation infrastructure.
The samples of the sub-campaigns are produced with different event seeds, and can thus be combined to give a total prediction for the Run~2 data-taking period between 2015 and 2018.

This analysis uses two types of samples to estimate signal and background contributions with final-state photons: samples where final-state photons are generated in the simulation of the hard interaction, and those where photons are not explicitly requested for the final state.
The first are referred to as \emph{dedicated} samples, the latter as samples \emph{inclusive in photons}.
Dedicated samples with photons were generated for the signal processes, described in detail in \cref{sec:simulation-dedicated}.
Dedicated samples were also produced for \plusjets{\Vy} final states, where $V$ refers to both \Wbosons and \Zbosons.
These and all other photon-inclusive background samples are summarised in \cref{sec:simulation-background}.
Although not simulated in the matrix-element generation step, photon-inclusive samples might still contain photons from the hard interaction as photon radiation is also simulated by the showering algorithms, irrespective of the generated matrix element.
This causes problems if a corresponding dedicated sample with photons from the matrix element is used simultaneously.
To avoid possible double-counting of events with photons, an overlap-removal procedure is applied between dedicated and photon-inclusive samples, as detailed in \cref{sec:simulation-overlap}.

To distinguish photons from the hard-scattering event, referred to as \emph{prompt} photons, from non-prompt photons and from other objects faking photon signatures, \MC generator information is used to identify the true source of a reconstructed photon candidate.
This is further detailed in \cref{sec:simulation-categorisation}.

\section{Dedicated simulations of the signal processes}
\label{sec:simulation-dedicated}

The fixed-order theory calculations by \citeauthor{Bevilacqua:2018woc}~\cite{Bevilacqua:2018woc,Bevilacqua:2018dny}, used as a reference for this analysis, make \NLO predictions for \tty \xsecs in the \emu final state, and all off-shell effects are taken into account in this calculation.
Thus, more precisely, the simulated process is \WbWby production, which also includes singly-resonant and non-resonant diagrams with one or no top-quark mass resonances.
In an attempt to classify \WbWby production into two separate contributions, the doubly-resonant diagrams would correspond to pure $\pp \to \tty$ production, and the singly-resonant diagrams to $\pp \to \tWby$ production -- for now neglecting any interference effects between the two.
A na{\"i}ve approach for simulating doubly-resonant contributions to the \WbWby final state would be to generate $\pp \to \tty$ events at leading order in \QCD, for example with the \Madgraph framework~\cite{Alwall:2014hca}, interfaced with a parton shower (\LOPS).
However, as diagrams with fewer resonances and associated interference effects are neglected, such a simulation would be insufficient to estimate the \WbWby fraction in data if a genuine comparison with the fixed-order calculation is desired.
An additional \LOPS simulation of $pp \to \tWby$ would add singly-resonant diagrams and could help to complete the picture if combined with the \tty simulation.
Examples of doubly-resonant and singly-resonant diagrams contributing to \WbWby production were shown in the previous chapter in \cref{fig:theory-tty-resonances}.
Unfortunately, generating large numbers of \tWby events has proven technically challenging and was not feasible for this analysis.

\begin{figure}
  \centering
  \includegraphics[scale=1.2]{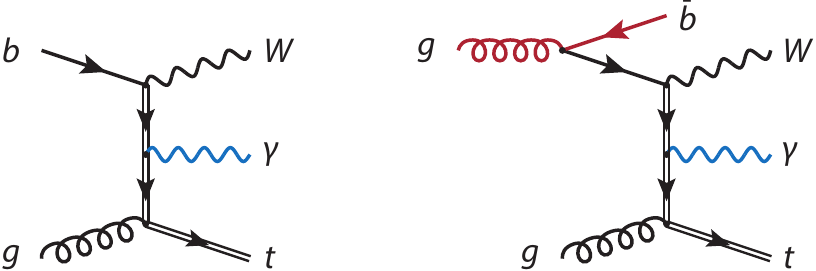}
  \caption[Feynman diagrams for \tWy and \tWby production]{%
    Representative tree-level Feynman diagrams for \tWy production on the left-hand side and \tWby production on the right-hand side.
    Diagrams such as the one on the left, with a \bquark in the initial state, can only be generated in the five-flavour scheme.
    In the four-flavour scheme, an additional vertex is needed to produce the \bquark.
    Thus, the right-hand side diagram is also the simplest possible diagram (i.e. leading order) for \tWy production in the four-flavour scheme.
  }
  \label{fig:sim-tWy-4flav-5flav}
\end{figure}

Single top-quark production in association with a \Wboson and a photon (\tWy), on the other hand, can be generated easily.
\MC simulations of \tWy are used for this analysis to estimate singly-resonant contributions to the \WbWby process.
They were computed at \LO in \QCD in the \emph{five-flavour scheme}, ignoring any mass effects of the \bquark and treating it as massless like the four lightest quarks.
In this scheme, the \bquark is described by parton distribution functions and it is considered in the perturbative evolution of \QCD in the initial state, described by the \DGLAP equations~\cite{Altarelli:1977zs,Dokshitzer:1977sg,Gribov:1972ri}.
As a consequence, and as opposed to the four-flavour scheme, the \bquark can enter the matrix element as an initial-state particle directly.
Hence, the leading-order partonic process for \tWy production is simply $bg \to \tWy$, and problems in the definition of the \tWy \xsec due to interference with \tty, as described in \cref{sec:theory-tty}, do not occur.
However, the singly-resonant diagrams of \tWy production only contribute to a \WbWby final state if an extra \bquark is present.
To add this \bquark to the matrix-element calculation, one would have to include \NLO real-emission corrections, which add $\pp \to \tWy [+X]$ final states to the computation.
This is technically even more challenging than only a \LOPS prediction of $\pp \to \tWby$.
Representative Feynman diagrams of \tWy and \tWby production are shown in \cref{fig:sim-tWy-4flav-5flav}.

However, even in a $bg \to \tWy$ simulation, the initial-state \bquark needs to be generated through the \PDFs by simulating a $g \to b\bar{b}$ split.
The second \bquark from this split acts as a spectator particle and corresponds to the missing \bquark of the \WbWby final state.
An estimate of this split through the \PDFs is far less precise than including the vertex of the split into the matrix element directly.
But, nonetheless, the \LOPS \tWy simulation can be used to estimate contributions by singly-resonant diagrams to a \WbWby final state.
Therefore, as a \enquote{best estimate} of the \WbWby final state, the \LOPS predictions for \tty and for \tWy are combined and treated as a single signal process in this analysis.
They are described in more detail in the following paragraphs.

\paragraph{doubly-resonant production.}
\ttbar production in association with a photon in the matrix element is simulated in a dedicated sample using the \Madgraph generator~\cite{Alwall:2014hca} at \LO in \QCD and the \NNPDFLO \PDF set~\cite{Ball:2012cx}.
The matrix-element generation includes the decays of the top quarks and uses a filter to only produce final states with at least one charged lepton.
The events are generated as an inclusive, doubly-resonant $2\to7$~process, that is, the generator produces final states such as $\pp \to q\bar{q}bl\nu b\gamma$ (\ljets channels) and $\pp \to b\bar{b}l\nu_ll\nu_l\gamma$ (dilepton channels).
Thus, all diagrams where the photon is radiated by any of the top quarks, by the \bquarks, by the \Wbosons as well as by the decay products of the \Wbosons are included in the simulation.
Although not the dominant production mode, the final state can be generated through quark-antiquark annihilation.
Therefore, the $2\to 7$~process also includes photons radiated by initial-state partons (\ISR).
To avoid infrared and collinear singularities due to the photon radiation, kinematic cuts are applied on matrix-element level.
Photons and charged leptons are required to have minimal transverse momenta of \SI{15}{\GeV}, and their pseudorapidities must fulfil $|\eta| < 5.0$.
In addition, the generated photons must not be in the vicinity of any other charged particle of the final state and at least be separated with $\Delta R > 0.2$.
The top-quark mass in this and all other simulations is set to \SI{172.5}{\GeV}.
The event generation is interfaced to \Pythia using the \emph{A14} parameter tune~\cite{ATL-PHYS-PUB-2014-021} and the \NNPDFLO \PDF set to model parton shower, hadronisation, fragmentation and the underlying event.
\Evtgen~\cite{Lange:2001uf} is used to simulate heavy-flavour decays, such as those of $B$~and $D$~hadrons.
The renormalisation and factorisation scales are dynamic and correspond to half the sum of all \emph{transverse masses} of the final-state particles:
\begin{align*}
  \mu_R = \mu_F = \frac{1}{2} \sum_f \sqrt{ m_f^2 + p_{\mathrm{T},f}^2 } \, ,
\end{align*}
where $f$~runs over all final-state particles, and $m_f$ and $p_{\mathrm{T},f}$ are the rest mass and the transverse momentum of particle $f$, respectively.

\paragraph{singly-resonant production.}
Associated \tW production with an additional photon in the matrix element is simulated in two sets of dedicated samples using the \Madgraph generator at \LO in \QCD and the \NNPDFLO \PDF set.
One set of events is generated as a $2 \to 3$ process and assumes a stable top quark and \Wboson.
Consequently, only diagrams of radiative \tW production are considered, where a photon is radiated during the production of the top quark or the \Wboson, or from the initial state.
The decays of the top quark and the \Wboson are then simulated subsequently in the \Madgraph framework.
A second set of events is generated as a $2 \to 2 \to 6$ process, where a top quark and a \Wboson are first generated without extra photon.
Then, the top quark and \Wboson are decayed, and the photon is only radiated in this second step of the generation process.
Hence, this set only contains diagrams where the photon originates from the decaying top quark, from the decaying $W$~boson or from any of the charged decay products of the two.
The two sets of \tWy events are complementary and, once combined, provide a full simulation of the process.
As done for the \tty simulation, the photons and charged leptons are required to have minimal transverse momenta of \SI{15}{\GeV}, and their pseudorapidities must fulfil $|\eta| < 5.0$.
The generated photons must again be separated from any other charged particle in the final state with $\Delta R > 0.2$.
In both sets of events, the matrix-element generator is interfaced to \Pythia, which uses the \emph{A14} parameter tune and the \NNPDFLO \PDF set.
\Evtgen is used to simulate heavy-flavour decays.

\section{Background simulations}
\label{sec:simulation-background}

Events with \plusjets{\Wy}~and \plusjets{\Zy}~final states are generated in dedicated samples using different versions of the \Sherpa programme~\cite{Gleisberg:2008ta,Hoeche:2009rj}.
\plusjets{\Wy} processes are simulated with \Sherpa 2.2.2 at \NLO in \QCD using the \NNPDFNNLO \PDF set.
\plusjets{\Zy} events are generated with \Sherpa 2.2.4 at leading order in \QCD with the same \PDF set.
All samples are normalised to the cross-sections given by the corresponding \MC simulation.
The simulation in \Sherpa includes all steps of the event generation, from the hard process to the observable particles.
All samples are matched and merged to the \Sherpa-internal parton showering based on Catani-Seymour dipoles~\cite{Gleisberg:2008fv,Schumann:2007mg} using the \textsc{meps}@\NLO prescription~\cite{Hoeche:2011fd,Catani:2001cc,Hoeche:2012yf}.
Virtual corrections for the \NLO accuracy in \QCD in the matrix element are provided by the \textsc{openloops} library~\cite{Cascioli:2011va,Denner:2016kdg}.
These samples with matrix-element photons are paired with simulations of \Wjets and \Zjets final states, simulated with \Sherpa 2.2.1 at \NLO in \QCD.
The \NNPDFNNLO \PDF set is used in conjunction with a dedicated tune provided by the \Sherpa authors.
These photon-inclusive samples are normalised to \NNLO cross-sections in \QCD~\cite{ATLAS-CONF-2015-039}.

Inclusive \ttbar production processes are simulated on matrix-element level at \NLO in \QCD using \Powheg{}~v2~\cite{Nason:2004rx,Frixione:2007vw,Alioli:2010xd}.
The matrix-element generator is interfaced to \Pythia to simulate parton shower, hadronisation, fragmentation and the underlying event.
Heavy-flavour decays are modelled with \Evtgen.
The matrix-element calculation uses the \NNPDFNLO \PDF set~\cite{Ball:2014uwa}.
The internal parameter \emph{hdamp} to control the threshold of the hardest emission is set to 1.5 times the top-quark mass following \ATLAS standards.
The showering in \Pythia uses the \emph{A14} tune in conjunction with the \NNPDFLO \PDF set.
By applying a \kfactor, the events are normalised to a \xsec value calculated with the \Topplusplus programme at \NNLO in \QCD, including soft-gluon resummation to next-to-next-to-leading-log order (see Ref.~\cite{Czakon:2011xx} and references therein), assuming a top-quark mass of \SI{172.5}{\GeV}.
The resulting \xsec for \ttbar production at $\sqrt{s} = \SI{13}{\TeV}$ amounts to $\sigma(pp \to \ttbar) = \SI{832}{\pb}$ with remaining uncertainties due to scale and \PDF choice of approximately 3\% and 4\%, respectively.

Single-top-quark processes are modelled separately for three possible production modes, namely \schannel, \tchannel and \tW production, each of which are generated for top-quark and anti-top-quark production separately.
The three production modes are simulated on matrix-element level at \NLO in \QCD with \Powheg and the \NNPDFLO \PDF set.
The matrix-element generator is interfaced to \Pythia with the \emph{A14} tune and the \PDF set as before.
Again, heavy-flavour decays are modelled with \Evtgen.
The sample cross-sections are normalised to \NNLO in \QCD using \kfactors~\cite{Kidonakis:2010tc,Kidonakis:2010ux,Kidonakis:2011wy}.
For the \tW production samples, the \emph{diagram removal scheme} is implemented to remove higher-order interference effects between \ttbar and \tW~production.

Events with two vector bosons, \ie $\mathit{WW}$, $\mathit{WZ}$ and $\mathit{ZZ}$, are generated with \Sherpa~2.2.2 (for the purely leptonic decays) and with \Sherpa~2.2.1 (for all other decays) at leading order in \QCD.
The \NNPDFNNLO \PDF set is used in conjunction with a dedicated tune provided by the \Sherpa authors.
The simulation includes all steps of the event generation, from the hard process to the observable particles.
The samples are normalised to \NLO \xsecs in \QCD~\cite{Campbell:1999ah}.

Events with a \ttbar pair and an associated \Wboson or \Zboson ($t\bar{t}V$) are simulated at \NLO in \QCD on matrix-element level with \Madgraph using the \NNPDFNLO \PDF set.
The matrix-element generator is interfaced to \Pythia, for which the \emph{A14} tune is used in conjunction with the \NNPDFLO \PDF set.
The samples are normalised to \NLO in both \QCD and electroweak theory~\cite{deFlorian:2016spz}.

\clearpage

\section{Sample overlap-removal strategy}
\label{sec:simulation-overlap}

Generally speaking, parton-shower algorithms lack formal accuracy in the simulation of \emph{hard} photon emissions, that is, emissions under large angles that carry considerable momentum fractions, due to the nature of their splitting function evolution.
Therefore, the radiation of high-\pT photons is simulated a lot more precisely in samples where these photons are generated at matrix-element level.
However, photons in the infrared/collinear limit, where their emission angles and momentum fractions become small, cannot be included in matrix-element calculations due to the occurring singularities.
Kinematic cuts are placed on the photons in the matrix-element generation step to avoid these limits.
As a consequence, events with photons below the kinematic thresholds can only be estimated from photon-inclusive samples.
To avoid double-counting of events with photons \emph{above} the thresholds, sample overlap-removal techniques need to be applied between dedicated photon samples and photon-inclusive samples.
In particular, the removal procedures are applied for the two signal processes, \tty and \tWy, and their photon-inclusive counterparts, \ttbar and associated \tW production.
In addition, sample overlap is removed between \plusjets{\Wy} and \plusjets{\Zy} samples and the photon-inclusive \Wjets and \Zjets samples, respectively.

The recipe for removing sample overlap is the following:
firstly, all events from the dedicated $X+\gamma$ sample are accepted as the radiation of hard photons is simulated with much higher accuracy in these samples.
Additionally, the populated phase space of photon-inclusive samples is much larger than that of dedicated photon samples, and only a fraction of events includes photons.
In practice, if only events with a photon are selected, the photon-inclusive samples would have to be generated with many more events to reach the same level of statistical accuracy.
Secondly, events of the photon-inclusive $X$ samples are removed if they fall in the phase-space region that is populated by the $X+\gamma$ samples.
This \emph{overlap} region is defined through the sets of cuts applied to the $X+\gamma$ samples on matrix-element level, and consists of the following two requirements for all samples:
\begin{enumerate}
\item $\pT (\gamma) > \SI{15}{\GeV}$ and
\item $\Delta R(\ell, \gamma) > 0.2$,
\end{enumerate}
where $\Delta R$ is defined in the \etaphi plane in the \ATLAS coordinate system.

Based on \MC-truth information, the sample overlap-removal algorithm first compiles lists of photon and charged-lepton candidates generated at matrix-element level.
The candidates need to be genuine photons and charged leptons (requested through their \PDGID values).
They must not originate from any interaction with the detector or from hadronic activity, but come from the primary interaction (required through their \emph{barcode} and \emph{truth origin}).
Additional criteria ensure that the found candidates are in their last and stable simulation states, or right before their decays in the case of muons and \tauleptons (required through their \emph{status code}).
Leptonically decaying \tauleptons are considered as such and their decay products are vetoed on the candidate lists.
After the lists are compiled, photon candidates are dropped if they fail the above criterion on their transverse momenta.
For the remaining photon candidates, the $\Delta R$ criterion is tested with all charged-lepton candidates, and the photon candidate is discarded as soon as it overlaps with any charged-lepton candidate.
If any photon candidates remain, the event is considered to fall in the overlap region, and it is vetoed for \ttbar, \tW, \Wjets and \Zjets samples.


\section{Categorisation of photon candidates}
\label{sec:simulation-categorisation}

The main interest of this analysis are \ttbar events where an additional photon is generated in the hard-scattering event.
These are referred to as \emph{prompt photons}.
However, photons and other photon-like objects may occur at many stages of the simulation process, both in \MC simulation and in the simulation of the detector response.
Some of the reconstructed photon candidates may also be other objects or activity recorded in the \ATLAS detector.
Among the candidates detected and reconstructed with \ATLAS, this analysis distinguishes three classes:
\begin{enumerate}
\item \emph{Prompt photons} from the hard-scattering event,
\item \emph{\efakephotons}, that is, electronic activity in the detector that fakes photon signatures in the calorimeters,
\item \emph{\hfakephotons}, that is, hadronic energy depositions in the calorimeters that fake photon signatures. This category also includes real, but non-prompt photons from hadronic decays of other particles, in particular from $\pi^0 \to \gamma\gamma$ decays.
\end{enumerate}
To estimate the fractions of these categories in \ATLAS data, \MC-truth information is used to categorise photon candidates in simulation.
The \ATLAS simulation infrastructure provides means to maintain an association between particles from the \MC generation step and reconstructed object candidates in the detector.
Corresponding \enquote{\MC-truth particles} of photon candidates can be retrieved from the \MC generator record, and quantities of these particles are used to categorise the photon candidate into the above classes.
The \ATLAS software uses an internal classification scheme for \MC-truth particles and introduces two parameters, \emph{truth type} and \emph{truth origin}, for this classification.

For a photon candidate to be classified as an \efakephoton, the associated \MC-truth particle needs to be an electron or overlap with an electron with $\Delta R < 0.1$.
For the latter criterion, a list of \MC-truth electrons is compiled similarly to the charged-lepton candidate list described in the previous section, based on requirements on the \PDGID, on the transverse momentum, on the centrality (\abseta) and on the \emph{barcode} of the \MC-truth particle.
If any of the two \efakephoton criteria are met, the candidate is categorised as such and is not tested for any of the other two categories to avoid possible double-classification.

The remaining photon candidates are then tested against three criteria for \hfakephotons, and any of the three is sufficient for the candidate to be accepted for this category.
Based on \emph{truth type} and \emph{truth origin} information, the photon candidate is categorised as an \hfakephoton if the associated \MC-truth particle
(1) is a photon and originates from a baryon or meson,
(2) is a photon and originates from a $\pi^0 \to \gamma\gamma$ decay or
(3) is a hadronic energy deposition.
Only if photon candidates meet none of the criteria for \efakephotons or \hfakephotons, they are categorised as prompt photons.



\chapter{Event selection}
\label{cha:selection}

With \MC simulations for all major signal and background processes prepared, their predictions may now be compared with \ATLAS data.
This analysis uses proton-proton collision data taken with the \ATLAS detector in the years 2015 to 2018 during \runii of the \LHC.
As summarised in \cref{sec:exp}, the total data delivered by the \LHC amounts to an integrated luminosity of \SI{156}{\ifb}, of which \SI{147}{\ifb}, more than \SI{94}{\percent}, were recorded with the \ATLAS detector.
The dataset is split in four subsets, one for each year of data-taking.
The subsets consist of data-taking \emph{runs} that are often, but not always identical to the periods the \LHC was refilled with proton beams.
Runs are divided in small, equal-length time units called \emph{luminosity blocks}, for which the instantaneous luminosity is averaged to calculate a luminosity-over-time profile.
Depending on the internal clock signal of the \ATLAS detector, one luminosity block corresponds to approximately one minute of data-taking.
Luminosity blocks are flagged if the \LHC beams were unstable, if the \ATLAS detector was not fully operational, or if the data recorded with \ATLAS did not fulfil a set of quality criteria.
The remaining luminosity blocks of all runs are considered to be \enquote{good} for physics analyses and enter the \emph{good-run lists}, one of which is compiled for each year of data-taking.
The dataset on the four good-run lists combined corresponds to an integrated luminosity of \SI{139}{\ifb}.
More details for individual years are given in \cref{tab:selection-grls}.
Unless analyses do not require all detector components to be online or have otherwise looser requirements, all \ATLAS analyses base their event selection on these good-run lists.

\begin{table}
  \centering
  \caption[Integrated luminosity for each year of \runii of the \LHC]{%
    Delivered integrated luminosity of the \LHC and the integrated luminosity of the \ATLAS good-run lists for each year of data-taking during \runii.
  }
  \label{tab:selection-grls}
  \footnotesize
  \begin{tabular}{lSS}
    \toprule
    Year & {Delivered $[\si{\ifb}]$} & {Good-run lists $[\si{\ifb}]$} \\
    \midrule
    2015  &   4.2 &   3.2 \\
    2016  &  38.5 &  33.0 \\
    2017  &  50.2 &  44.3 \\
    2018  &  63.3 &  58.5 \\
    \midrule
    Total & 156.2 & 139.0 \\
    \bottomrule
  \end{tabular}
\end{table}

The aim of the analysis presented in this thesis is to select phase-space regions highly enriched in \tty events.
However, before tightening the selection to signal-like events in the \emu final state, a more generic pre-selection of events is applied.
This reduces the amount of data drastically and only leaves events that are interesting for control studies and the measurement itself.
Among the most prominent features of the \emu final state are the two charged leptons with high transverse momenta.
Therefore, events are required to have fired a single-lepton trigger in the \ATLAS \HLT trigger system.
For both electrons and muons, not single triggers, but chains of triggers are defined, out of which only one must have fired.
The triggers in a chain differ in their \pT thresholds and identification and isolation requirements:
those with lower \pT thresholds require the trigger objects to be more tightly identified and isolated, whereas high-\pT triggers have looser requirements.
The full list of the considered triggers is given in \cref{tab:selection-triggers} following the convention:
\begin{align*}
  \mathtt{HLT\_<type><pT>\_<ID>\_<iso>} \, ,
\end{align*}
where the \texttt{<pT>} tag defines the \pT threshold and the following tags describe the identification and isolation criteria.
Due to the increased average pile-up in later years of data-taking, the trigger chains were adjusted accordingly to not overload the \TDAQ system with too high event rates.
In addition to the trigger-fire requirements, an electron or muon candidate must be reconstructed according to the criteria defined in \cref{sec:exp_objects} and matched to the candidate that fired the trigger.
To omit regions of low trigger efficiency%
\footnote{Due to detection inefficiencies and uncertainties in the energy/momentum measurements of the involved \ATLAS components, triggers show a turn-on behaviour in trigger efficiency as the object candidate's transverse momentum increases.}%
, the matched electron and muon candidates must fulfil the \pT requirements also listed in \cref{tab:selection-triggers}: at least \SIlist{25;27;28;28}{\GeV} for 2015 to 2018, respectively.
The identical requirements are imposed on \ATLAS data and \MC simulation.

\begin{table}
  \caption[List of single-lepton triggers used in the event pre-selection]{%
    List of single-lepton triggers used in the event pre-selection.
    Events need to have either one of the single-electron or single-muon triggers fired.
    Then, a lepton must be selected above the listed \pT thresholds and matched to that trigger.
  }
  \label{tab:selection-triggers}
  \centering
  \footnotesize
  \begin{tabular}{llll}
    \toprule
    Year & Single-electron triggers & Single-muon triggers & Lepton \pT \\
    \midrule
         & HLT\_e24\_lhmedium\_L1EM20VH       & HLT\_mu20\_iloose\_L1MU15 &                  \\
    2015 & HLT\_e60\_lhmedium                 & HLT\_mu50                 & $> \SI{25}{GeV}$ \\
         & HLT\_e120\_lhloose                 &                           &                  \\
    \midrule
         & HLT\_e26\_lhtight\_nod0\_ivarloose & HLT\_mu26\_ivarmedium     &                  \\
    2016 & HLT\_e60\_lhmedium\_nod0           & HLT\_mu50                 & $> \SI{27}{GeV}$ \\
         & HLT\_e140\_lhloose\_nod0           &                           &                  \\
    \midrule
         & HLT\_e26\_lhtight\_nod0\_ivarloose & HLT\_mu26\_ivarmedium     &                  \\
    2017 & HLT\_e60\_lhmedium\_nod0           & HLT\_mu50                 & $> \SI{28}{GeV}$ \\
         & HLT\_e140\_lhloose\_nod0           &                           &                  \\
    \midrule
         & HLT\_e26\_lhtight\_nod0\_ivarloose & HLT\_mu26\_ivarmedium     &                  \\
    2018 & HLT\_e60\_lhmedium\_nod0           & HLT\_mu50                 & $> \SI{28}{GeV}$ \\
         & HLT\_e140\_lhloose\_nod0           &                           &                  \\
    \bottomrule
  \end{tabular}
\end{table}

In addition, all events must have at least one primary vertex with at least two tracks with $\pT > \SI{400}{\MeV}$ matched to that vertex.
Then, the object candidates are reconstructed with the criteria listed in \cref{sec:exp_objects}, but without imposing any identification or isolation requirements yet.
With these lists of candidates, the overlap-removal steps described in the same section are applied.
What follows is a procedure called \emph{event cleaning}, in which all reconstructed jet candidates are tested against a set of quality criteria to distinguish them from non-collision background.
This includes beam-induced background due to upstream proton losses in the beam, showers in the detector induced by cosmic rays, and large-scale coherent calorimeter noise.
The \emph{BadLoose} criteria, further described in Ref.~\cite{ATLAS-CONF-2015-029}, are applied.
If any of the reconstructed jet candidates fails the requirements, the entire event is discarded because even low-\pT candidates would spoil the \MET calculation.

\paragraph{definition of the signal region.}
After pre-selecting events, the identification and isolation requirements described in \cref{sec:exp_objects} are imposed on all reconstructed object candidates.
Then, the \tty \emu signal region is selected as follows:
exactly one electron and one muon are required, each of which must carry $\pT > \SI{25}{\GeV}$.
Note that, regardless, one of the two charged leptons must fulfil the trigger-match requirements of the pre-selection and pass the corresponding \pT threshold.
Electron and muon must be of opposite electric charge.
In addition, the event is required to have at least two jets with $\pT > \SI{25}{\GeV}$, at least one of which must be \btagged with the \MVtwo discriminant at the \SI{85}{\percent} operating point.
The event must have exactly one photon with $\pT > \SI{20}{\GeV}$.
Although the \tty \emu final state contains two neutrinos and large missing transverse momentum is expected, no explicit cut on \MET is imposed due to the low background contamination of the selected signal region.

The selected events from \MC simulation are then grouped based on their origin:
the \catttyemu and \cattWyemu categories contain events from simulated \tty and \tWy events in the \emu decay channel, respectively.
The decay channels are identified through \MC-truth information for each event.
Dilepton events involving \tauleptons, which subsequently decayed into other leptons, are counted separately as they are not included in the reference theory computation~\cite{Bevilacqua:2018woc,Bevilacqua:2018dny}.
Minor contributions are also expected from \ljets events with an additional lepton faked by hadronic activity.
They are combined with the dilepton events involving \taulepton decays and counted in the category \catother.
Selected events from background simulation are categorised based on the origin of the reconstructed photon, which is identified as described in \cref{sec:simulation-categorisation}.
The categories are:
\begin{enumerate}
\item%
  The \cathfake category with any type of \hfakephoton that passed the photon selection criteria.
  The category is dominated by \ttbar events with an additional \hfakephoton in the final state, but it may also contain events with a prompt photon in the simulation, that failed detection or reconstruction, and an additionally reconstructed \hfakephoton.
\item%
  The \catefake category with events that have no prompt photon, but an additional electron misreconstructed as a photon (\efakephoton).
  Again, this may also include events with a prompt photon in the simulation, that failed detection or reconstruction, and an additionally reconstructed \efakephoton.
\item%
  The \catprompt category with background-like events, where the reconstructed photon candidate is a prompt photon.
\end{enumerate}
The predicted event yields of all categories and the numbers of reconstructed events in \ATLAS data are listed in \cref{tab:selection-statonly-yields}.
To allow a better comparison, the numbers are also given separately for 2015/16, 2017 and 2018.
The quoted uncertainties on the predictions are \MC-statistical uncertainties.
While all background processes are simulated at \NLO in \QCD, or their \xsecs are reweighted to that order through \kfactors, the estimation of the signal processes is only based on the \LO \xsecs calculated in \Madgraph and, thus, is not expected to be very accurate in its total rate.%
\footnote{%
The \ATLAS \tty analysis using \SI{36}{\ifb} performed dedicated calculations of \LO and \NLO \xsecs in the fiducial volume of the measurement, which yielded a \kfactor as high as 1.44 in the \emu channel~\cite{TOPQ-2017-14}.
}
Therefore, for aesthetic reasons in control plots and to allow a better assessment of the event composition in \MC simulation when comparing to data, the predictions of the \tty and \tWy categories marked with (*) were scaled in such a way that the total \MC prediction matches the numbers of reconstructed events in data in each column of the table.
However, this has no effect on the results of the measurement as the fiducial inclusive \xsec is extracted independently of the predicted \xsec as detailed in \cref{cha:strategy}.

\begin{table}
  \centering
  \caption[Predicted event yields with \MC-statistical uncertainties only]{%
    Predicted event yields for all \MC categories and numbers of reconstructed events in \ATLAS data in the \emu signal region.
    All categories are estimated based on \MC simulation only.
    The listed uncertainties for the predictions are \MC-statistical uncertainties.
    The predictions of the \tty and \tWy categories marked with (*) were scaled to match the numbers of reconstructed events in data in each column.
    The scaling factors are \num{1.358}, \num{1.445}, \num{1.414} and \num{1.408}, respectively.
  }
  \label{tab:selection-statonly-yields}
  \sisetup{round-mode=places, round-precision=1}
  \begin{tabular}{
    l
    S[table-format=3.1] @{${}\pm{}$} S[table-format=1.1]
    S[table-format=3.1] @{${}\pm{}$} S[table-format=1.1]
    S[table-format=4.1] @{${}\pm{}$} S[table-format=1.1]
    S[table-format=4.1] @{${}\pm{}$} S[table-format=2.1]
    }
    \toprule
    & \multicolumn{2}{c}{2015/16}
    & \multicolumn{2}{c}{2017}
    & \multicolumn{2}{c}{2018}
    & \multicolumn{2}{c}{full dataset}\\
    \midrule
    \catttyemu{}* & 642.7 & 5.3  & 759.6 & 6.1 & 989.2 & 6.9 & 2391.4 & 10.6 \\
    \cattWyemu{}* & 42.9  & 1.0  & 49.9  & 1.1 & 62.9  & 1.3 & 155.7  & 2.0 \\
    \catother{}*  & 75.6  & 1.8  & 88.2  & 2.0 & 115.2 & 2.3 & 279.0  & 3.6 \\
    \cathfake     & 22.9  & 1.0  & 23.1  & 1.0 & 31.4  & 1.1 & 77.5   & 1.8 \\
    \catefake     & 5.63  & 0.33 & 7.0   & 0.4 & 10.3  & 0.5 & 23.0   & 0.7 \\
    \catprompt    & 19.3  & 1.0  & 30.2  & 1.2 & 38.0  & 1.4 & 87.5   & 2.1 \\
    \midrule
    Total \MC & 809.0 & 5.9 & 958.0 & 6.7 & 1247.0 & 7.6 & 3014.0 & 11.7 \\
    \midrule
    Data
      & \multicolumn{2}{l}{\num{809}}
      & \multicolumn{2}{l}{\num{958}}
      & \multicolumn{2}{l}{\hspace{0.16em}\num{1247}}
      & \multicolumn{2}{l}{\hspace{0.14em}\num{3014}} \\
    \bottomrule
  \end{tabular}
\end{table}

\Cref{tab:selection-contributions} gives a more detailed overview of the composition of the background categories:
individual contributions of all \MC simulations to the categories are listed separately.
The same scaling as in \cref{tab:selection-statonly-yields} to the data yields is applied to the \tty and \tWy categories marked with (*).
The \Wjets and \Zjets processes marked with ($\dagger$) are found to have negligible contributions to all categories and are therefore discarded in further steps of the analysis.
Negligible contributions are expected as any overlap with the $\plusjets{\Wy}$ and $\plusjets{\Zy}$ samples is removed from the \Wjets and \Zjets samples, \cf \cref{sec:simulation-overlap}.
Thus, the remaining cases to cover would only be \Wjets and \Zjets with \hfakephotons or \efakephotons in the final state or phase-space regions not simulated in the $\plusjets{\Wy}$ and $\plusjets{\Zy}$ samples, which are expected to only contribute weakly to the chosen signal region.

\begin{table}[p]
  \centering
  \caption[Breakdown of the predicted event yields for all \MC categories]{%
    Breakdown of the predicted event yields for all \MC categories in the \emu signal region.
    The categories are identical to those listed in \cref{tab:selection-statonly-yields}, but their individual process contributions are listed.
    The predictions of the \tty and \tWy categories marked with (*) were scaled to match the numbers of reconstructed events in data in each column.
    Simulations marked with ($\dagger$) have negligible contributions to all categories and are therefore discarded in further steps of the analysis.
  }
  \label{tab:selection-contributions}
  \sisetup{round-mode=places, round-precision=1}
  \begin{tabular}{
    l
    S[table-format=3.1] @{${}\pm{}$} S[table-format=1.1]
    S[table-format=3.1] @{${}\pm{}$} S[table-format=1.1]
    S[table-format=3.1] @{${}\pm{}$} S[table-format=1.1]
    S[table-format=4.1] @{${}\pm{}$} S[table-format=2.1]
    }
    \toprule
    & \multicolumn{2}{c}{2015/16}
    & \multicolumn{2}{c}{2017}
    & \multicolumn{2}{c}{2018}
    & \multicolumn{2}{c}{full dataset}\\
    \midrule
    \catttyemu{}*                     & 642.7 & 5.3  & 759.6 & 6.1  & 989.2 & 6.9  & 2391.4 & 10.6 \\
    \midrule
    \cattWyemu{}*                     & 42.9  & 1.0  & 49.9  & 1.1  & 62.9  & 1.3  & 155.7  & 2.0  \\
    \midrule
    \catother{}*                      & 75.6  & 1.8  & 88.2  & 2.0  & 115.2 & 2.3  & 279.0  & 3.6  \\
    $\rightarrow$ from \tty           & 70.6  & 1.8  & 82.1  & 2.0  & 107.6 & 2.3  & 260.4  & 3.5  \\
    $\rightarrow$ from \tWy           & 5.03  & 0.34 & 6.1   & 0.4  & 7.5   & 0.4  & 18.6   & 0.7  \\
    \midrule
    \cathfake                         & 22.9  & 1.0  & 23.1  & 1.0  & 31.4  & 1.1  & 77.5   & 1.8  \\
    $\rightarrow$ from \tty           & 0.17  & 0.09 & 0.30  & 0.13 & 0.43  & 0.13 & 0.90   & 0.19 \\
    $\rightarrow$ from \ttbar         & 21.1  & 0.9  & 21.5  & 0.9  & 29.9  & 1.1  & 72.5   & 1.7  \\
    $\rightarrow$ from \tWy           & 1.4   & 0.4  & 0.8   & 0.4  & 0.6   & 0.4  & 2.8    & 0.6  \\
    $\rightarrow$ from \Wy            & \emptyvalue  & \emptyvalue  & \emptyvalue  & \emptyvalue   \\
    $\rightarrow$ from \daggered{\Wjets} & \emptyvalue & \emptyvalue & \emptyvalue & \emptyvalue   \\
    $\rightarrow$ from \Zy            & \emptyvalue  & \emptyvalue  & \emptyvalue  & \emptyvalue   \\
    $\rightarrow$ from \daggered{\Zjets} & \emptyvalue & \emptyvalue & \emptyvalue & \emptyvalue   \\
    $\rightarrow$ from diboson        & \emptyvalue  & 0.21  & 0.09 & 0.10  & 0.05 & 0.36   & 0.10 \\
    $\rightarrow$ from \ttV           & 0.29  & 0.05 & 0.22  & 0.08 & 0.45  & 0.08 & 0.96   & 0.12 \\
    \midrule
    \catefake                         & 5.63  & 0.33 & 7.0   & 0.4  & 10.3  & 0.5  & 23.0   & 0.7  \\
    $\rightarrow$ from \tty           & 0.49  & 0.13 & 1.05  & 0.20 & 1.63  & 0.24 & 3.18   & 0.33 \\
    $\rightarrow$ from \ttbar         & 1.35  & 0.24 & 1.69  & 0.27 & 2.96  & 0.33 & 6.0    & 0.5  \\
    $\rightarrow$ from \tWy           & 0.12  & 0.27 & \emptyvalue  & \emptyvalue  & 0.12   & 0.27 \\
    $\rightarrow$ from \Wy            & \emptyvalue  & \emptyvalue  & \emptyvalue  & \emptyvalue   \\
    $\rightarrow$ from \daggered{\Wjets} & \emptyvalue & \emptyvalue & \emptyvalue & \emptyvalue   \\
    $\rightarrow$ from \Zy            & \emptyvalue  & \emptyvalue  & \emptyvalue  & \emptyvalue   \\
    $\rightarrow$ from \daggered{\Zjets} & \emptyvalue & \emptyvalue & \emptyvalue & \emptyvalue   \\
    $\rightarrow$ from diboson        & 1.85  & 0.12 & 2.06  & 0.14 & 2.71  & 0.16 & 6.63   & 0.25 \\
    $\rightarrow$ from \ttV           & 1.81  & 0.10 & 2.22  & 0.14 & 3.00  & 0.13 & 7.03   & 0.22 \\
    \midrule
    \catprompt                        & 19.3  & 1.0  & 30.2  & 1.2  & 38.0  & 1.4  & 87.5   & 2.1  \\
    $\rightarrow$ from \ttbar         & 11.9  & 0.7  & 21.6  & 0.9  & 25.8  & 1.0  & 59.3   & 1.5  \\
    $\rightarrow$ from \Wy            & 1.0   & 0.6  & \emptyvalue  & 0.10  & 0.11 & 1.1    & 0.6  \\
    $\rightarrow$ from \daggered{\Wjets} & \emptyvalue & \emptyvalue & \emptyvalue & \emptyvalue   \\
    $\rightarrow$ from \Zy            & 0.59  & 0.16 & 2.0   & 0.8  & 3.4   & 1.0  & 6.0    & 1.2  \\
    $\rightarrow$ from \daggered{\Zjets} & \emptyvalue & \emptyvalue & \emptyvalue & \emptyvalue   \\
    $\rightarrow$ from diboson        & 1.16  & 0.22 & 1.29  & 0.14 & 1.63  & 0.21 & 4.09   & 0.33 \\
    $\rightarrow$ from \ttV           & 4.63  & 0.24 & 5.32  & 0.28 & 7.06  & 0.34 & 17.0   & 0.5  \\
    \bottomrule
  \end{tabular}
\end{table}

The predicted event yields for the full \runii dataset show that the defined signal region is expected to be dominated by \tty and \tWy contributions.
The \catttyemu and \cattWyemu signal categories with a total of \num{2547} expected events account for \SI{84.5}{\percent} of the total \MC estimate, which corresponds to a signal-to-background ratio of \num{5.45}.
This results in a statistical significance of the signal of $\sfrac{S}{\sqrt{B}} = \num{117.9}$.
The combined total \tty and \tWy prediction is composed of \SI{90.1}{\percent} \emu-channel events and \SI{9.9}{\percent} non-\emu-channel events.
Within these numbers, the \tty-to-\tWy ratio is \num{15.4} and \num{14.0} for \emu-channel and non-\emu-channel events, respectively.
The non-\emu-channel contributions from \tty and \tWy also constitute the largest background:
within the \SI{15.5}{\percent} background events, some \SI{60}{\percent} are classified as \catother.
The remaining background events are composed of \SI{17}{\percent} from the \cathfake category, \SI{5}{\percent} from the \catefake category, and \SI{19}{\percent} from the \catprompt category.

Within the \cathfake category, the vast majority of events comes from \ttbar simulation:
the \ttbar pair decays in the \emu channel, and hadronic activity originating from an additionally radiated gluon mimics a photon signature in the \ATLAS detector.
Of the \num{77.5} expected events in the \cathfake category, more than \SI{93}{\percent} are of this type.
The \catefake category with \num{23} expected events has its largest contributions from \ttbar, diboson and \ttV simulation.
The latter two types of processes produce final states such as $\plusjets{e\mkern-1.5mu\emu}$, where one of the two electrons is misreconstructed as a photon.
The \num{87.5} expected events in the \catprompt category are dominated by a \SI{68}{\percent} contribution from \ttbar simulation.
Na{\"i}vely, these events would be removed in the sample overlap-removal strategy described in \cref{sec:simulation-overlap}, because they appear signal-like and should be estimated from \tty simulation.
However, these particular events from \ttbar simulation do not fall in the defined overlap region and therefore passed the overlap-removal checks.
This means that either the \pT of the simulated photon is below the defined threshold, or that the simulated photon is too close to a lepton.
In both cases, this type of event is not what is considered \emph{signal} in this analysis, and it should therefore be sorted into the \catprompt category instead.
The remaining events in the \catprompt category are dominated by a \SI{19}{\percent} contribution from \ttV simulation, with smaller fractions coming from \Wy, \Zy and diboson simulation.

\begin{figure}[p]
  \centering
  \includegraphics[width=0.48\textwidth]{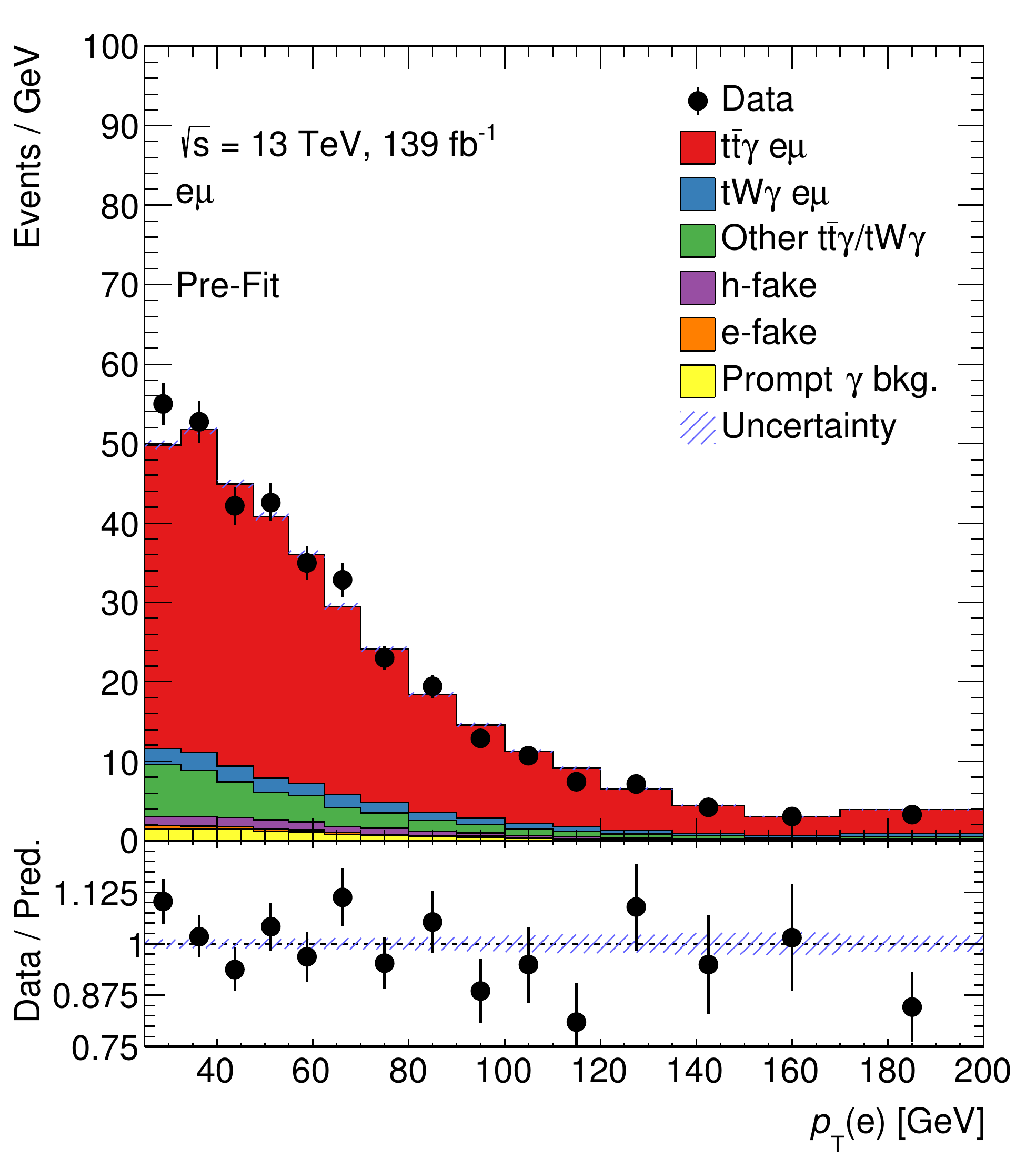}%
  \includegraphics[width=0.48\textwidth]{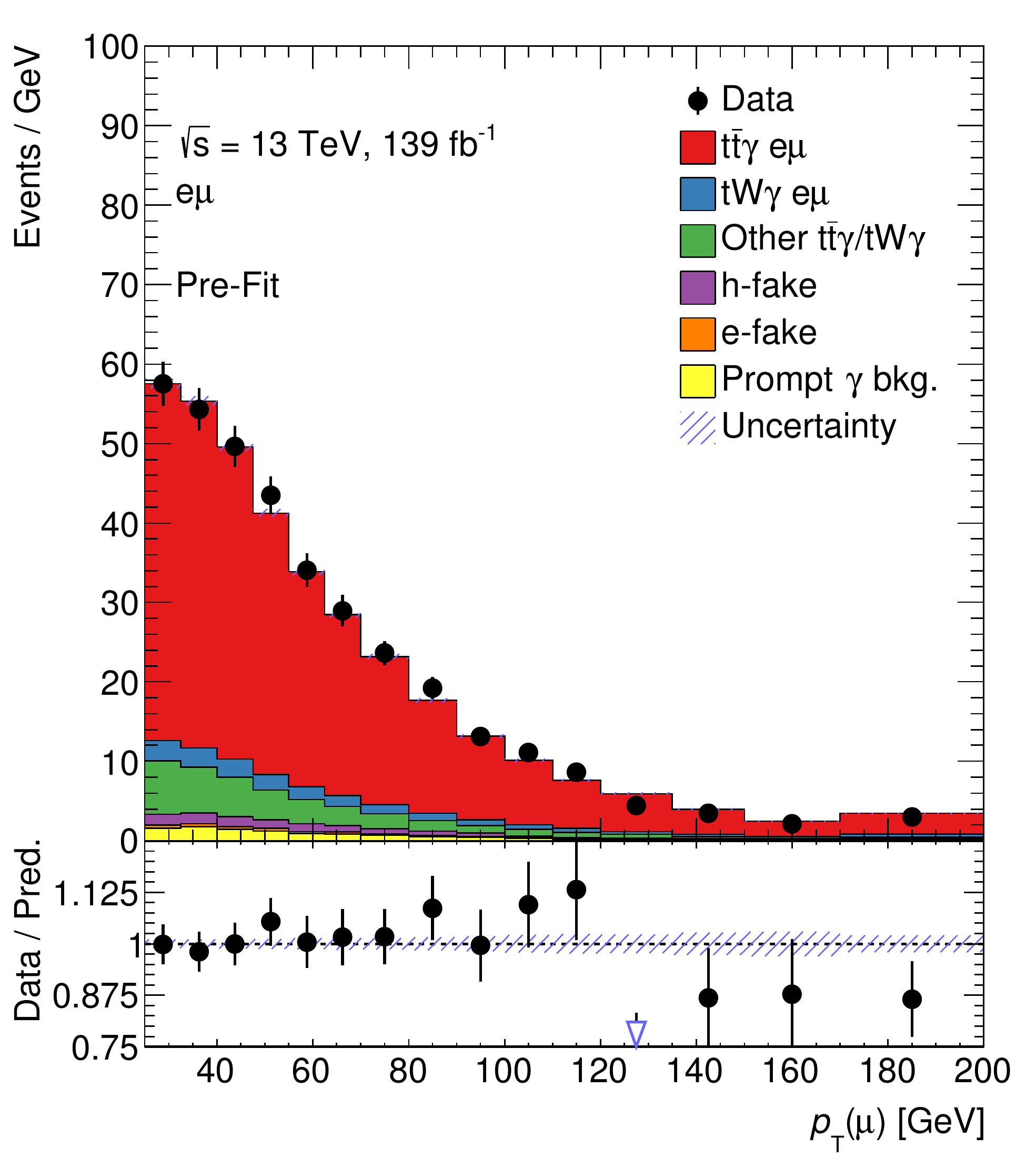}%
  \\
  \includegraphics[width=0.48\textwidth]{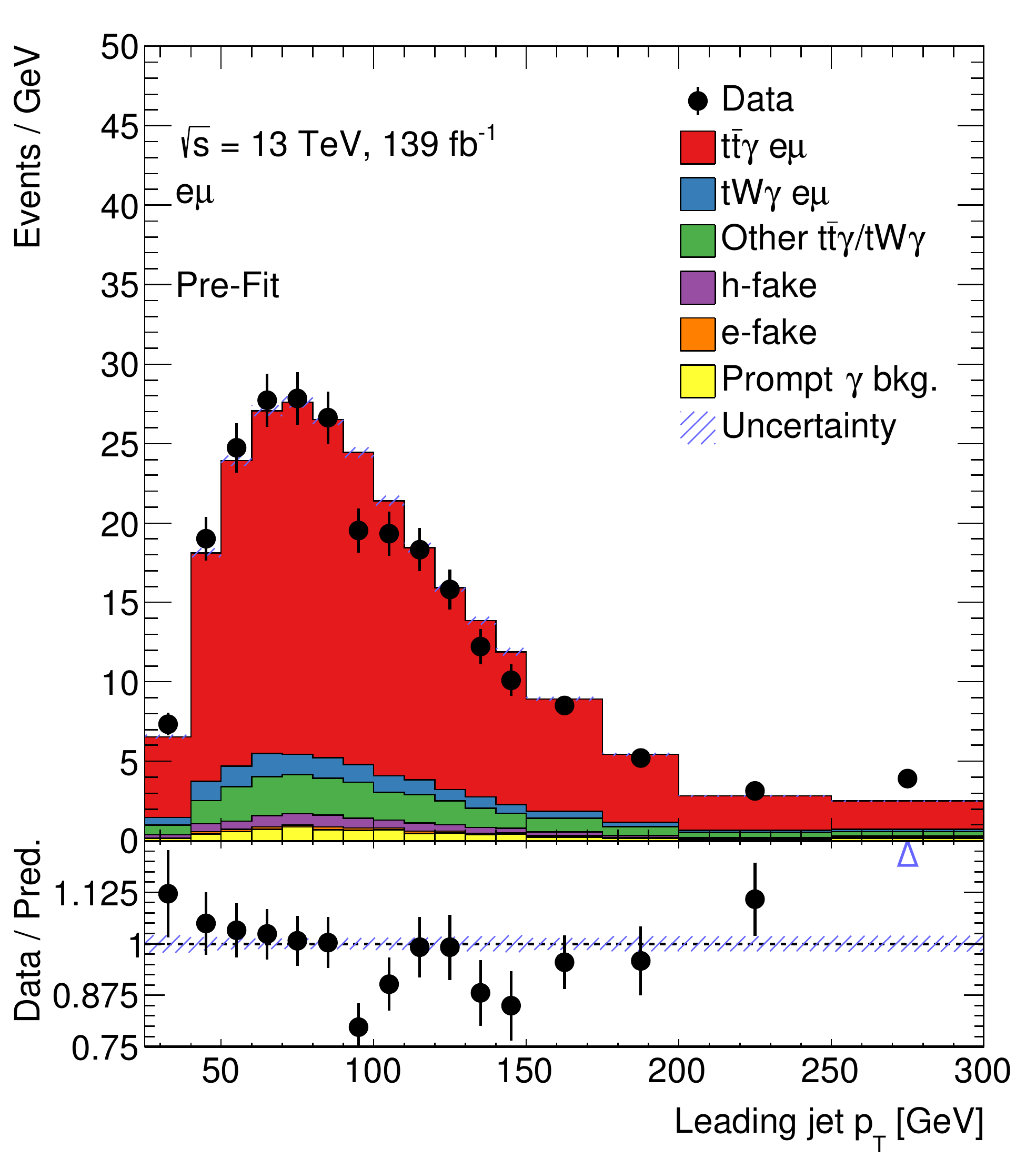}%
  \includegraphics[width=0.48\textwidth]{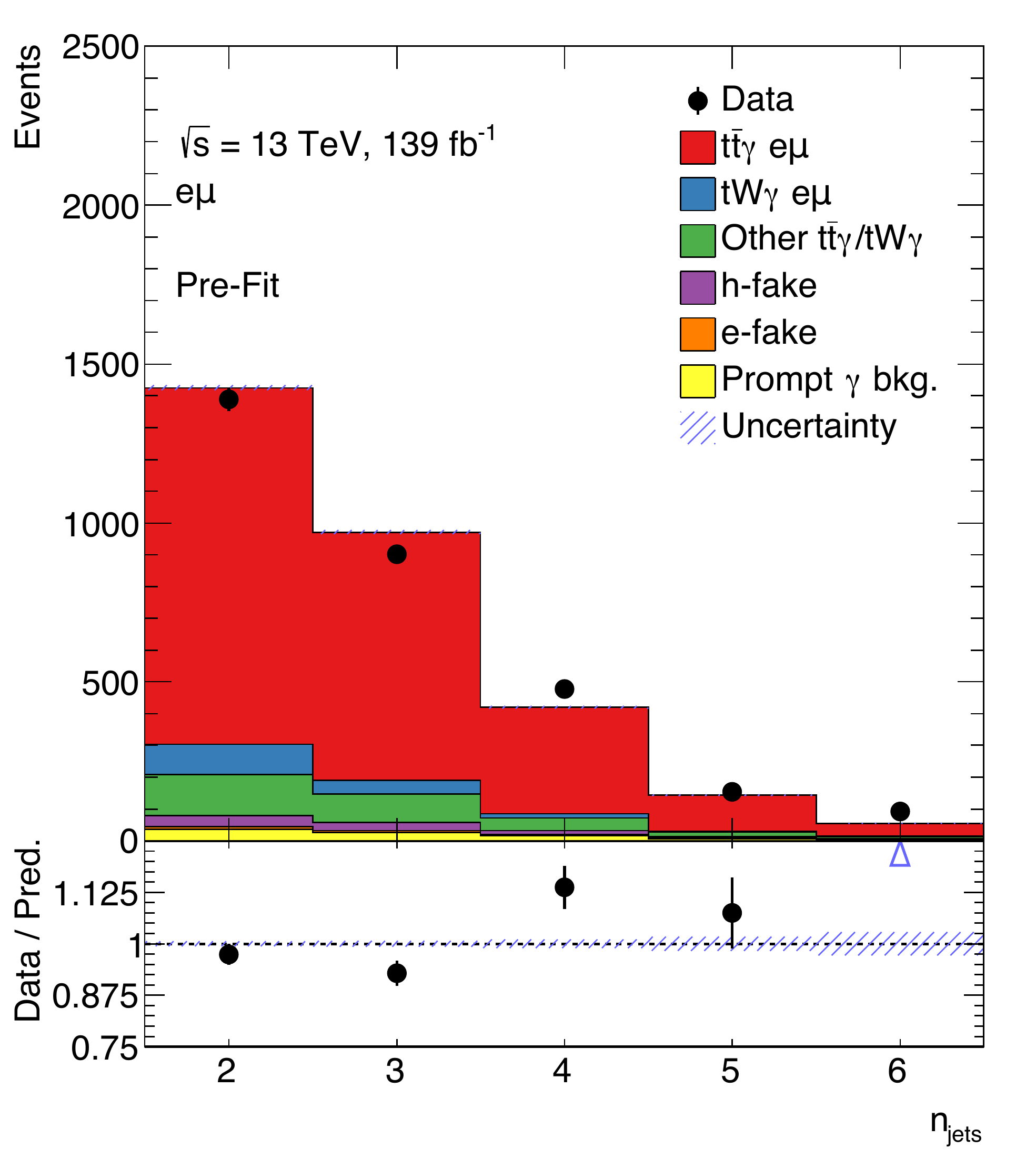}%
  \caption[Control plots with \MC-statistical uncertainties only (1)]{%
    Control plots for a data/\MC comparison in the \emu signal region.
    The shaded error bands of the prediction are \MC-statistical uncertainties only.
    As in \cref{tab:selection-statonly-yields}, the predictions of the \tty and \tWy categories were scaled to match the numbers of reconstructed events in data.
    The shown observables are the transverse momenta of the electron, of the muon and of the leading jet, as well as the jet multiplicity.
  }
  \label{fig:selection-controlplots-1}
\end{figure}

\begin{figure}[p]
  \centering
  \includegraphics[width=0.48\textwidth]{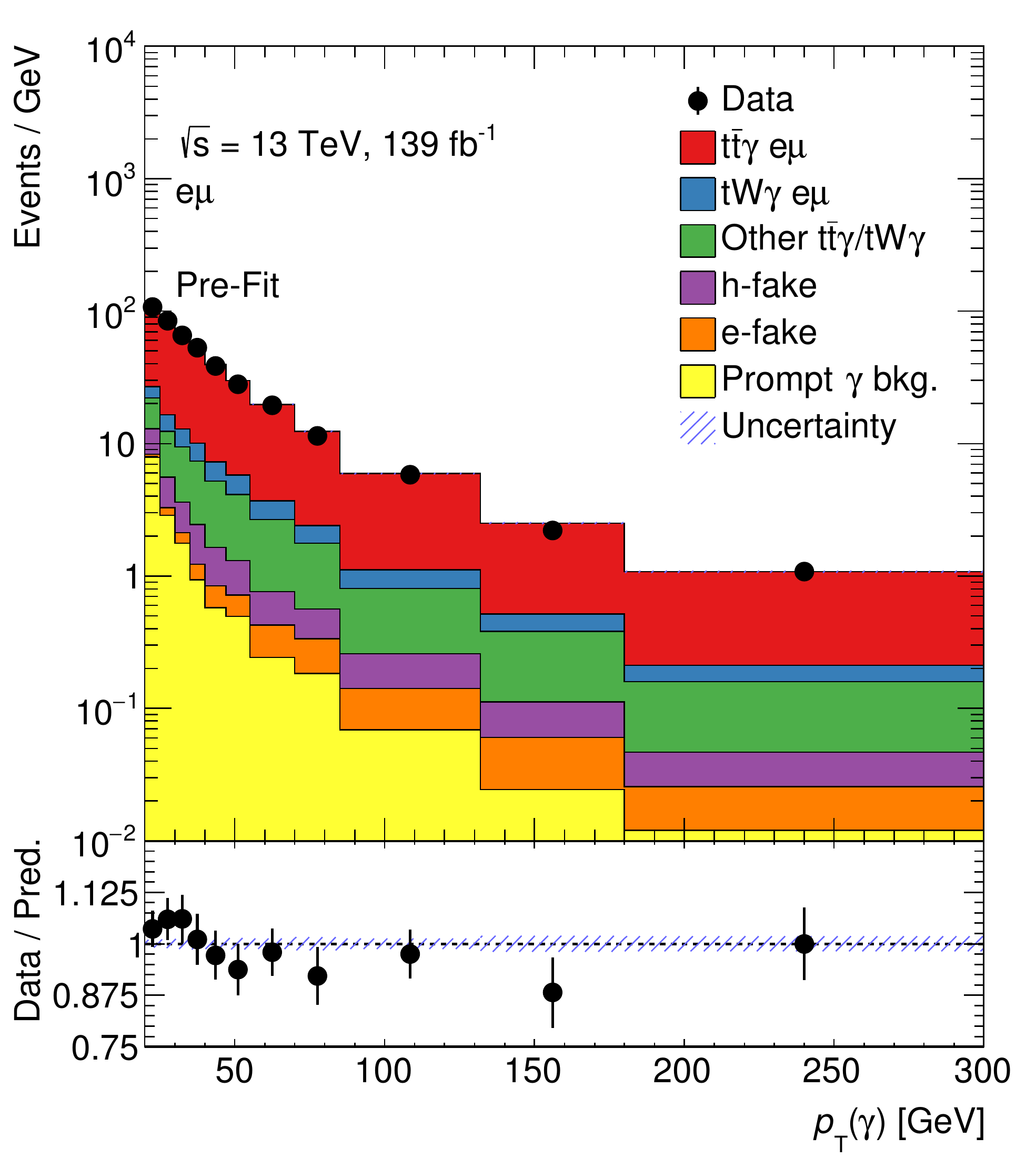}%
  \includegraphics[width=0.48\textwidth]{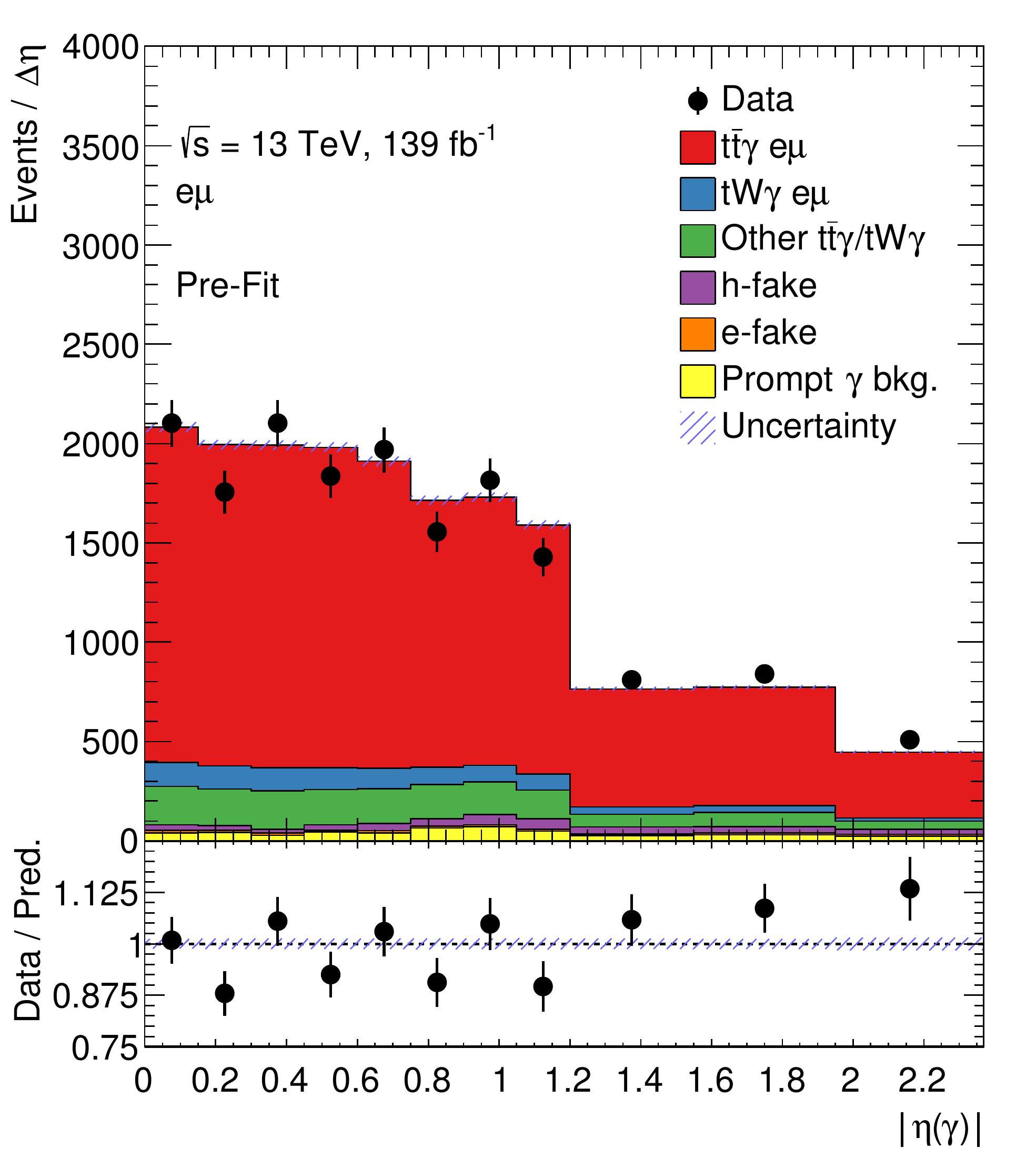}%
  \\
  \includegraphics[width=0.48\textwidth]{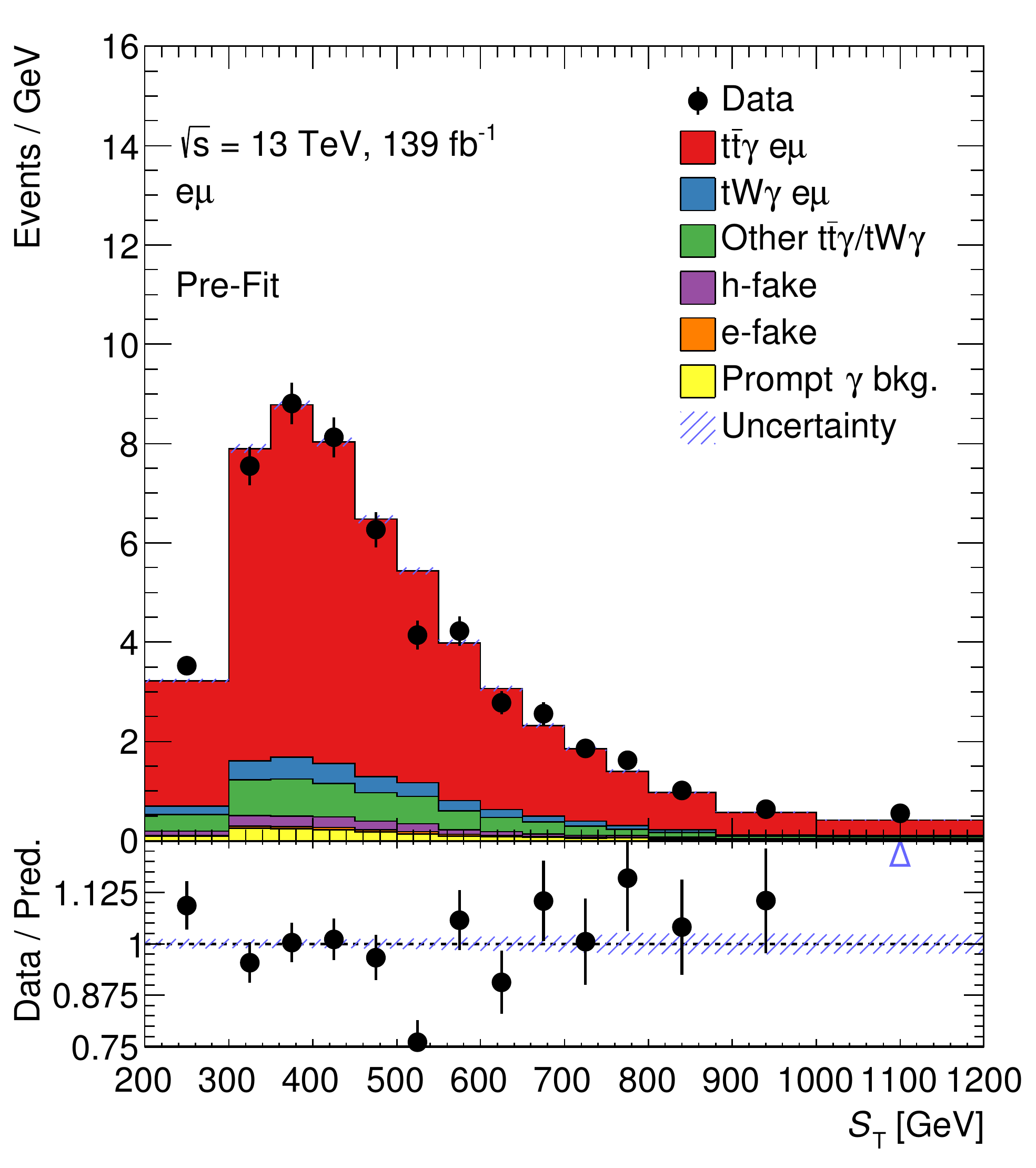}%
  \includegraphics[width=0.48\textwidth]{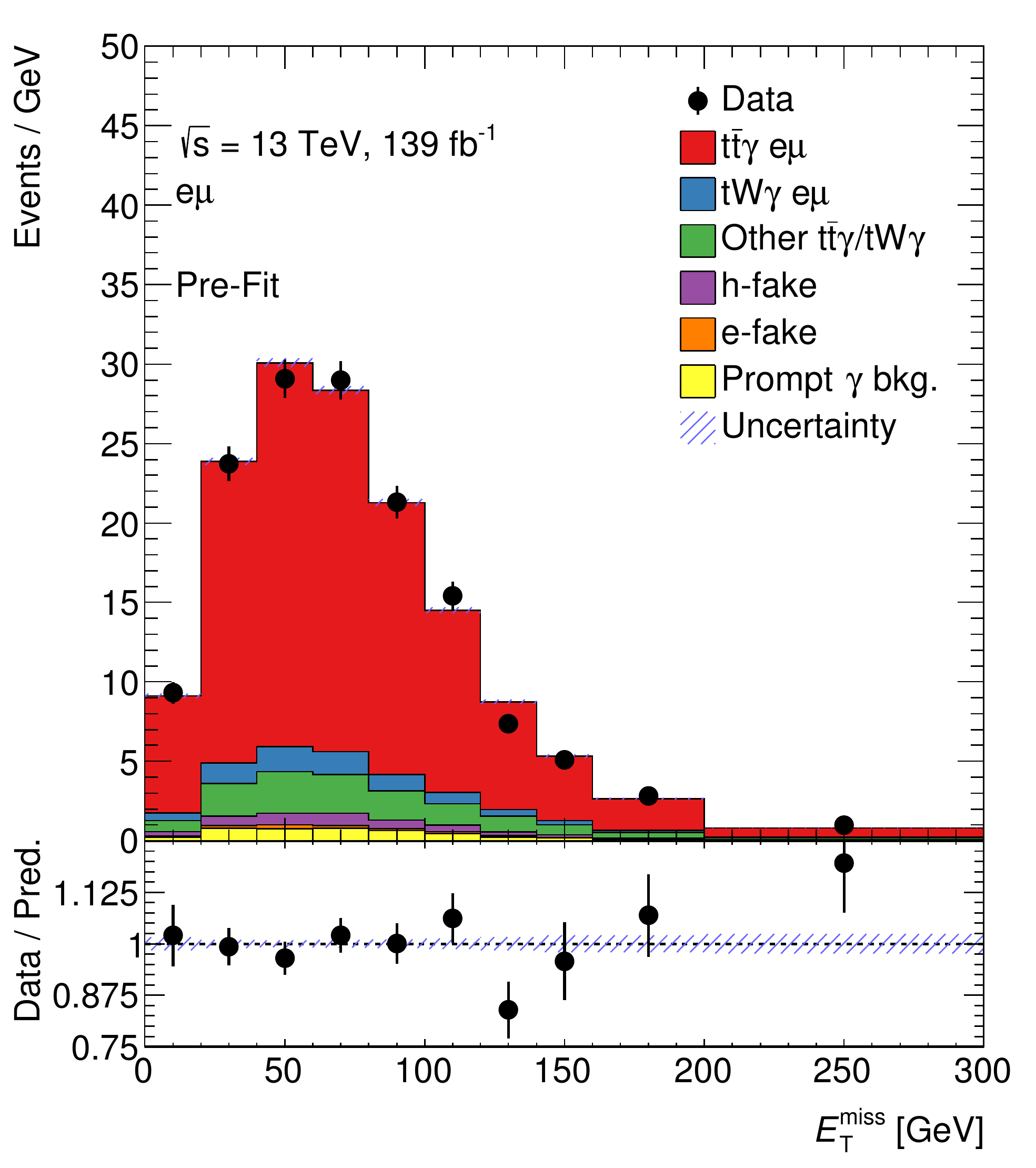}%
  \caption[Control plots with \MC-statistical uncertainties only (2)]{%
    Control plots for a data/\MC comparison in the \emu signal region.
    The shaded error bands of the prediction are \MC-statistical uncertainties only.
    As in \cref{tab:selection-statonly-yields}, the predictions of the \tty and \tWy categories were scaled to match the numbers of reconstructed events in data.
    The shown observables are the transverse momentum and absolute pseudorapidity of the photon, the missing transverse momentum \MET, and the scalar sum \ST of all transverse momenta of the event, including \MET.
  }
  \label{fig:selection-controlplots-2}
\end{figure}

Data/\MC control plots for various observables of the \emu final state are presented in \cref{fig:selection-controlplots-1,fig:selection-controlplots-2}.
\Cref*{fig:selection-controlplots-1} shows the transverse momenta of the electron, of the muon and of the leading jet as well as the jet multiplicity.
\Cref*{fig:selection-controlplots-2} shows the transverse momentum and absolute pseudorapidity of the photon, the missing transverse momentum \MET, and the scalar sum \ST of all transverse momenta of the event.
\ST includes transverse momenta of the electron, of the muon, of the photon, of all jets as well as \MET.
The shaded uncertainty bands represent \MC-statistical uncertainties on the prediction, whereas the black error bars are Poissonian statistical uncertainties of the data points.
As done in \cref{tab:selection-statonly-yields,tab:selection-contributions}, the combined integrals of the \tty and \tWy categories were scaled in such a way that the total integral of the \MC prediction matches the data yields for each plot.
Some of the observables use non-equidistant binning to reduce statistical fluctuations in the less populated areas of the distributions.
To avoid shape distortions of the observables, the histograms do not show the numbers of events per bin, but the bin content is scaled to the inverse of the bin width.
For example, the non-equidistant bins in the transverse momenta distributions show events per \si{\GeV}, not events per bin.
Underflow and overflow events, \ie events below or above the shown observable range, are included in the leftmost and rightmost bins of all distributions, respectively.

Good shape agreement within statistical and \MC-statistical uncertainties is observed for the electron and muon transverse momenta.
The prediction undershoots data in the high-momentum tail of the leading jet \pT distribution for $\pT > \SI{200}{\GeV}$.
Some discrepancies between prediction and data are observed in the 3-jet and 4-jet bins of the jet multiplicity distribution, where the prediction first overshoots, then undershoots data.
The photon transverse momentum and absolute pseudorapidity show good agreement between prediction and data across the entire plotted range.
The \ST and \MET observables show minor discrepancies in the rightmost bins of their distributions, but are otherwise well modelled in \MC simulation when compared to data.


\chapter{Analysis strategy}
\label{cha:strategy}

The previous chapter introduced the \ATLAS dataset and the signal region used to perform the measurement.
The following sections detail the necessary steps and prerequisites to extract a value for the \xsec of the signal process.
As summarised in \cref{chap:simulation}, this measurement uses \LOPS predictions for \tty and \tWy final states in the \emu channel to estimate the production of \tty including off-shell and interference effects, and both \MC simulations combined are treated as the signal of this \xsec measurement.

Many analyses with contributions from fake photons spend significant effort on the estimation of these fake processes using data-driven techniques.
Examples of these are the \tty analysis in the \ljets channels~\cite{TOPQ-2017-14} discussed in \cref{chap:PPT} in the context of photon identification.
For the analysis in the \emu channel presented here, the event yields in \cref{cha:selection} revealed low predicted contributions in the \cathfake and \catefake categories when estimated from \MC simulation directly.
With less than \SI{3}{\percent} and less than \SI{1}{\percent} of the total predicted events, respectively, they only constitute a minor background with little expected impact on the result.
To assess whether data-driven techniques are needed for this analysis, control studies in the \ljets channels were performed, where possible simulation-to-data scale factors for the fake contributions were largely compatible with 1 within uncertainties.
In addition, it was found that such scale factors would have a negligible effect on the shape of all relevant observables, even when determined as a function of the photon transverse momentum and pseudorapidity.
As a consequence, the contributions in the \cathfake and \catefake categories are estimated from \MC simulation directly and only get assigned a very conservative global rate uncertainty of \SI{50}{\percent}.
The uncertainty model is summarised in \cref{chap:systematics}.

\Cref{sec:strategy-fid-phase-space} of this chapter starts by defining the fiducial phase-space volume at parton level, in which the \xsec measurement is performed.
The volume is chosen to be as close as possible to the signal region at the level of reconstruction.
\Cref{sec:strategy-fit} then highlights the idea of likelihood fits and, in particular, of the maximum likelihood estimate for the extraction of the \xsec from \ATLAS data.
It also explains the template method and the profiling technique for the treatment of uncertainties of the measurement.
The section concludes with how the fiducial inclusive \xsec is extracted from the fit results.
Although measurements of differential \xsecs were not the main focus of the author's work, they are included in Ref.~\cite{TOPQ-2020-03} as complementary results to the ones presented here.
\Cref{sec:strategy-differential} briefly summarises the methods behind them.
\Cref{sec:strategy-prediction} introduces the theory prediction used as a reference for this measurement in more detail.

\section{Definition of the fiducial phase space}
\label{sec:strategy-fid-phase-space}

To extract a \xsec value from a measurement that is purely based on reconstructed events, the fiducial phase-space volume at parton level of such a \xsec needs to be defined first.
The volume used for this measurement is constructed as close as possible to the signal region defined in \cref{cha:selection}.
Its selection criteria are looser than those used in the fixed-order theory predictions in Refs.~\cite{Bevilacqua:2018woc,Bevilacqua:2018dny} used as a reference for this measurement, but the authors performed a dedicated recalculation in the volume defined in this section.
More details on the theory prediction are given in \cref{sec:strategy-prediction}.
The tricky part with defining the fiducial volume is that the results of this measurement are evaluated based on the \LOPS \MC simulations of the \tty and \tWy signal processes, which already include modelling the parton shower and hadronisation.
However, the volume needs to be defined at \emph{parton level}, that is, with objects originating from the hard interaction before showering and hadronisation, to be compatible with the fixed-order theory predictions.
This requires a careful extraction of those objects from \MC-truth information of the \LOPS simulations and including those effects that are also modelled at \NLO in \QCD, but excluding anything that goes beyond that.
The physics objects are defined as follows.

Leptons are required to have $\pT > \SI{25}{\GeV}$ and $\abseta < 2.5$.
In addition, as the \LOPS samples also simulate soft emissions of photons in the showering, the leptons are combined with close-by photons in a procedure called \emph{dressing} if the photons lie within a cone of $\Delta R = 0.1$ in the \etaphi plane around the lepton.
Instead of using the bare \bquarks from \MC-truth information, they are reconstructed as \bjets using the \antikt clustering algorithm with a distance parameter of $R = 0.4$.
This is done to include possible soft radiation into the cones around the two original \bquarks.
The \bjets are required to have $\pT > \SI{25}{\GeV}$ and $\abseta < 2.5$, and they must contain a \bquark originating from the decay of one of the top quarks of the hard interaction.
Photons must carry $\ET > \SI{20}{\GeV}$ and be within $\abseta < 2.37$.
In addition, $\Delta R > 0.4$ is required between the two \bjets, between the two charged leptons, and between each of the two \bjets and each of the charged leptons.
These conditions reflect the experimental cuts that are used to define objects within the \ATLAS detector and to select the reconstruction-level signal region of the measurement.

The authors of Refs.~\cite{Bevilacqua:2018woc,Bevilacqua:2018dny} require the photons to be Frixione-isolated~\cite{Frixione:1998jh}.
Instead of placing a hard cut on $\Delta R$ between photons and close-by objects, this implements a soft, smooth isolation, which is also required for the photons in the fiducial volume to align the definitions in the measurement with those in the theory prediction.
For a given maximum value of $R_0 = 0.4$, the isolation of the photon is tested at all points $R \leq R_0$.
For each of them, the sum of transverse momenta of charged leptons and jets contained in a cone of radius $R$ in the \etaphi plane around the photon must be below a predefined threshold.
Using the Heaviside function $\Theta(R - \Delta R_{\gamma i})$, the Frixione isolation criteria can be denoted as
\vspace*{0pt minus 2pt}  
\begin{align}
  \label{eq:strategy-Frixione}
  \sum_i \pT(i) \, \Theta (R - \Delta R_{\gamma i}) \leq \ET(\gamma) \left( \frac{1 - \cos(R)}{1 - \cos(R_0)} \right) \, ,
\end{align}
which must be fulfilled for all $R \leq R_0$ and for all charged leptons and clustered jets $i$ with transverse momentum $\pT(i)$ and with distance $\Delta R_{\gamma i}$ to the photon.

For an event to pass the selection for the fiducial volume, exactly one electron and one muon with the above definitions are required.
Electrons and muons not originating from the top quarks directly, but from intermediate \taulepton decays, are rejected based on \MC-truth information.
This is done to be in alignment with the phase space calculated in the theory computation.
In addition, the event must have two \bjets, each of which must contain either the \bquark or the anti-\bquark that originated from the hard interaction.
Exactly one Frixione-isolated photon is required.

After selection, the total number of signal events in the fiducial volume is predicted to be \num{3913.2}.
This corresponds to a fiducial acceptance, that is, the fraction of total generated $\emu\gamma$~events that enter into the fiducial volume, of
\vspace*{0pt minus 2pt}  
\begin{align}
  \label{eq:strategy-acceptance}
  A_{\text{fid}} = \SI{13.07}{\percent} \, .
\end{align}
The number of events in the fiducial volume at parton level can be compared to that in the reconstruction-level signal region for the \catttyemu and \cattWyemu signal categories.
Without the scaling applied in the previous chapter in \cref{tab:selection-statonly-yields}, the reconstruction-level selection yields \num{1808.6} signal events.
Based on \MC-truth information, fiducial volume and reconstruction level can be matched on an event-by-event basis to understand the differences in the event numbers.
One figure used for that is the signal efficiency $\epsilon$, that is, the number of reconstructed events passing the fiducial-volume cuts over the total number of events in the fiducial volume.
Another one is the fraction $\fout$ of events that migrated into the reconstruction-level selection from outside the fiducial volume, in other words, the fraction of reconstructed events that did \emph{not} pass the fiducial cuts.
The values are found to be:
\vspace*{0pt minus 2pt}  
\begin{align}
  \label{eq:strategy-efficiency-migration}
  \epsilon &= \frac{N(\text{reco}\,|\,\text{fid})}{N(\text{fid})} = \SI{29.73}{\percent} &
  \fout &= \frac{N(\text{reco}\,|\,!\,\text{fid})}{N(\text{reco})} = \SI{35.67}{\percent}
\end{align}
with $N(\text{reco})$ the number of reconstruction-level events and $N(\text{fid})$ the number of events in the fiducial volume.
The conditional notation $N(\text{reco}\,|\,\text{fid})$ refers to those events from the fiducial phase-space volume that were reconstructed.
The two total numbers, $N(\text{reco})$ and $N(\text{fid})$, at reconstruction and parton level can be related by calculating the ratio between the efficiency $\epsilon$ and the fraction of \emph{non}-migrated events, $1 - \fout$:
\vspace*{0pt minus 2pt}  
\begin{align}
  \label{eq:strategy-correction-factor}
  C \equiv \frac{\epsilon}{1 - \fout} = \frac{N(\text{reco})}{N(\text{fid})} = 0.4622 \, ,
\end{align}
where $C$~is known as the fiducial \emph{correction factor} that relates the number of signal events on reconstruction level to the number of events in the fiducial volume.
The first equivalence merely shows how the correction factor is constructed, but it is calculated from the two event counts directly.
After performing a maximum likelihood fit, as detailed in the following section, the value quoted in \cref{eq:strategy-correction-factor} is used to calculate the observed number of events in the fiducial volume from the reconstruction-level \ATLAS data.

\section{Extraction of the \xsec value}
\label{sec:strategy-fit}

The determination of a \xsec value from a measured set of particle physics data is a highly non-trivial process due to the many parameters involved.
The default strategy is to perform an estimation of the parameters from an observed data distribution (or distributions) during the \emph{fitting} process.
A good estimator \emph{\estimate{\theta}} of a parameter ought to be \emph{consistent}, \emph{unbiased} and \emph{efficient}.
In other words, it should converge to the true value $\theta$ as the number of data points $N$ increases, the expectation value of the estimator should equal the true value $\theta$, and the estimator's variance should be small (ideally fulfilling the Cram{\'e}r-Rao minimum-variance bound~\cite{Cramer:1946aa,Darmois:1945aa,Frechet:1943aa,Rao:1945aa}).
In practice, the model involves a whole set of parameters $\vec{\theta}$, all of which need to be estimated through a set of estimators $\vec{{\estimate{\theta}}}$ in one go.
The method used in this measurement is the maximum-likelihood estimate, briefly summarised in the following paragraphs.
Systematic uncertainties of the measurement, which are introduced in detail in \cref{chap:systematics}, are included in the fit directly using the template method and the profile likelihood technique.
This provides a coherent statistical interpretation of the result and allows an estimate of the result's uncertainties.
The implementation of the maximum-likelihood estimate, the template method and the profiling is done with the \HistFactory{}~\cite{Cranmer:2012sba} software, which utilises the \textsc{roofit}~\cite{Verkerke:2003ir} and \textsc{roostats}~\cite{Moneta:2010pm} libraries of the software framework \textsc{root}~\cite{Brun:1997pa}.

The general idea of the parameter estimation is that a measurement consists of $N$ statistically independent data points $\{x_i\}$, and that each of these data points follows a (unknown) probability density function $P(x | \vec{\theta})$ described through a set of model parameters $\vec{\theta}$ that are to be determined.
In the case of unbinned data, the joint probability density function for the set $\vec{x} = \{x_i\}$ is then given by the likelihood function
\vspace*{0pt minus 5pt}  
\begin{align}
  \label{eq:strategy-likelihood}
  L(\vec{x} | \vec{\theta}) = \prod_{i=1}^N P(x_i | \vec{\theta}) \, .
\end{align}
With all data points $\{x_i\}$ already measured, the maximum likelihood estimate for the parameters $\vec{\theta}$ is then the set of values $\vec{\estimate{\theta}}$, for which the likelihood in \cref{eq:strategy-likelihood} reaches its global maximum.
This, however, does not consider statistical uncertainties of the absolute rate of the data points.
Repeating the measurement with identical conditions would show that the number of observed events $N$ fluctuates according to a Poisson distribution around the (unknown) expectation value~$\nu$.
Incorporating this Poisson term leads to the \emph{extended} maximum likelihood estimate.
In practice, for numerical reasons, it is common to minimise the negative logarithm of the likelihood instead of trying to find the global maximum.
When combining \cref{eq:strategy-likelihood} with the Poisson term, the extended log-likelihood is
\vspace*{0pt minus 5pt}  
\begin{align}
  \label{eq:strategy-likelihood-extended}
  \ln L (\vec{x} | \nu, \vec{\theta}) = \sum_{i=1}^N \ln P(x_i | \vec{\theta}) + N \ln \nu - \nu \, ,
\end{align}
where all constant terms have already been dropped.

The \xsec of a particle physics process can be determined by comparing the observed process rate in data to the predicted rate from \MC simulation.
However, only in the rarest cases the data is pure and only sampled from one single probability density distribution $P(x_i | \vec{\theta})$.
In the measurement presented here, \MC simulations predict contributions from several physics processes in the selected region, as listed in \cref{tab:selection-statonly-yields}, each of which follow slightly different distributions.
Differences in these distributions, in particular between those of signal-like and background-like events, $P_S(x_i|\vec{\theta})$ and $P_B(x_i|\vec{\theta})$, can be exploited to determine their fractions in the dataset $\{x_i\}$.
To parameterise the fraction of signal-like events in the model, the signal strength $\mu$ is introduced, which is the \emph{parameter of interest} of the measurement.
$\mu$~scales the rate of the signal prediction, that is, $\mu = 0$ corresponds to the background-only hypothesis, and $\mu = 1$ corresponds to the signal-plus-background hypothesis with the strength of the signal as predicted by \MC simulation.
Thus, the total number of reconstructed events in a measurement is $N = \mu \, S + B$, with $S$ and $B$ the numbers of predicted signal and background events, respectively.
The other parameters $\vec{\theta}$ of the model are then usually referred to as \emph{nuisance} parameters.
Replacing the expectation value of \cref{eq:strategy-likelihood-extended} with $\mu \, S + B$, leads to the extended log-likelihood~\cite{Cranmer:2012sba}
\vspace*{0pt plus 2pt}  
\begin{align}
  \label{eq:strategy-likelihood-extended2}
  \ln L (\vec{x} | \mu, \vec{\theta}) = \sum_{i=1}^N \ln \left[   \mu \, S \, P_S(x_i | \vec{\theta}) + B \, P_B(x_i | \vec{\theta}) \right] - (\mu \, S + B) \, ,
\end{align}
which now additionally depends on the signal strength.
For large numbers of data points $N$, it is common to bin the data to increase computational efficiency.
Each bin $b$ of the distribution is then identified with the number of events $N_b$ of that bin, and the probability density functions $P_S(x_i|\vec{\theta})$ and $P_B(x_i|\vec{\theta})$ are replaced with bin-wise probabilities.
It can be shown that the binned, extended log-likelihood can simply be expressed as
\vspace*{0pt plus 2pt}  
\begin{align}
  \label{eq:strategy-binned-likelihood}
  L(N | \mu, \vec{\theta}) = \prod_{b \mkern3mu \in \text{bins}} \text{Poisson} \left( N_b \, \left| \, \mu \, \nu_b^S(S, \vec{\theta}) + \nu_b^B(B, \vec{\theta}) \right.\right) \, ,
\end{align}
that is, as the product of Poisson probabilities to observe $N_b$ events in bin $b$.
$\nu_b^S$ and $\nu_b^B$ are the expected numbers of events for bin $b$ for signal-like and background-like events, respectively, and they both depend on the nuisance parameters $\vec{\theta}$.

The \HistFactory software allows the inclusion of nuisance parameters in various ways.
Generally, there are three different types:
firstly, the prediction for each bin is affected by statistical uncertainties of the \MC simulation.
The corresponding nuisance parameters $\gamma_{bp}$ are unique for each bin~$b$ and uncorrelated, but they are constrained through the overall rate of the process~$p$.
To avoid a larger number of $\gamma_{bp}$-type nuisance parameters in the fit model, the statistical bin-by-bin uncertainties are evaluated for the sum of all processes $\{p_i\}$ only -- which is a good approximation unless the different \MC simulations come with very different statistics.
Secondly, changes of the overall rate of a process $p$ are parameterised through unconstrained nuisance parameters $\phi_p$, like the signal strength $\mu$ for the signal process.
These are used if the overall rate is unknown and to be determined during the fit.
Thirdly, nuisance parameters reflecting systematic uncertainties of the predictions usually change both rate and shape of the distribution simultaneously, but are constrained by an auxiliary measurement or some prior knowledge.
Technically, they are split in normalisation-only variations $\eta_p(\vec{\theta})$ and shape-only variations $\sigma_{pb}(\vec{\theta})$, the latter of which affect each bin $b$ differently, and these two variations are correlated in the fitting procedure.
Combining all of these types, the expected number of events for bin $b$ for a process $p$ can be written as:
\vspace*{0pt plus 2pt}  
\begin{align}
  \label{eq:strategy-bin-expectation}
  \nu_{bp} (\gamma_b, \phi_p, \vec{\theta}) &= \gamma_b \phi_p(\vec{\theta}) \eta_p(\vec{\theta}) \sigma_{bp}(\vec{\theta}) \, .
\end{align}
The systematic nuisance parameters $\vec{\theta}$ are treated as free parameters of the fit, but the binned, extended log-likelihood in \cref{eq:strategy-binned-likelihood} must be augmented with additional terms for each parameter that constrain them.
These can either be auxiliary measurements with high sensitivity to the parameters, or one may add prior probability density functions that limit the parameter range.

In the latter approach, which is used here, the prior distributions are constructed using the \emph{template method}:
systematic uncertainties of the measurement are estimated from alternative \MC models, either from independently sampled sets of data points or through reweighting the nominal set of data points.
These alternative models yield histograms of the fitted distribution referred to as \emph{templates}.
The \ATLAS uncertainty model provides either one or two alternative models in addition to the nominal prediction, thus, resulting in two-point or three-point systematic uncertainties.
The priors of the parameters $\vec{\theta}$ are constructed in such a way that the alternative models correspond to the one-standard-deviation prediction of a Gaussian prior, and the nominal prediction is the central value.
\HistFactory employs exponential interpolation for all normalisation variations $\eta(\vec{\theta})$ to create continuous distributions.
Piece-wise linear interpolation is used for the shape-only bin-by-bin variations $\sigma_b(\vec{\theta})$.

\begin{figure}
  \centering
  \includegraphics[%
  width=0.46\textwidth,clip=true,trim=0 100pt 0 0]{%
    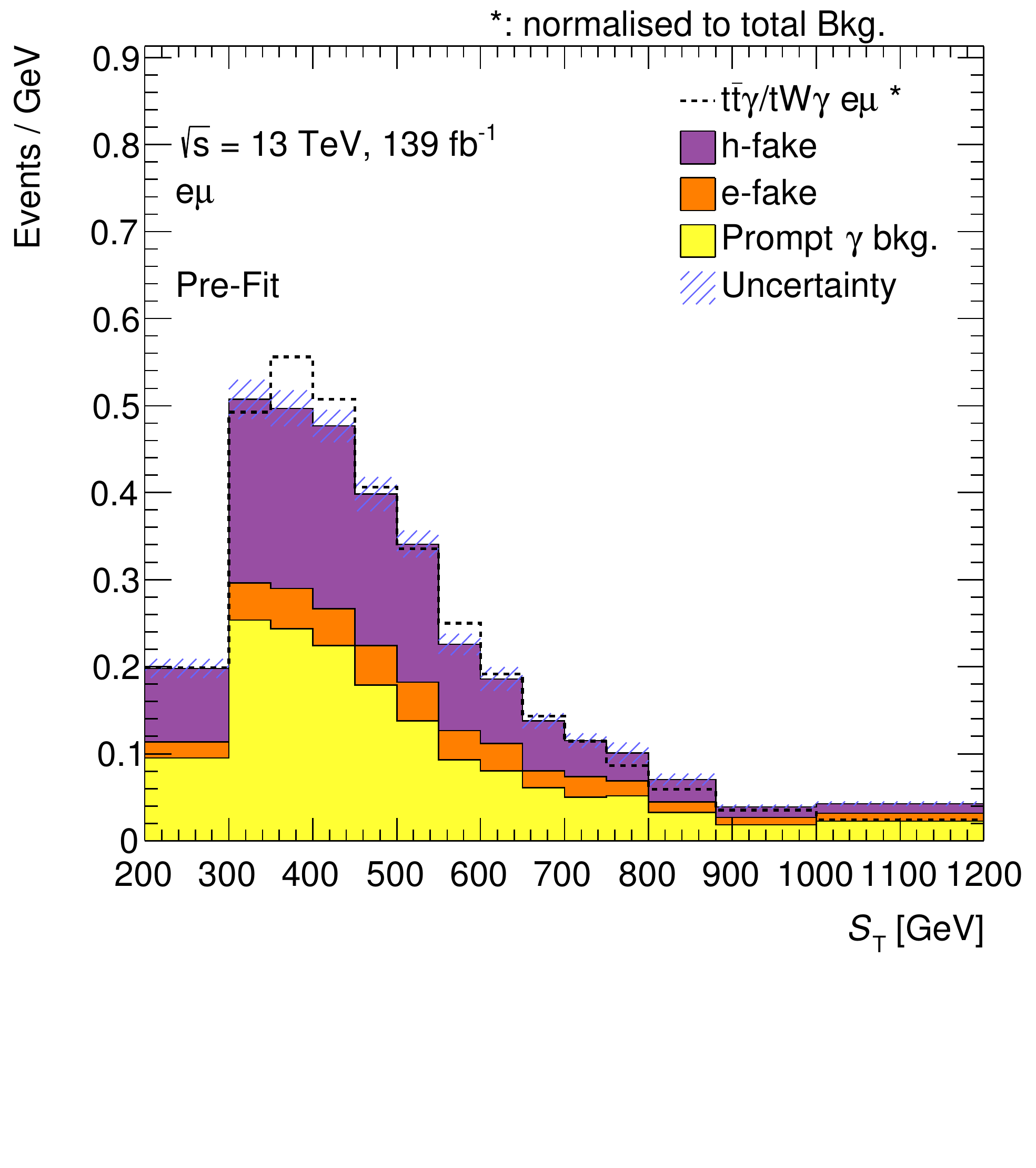}
  \includegraphics[%
  width=0.46\textwidth,clip=true,trim=0 100pt 0 0]{%
    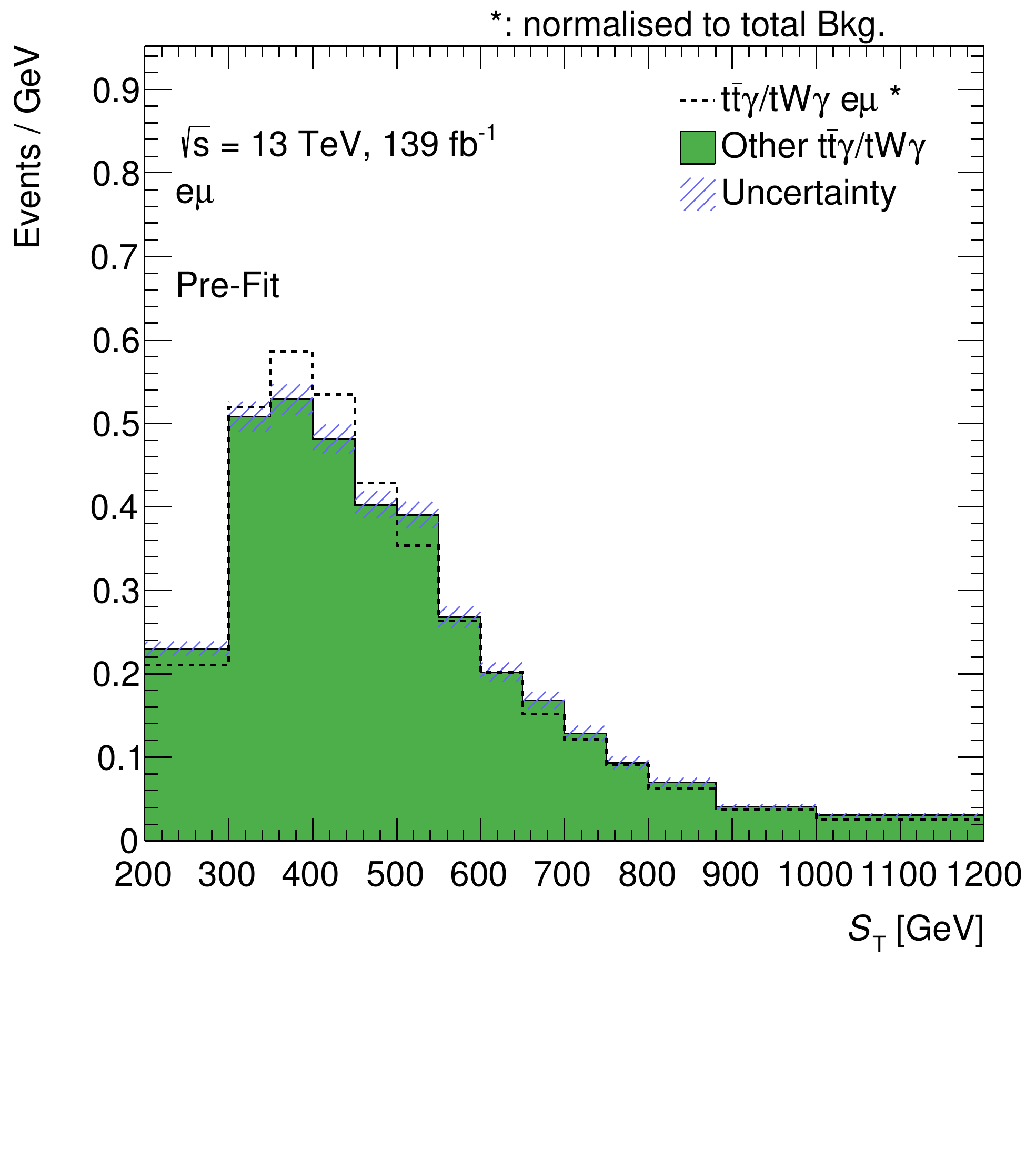}
  \caption[Shape comparison of signal and background categories]{%
    Shape comparison for the combined \tty/\tWy \emu signal categories against (1) the \cathfake, \catefake and \catprompt background categories, and (2) the \catother category.
    The dashed lines represent the shape of the signal and they are normalised to the integrals of the combined background categories and of the \catother category in the two plots, respectively.
  }
  \label{fig:strategy-signal-bkgd-shape}
  \vspace*{-3pt}  
\end{figure}

The measurement of the \tty \xsec presented in this thesis is based on the minimisation of the binned extended log-likelihood of \cref{eq:strategy-binned-likelihood} with additional Gaussian prior constraints of the nuisance parameters.
The distribution used to discriminate signal-like and background-like events and to extract the signal strength $\mu$ is the \ST distribution as introduced before:
\ST is the scalar sum of all transverse momenta of the event, which includes the transverse momenta of the electron, of the muon, of the photon, of all jets, and \MET.
\Cref{fig:strategy-signal-bkgd-shape} shows two shape comparisons in the \ST distribution:
on the left-hand side, the combined signal categories are compared with the \cathfake, \catefake and \catprompt background categories;
on the right-hand side, the combined signal categories are compared with the \catother category.
The dashed lines represent the combined spectrum of the \catttyemu and \cattWyemu categories, normalised to the integral of the spectra they are compared with, respectively.
Differences to the background categories are observed in the most populated areas of the spectrum around \SI{400}{\GeV} and around the tail of the distribution for $\ST > \SI{1000}{\GeV}$.
The comparison with the \catother category reveals shape differences in the bins up to $\ST < \SI{550}{\GeV}$.

To reduce the complexity of the fit of the \ST distribution with some hundred nuisance parameters, the \emph{profiling} technique is used.
The parameters of the systematic uncertainties are profiled as functions of the parameter of interest, that is, $\vec{\theta} \to \vec{\theta}(\mu)$.
This is done by calculating the maximum likelihood estimate $L(n_b | \mu, \mkern3mu \estimate{ \estimate{ \mkern-3mu \vec{\theta}}} (\mu))$ of the parameters~$\vec{\theta}(\mu)$ for different values of $\mu$, also known as the profile likelihood estimate for $\vec{\theta}(\mu)$.
The ratio of this likelihood and the global maximum likelihood estimate $L(n_b | \estimate{\mu}, \estimate{\vec{\theta}})$, without constraining $\mu$ to a fixed value, is then a \emph{profile likelihood} to estimate the effect of the nuisance parameters on $\mu$, and it is used to evaluate uncertainties on the signal strength.
\HistFactory and its underlying packages use the \textsc{migrad} and \textsc{minos} minimisation techniques implemented in the \textsc{minuit} framework~\cite{James:1975dr} to perform the minimisation.

Once the negative profile log-likelihood is minimised, the fit yields estimates of the signal strength and its uncertainties.
This takes possible correlations of the nuisance parameters, calculated during the profiling, into account.
The post-fit values for $\mu$ and $\vec{\theta}$ can then be used to calculate post-fit event yields for the signal and background categories, from which the number of signal events $N_{\text{meas}}(\text{fid})$ in the fiducial volume is extracted using \cref{eq:strategy-correction-factor}.
The fiducial \xsec is calculated using the correction factor~$C$ and the well-known relation between \xsec and luminosity $\mathcal{L}$
\begin{align}
  \label{eq:strategy-fidxsec-prelim}
  \sigma^{\text{fid}}_{\text{meas}} =
  \frac{N_{\text{meas}}(\text{fid})}{\mathcal{L}} =
  \frac{N(\text{data}) - N_{\text{meas}}(\text{non-\emu signal}) - N_{\text{meas}}(\text{bkgd})}{\mathcal{L} \cdot C} \, ,
\end{align}
where $N(\text{data})$ are the event yields in data.
$N_{\text{meas}}(\text{non-\emu signal})$ and $N_{\text{meas}}(\text{bkgd})$ are the post-fit predicted event yields for the \catother and all other background categories, respectively.
However, the \catother category of the measurement is simulated with the same \MC samples as the \catttyemu and \cattWyemu signal categories and is, thus, also scaled with the signal strength~$\mu$ in the fit.
To avoid a direct dependency on the signal strength and the \LO \xsec associated with the \tty and \tWy simulations, one can rearrange \cref{eq:strategy-fidxsec-prelim} to contain the fraction $f_{\emu}$ of \emu events in all signal events.
The resulting formula for the \xsec,
\begin{align}
  \label{eq:strategy-fidxsec-final}
  \sigma^{\text{fid}}_{\text{meas}} = \frac{N(\text{data}) - N_{\text{meas}}(\text{bkgd})}{\mathcal{L} \times C} \times f_{\emu}
  \quad \text{with}~~ f_{\emu} = \frac{N_{\text{meas}}(\text{\emu signal})}{N_{\text{meas}}(\text{all signal})} \, ,
\end{align}
now only contains fractions of \tty and \tWy contributions:
$N_{\text{meas}}(\text{all signal})$ are the combined post-fit predicted event yields for the \catttyemu, \cattWyemu and \catother categories.

\section{Measurement of differential distributions}
\label{sec:strategy-differential}

The idea of taking reconstructed \ATLAS data and comparing it with predictions in a fiducial volume can be extended to entire distributions -- yielding differential \xsec values with respect to a chosen observable.
Although the approach is similar to what is done with post-fit event yields to obtain a fiducial inclusive \xsec value, it requires a few more steps and comes with additional caveats, for example necessary tests of the method's stability.
In order to compare observable distributions of \ATLAS data to theory predictions, such as those made in Refs.~\cite{Bevilacqua:2018woc,Bevilacqua:2018dny}, the data distributions need to be corrected to fiducial parton level using a technique known as \emph{unfolding}:
any effects originating from instrumental aspects of the measurement need to be removed, such as limitations of the detector acceptance, smearing due to limited detector resolution etc.
With all such influences gone, the data can be compared not only to theory predictions, but also to (unfolded) measurements of other experiments.
Differential measurements were not the main focus of the author's work, but they provide valuable complementary results to the fiducial inclusive measurement.
Therefore, their strategy used in Ref.~\cite{TOPQ-2020-03} is summarised briefly in the following paragraphs.

\begin{figure}
  \centering
  \includegraphics[width=0.49\textwidth]{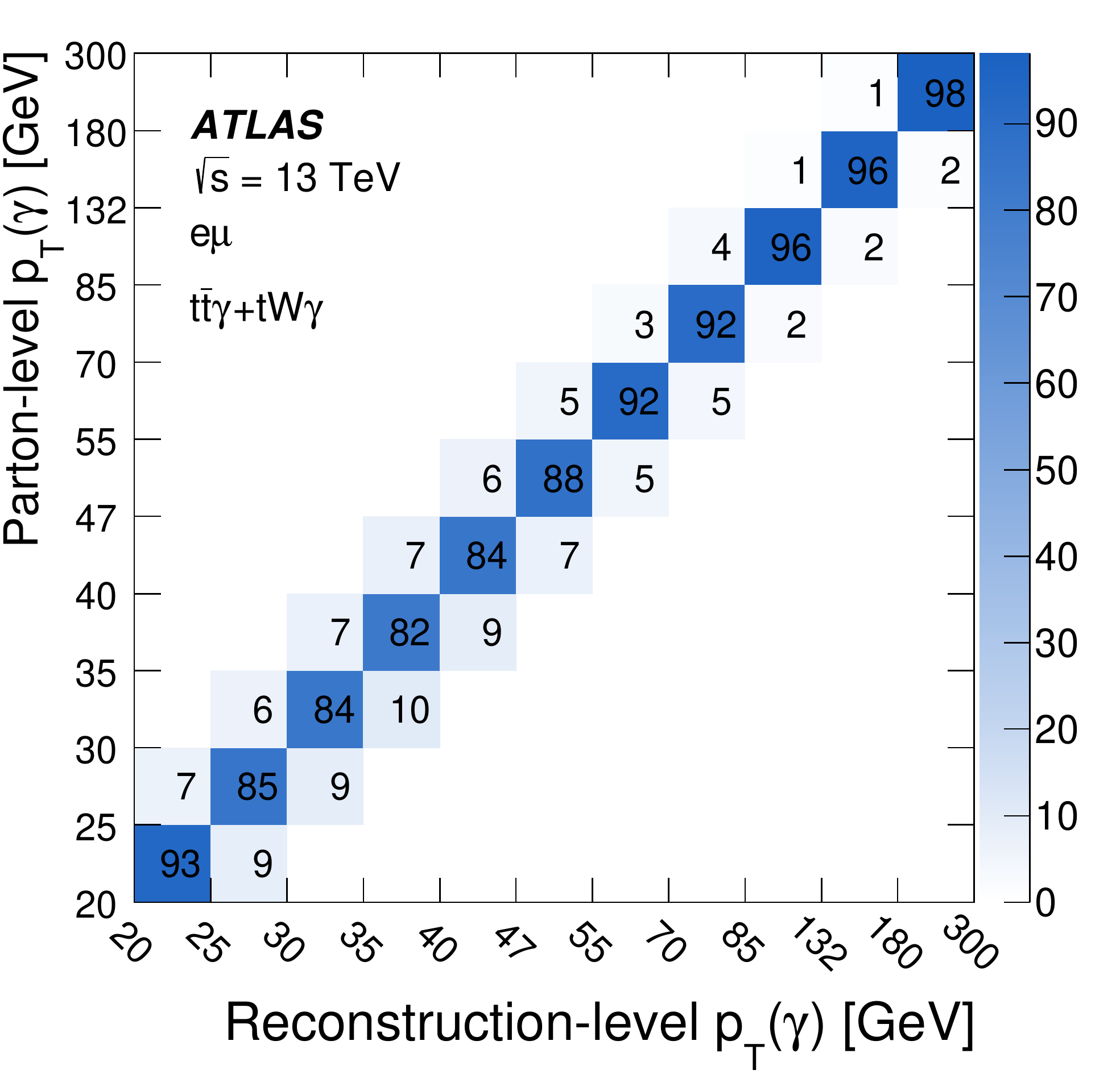}
  \includegraphics[width=0.49\textwidth, trim=0 10pt 0 -10pt,clip=true]{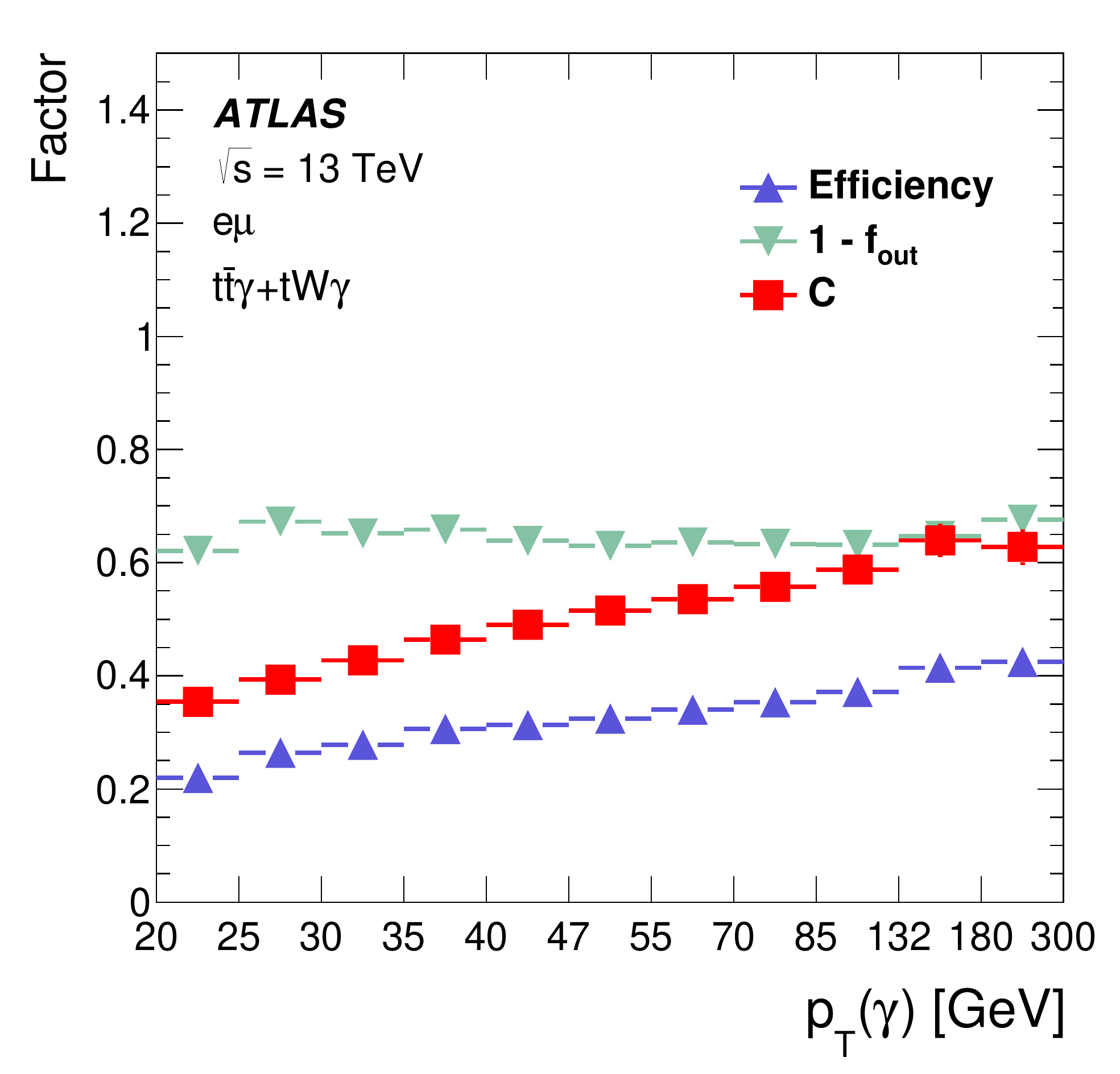}
  \caption[Migration matrix, migration fraction \fout and efficiency~$\epsilon$ of $\pT(\gamma)$]{%
    On the left: migration matrix for the photon transverse momentum that shows migration effects from parton-level bins to various bins at reconstruction level.
    The values are normalised per column and shown in percentages.
    On the right: migration fraction \fout, signal efficiency $\epsilon$ and correction factor~$C$ for the bins of the photon \pT distribution.
    Figures taken from Ref.~\cite{TOPQ-2020-03}.
  }
  \vspace*{3pt} 
  \label{fig:strategy-diff-matrix-effmig}
\end{figure}

The groundwork for unfolding was already laid in the descriptions of the fiducial phase-space volume in \cref{sec:strategy-fid-phase-space}:
by comparing the fiducial parton level with the reconstruction level in \MC simulations of signal-like events, the relations between the two levels can be evaluated.
However, instead of using the overall event numbers, the comparisons are done on a bin-by-bin basis for differential measurements.
In correspondence to the definitions given in \cref{eq:strategy-efficiency-migration}, signal efficiencies for each parton-level bin~$k$ and migration fractions for each reconstruction-level bin~$j$ can be defined in the following way:
\vspace*{0pt minus 2pt}  
\begin{align}
  \label{eq:strategy-diff-eff-mig}
  \epsilon_k &= \frac{\sum_j N_j(\text{reco}\,|\,\text{fid.\,bin\,$k$})}{N_k(\text{fid})} &
  f_{\text{out},j} &= \frac{N_j(\text{reco}\,|\,!\,\text{fid})}{N_j(\text{reco})} \, ,
\end{align}
that is, the efficiency $\epsilon_k$ is the fraction of events in bin~$k$ of the parton-level distribution that enters one of the reconstruction-level bins, and $f_{\text{out},j}$ is the fraction of events in bin~$j$ on reconstruction level that does not fall into the fiducial volume.
In a first step, and as done for the fiducial inclusive \xsec, the prediction for the \emu signal is obtained by subtracting all non-\emu contributions from data.
The obtained bin values can then be corrected for migration effects by multiplying with the non-migrated fraction of events:
\vspace*{0pt minus 2pt}  
\begin{align}
  \label{eq:strategy-diff-migcorrection}
  N_{\text{corr},j}(\emu) = \left[ N_j(\text{data}) - N_j(\text{non-\emu signal}) - N_j(\text{bkgd}) \right] \cdot \left( 1 - f_{\text{out},j} \right) \, .
\end{align}
The migration-corrected \emu signal distribution is then related to the differential \xsec~$\sigma^{\text{fid}}_k$ in bin~$k$ through
\vspace*{0pt minus 2pt}  
\begin{align}
  \label{eq:strategy-diff-migmatrix}
  N_{\text{corr},j}(\emu) = \mathcal{L} \cdot \sum_k \sigma^{\text{fid}}_k \epsilon_k \mathcal{M}_{kj} \, ,
\end{align}
where $\mathcal{M}$ is the \emph{migration matrix}.
Its elements $\mathcal{M}_{kj}$ represent the probability of an event generated in bin~$k$ in the fiducial volume at parton level to be observed in bin $j$ on reconstruction level.
The fractions are estimated from \MC-simulated signal-like events that enter both the fiducial volume and the reconstruction-level signal region.
Examples of a migration matrix, signal efficiency values, migration fractions and resulting correction factors are shown in \cref{fig:strategy-diff-matrix-effmig} for the distribution of the photon transverse momentum.
The chosen binning shows little migration from off-diagonal bins over the entire spectrum of the distribution.
The migration fraction \fout is almost constant for all bins, whereas the signal efficiency~$\epsilon$ and the correction factor~$C$ grow as the transverse momentum increases.

The unfolding comes into play when solving \cref{eq:strategy-diff-migmatrix} for the differential \xsec: the migration matrix needs to be inverted.
Various methods exist, and the data of the \xsec measurement presented here is unfolded using an iterative matrix-unfolding technique based on Bayes' theorem developed by \citeauthor{DAgostini:1994fjx}~\cite{DAgostini:1994fjx,DAgostini:2010hil}.
The technique is implemented in the \textsc{roounfold} package~\cite{Adye:2011gm}.
The iterative approach interprets the migration matrix as a conditional probability $P(E_j|C_k)$ to observe effect $E_j$ in bin~$j$ of the reconstruction-level distribution, given a cause~$C_k$ in bin~$k$ at parton level.
Following this interpretation, the elements of the inverted matrix can be expressed as conditional probabilities $P(C_k|E_j)$ to have a cause in bin~$k$ given an observation of an effect in bin~$j$, and they can be calculated using Bayes' theorem~\cite{DAgostini:1994fjx}
\begin{align}
  \label{eq:strategy-diff-bayes}
  P(C_k|E_j) = \frac{ P(E_j|C_k) \cdot P_0(C_k) }{ \sum_l P(E_j|C_l) \cdot P_0(C_l) } \, ,
\end{align}
where $P_0(C_k)$ is an initial, prior probability density distribution of causes $C_k$.
While the conditional probabilities $P(E_j|C_k)$ are determined from \MC simulation and are assumed constant, the distribution of $P(C_k)$ can be updated \emph{iteratively}.
The observed distribution of effects $E_j$ is then unfolded with the obtained $P(C_k|E_j)$ and used as the prior of the next step of the iteration.
The closer the initial $P_0(C_k)$ to real data, the better the agreement of $\estimate{P}(C_k)$ with the true distribution.
Finding the optimal number of iterations to obtain the inverted matrix is a delicate process, and it requires balancing closure with the true distribution from \MC simulation and statistical uncertainties associated with the technique.
Closure is tested with a Pearson $\chi^2$ test.
Additional stability tests of the unfolding procedure, such as pull and stress tests, are performed using pseudo-datasets obtained from \MC simulation using the bootstrap method, \cf \cref{sec:results-differential} for details.

Once calculated, the inverse of the migration matrix can be used to unfold \ATLAS data and to calculate a differential \xsec distribution.
As before, by including the fraction $f_{\emu,j}$ of \emu events in all signal events per bin, the dependency on the \LO \xsec associated with the \tty and \tWy simulations can be avoided.
Then, the differential \xsec in bin~$k$ with respect to observable $X$ reads
\begin{align}
  \label{eq:strategy-diff-xsec}
  \frac{\mathrm{d}\sigma^{\text{fid}}}{\mathrm{d}X_k} =
  \frac{1}{\mathcal{L} \cdot \Delta X_k \cdot \epsilon_k} \,
  \sum_j \mathcal{M}_{jk}^{-1} \times (N_j(\text{data}) - N_j(\text{bkgd})) \, f_{\emu,j} \, (1 - f_{\text{out},j}) \, ,
\end{align}
where $\Delta X_k$ is the bin width of bin~$k$ at parton level, and $\mathcal{M}^{-1}$ is the inverted migration matrix obtained through the unfolding.
The unfolded results included in Ref.~\cite{TOPQ-2020-03} comprise differential distributions of five observables:
the transverse momentum and absolute pseudorapidity of the photon,
the distance \DRlph in the \etaphi plane between the photon and the closer of the two charged leptons,
and the absolute differences in pseudorapidities \Detall and in azimuthal angles \Dphill between the two charged leptons.

\section{Reference of the measurement: theory prediction}
\label{sec:strategy-prediction}

As a reference point for the measurements of both the fiducial inclusive and differential \xsecs, a dedicated theory prediction is used.
The calculation is similar to that presented in Refs.~\cite{Bevilacqua:2018woc,Bevilacqua:2018dny} and is performed by the same authors (Bevilacqua, Hartanto, Kraus, Weber and Worek), but uses the fiducial volume defined in \cref{sec:strategy-fid-phase-space}.
The authors make precise predictions for the \xsec of the $\pp \to e^+\nu_e\mu^-\bar{\nu}_\mu b\bar{b}\gamma + X$ final state at a \com energy of \SI{13}{\TeV}.
The computations are done at $\mathcal{O}(\alpha_s^3\alpha^5)$ and use the following values for the \SM input parameters:
\vspace*{0pt plus 2pt}  
\begin{subequations}
\begin{align}
  \label{eq:strategy-prediction-settings}
  M_W &= \SI{80.385}{\GeV}  & \Gamma_W &= \SI{2.0988}{\GeV} \\[0.6ex]  
  M_Z &= \SI{91.1876}{\GeV} & \Gamma_Z &= \SI{2.50782}{\GeV} \\[0.6ex]  
  m_t &= \SI{173.2}{\GeV}   & \Gamma_t^{\NLO} &= \SI{1.351579}{\GeV} \\[0.6ex]  
  G_F &= \SI{1.166378e-5}{\GeV\tothe{-2}}
\end{align}
\end{subequations}
While the electroweak coupling is evaluated at the Fermi constant and uses the coupling strength $\alpha(G_F) \approx \sfrac{1}{132}$, the leading emission is done with $\alpha(0) \approx \sfrac{1}{137}$.
This leads to a decrease of the calculated \xsecs by about \SI{3}{\percent}, already included in the values presented in the following.
The hard-scattering event, initiated through either two gluons or quark-antiquark annihilation, is performed at \NLO in \QCD with all off-shell and interference effects included using automated off-shell algorithms~\cite{Papadopoulos:2005ky} implemented in the \textsc{helac-dipoles}~\cite{Czakon:2009ss} package and the \textsc{helac-phegas}~\cite{Cafarella:2007pc} \MC program.
For more details on the computational framework, especially on including \NLO virtual and real-emission corrections, see the original references of the authors~\cite{Bevilacqua:2018woc,Bevilacqua:2018dny} and references therein.

Results for the \xsecs are given for two possible choices of the renormalisation and factorisation scales: with fixed scales set to half the top mass, $\mu_R = \mu_F = m_t/2$, and with dynamical scales, $\mu_R = \mu_F = \ST/4$.
Both fiducial inclusive and differential \xsecs show reduced scale dependencies when computed with dynamical scales, thus, the latter are chosen here for the comparison with the presented measurement.
The observable \ST of the calculation is the theory equivalent of the experimental \ST, the variable used in this measurement for the profile likelihood fit.
In the theory calculation, it is defined as the scalar sum of all transverse momenta, \ie the transverse momenta of the electron, of the muon, of the photon, of the two \bjets, and the missing transverse momentum due to the neutrinos.
The two \bjets are obtained by clustering final-state partons with the \antikt algorithm with a distance parameter of $R = 0.4$ and pseudorapidities $\abseta < 5.0$.

The dedicated theory computation applies the same cuts on the transverse momenta and absolute pseudorapidities of each particle as listed in \cref{sec:strategy-fid-phase-space}.
The identical distance requirements in the \etaphi space are imposed as well.
Using the \textsc{ct14}~\cite{Dulat:2015mca}, the \textsc{mmht14}~\cite{Harland-Lang:2014zoa} and the \NNPDF{}\,3.0 \PDF sets, \xsec values listed in \cref{tab:strategy-prediction-xsecs} are obtained.
The table gives \xsec values for both \LO and \NLO calculations in \QCD, as well as the resulting \kfactors.
Uncertainties on the scale choice are estimated by varying both scales independently and simultaneously to twice and half their values, resulting in a total of six pairs of alternative scale values.
Uncertainties owing to the choice of \PDF are evaluated following the recommendations for the respective \PDF sets; they are explained in detail by the authors of the theory computation in Ref.~\cite{Bevilacqua:2016jfk}.
The \textsc{ct14} \PDF uncertainties, usually given at the \SI{90}{\percent} confidence level, were rescaled to the \SI{68}{\percent} confidence level in order to be consistent with the uncertainty estimates of the other two sets.

\begin{table}
  \centering
  \caption[Predicted fiducial \xsec values of the measurement]{%
    Cross-section values for the fiducial volume defined in \cref{sec:strategy-fid-phase-space} computed by the authors of Refs.~\cite{Bevilacqua:2018woc,Bevilacqua:2018dny}.
    For comparison, \xsecs are listed for both \LO and \NLO calculations in \QCD, and using three different \PDF sets.
    For all three sets, the uncertainties on the scale choice are reduced drastically when including \NLO terms in the calculation.
    In all cases, the nominal scale values are chosen dynamically with $\mu_R = \mu_F = \ST/4$.
  }
  \label{tab:strategy-prediction-xsecs}
  \begin{tabular}{l S S S S S S S}
    \toprule
    \PDF set
    & {$\sigma^{\LO}\,[\si{\fb}]$}
    & {$\delta_{\text{scale}}\,[\si{\fb}]$}
    & {$\sigma^{\NLO}\,[\si{\fb}]$}
    & {$\delta_{\text{scale}}\,[\si{\fb}]$}
    & {$\delta_{\PDF}\,[\si{\fb}]$}
    & {\kfactor} \\
    \midrule
    \textsc{ct14}   & 18.60 & $^{+6.11}_{-4.30}$ & 19.25 & $^{+0.28}_{-1.09}$ & $^{+0.52}_{-0.59}$ & 1.03 \\[2.0ex]
    \textsc{mmht14} & 20.36 & $^{+7.36}_{-5.03}$ & 19.46 & $^{+0.28}_{-1.09}$ & $^{+0.41}_{-0.38}$ & 0.96 \\[2.0ex]
    \NNPDF{}\,3.0   & 18.98 & $^{+6.67}_{-4.59}$ & 20.00 & $^{+0.28}_{-1.09}$ & $^{+0.28}_{-0.28}$ & 1.05 \\
    \bottomrule
  \end{tabular}
  \vspace*{4pt} 
\end{table}

To have a direct reference point for the presented \ATLAS measurement in the \emu channel, the results obtained in the $e^+\mu^-$ final state need to be doubled.
Using the default \PDF set of the authors of the computation, \textsc{ct14}, the fixed-order \NLO calculation yields a \xsec of the \emu final state in the fiducial phase space described in \cref{sec:strategy-fid-phase-space} of
\begin{align}
  \label{eq:strategy-prediction-xsec}
  \sigma^{\text{fid}}_{\NLO} = 38.50 \, ^{+0.56}_{-2.18} \, (\text{scale}) \, ^{+1.04}_{-1.18} \, (\text{\PDF}) \, \si{\fb} \, ,
\end{align}
with total relative uncertainties of $^{+3.1\,\si{\percent}}_{-6.4\,\si{\percent}}$.
Again, this result uses the dynamical scale choice of $\mu_R = \mu_F = \ST/4$ and not fixed values for the scales to reduce the scale dependence of the result.
The equivalent calculation at \LO in \QCD is subject to relative uncertainties of more than \SI{20}{\percent}, dominated by uncertainties on the scale choice.
The theory predictions in \cref{eq:strategy-prediction-xsec} are used as a reference for the fiducial inclusive \xsec measurement.
In addition, the authors provide binned histograms of their computation for the five observables that are measured differentially.
These binned theory predictions are included directly in the plots in \cref{chap:results} to be able to compare them with unfolded \ATLAS data.


\chapter{Systematic uncertainties}
\label{chap:systematics}

Various sources of uncertainties need to be considered for a \xsec measurement of the \tty process.
Apart from the statistical uncertainty of the measurement, determined by the number of observed events, every measurement is subject to systematic uncertainties.
This fairly generic term describes a broad range of uncertainty sources, but they can be sub-classified in two distinct categories of different origin: \emph{experimental} uncertainties and \emph{modelling} uncertainties.
The first category includes all uncertainty sources associated with the experimental setup and its \enquote{deficiencies}, such as inefficiencies of the \ATLAS detector in the detection or reconstruction of particle signatures, limitations of the calibration of detector components or a finite knowledge of the exact amount of collected data.
The other category, modelling uncertainties, comprises all uncertainties on the \MC simulation of signal and background physics processes.
Phenomenological inputs, such as the \PDFs of gluons and quarks, are only known to a certain precision, and predictions rely on model-specific parameters, such as the choice of the factorisation scale $\mu_F$ of a \MC simulation.
While the nominal prediction for the signal and background processes makes a choice for each of these inputs and parameters, alternative models need to be used to evaluate the impact on the measurement when one of these inputs is varied.
In addition, \MC samples are only generated with a limited number of events.
Statistical uncertainties associated with the \MC simulation also affect the precision of the measurement and are considered as part of the modelling uncertainties.

Systematic uncertainties are included in the measurement of the fiducial inclusive \xsec with the template method, as described in \cref{sec:strategy-fit}:
different scenarios for the signal and background predictions are evaluated and reconstructed, leading to different predicted event yields and shapes of observable distributions.
These alternative predictions are filled into histograms called \emph{systematic templates}.
The templates are then compared against the nominal prediction and their differences are a measure for the impact of this uncertainty on the measurement.
The treatment of template-based systematics and how they enter the profile likelihood fit is further detailed in \cref{sec:systematics-templates}.
All experimental uncertainties considered in the \tty \xsec measurement are introduced in \cref{sec:systematics-exp}, while the modelling uncertainties are described in \cref{sec:systematics-modelling}.
Modelling uncertainties can also be evaluated by looking at their impact on the correction factor~$C$ that relates the reconstruction-level signal region to the fiducial phase-space volume at parton level.
However, the evaluation on~$C$ can only be done pre-fit and, hence, cannot consider correlations between individual uncertainties or constraints of uncertainties obtained in the profile likelihood fit.
It is therefore inferior to the profile likelihood estimate.
Nonetheless, it provides a good cross-check and helps to understand the effect of the modelling uncertainties.
This complementary evaluation is documented in \cref{sec:systematics-acceptance}.

\section{Systematic templates in the profile likelihood fit}
\label{sec:systematics-templates}

Systematic variations may affect both the number of events after event selection and the distributions of observables.
Some of these variations are obtained by applying reweighting techniques to a set of generated events, for example when input parameters to the \MC simulation are varied or when the simulated data is reweighted to a pile-up profile observed in data.
Other variations need to be stored as a separate set of events, for example when energy calibrations are varied that have an immediate effect on kinematic distributions of particles.
In all cases, the initial sum of weighted events before event selection is stored to allow all variations to be scaled to the same nominal predicted \xsec for a process.
Thus, any differences observed after event selection are due to different selection efficiencies and can be attributed to the systematic variations themselves.

Systematic variations enter the profile likelihood fit via nuisance parameters as detailed in \cref{cha:strategy}.
To create these nuisance parameters, template distributions of the \ST observable are created for each of these variations and for each of the \catttyemu, \cattWyemu, \catother, \cathfake, \catefake and \catprompt event categories separately.
Some variations only have templates for some of the categories -- for example, \tty scale variations only affect the \catttyemu and \catother categories.
But others, such as the pile-up reweighting, have templates for all of them.
However, the templates attributed to one variation are described through one single nuisance parameter and are correlated across all event categories.

\paragraph{smoothing.}
To avoid high sensitivities to statistical fluctuations in the templates, \emph{smoothing} techniques are applied to some of them:
statistics are averaged across bins to prevent large spikes in the template distributions.
As a first step, a histogram with the relative differences between the templates of the systematic variation and the nominal prediction is created.
The algorithm then searches for neighbouring bins with large differences in bin content, but also large statistical uncertainties on this difference.
If the relative uncertainty exceeds a predefined threshold, the two neighbouring bins are merged to increase bin statistics.
The merge threshold is chosen according to the global template statistics of the systematic variation and of the nominal prediction.
Once all bin statistics are sufficiently high, the smoothing algorithm 353\kern0.05em\textsc{qh twice} is applied%
\footnote{%
  353\kern0.05em\textsc{qh twice} denotes a sequential smoothing algorithm which consists of \enquote{running medians of 3}, followed by \enquote{running medians of 5} and another \enquote{running medians of 3}, with special conditions for the next-to-end and end points.
  This produces smooth, but flattened maxima and minima which can be cured by an additional quadratic interpolation (the \enquote{Q} step).
  Non-smooth monotonic sequences are dealt with by \emph{Hann} smoothing (\enquote{H}) as a final step.
  The \emph{353QH} sequence is applied a second time on the residuals of the smoothed distribution, the results of which are combined afterwards.
  For more details, c.f. Ref.~\cite{Friedman:1974vj}.
}
to create smooth transitions between the merged bins.
In particular, smoothing algorithms are applied to those systematic variations that are simulated with a separate event seed.
These are prone to large statistical fluctuations with respect to the nominal prediction.
\Cref{tab:syst_smoothing_symm} gives an overview of all systematic variations and whether smoothing is applied to their templates.

\paragraph{symmetrisation.}
Many of the systematic variations come in pairs in the \ATLAS uncertainty model, that is, one up variation and one down variation of a parameter from its nominal value.
Ideally this creates a three-point uncertainty with the nominal template as the central prediction and the two systematic templates providing two opposing variations.
However, in some cases these variations are highly non-symmetric.
To centre the two systematic variations around the nominal template, \emph{symmetrisation} techniques may be applied.
For three-point uncertainties, \emph{two-sided} symmetrisation is the default strategy to cure asymmetries:
in a first step, the relative differences to the nominal template are calculated for both variations.
These relative differences are then centred around their arithmetic mean.
Hence, the positive and negative of the following expression are taken as the new relative up and down variations:
\begin{align}
  \text{new relative up/down}
  = \pm ~ \left\lvert \frac{\text{absolute up}-\text{absolute down}}{2 \times \text{nominal}}\right\rvert \, .
\end{align}
If the two variations are symmetric around the nominal template by construction, this procedure will have no impact.
A more conservative alternative is to use \emph{maximum} symmetrisation:
this approach takes the larger of the two relative variations bin-by-bin and mirrors its values around the nominal prediction.
This conserves both amplitude and direction of the larger of the two relative variations in each bin, while providing symmetric up and down templates.
Maximum symmetrisation is used for templates with large statistical fluctuations where a conservative estimate of systematic variations is needed.
In particular in those cases, where both relative variations occur with the same sign in a few bins, maximum symmetrisation is superior to two-sided symmetrisation:
while maximum symmetrisation provides a symmetric three-point uncertainty based on the larger of the two relative variations, two-sided symmetrisation would cancel the effects of the two systematic templates.

\emph{One-sided} symmetrisation can be used when only a single variation is provided.
In this case, the variation is mirrored around the nominal prediction to provide a supplementary variation.
\Cref{tab:syst_smoothing_symm} lists all systematic variations of the analysis and the symmetrisation techniques applied to them.

\begin{table}
  \centering
  \caption[List of systematics with smoothing and symmetrisation techniques applied]{%
    Smoothing and symmetrisation applied to the templates of systematic variations.
    All variations are introduced in detail in \cref{sec:systematics-exp,sec:systematics-modelling}.
    Systematic variations with both up and down variation are usually symmetrised via two-sided symmetrisation, or via maximum symmetrisation in the case of modelling uncertainties.
    One-sided symmetrisation is used for single systematic variations to create a three-point uncertainty.
    }
  \label{tab:syst_smoothing_symm}
  \begin{tabular}{lccc}
    \toprule
    Systematic                     & Smoothing  & Symmetrisation \\
    \midrule
    all experimental uncertainties, except: & \checkmark & two-sided \\
    ~--~all scale-factor systematics      & ---       & two-sided \\
    ~--~\textsc{jes} data vs. \MC (\FS/\atlfast)   & \checkmark & one-sided \\
    ~--~\textsc{met} soft-track parallel res.      & \checkmark & one-sided \\
    ~--~\textsc{met} soft-track perpendicular res. & \checkmark & one-sided \\
    \midrule
    \tty $\mu_R$                   & ---        & --- \\
    \tty $\mu_F$                   & ---        & --- \\
    \tty \PS model                  & \checkmark & one-sided \\
    \tty \Pythia \emph{A14 var3c}  & \checkmark & maximum \\
    \tty \PDF                       & ---        & one-sided \\
    \midrule
    \tWy $\mu_R$                   & ---        & --- \\
    \tWy $\mu_F$                   & ---        & --- \\
    \tWy \PS model                  & \checkmark & one-sided \\
    \midrule
    \ttbar $\mu_R$ (shape)         & ---        & --- \\
    \ttbar $\mu_F$ (shape)         & ---        & --- \\
    \ttbar \PS model (shape)        & \checkmark & one-sided \\
    \ttbar \Pythia \emph{A14 var3c} (shape) & --- & --- \\
    \ttbar \emph{hdamp} (shape)    & \checkmark & one-sided \\
    \bottomrule
  \end{tabular}
\end{table}

\paragraph{pruning.}
Systematic variations can generally show two types of differences to the nominal template: a shift in the overall rate and an altered shape of the observable distribution, or combinations of both.
While some variations might show large rate differences with a shape very similar to that of the nominal template, others could have very different shapes and only a small shift in the rate.
These two effects can be disentangled by comparing the overall rate to that of the nominal template and, for a pure shape comparison, by normalising the systematic template to the integral of the nominal prediction.

\begin{figure}[p]
  \centering
  \adjincludegraphics[%
  width=0.33\textwidth,%
  trim=0 {.66\height} 0 0,%
  clip]{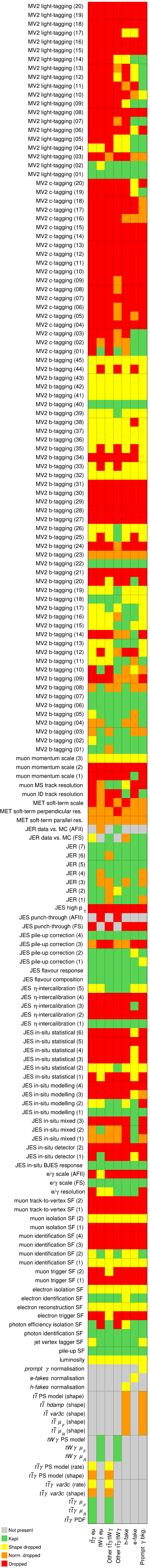}%
  \adjincludegraphics[%
  width=0.33\textwidth,%
  trim=0 {.33\height} 0 {.33\height},%
  clip]{figures/results/pruning-plot}%
  \adjincludegraphics[%
  width=0.33\textwidth,%
  trim=0 0 0 {.66\height},%
  clip]{figures/results/pruning-plot}%
  \caption[Overview of pruned systematic uncertainties]{%
    Pruning of systematic uncertainties.
    The different columns indicate whether a systematic was pruned for the respective event category.
    The colour code is: green for both normalisation and shape kept; yellow for normalisation only, but shape dropped; orange for shape only, but normalisation dropped; and red for all components dropped.
    Grey means that the corresponding templates do not exist.
    The plot is cut off at the bottom of each column and continues at the top of the following column.
  }
  \label{fig:syst-pruning}
\end{figure}

Adding all systematic variations with both rate and shape components as nuisance parameters to the profile likelihood fit, as theorised in \cref{sec:strategy-fit}, would lead to a high-dimensional phase space and an unstable fit procedure.
Therefore, \emph{pruning thresholds} are defined and systematic variations are dropped for an event category if their impact on the nominal prediction of that category remains below these thresholds.
This reduces the number of local minima in the multi-dimensional phase space and generally increases the stability of the fit procedure.
If the rate shift is below the \emph{normalisation pruning threshold}, the rate component of a systematic template is dropped and it is normalised to the integral of the nominal prediction.
If, without considering any global rate shift, the maximum difference between the systematic variation and the nominal template in all of the bins is below the \emph{shape pruning threshold}, the shape component of a template is dropped%
\footnote{%
  Note that the pruning is only applied after smoothing and symmetrisation -- thus, the pruning should not be sensitive to bin-by-bin statistical fluctuations in the templates.
  }.
For this analysis, the normalisation and shape pruning thresholds are chosen to be \SI{0.05}{\percent} and \SI{0.2}{\percent}, respectively.
Different thresholds were tested to evaluate the impact of pruning on the result.
As a figure of merit, the expected uncertainty on the signal strength in a fit to \emph{Asimov pseudo-data} was evaluated and compared against a fit scenario without pruning.
The Asimov dataset is created from the predicted number of total events in \MC simulation in each bin of the \ST distribution and, hence, constitutes a dataset that matches the \MC predictions perfectly.
The results of these tests are detailed in \cref{tab:pruning_tests} and show the relative change of the expected uncertainty with respect to the no-pruning scenario.
The chosen thresholds drop approximately half of the rate components and half of the shape components, while the uncertainty is expected to be affected by less than \SI{0.2}{\percent}, which is far below the quoted precision of the final result.
\Cref{tab:syst_table} shows all systematics considered before any pruning is applied.
\Cref{fig:syst-pruning} gives an overview of the pruned systematic components according to the defined thresholds.
They are introduced in the two following sections in more detail.

\begin{table}
  \centering
  \caption[Normalisation and shape pruning thresholds and their impact]{%
    Effect of different normalisation and pruning thresholds in Asimov fit scenarios.
    Thresholds with approximately \sfrac{1}{2}, \sfrac{3}{4} and \sfrac{4}{5} of the systematic variations dropped were tested and evaluated
    (corresponding to normalisation pruning thresholds of \SI{0.05}{\percent}, \SI{0.1}{\percent} and \SI{0.2}{\percent} and shape pruning thresholds of \SI{0.2}{\percent}, \SI{0.5}{\percent} and 1\%).
    As a figure of merit, the relative change of the expected uncertainty on the signal strength in percent with respect to the no-pruning scenario is listed.
    The tests show how the expected uncertainty on the signal strength decreases with more rigid pruning in almost all cases.
    The thresholds \SI{0.05}{\percent} and \SI{0.2}{\percent}, marked with (*), are chosen for this analysis.
    In combination, they reduce the expected uncertainty on the signal strength by less than \SI{0.2}{\percent}, which is far below the quoted uncertainty of the measurement.
  }
  \label{tab:pruning_tests}
  \sisetup{table-format=2.2,round-precision=2,round-mode=places,table-sign-mantissa}
  \begin{tabular}{l S@{\,} l S@{\,} l S@{\,} l S@{\,} l}
    \toprule
    {\diagbox{shape}{rate}} & \multicolumn{2}{c}{no pruning} & \multicolumn{2}{c}{\sfrac{1}{2} pruned*} & \multicolumn{2}{c}{\sfrac{3}{4} pruned} & \multicolumn{2}{c}{\sfrac{4}{5} pruned} \\
    \midrule
    {no pruning}   & \multicolumn{2}{c}{---} & -0.0448 & \% & -0.1970 & \% & -0.5844 & \% \\
    \sfrac{1}{2} pruned* & -0.0803 & \% & -0.1300 & \% & -0.2935 & \% & -0.6672 & \% \\
    \sfrac{3}{4} pruned  & -0.4469 & \% & -0.5009 & \% & -0.6775 & \% & -1.0691 & \% \\
    \sfrac{4}{5} pruned  & -0.1939 & \% & -0.2548 & \% & -0.4517 & \% & -0.8909 & \% \\
    \bottomrule
  \end{tabular}
\end{table}

 \begin{table}[p]
 \centering
 \caption[List of all systematics considered before pruning]{%
   Complete list of systematic templates considered before pruning is applied.
   The type \emph{N} indicates that the normalisation component of the systematic templates is considered.
   The type \emph{S} means that the shape component is considered.
   In addition, the numbers of nuisance parameters are listed.
 }
 \label{tab:syst_table}
 \small
 \begin{tabular}{lcc}
   \toprule
   Systematic                               & Type   & \# Components \\
   \midrule
   \tty $\mu_R$ scale                       & SN     & 1 \\
   \tty $\mu_F$ scale                       & SN     & 1 \\
   \tty \PS model                           & SN     & 1 \\
   \tty \Pythia \emph{A14 var3c}            & SN     & 1 \\
   \tty \PDF                                & SN     & 1 \\
   \midrule
   \tWy $\mu_R$ scale                       & SN     & 1 \\
   \tWy $\mu_F$ scale                       & SN     & 1 \\
   \tWy \PS model                           & SN     & 1 \\
   \midrule
   \ttbar $\mu_R$ scale (shape)             & S      & 1 \\
   \ttbar $\mu_F$ scale (shape)             & S      & 1 \\
   \ttbar \PS model (shape)                 & S (3x) & 1 \\
   \ttbar \Pythia \emph{A14 var3c} (shape)  & S      & 1 \\
   \ttbar \emph{hdamp} (shape)              & S (3x) & 1 \\
   \midrule
   \cathfake (50\% normalisation)           & N      & 1 \\
   \catefake (50\% normalisation)           & N      & 1 \\
   \catprompt (50\% normalisation)          & N      & 1 \\
   \midrule
   Luminosity                               & N      & 1  \\
   Pile-up                                  & SN     & 1  \\
   \midrule
   Muons (trigger, reconstruction, identi-  & \multirow{3}{*}{SN} & \multirow{3}{*}{15} \\
   \hspace*{2em}fication, isolation, momentum &      &    \\
   \hspace*{2em}resolution, momentum scale) &        &    \\
   Electrons (trigger, reconstruction,      & \multirow{2}{*}{SN} & \multirow{2}{*}{4} \\
   \hspace*{2em}identification, isolation)  &        &    \\
   $e/\gamma$ (resolution, scale)           & SN     & 3  \\
   MET (resolution, scale)                  & SN     & 3  \\
   Photons (efficiency, isolation)          & SN     & 2  \\
   \midrule
   Jet energy scale (\textsc{jes})          & SN     & 30 \\
   Jet energy resolution (\textsc{jer})     & SN     & 8  \\
   Jet vertex tagger (\textsc{jvt})         & SN     & 1  \\
   \MVtwo: $b$-tagging efficiency           & SN     & 45 \\
   \MVtwo: $c$-mistagging rate              & SN     & 20 \\
   \MVtwo: light-mistagging rate            & SN     & 20 \\
   \bottomrule
 \end{tabular}
\end{table}

\section{Experimental uncertainties}
\label{sec:systematics-exp}

The analysis considers various uncertainties originating from the experimental setup, ranging from inefficiencies in particle detection to uncertainties on the calibration of energy measurements and of the luminosity monitors.
Many of these originate from the reconstruction, identification and isolation methods, detailed in \cref{sec:exp_objects}, or their calibration techniques.
Unless stated otherwise, the considered experimental uncertainties come with an up variation and a down variation in the \ATLAS uncertainty model and enter the profile likelihood fit as a three-point uncertainty.
To reduce statistical limitations and to provide symmetric templates, two-sided symmetrisation and smoothing algorithms are applied to those variations with separate event seed.
If the variations are obtained through reweighting, only two-sided symmetrisation, but no smoothing is applied.
While the following paragraphs introduce all considered systematic variations, only a few example plots of the systematic templates are shown.
\Cref{cha:app-red-blue-plots} contains additional distributions.

\begin{figure}
  \centering
  \includegraphics[width=0.48\textwidth]{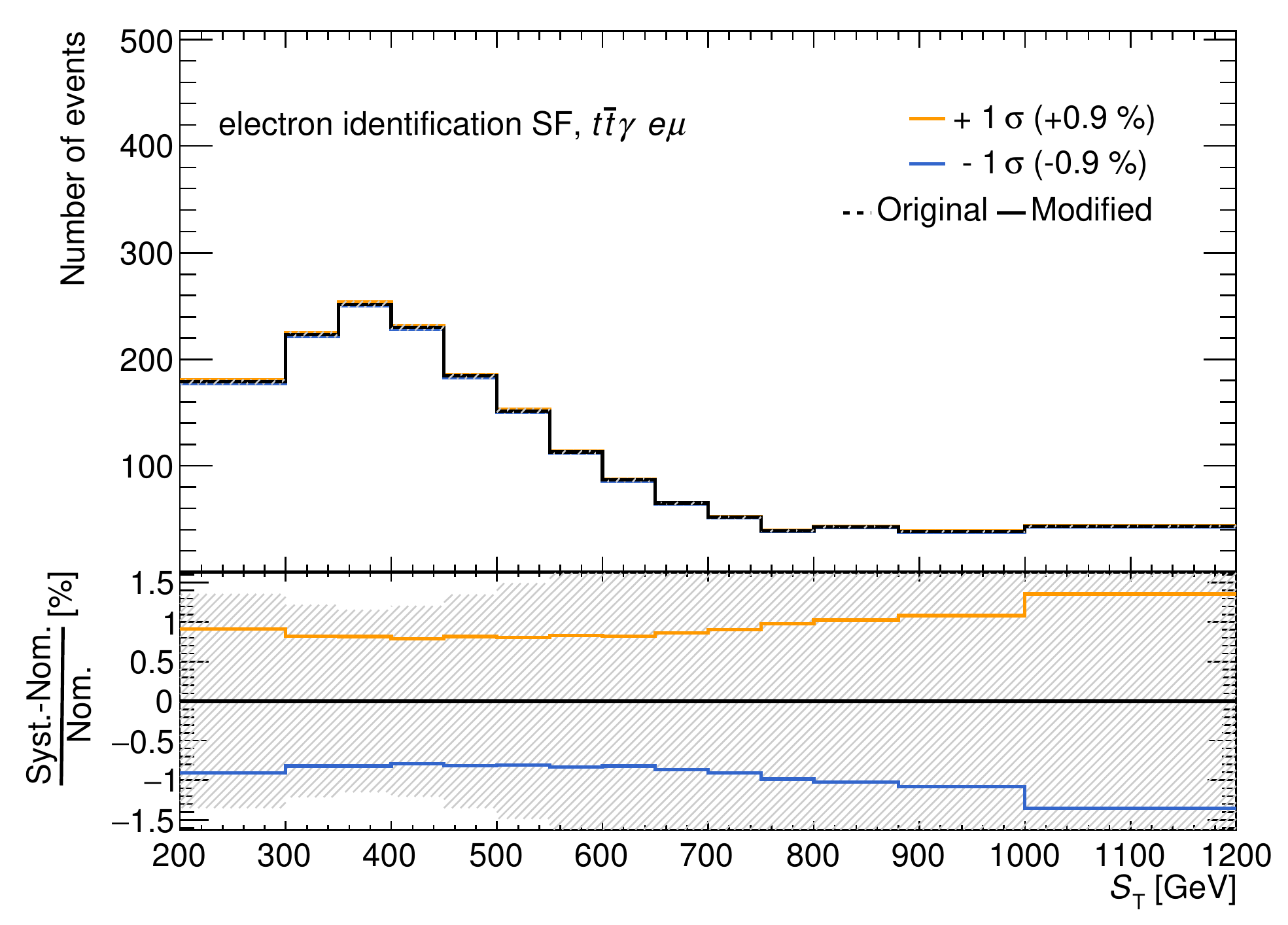}
  \includegraphics[width=0.48\textwidth]{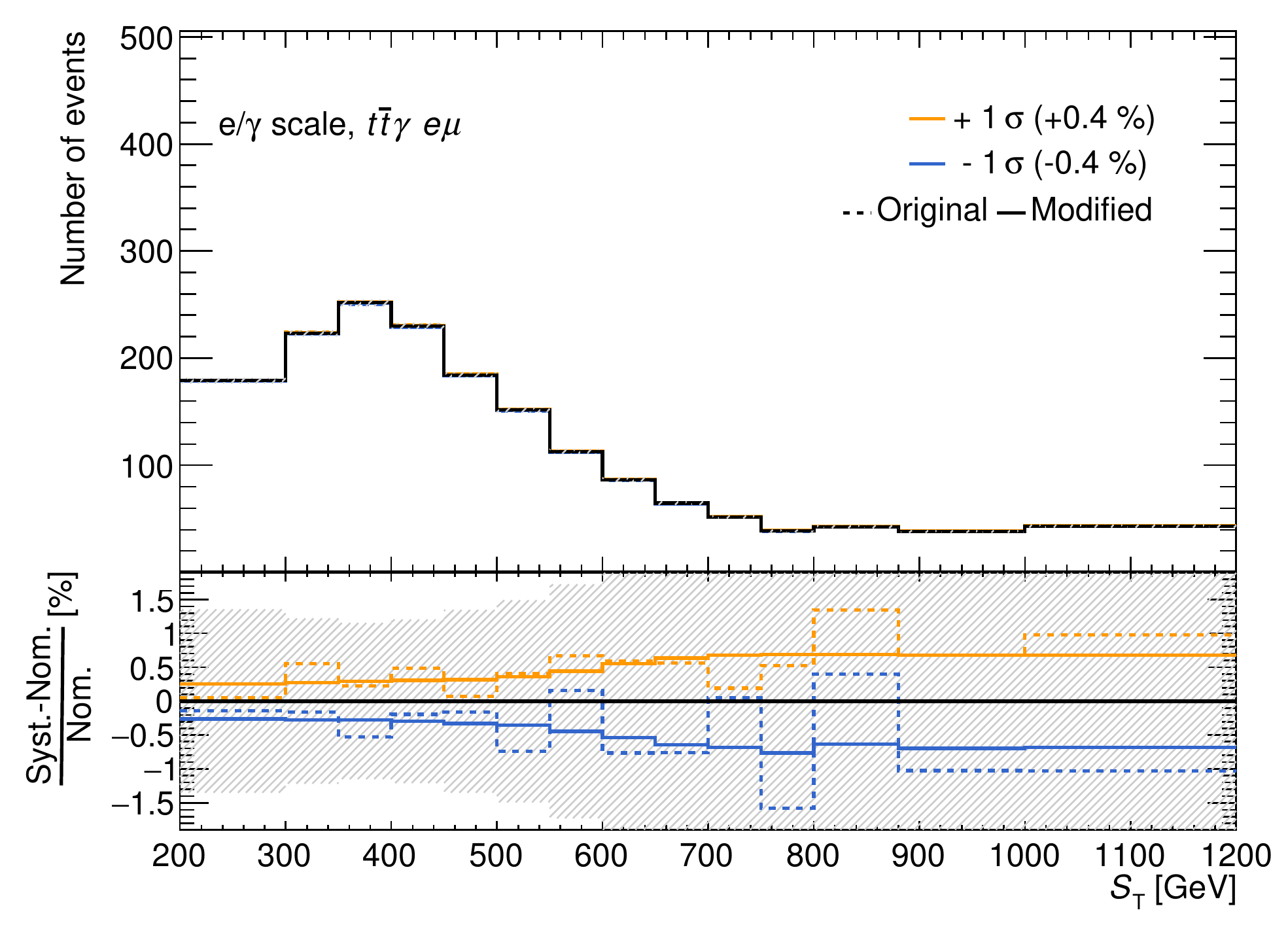}
  \caption[Templates for electron ID and $e/\gamma$ energy scale systematics]{%
    Systematic uncertainty templates for the electron identification scale factors and the electron/photon energy scale for the fit variable \ST.
    The final templates are shown in solid orange and blue to be compared against the nominal prediction in black.
    The shaded uncertainty bands represent \MC-statistical uncertainties on the nominal prediction.
  }
  \label{fig:syst_electrons}
\end{figure}

\paragraph{charged leptons.}
For charged leptons, two classes of systematic uncertainties are considered.
The first class are uncertainties on the efficiencies of charged-lepton triggers, and on the efficiencies of their reconstruction and identification.
Scale factors, measured using the tag-and-probe method in $Z \to \ell\ell$ and $J/\Psi \to \ell\ell$ events, are applied to \MC simulation to correct to efficiencies observed in data~\cite{PERF-2015-10,ATLAS-CONF-2016-024}.
The values of these scale factors are varied within their uncertainties to study their impact on the result.
The uncertainty templates for the electron identification scale factors are shown on the left-hand side in \cref{fig:syst_electrons}.

The second class concerns uncertainties on the charged-lepton energy and momentum calibration.
The muon momentum is studied in $Z \to \mu\mu$ and $J/\Psi \to \mu\mu$ decays, and correction factors are derived to correct the muon momentum scale and resolution in \MC simulation to match those observed in data~\cite{PERF-2015-10}.
These calibration constants are varied within their uncertainties.
For electrons, methods detailed in Refs.~\cite{PERF-2017-03,EGAM-2018-01} are used to calibrate the electron energy scale and resolution.
The full systematic model is reduced and merged into one single nuisance parameter for the energy scale and one for the energy resolution.
As their detection methods and uncertainty models overlap, these nuisance parameters are derived together for electrons and photons to reflect the uncertainties on both their calibrations.
As an example, the systematic templates for the uncertainties on the electron/photon energy scale are shown on the right-hand side in \cref{fig:syst_electrons}.

\begin{figure}
  \centering
  \includegraphics[width=0.48\textwidth]{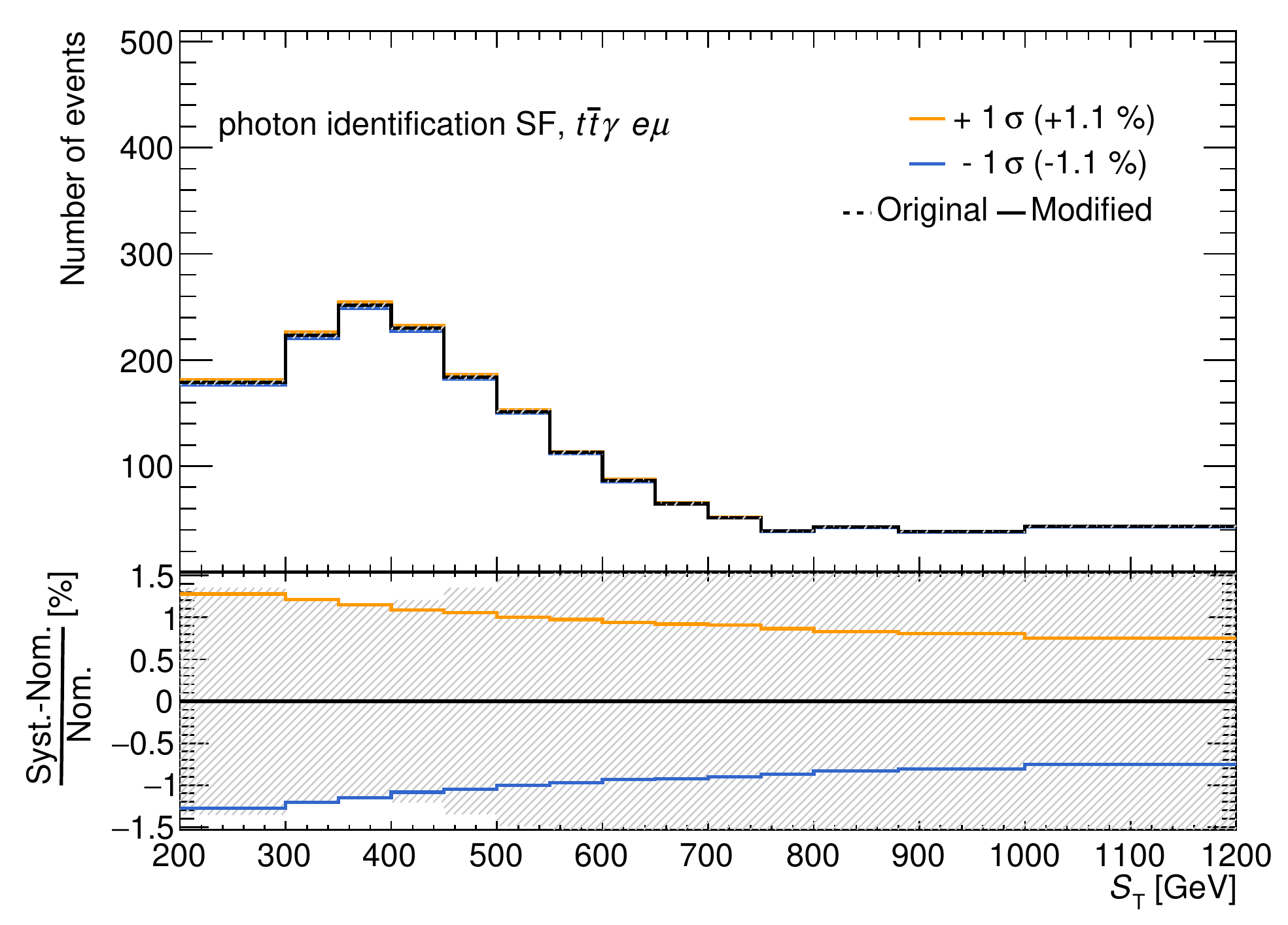}
  \includegraphics[width=0.48\textwidth]{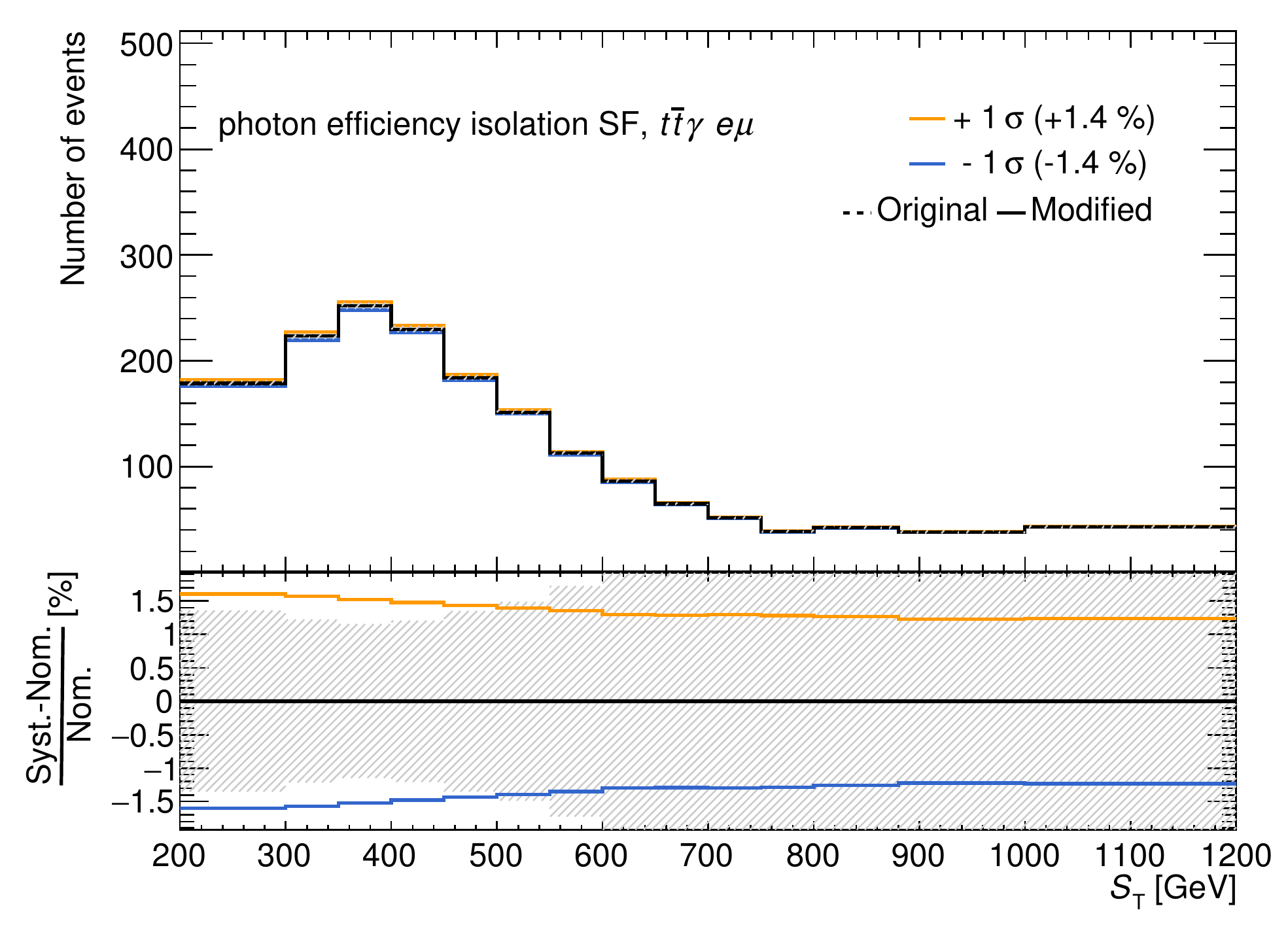}
  \caption[Templates for photon ID and photon isolation systematics]{%
    Systematic templates for the scale-factor uncertainties on the photon identification and the photon isolation efficiencies.
    The final templates are shown in solid orange and blue to be compared against the nominal prediction in black.
    The shaded uncertainty bands represent \MC-statistical uncertainties on the nominal prediction.
  }
  \label{fig:syst_photonSF}
\end{figure}

\paragraph{photons.}
Apart from the aforementioned energy scale and resolution uncertainties, two types of simulation-to-data scale factors are used for photons.
Identification and isolation efficiency scale factors are applied to photons in \MC simulation to correct them to efficiencies measured in data.
The identification scale factors were derived with three techniques in different energy ranges~\cite{PERF-2017-02,EGAM-2018-01}:
firstly, the \emph{radiative \Zboson} method based on low-\pT photons radiated during a $Z \to \ell\ell$ decay,
secondly, the \emph{electron extrapolation} method using $Z \to ee$ events where the similarity between electrons and photons in the detector is exploited with a tag-and-probe method,
and thirdly, the \emph{inclusive photon} method using events with isolated, high-\pT photons.
The last method exploits the weak correlation between the narrow-strip variables used for identifying these photons and the photon isolation.
The sets of scale factors of the three techniques are combined into one set used to correct the efficiencies in \MC simulation.
Photon isolation scale factors were determined using the radiative \Zboson and inclusive photon methods~\cite{EGAM-2018-01} and are combined into one set of scale factors.
These combined identification and isolation scale factors are both varied within their uncertainties to estimate their impact on the analysis results.
Their templates are shown in \cref{fig:syst_photonSF}.

\begin{figure}
  \centering
  \includegraphics[width=0.48\textwidth]{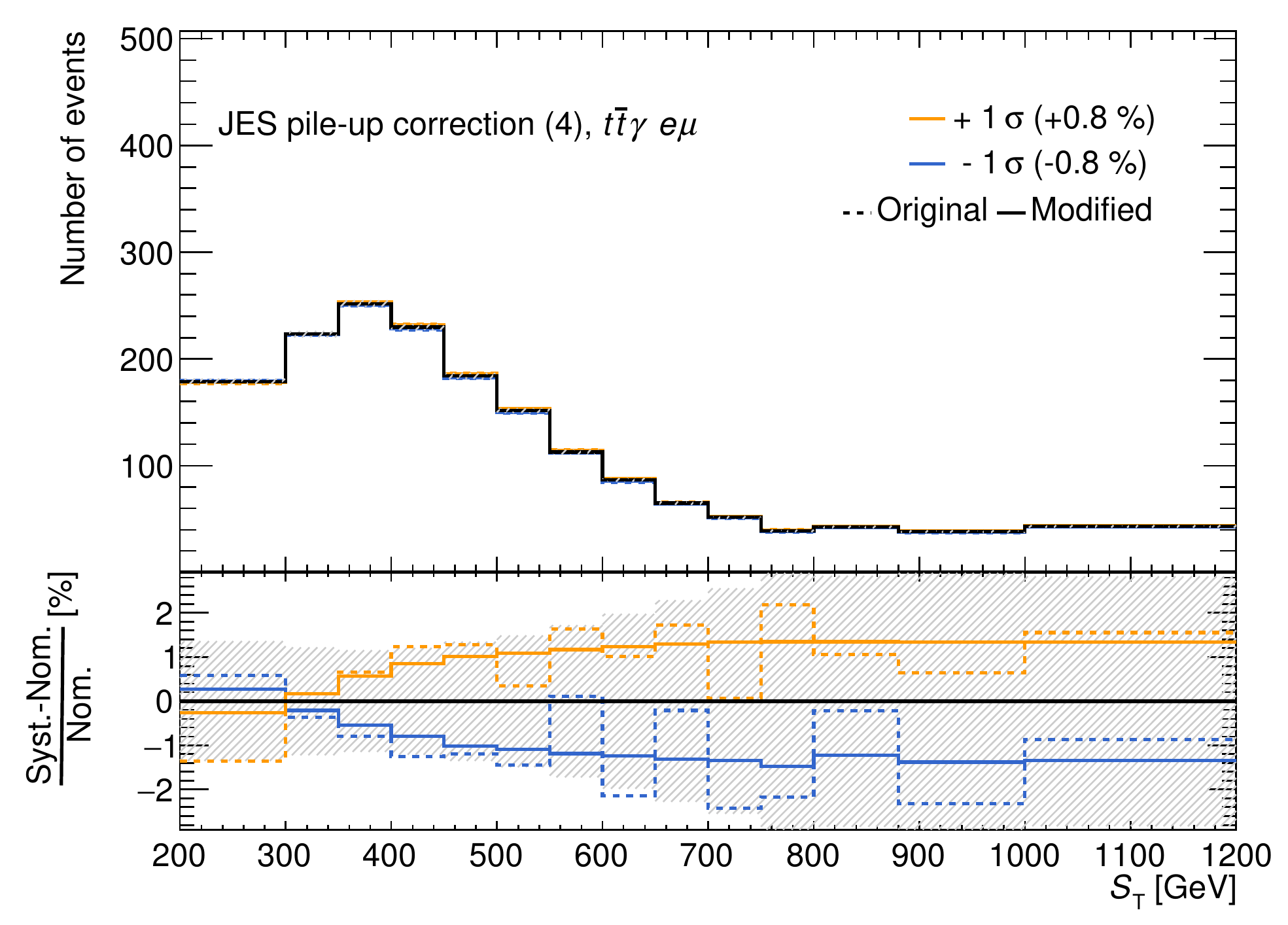}
  \includegraphics[width=0.48\textwidth]{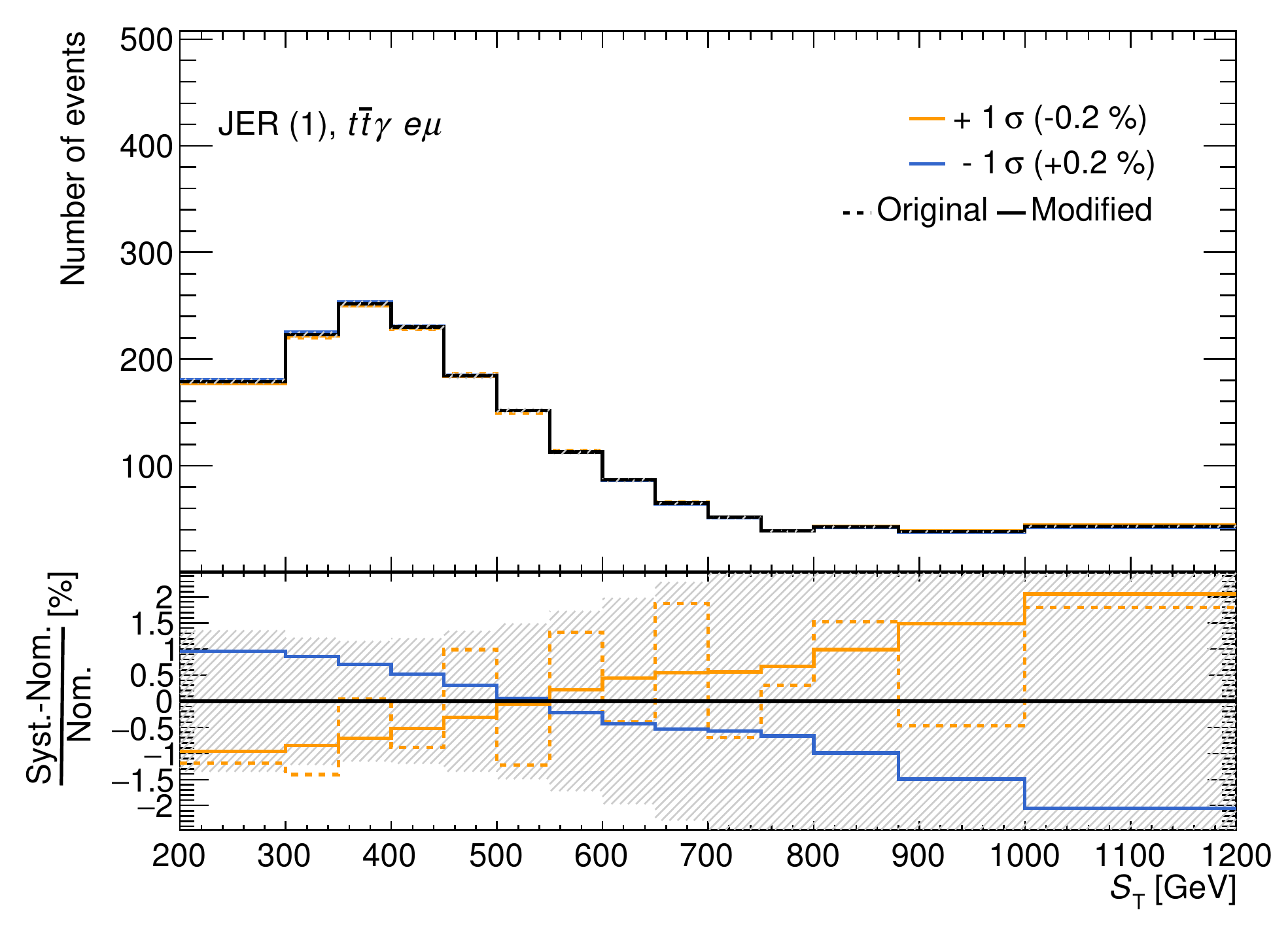}
  \caption[Templates for JES $\rho$~topology and JER systematics]{%
    Systematic templates for two jet uncertainties:
    one of the nuisance parameters of the JES pile-up correction concerning the topology of the jet transverse momentum density~$\rho$, and
    one of the effective nuisance parameters of the uncertainty on the jet energy resolution.
    Two-sided symmetrisation is applied to the templates on the left, one-sided symmetrisation to that on the right.
    The final templates are shown in solid orange and blue to be compared against the nominal prediction in black.
    The shaded uncertainty bands represent \MC-statistical uncertainties on the nominal prediction.
  }
  \vspace*{3pt} 
  \label{fig:syst_jets}
\end{figure}

\paragraph{jets.}
As detailed in \cref{sec:exp_objects}, the calibration of the jet energy scale is performed in several steps, combining measurements in \MC simulation and \emph{in-situ} calibration on data~\cite{PERF-2016-04}.
The individual steps correct various effects, such as improvements of the jet's $\eta$ resolution through jet-origin correction, the removal of excess energy due to pile-up, four-momentum corrections based on \MC simulation, and in-situ corrections applied to data by using high-resolution reference objects.
The uncertainties on the calibration have many sources and are reduced to a set of 30 effective nuisance parameters through eigenvector decomposition, out of which 29 are active on an event-by-event basis as the uncertainties on the modelling of punch-through jets%
\footnote{%
  Punch-through jets penetrate both \ATLAS calorimeters without depositing their entire energy.
}
are derived independently for \atlfast and full simulation.
The effective nuisance parameters can be classified in several categories:
uncertainties on the pile-up corrections (4 \NPs), uncertainties on the jet $\eta$-intercalibration%
\footnote{%
  $\eta$-intercalibration is a technique to use well-measured energies of central jets to obtain calibrations for jets with high $\eta$, where the calorimeter response is more complex and less understood.
}
(5 \NPs), uncertainties on the jet flavour composition (2 \NPs), uncertainties on the in-situ calibration (16 \NPs), as well as uncertainties on the modelling of punch-through jets (1 \NP) and of high-\pT jets (1 \NP).
Systematic templates for one of the largest uncertainties are shown in \cref{fig:syst_jets} on the left.
The shown nuisance parameter reflects uncertainties on the topology of the jet transverse momentum density~$\rho$ used in the pile-up corrections.

\Cref{sec:exp_objects} summarises the determination of the jet energy resolution using the \emph{dijet imbalance} and \emph{bi-sector} methods.~\cite{PERF-2011-04}.
Eigendecomposition is performed to obtain a reduced set of 7 uncorrelated nuisance parameters for the jet energy resolution, one of which is shown in \cref{fig:syst_jets}.
Resolution distributions in \MC simulation are smeared to match those in data.
An additional nuisance parameter accounts for uncertainties on this smearing and is derived separately for \MC samples with full detector simulation and \atlfast simulation.

The efficiencies of the jet vertex tagging algorithm in data are measured with $Z \to \mu\mu + \text{jets}$ events and compared to those obtained from \MC simulation~\cite{ATLAS-CONF-2014-018}.
Scale factors are derived with a tag-and-probe method from these events to correct the efficiencies in simulation.
These scale factors are varied within their uncertainties to account for possible mis-modelling in the simulations.

\paragraph{flavour-tagging.}
Efficiencies of the \MVtwo flavour-tagging algorithm used for \btag are measured in both data and \MC simulation and simulation-to-data scale factors are derived in multiple bins of jet transverse momenta~\cite{FTAG-2018-01}.
Various sources of uncertainties on these scale factors are evaluated, including limited data statistics, uncertainties on the jet energy scale and uncertainties on the modelling of \MC samples that are used to derive them.
Variations related to systematic uncertainties are fully correlated across all bins of the \pT spectrum.
A total of 45, 20 and 20 scale-factor variations are obtained through eigenvector decomposition and account for uncertainties associated with the \btag, $c$-mistagging and light-mistagging performances, respectively.
Reduced sets of nuisance parameters are available for each of the fixed-efficiency operating points, including the one used in this analysis with \SI{85}{\percent} efficiency.
But the reduced models are not used to maintain maximum flexibility for combining the results with those of other analyses.

\paragraph{missing transverse momentum.}
The missing transverse momentum is reconstructed from the vector sum of several terms corresponding to different types of reconstructed objects~\cite{ATLAS-CONF-2018-023}.
The respective uncertainties for electrons, muons, photons and jets enter the hard term of the missing transverse momentum, $\pT^{\mathrm{hard}}$, and are propagated into its uncertainty.
In addition, uncertainties on the track-based soft term $\pT^{\mathrm{soft}}$ are evaluated.
These uncertainties are computed from the maximal disagreement between data taken between 2015 and 2016 and \MC simulation and consist of three terms, all evaluated as projected quantities with respect to $\pT^{\mathrm{hard}}$:
firstly, the parallel resolution, defined as the root mean square of the parallel projection along the hard term, $p_{||}^{\mathrm{soft}}$;
secondly, the parallel scale, defined as the mean of $p_{||}^{\mathrm{soft}}$;
and thirdly, the perpendicular resolution, defined as the root mean square of the perpendicular component, $p_{\perp}^{\mathrm{soft}}$.
Scale and resolution terms are varied up and down by one standard deviation to study their impact on the analysis.

\paragraph{pile-up.}
Simulation-to-data scale factors are applied to match the \MC simulation to the pile-up profiles observed in data.
These scale factors are varied within their uncertainties and represent the uncertainty on this procedure.

\paragraph{luminosity.}
As quoted in \cref{cha:selection}, the total integrated luminosity of \ATLAS Run~2 data has an uncertainty of 1.7\%.
It is derived following a methodology similar to that detailed in Ref.~\cite{DAPR-2013-01} for $\SI{8}{\TeV}$ data.

\section{Modelling uncertainties}
\label{sec:systematics-modelling}

The \tty analysis considers a broad range of uncertainties on the modelling of signal and background processes.
These range from variations of the modelling setup, for example by changing input parameters to the \MC generators, to global uncertainties assigned to event categories or processes.
Variations of the modelling setup are considered for the three \MC simulations that contribute most to the predicted events: \tty, \tWy and \ttbar.
Some of these variations are obtained through reweighting the same sets of events used for the nominal prediction, but others use separate \MC samples, for example when a variation uses an entirely different generator setup.
Again, only a few examples of the templates are shown in figures; \cref{cha:app-red-blue-plots} contains additional plots of modelling uncertainty templates.

\paragraph{signal modelling.}
Various uncertainties on the \LOPS predictions of the \tty and \tWy signal processes are considered:
uncertainties on the scale choice in the nominal samples, uncertainties on the parton-shower model, uncertainties on the parameter choice to model initial-state and final-state radiation, and uncertainties owing to the choice of the \PDF sets.
All of these \MC samples are generated in a phase space much larger than the fiducial volume defined in \cref{sec:strategy-fid-phase-space}.
Although the total numbers of events in each sample were normalised to the same nominal \xsec for the generated phase space, not all modelling variations would yield identical acceptances for the fiducial volume.
Evaluating the templates of these uncertainties in a reconstruction-level selection would, thus, also include possible fiducial acceptance differences.
To avoid this, the acceptance values of all modelling variations are determined and their systematic templates are reweighted to match the nominal fiducial acceptance as quoted in \cref{eq:strategy-acceptance}.
Thus, remaining rate differences in the templates with respect to the nominal predictions only include migration and efficiency effects.
These remaining effects of the signal modelling are also summarised in \cref{sec:systematics-acceptance}.

\begin{figure}
  \centering
  \includegraphics[width=0.48\textwidth]{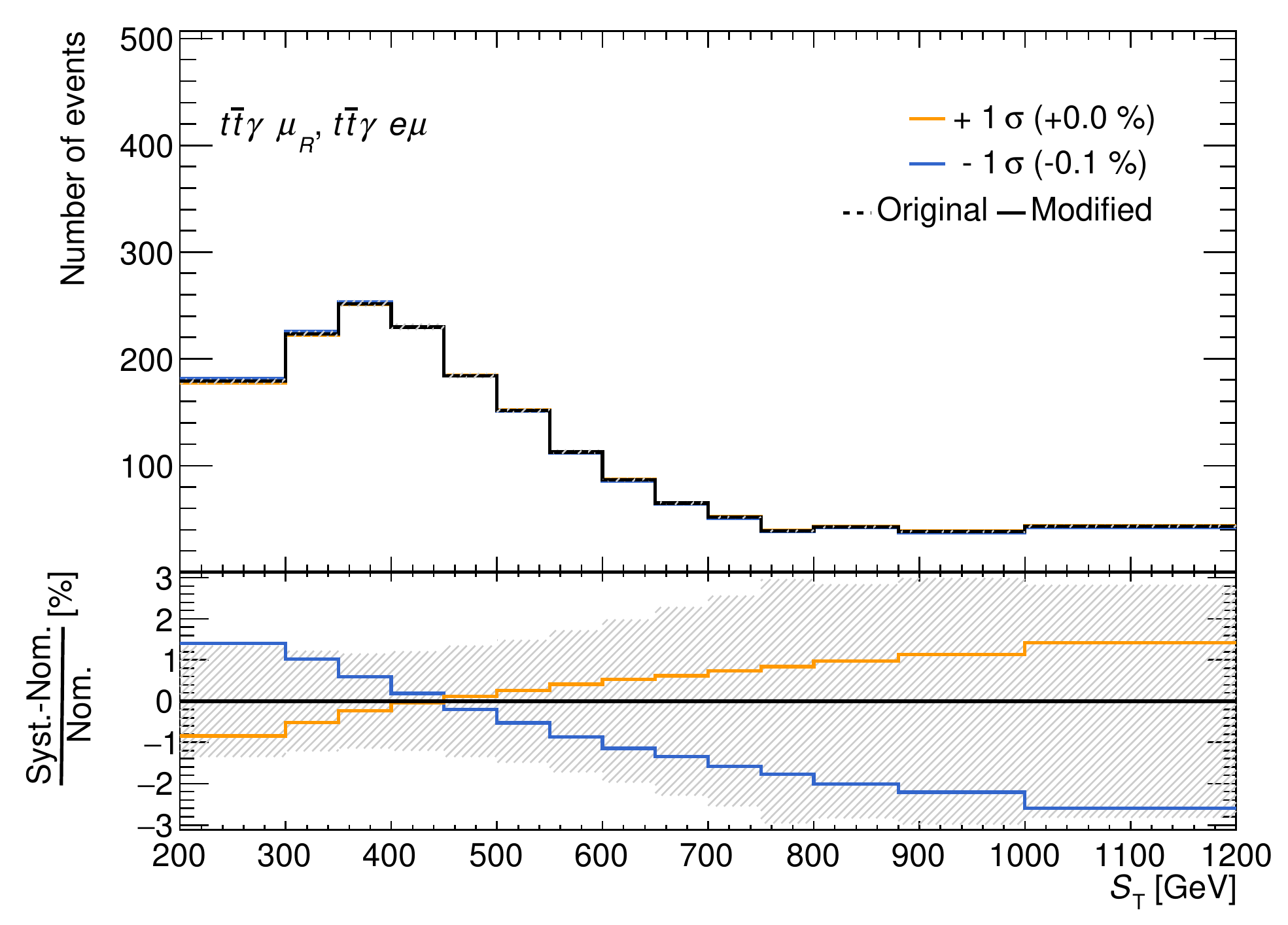}
  \includegraphics[width=0.48\textwidth]{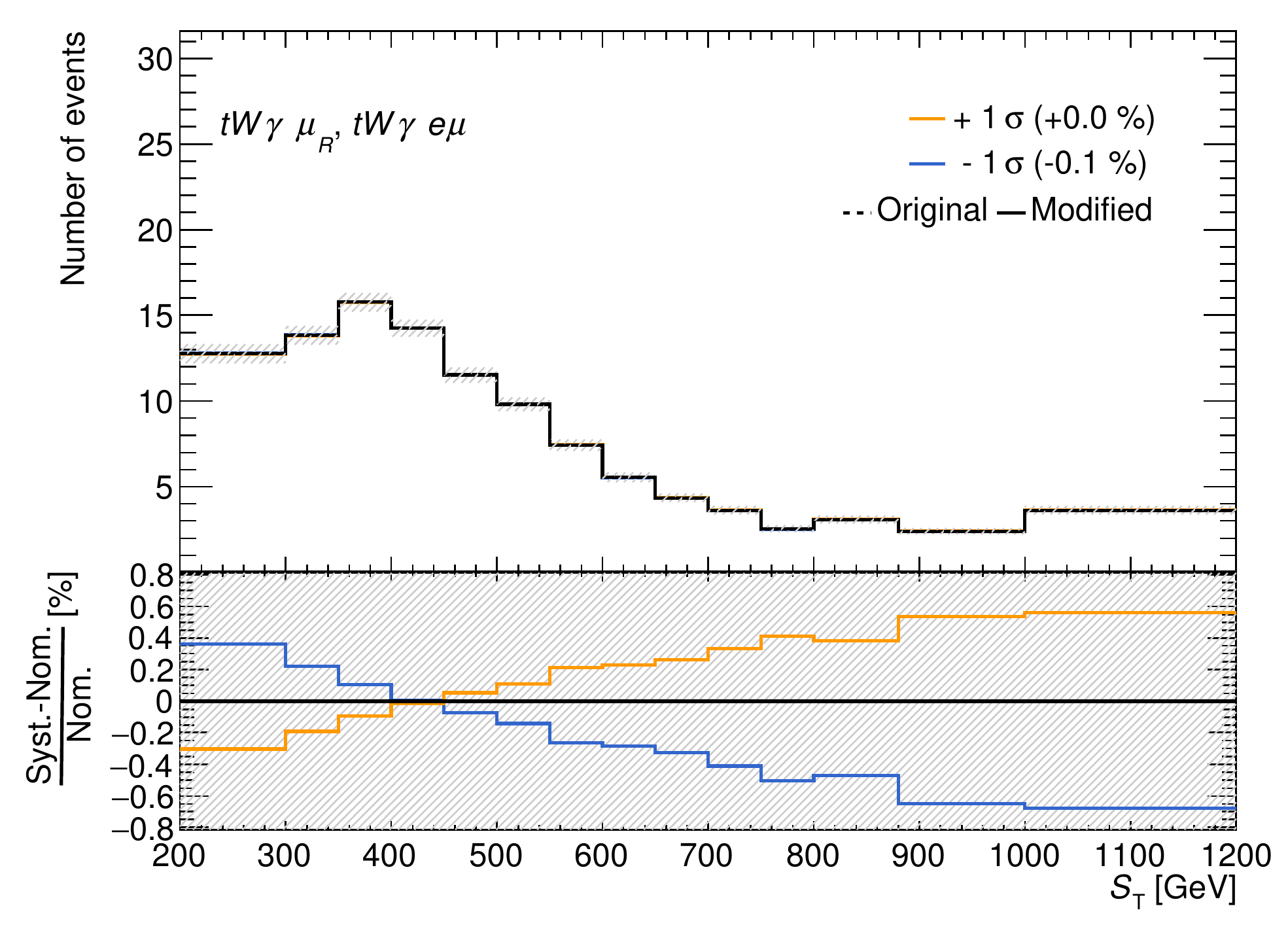}
  \caption[Templates for \tty and \tWy renormalisation scale systematics]{%
    Systematic templates of the uncertainty on the renormalisation scales for
    the \tty signal, shown in the \catttyemu category, and
    for the \tWy signal, shown in the \cattWyemu category.
    The templates are obtained through reweighting the nominal sets of events.
    The shaded uncertainty bands represent \MC-statistical uncertainties on the nominal prediction.
  }
  \label{fig:syst_muR}
\end{figure}

The dependency on the choice of the renormalisation and factorisation scales~$\mu_R$ and~$\mu_F$ is estimated by varying both scales up and down separately by a factor of~2 with respect to the value used in the nominal prediction.
Thus, three-point uncertainties for each of the scales are introduced as nuisance parameters to the fit, separately for \tty and \tWy.
To reduce the effect of statistical fluctuations, the variation of the scales is done by reweighting the nominal set of events.
As a consequence, no symmetrisation or smoothing techniques need to be applied to the templates.
As an example, the resulting templates for the renormalisation scale uncertainties are shown in \cref{fig:syst_muR}.

\begin{figure}
  \centering
  \includegraphics[width=0.48\textwidth]{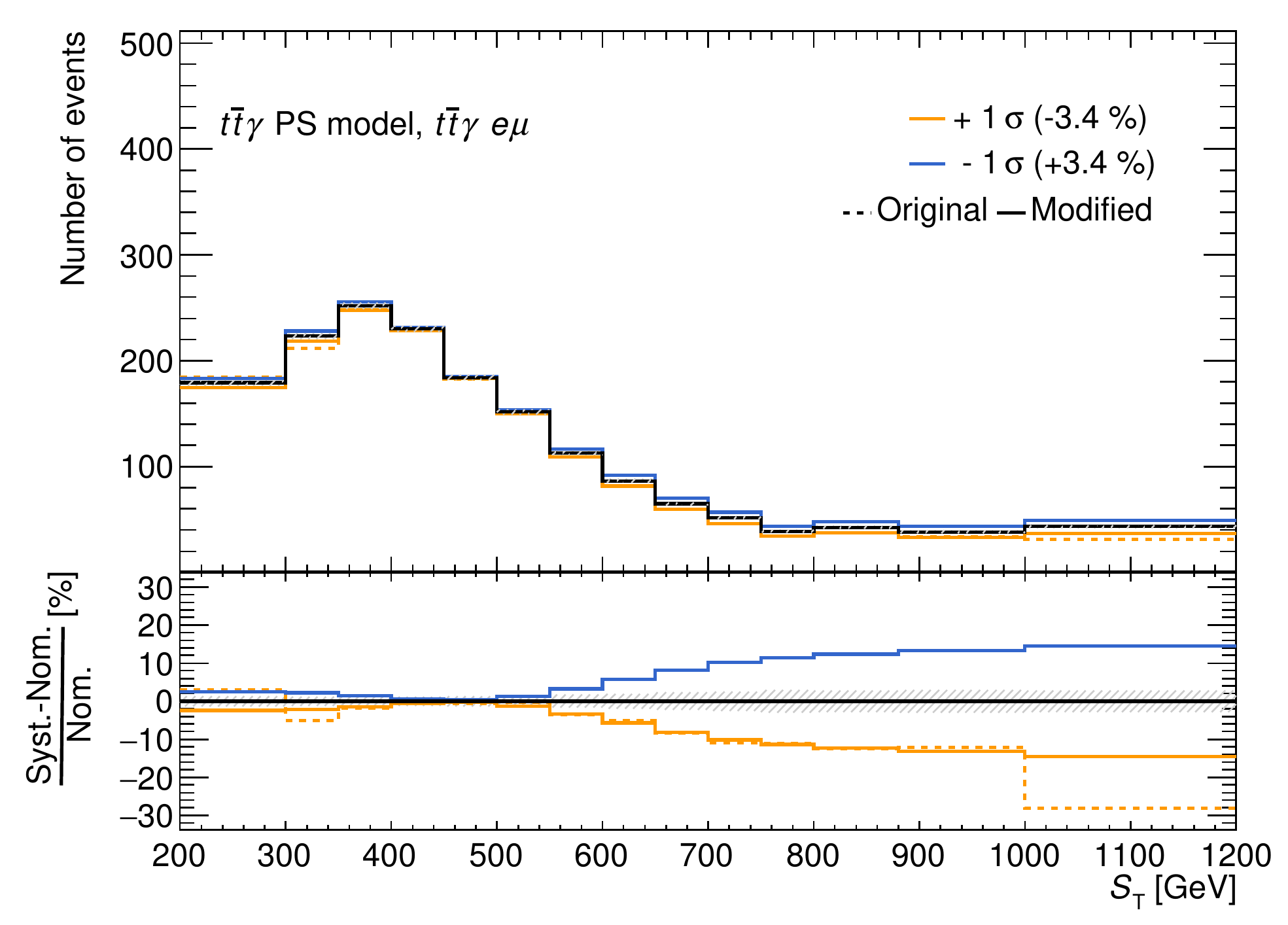}
  \includegraphics[width=0.48\textwidth]{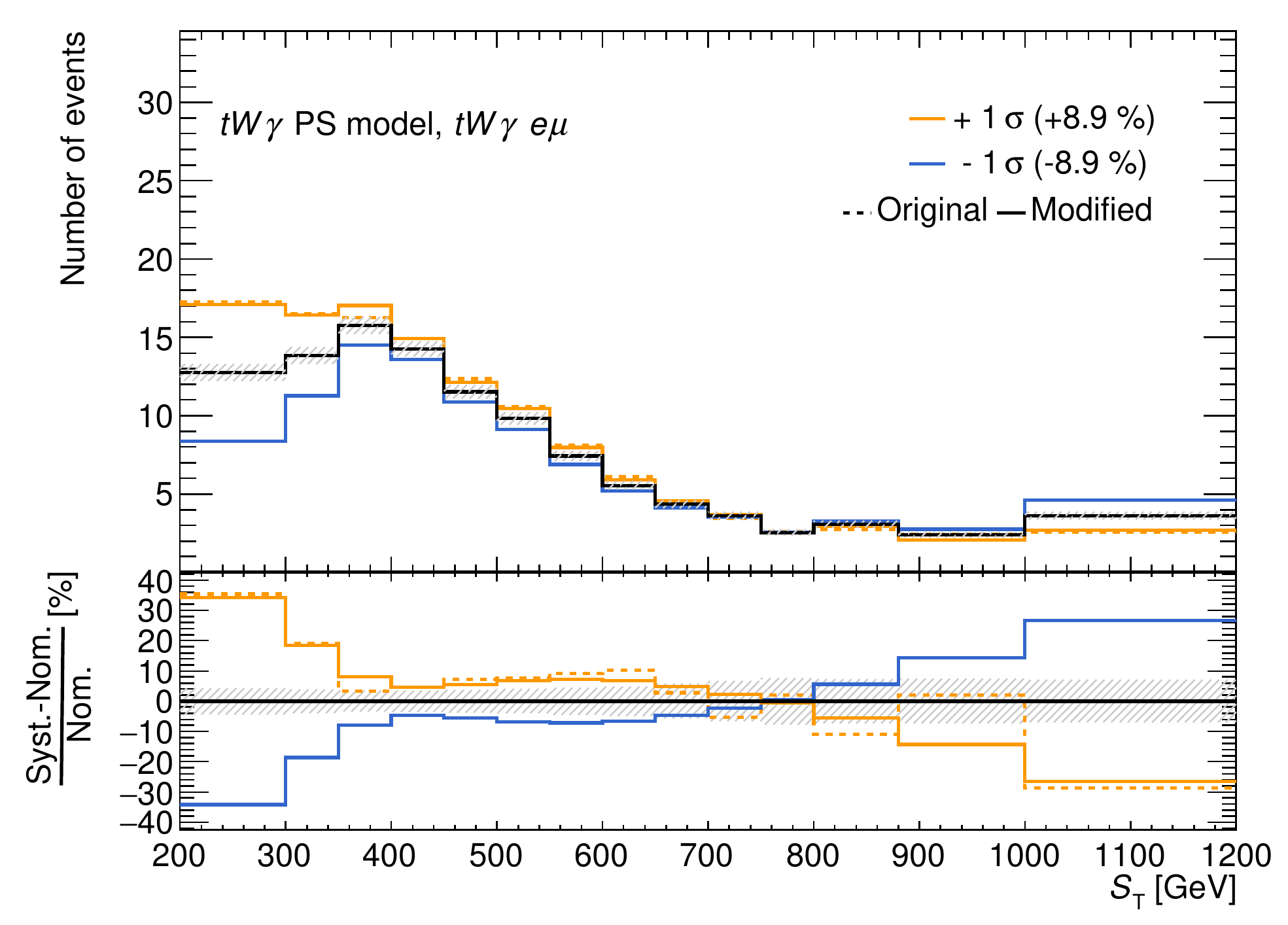}
  \caption[Templates for \tty and \tWy \PS model systematics]{%
    Systematic templates of the uncertainty on the choice of the parton-shower model
    for the \tty signal, shown in the \catttyemu category, and
    for the \tWy signal, shown in the \cattWyemu category.
    The dashed lines are the non-smoothed templates; the final templates after smoothing and symmetrisation are shown in solid orange and blue.
    The shaded uncertainty bands represent \MC-statistical uncertainties on the nominal prediction.
  }
  \vspace*{3pt} 
  \label{fig:syst_PS_model}
\end{figure}

To estimate the uncertainty on the parton shower and hadronisation generated with \Pythia, the same sets of events used for the nominal prediction are showered with \Herwig~\cite{Bahr:2008pv} in alternative \tty and \tWy simulations.
While the first generator is based on the string fragmentation model~\cite{Andersson:1983ia}, the latter utilises the cluster fragmentation model~\cite{Webber:1983if}, and possible differences between these approaches provide an estimate of the uncertainty on the model choice.
To avoid two-point uncertainties in the fit, the \tty and \tWy templates of the parton-shower model uncertainty are symmetrised via one-sided symmetrisation and also smoothed to reduce statistical fluctuations.
The result are two symmetric three-point uncertainties, one each for \tty and \tWy.
The parton-shower uncertainty for \tty is generated with \atlfast simulation of the detector response, whereas the nominal \tty prediction uses full detector simulation.
To avoid any dependencies on possible differences between \atlfast and full simulation, a separate \atlfast simulation of the nominal \tty prediction is used as a reference such that differences between the nominal \atlfast and full simulations do not enter the systematic templates.
For \tWy this step is unnecessary as the nominal prediction already uses \atlfast.
The resulting templates for \tty and \tWy are shown in \cref{fig:syst_PS_model}.

\begin{figure}
  \centering
  \includegraphics[width=0.48\textwidth]{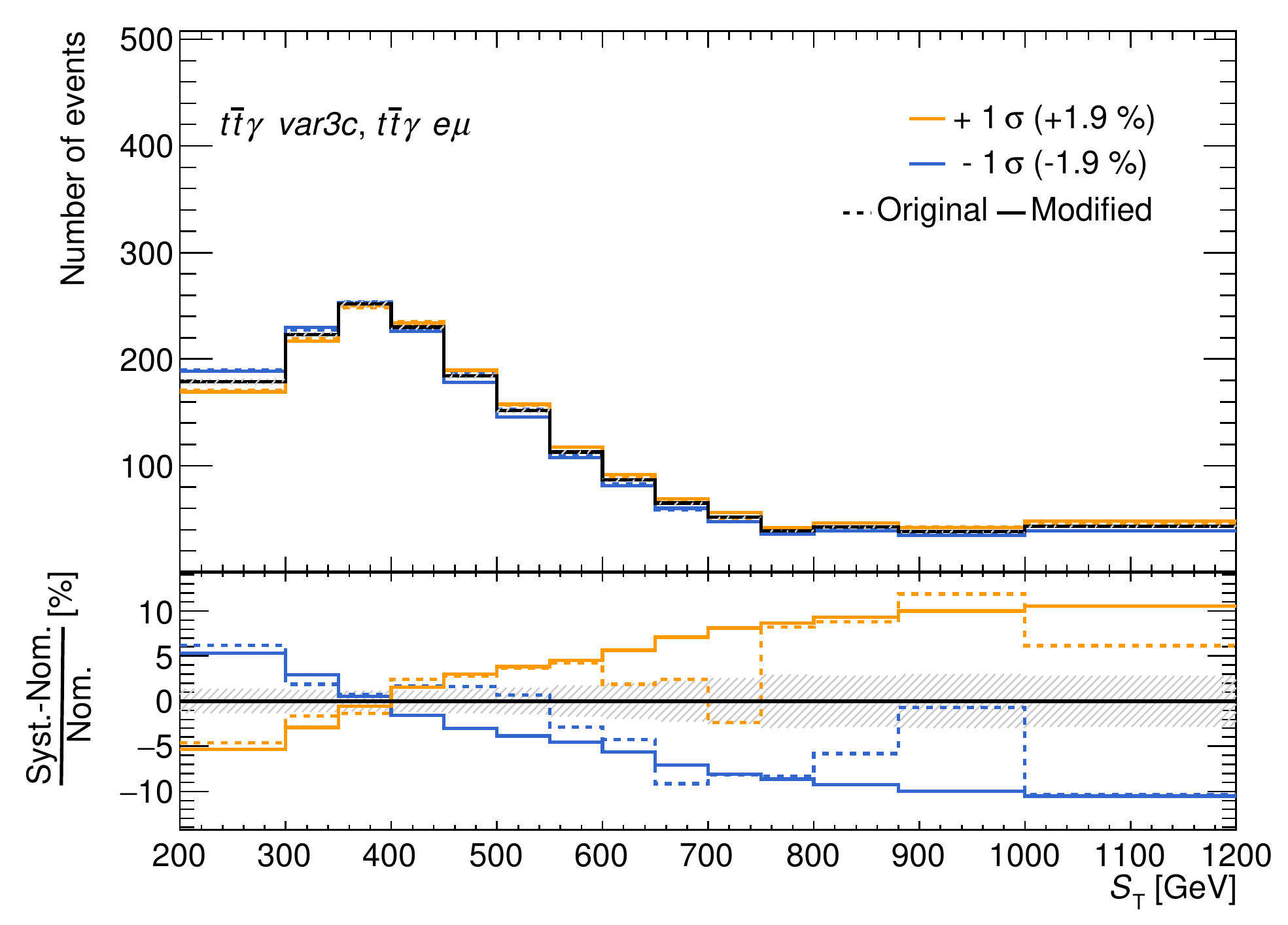}
  \includegraphics[width=0.48\textwidth]{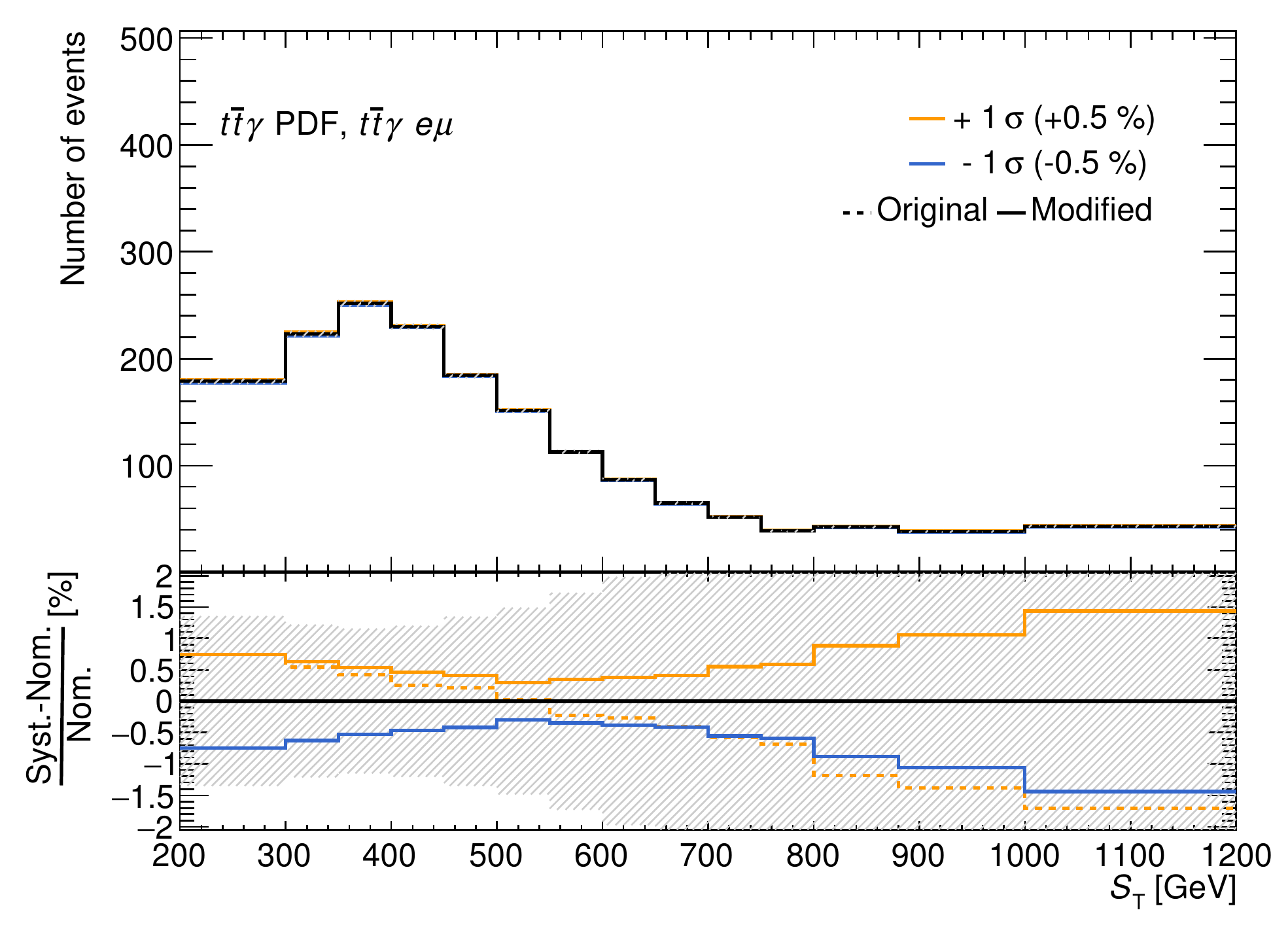}
  \caption[Templates for \tty and \tWy \Pythia eigentune systematics]{%
    On the left:
    systematic templates for the \tty radiation uncertainty (\Pythia \emph{var3c} eigentune) in the \catttyemu category.
    The templates are smoothed and use \emph{maximum} symmetrisation.
    The dashed lines are the non-smoothed templates; the final templates are shown in solid orange and blue.
    On the right:
    combined systematic template for the \tty \PDF uncertainty.
    The dashed line in the background shows one of the \NNPDF replicas.
    The shaded uncertainty bands represent \MC-statistical uncertainties on the nominal prediction.
  }
  \vspace*{3pt} 
  \label{fig:syst_tty_var3c_PDF}
\end{figure}

For \tty only, the uncertainty on initial-state and final-state radiation is estimated through two dedicated sets of events generated with \Madgraph + \Pythia, where a prescribed variation of the \Pythia \emph{A14} tune is implemented (named \emph{A14 var3c} eigentune).
The tune variations simulate high-radiation and low-radiation scenarios.
The templates are smoothed and \emph{maximum} symmetrisation is applied to provide a conservative estimate of the impact of this variation.
Again, the extra sets of events use \atlfast simulation, and to avoid any dependencies on this choice, the \atlfast simulation of the nominal \tty prediction is used as a reference.
The templates are shown in \cref{fig:syst_tty_var3c_PDF} on the left-hand side.

To evaluate uncertainties on the choice of the \PDF set, the \tty simulation comes with an additional one hundred sampled replicas of the \NNPDFLO distributions.
Through reweighting of the nominal set of events, systematic templates are first generated separately for each of the one hundred sets of event weights.
Then, the relative differences between all systematic templates and the nominal prediction are calculated in each bin of the observable distribution.
This yields ensembles of one hundred relative differences for each bin that are distributed around zero.
As prescribed by the \NNPDF Collaboration~\cite{Ball:2012cx}, the bin-by-bin standard deviation of these ensembles is then used to create a single combined template, which is mirrored around the nominal prediction through one-sided symmetrisation to provide a symmetric three-point uncertainty.
Both rate and shape components of these combined templates are considered.
The resulting templates are shown in \cref{fig:syst_tty_var3c_PDF} on the right-hand side.
This procedure is only done for \tty, but not for \tWy as the resulting uncertainty is expected to be about one order of magnitude smaller than that of \tty, and hence becomes negligible in this measurement.

While evaluating the matching between reconstruction-level signal region and fiducial volume, as explained in \cref{sec:strategy-fit}, it was noticed that a large fraction of \tWy events only contain the \bquark from the top quark in the \MC-truth information, but no second \bquark.
It is unclear whether the $g\to b\bar{b}$ split is not modelled accurately for these events, or whether the second \bquark is removed in the processing of the \MC sample in the \ATLAS simulation infrastructure.
Either way, the \tWy samples are found to contain a significant fraction of \emu-channel events that fail the cuts of the fiducial volume defined in \cref{sec:strategy-fid-phase-space} due to the missing \bquark.
Depending on the sample, this fraction is as high as \SI{50}{\percent}.
As this affects the value obtained for the correction factor~$C$, an additional \tWy parton-definition uncertainty is introduced:
in an alternative scenario, the number of parton-level events for \tWy is assumed to be twice as high as the determined value.
The value of the efficiency~$\epsilon$ is expected to remain constant, thus, only the (combined \tty and \tWy) migration fraction \fout and the correction factor~$C$ are changed.
As this additional uncertainty has no impact at reconstruction level, it cannot be included as a nuisance parameter into the fit.
Instead, it is assigned as a fixed uncertainty to the correction factor~$C$, which affects the calculation of the final \xsec according to \cref{eq:strategy-fidxsec-final}.
In that step, the \tWy parton-definition uncertainty is added to the systematic uncertainties obtained from the profile likelihood fit in quadrature.
The effect of this uncertainty is summarised in \cref{sec:systematics-acceptance}.

\paragraph{background modelling.}
Various processes and \MC simulations contribute to the background categories of the analysis.
Estimating individual modelling uncertainties for each of the contributing processes would complicate the analysis unnecessarily, given the small overall background fraction.
However, since \ttbar production is the dominant process that contributes to the three background categories, \ttbar modelling uncertainties are evaluated for the \cathfake and \catprompt categories, the two larger of the three.
To account for possible mis-modelling of all processes contributing, global rate uncertainties of $50\%$ are assigned to each of the three background categories.
To avoid correlation between these and the \ttbar modelling uncertainties, only shape components of the systematic templates of \ttbar are considered and their normalisation components are dropped for the fit.

As done for the signal processes, the dependency on the choice of the renormalisation and factorisation scales is estimated by varying both scales up and down separately by a factor of~2 with respect to the value used in the nominal prediction.
To reduce the effect of statistical fluctuations, the variation of the scales is done by reweighting the nominal set of \ttbar events.
As a consequence, no symmetrisation or smoothing techniques need to be applied to the templates, but their rate components are dropped before the fit.

\begin{figure}
  \centering
  \includegraphics[width=0.48\textwidth]{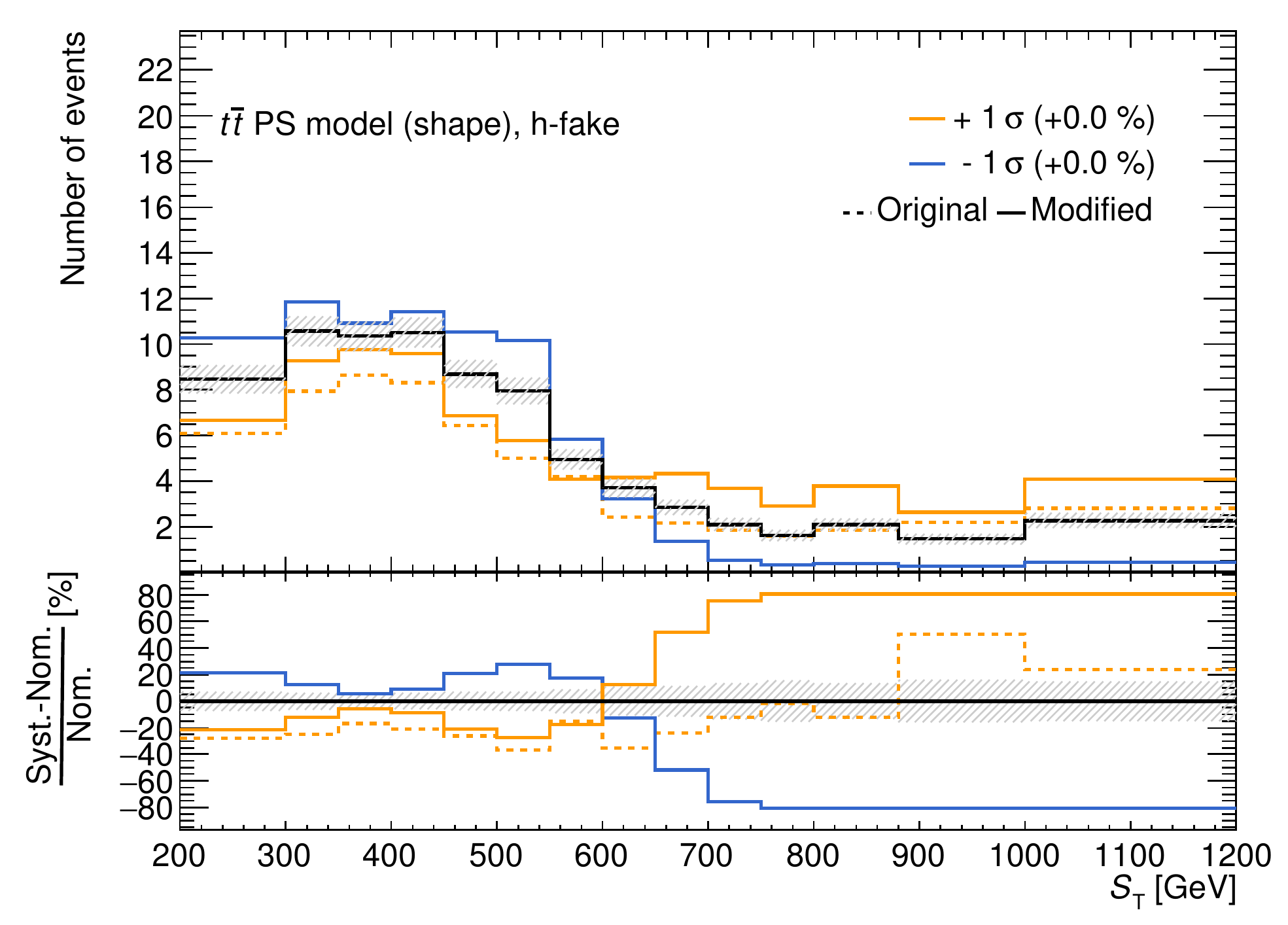}
  \includegraphics[width=0.48\textwidth]{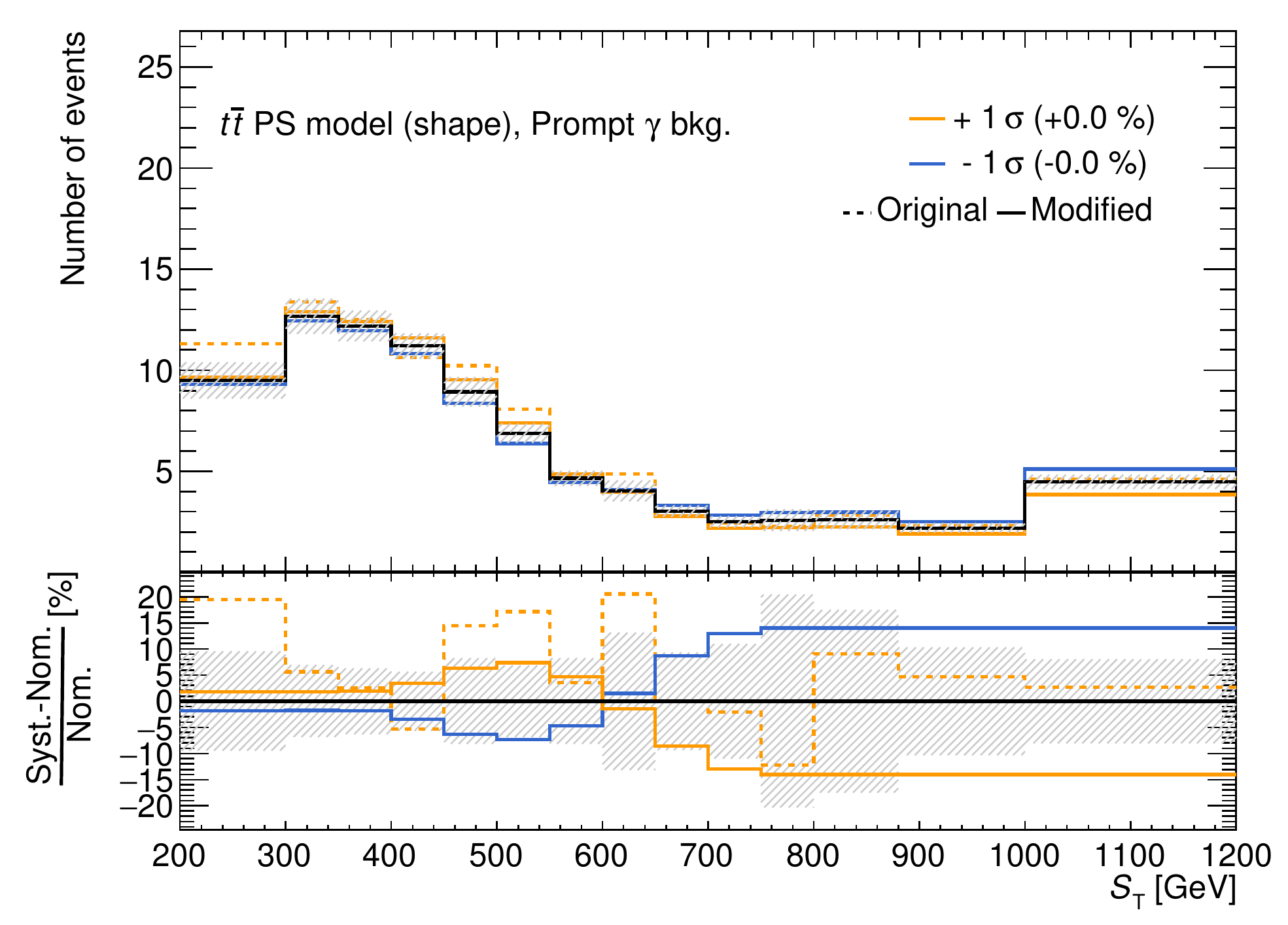}
  \caption[Templates for \ttbar \PS model systematics]{%
    Systematic templates for the \ttbar parton-shower model uncertainty in
    the \cathfake and
    the \catprompt categories.
    The templates are smoothed and one-sided symmetrisation is applied.
    Their rate components are dropped and their shape components are amplified by a factor of~3.
    The dashed lines are the non-smoothed templates; the final templates after smoothing and symmetrisation are shown in solid orange and blue.
    The shaded uncertainty bands represent \MC-statistical uncertainties on the nominal prediction.
  }
  \vspace*{3pt} 
  \label{fig:syst_ttbar_PS_model}
\end{figure}

To estimate the uncertainty on the parton shower and hadronisation generated with \Pythia, the same set of events used for the nominal prediction is showered with \Herwig~\cite{Bahr:2008pv} in alternative \ttbar simulations.
The systematic templates are smoothed and symmetrised via one-sided symmetrisation.
This results in two symmetric three-point uncertainties, one each for the \cathfake and \catprompt categories.
As done for \tty, since the parton-shower uncertainty is generated with \atlfast simulation, the templates are compared against an \atlfast simulation of the nominal \ttbar prediction.
The rate components of the templates are dropped.
As large bin-by-bin variations are observed, but removed partially by the smoothing algorithm, the smoothed shape components are amplified by a factor of~3 to avoid obscuring possible physics effects that caused these variations.
This provides a conservative estimate of this uncertainty.
The resulting templates are shown in \cref{fig:syst_ttbar_PS_model}.

The uncertainty on initial-state and final-state radiation is estimated with the \Pythia \emph{A14 var3c} eigentune.
Contrary to what is done for the signal processes, the variations are obtained through reweighting the nominal set of events.
Thus, no smoothing and symmetrisation are applied, but the rate components of the templates are dropped.

\begin{figure}
  \centering
  \includegraphics[width=0.48\textwidth]{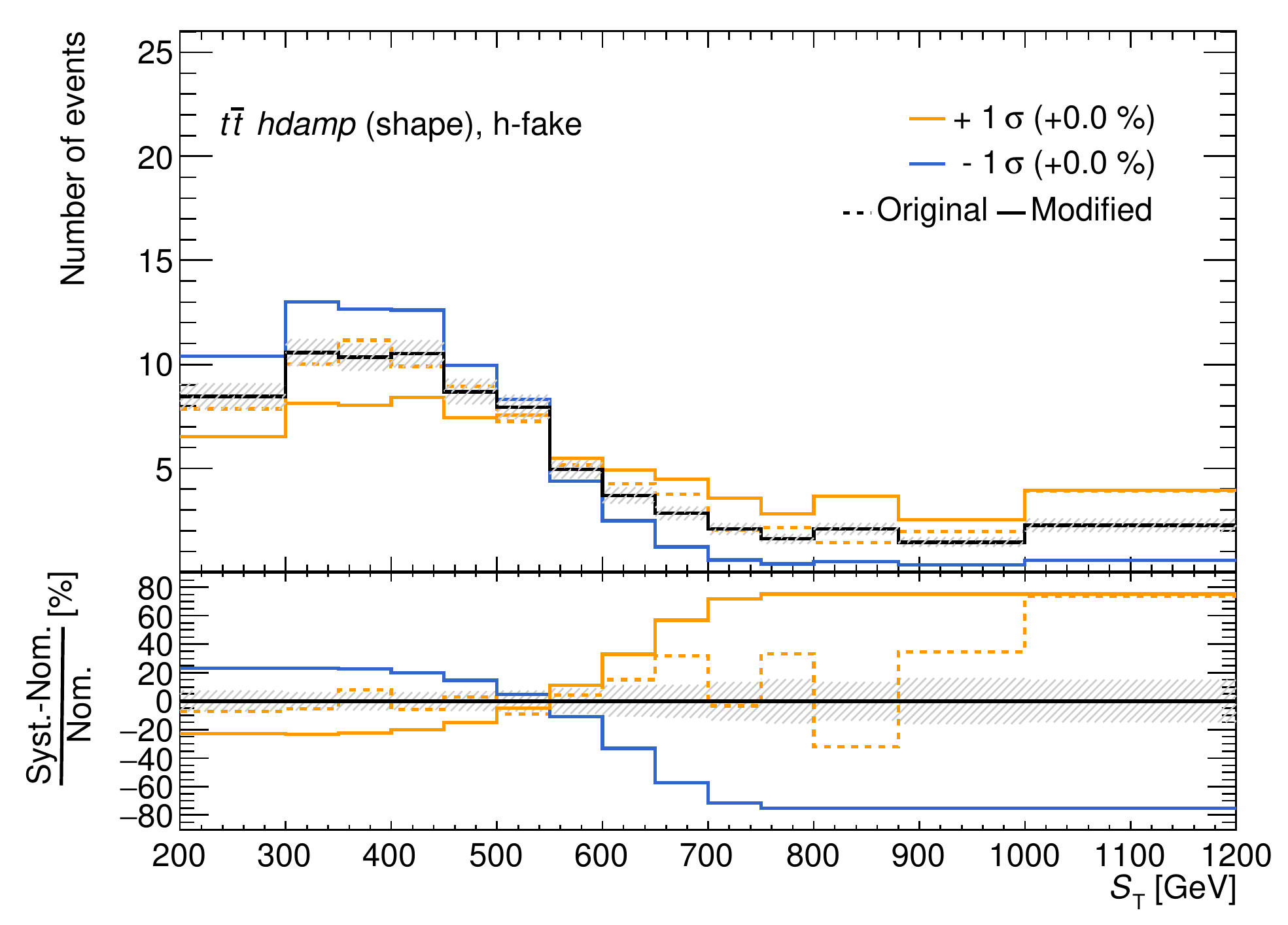}
  \includegraphics[width=0.48\textwidth]{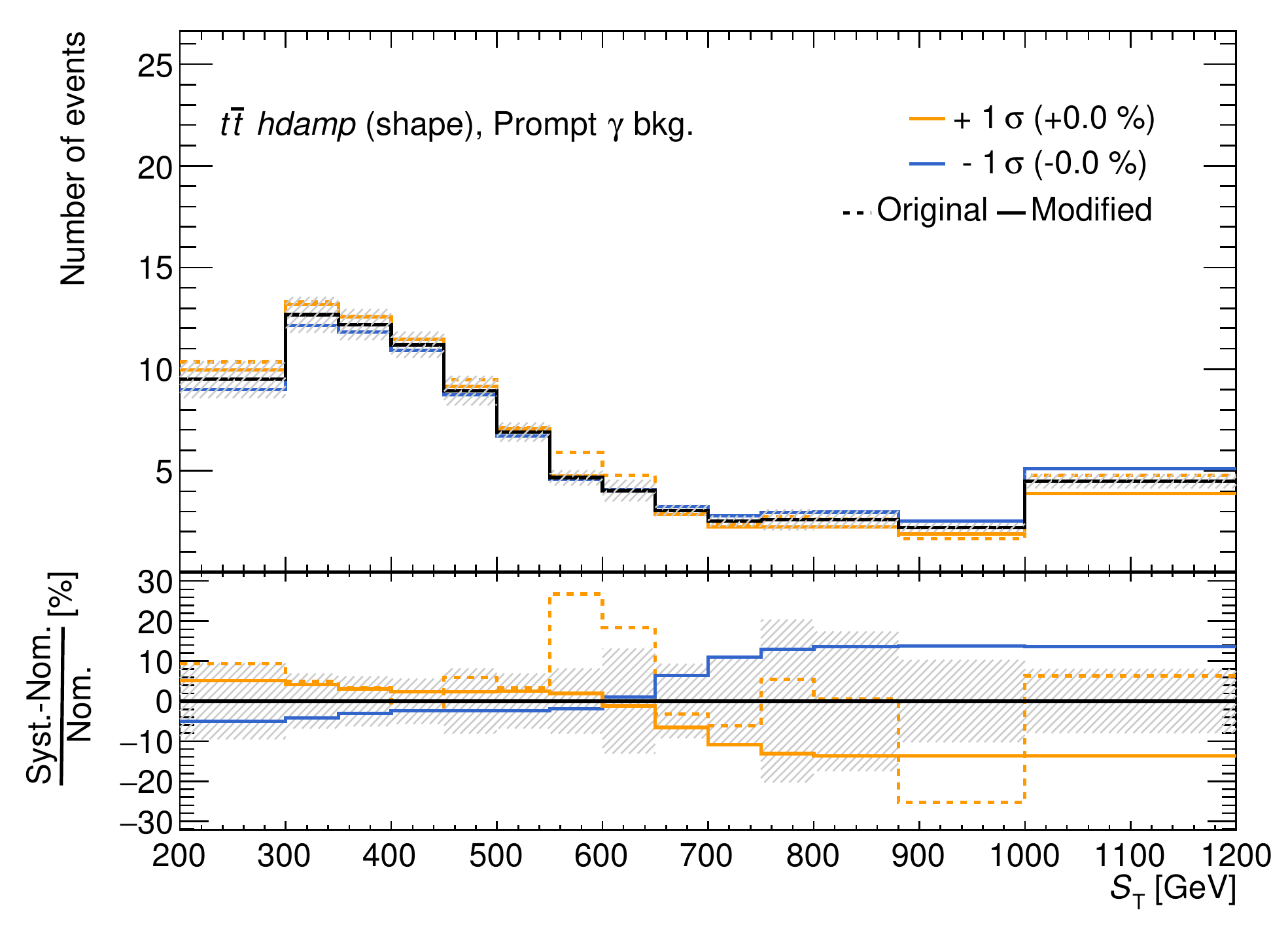}
  \caption[Templates for \ttbar hdamp systematics]{
    Systematic templates for the \ttbar \emph{hdamp} uncertainty in
    the \cathfake and
    the \catprompt categories.
    The templates are smoothed and one-sided symmetrisation is applied.
    Their rate components are dropped and their shape components are amplified by a factor of~3.
    The dashed lines are the non-smoothed templates; the final templates after smoothing, symmetrisation and amplification are shown in solid orange and blue.
    The shaded uncertainty bands represent \MC-statistical uncertainties on the nominal prediction.
  }
  \label{fig:syst_ttbar_hdamp}
\end{figure}

An additional uncertainty is considered for \ttbar, where the \emph{hdamp} parameter of \Powheg is varied, which controls the threshold of the hardest emission.
An alternative set of events is generated with the parameter set to twice the value used in the nominal prediction.
One-sided symmetrisation and smoothing are applied to the resulting template.
As the templates are generated with \atlfast simulation, they are compared against an \atlfast simulation of the nominal \ttbar prediction.
The rate components of the final templates are dropped, but the shape components are amplified by a factor of~3 to provide a conservative estimate of this uncertainty.
The resulting templates are shown in \cref{fig:syst_ttbar_hdamp}.

To validate the procedure of amplifying the shape of the parton-shower and \emph{hdamp} templates, the amplified scenario was compared with a scenario without shape amplification.
As a figure of merit, the expected uncertainty on the signal strength in an Asimov fit scenario was evaluated and compared against a fit without the \ttbar uncertainties in question included.
The results of these tests are listed in \cref{tab:syst_ttbar_shape} and show the relative change of the expected uncertainty with respect to the scenario without those systematics.
The values show a very minor increase of the expected uncertainty, even with amplified shape components of the systematic templates.
Thus, to provide conservative estimates for these uncertainties that are subject to large bin-by-bin variations, the amplified scenario was chosen for the fit.

\begin{table}
  \centering
  \caption[\ttbar shape amplification scenarios and their impact on the sensitivity]{%
    Impact of amplifying shape effects for the \ttbar \PS model and \emph{hdamp} variations.
    To test their impact on the fit, scenarios without the two systematics, with their shape components at their default amplitude and with their shape components amplified (3x) were evaluated.
    As a figure of merit, the relative change of the expected uncertainty on the signal strength in percent with respect to the dropped scenario is shown.
  }
  \sisetup{table-format=4.2,round-precision=2,round-mode=places,table-sign-mantissa}
  \label{tab:syst_ttbar_shape}
  \begin{tabular}{l c}
    \toprule
    Model & Relative change \\
    \midrule
    systematics dropped & --- \\
    shape only (1x)     & +0.03\,\% \\
    shape only (3x)     & +0.23\,\% \\
    \bottomrule
  \end{tabular}
\end{table}

\section{Uncertainties on the fiducial phase-space correction}
\label{sec:systematics-acceptance}

After having obtained a number of measured signal-like events in data, a fiducial \xsec is calculated at parton level as defined in \cref{eq:strategy-fidxsec-final} in \cref{cha:strategy}.
This involves correcting the obtained event yields at reconstruction level for migration and efficiency effects to obtain a value for the measured number of events in the fiducial volume at parton level.
This correction depends on the reconstruction efficiencies and on the fraction of migrated events, which pass the reconstruction-level selection, but do not lie in the fiducial phase-space volume.
Efficiency~$\epsilon$, migration fraction \fout and the correction factor~$C$ were already determined for the nominal prediction in \cref{sec:strategy-fid-phase-space}, which are the values used to determine the fiducial \xsec in \cref{chap:results}.

However, the modelling uncertainties introduced in the previous section might show different efficiency and migration effects.
By reweighting all signal modelling templates to the nominal fiducial acceptance, these differences are already included as rate differences in the templates when compared to the distributions of the nominal prediction.
Nonetheless, for each of these alternative models, the reconstruction efficiencies, migration fractions and correction factors can be calculated and can then be compared against the values obtained with the nominal prediction.
This gives additional insight into the importance of each modelling uncertainty, but is irrelevant for the profile likelihood fit or for the extraction and determination of the fiducial \xsec.
As \tty and \tWy were generated as separate processes, their systematic variations are evaluated individually, but always considering the impact on the combined values of $\epsilon$, \fout and~$C$.
As before, some of the systematic variations for the \tty process were not generated with a full simulation of the \ATLAS detector, but with \atlfast simulation.
To avoid any dependencies on possible differences between \atlfast and full simulation, a separate \atlfast simulation of the nominal \tty prediction is used as a reference for the systematic variations.
An overview of all modelling variations, their efficiencies, migration fractions and correction factors is given in \cref{tab:acceptance_systematics}.

\begin{table}
  \centering
  \caption[Efficiencies, outside-migration fractions and correction factors for the nominal signal predictions and all systematic variations]{%
    Obtained values for efficiencies~$\epsilon$, migration fractions~\fout and correction factors~$C$ for the nominal signal predictions and all systematic variations considered.
    These include variations of the renormalisation and factorisation scales $\mu_R$ and $\mu_F$, alternative modelling of the parton shower with \Herwig, variations of the \Pythia \emph{A14} tune and uncertainties originating from the choice of the \PDF set.
    For the latter, the mean values and standard deviations and the resulting relative uncertainties are quoted.
    In addition, the effect of the \tWy parton-definition uncertainty is listed.
  }
  \label{tab:acceptance_systematics}
  \begin{tabular}{
      l
      S[separate-uncertainty = false, table-format=1.4(1)]
      S[separate-uncertainty = false, table-format=1.4(1)]
      S[separate-uncertainty = false, table-format=1.4(1)]
      S[table-format=1.3, table-sign-mantissa]@{\,}l
    }
    \toprule
    Evaluated model & {$\epsilon$} & {\fout} & {$C$} & \multicolumn{2}{c}{rel. effect} \\
    \midrule
    Nominal (\tty FS + \tWy)          & 0.2973 & 0.3567 & 0.4622 & \multicolumn{2}{c}{---} \\
    Nominal (\tty \atlfast + \tWy)    & 0.2908 & 0.3584 & 0.4533 & \multicolumn{2}{c}{---} \\
    \midrule
    \tty $\mu_R \times 2.0$           & 0.2977 & 0.3562 & 0.4623 &  0.031 & \% \\
    \tty $\mu_R \times 0.5$           & 0.2967 & 0.3575 & 0.4618 & -0.072 & \% \\
    \tty $\mu_F \times 2.0$           & 0.2971 & 0.3563 & 0.4615 & -0.142 & \% \\
    \tty $\mu_F \times 0.5$           & 0.2977 & 0.3570 & 0.4630 &  0.180 & \% \\
    \midrule
    \tWy $\mu_R \times 2.0$           & 0.2973 & 0.3567 & 0.4622 &  0.006 & \% \\
    \tWy $\mu_R \times 0.5$           & 0.2973 & 0.3566 & 0.4621 & -0.008 & \% \\
    \tWy $\mu_F \times 2.0$           & 0.2973 & 0.3566 & 0.4621 & -0.023 & \% \\
    \tWy $\mu_F \times 0.5$           & 0.2974 & 0.3568 & 0.4623 &  0.031 & \% \\
    \midrule
    \tty \PS model (\Herwig)          & 0.2915 & 0.3352 & 0.4384 & -3.278 & \% \\
    \tty \Pythia \emph{A14 var3c up}  & 0.2908 & 0.3658 & 0.4585 &  1.152 & \% \\
    \tty \Pythia \emph{A14 var3c down}& 0.2940 & 0.3504 & 0.4526 & -0.150 & \% \\
    \tWy \PS model (\Herwig)          & 0.2975 & 0.3569 & 0.4626 &  0.094 & \% \\
    \midrule
    \tty \PDF & 0.2973 \pm 0.0003 & 0.3567 \pm 0.0009 & 0.4622 \pm 0.0007 &  0.141 & \% \\
    \tWy \PDF & 0.2973 \pm 0.0001 & 0.3567 \pm 0.0001 & 0.4622 \pm 0.0001 &  0.014 & \% \\
    \midrule
    \tWy parton definition            & 0.2973 & 0.3383 & 0.4495 & -2.750 & \% \\
    \bottomrule
  \end{tabular}
\end{table}

The dependency on the choice of the renormalisation and factorisation scales~$\mu_R$ and~$\mu_F$, estimated as described in \cref{sec:systematics-modelling}, shows little relative differences with respect to the nominal prediction.
Less than $\pm \SI{0.2}{\percent}$ difference are obtained when varying the \tty factorisation scale, and the effects of the renormalisation scale variations are even smaller.
The \tWy scale variations have low impact with relative uncertainties below $\pm \SI{0.05}{\percent}$.
Alternative showering with \Herwig is simulated with \atlfast and is, thus, compared against the nominal \atlfast simulation of the \tty and \tWy predictions.
The choice of the \tty parton-shower model shows a large impact with a relative difference of $- \SI{3.3}{\percent}$, while the alternative \tWy parton-shower model differs by $+ \SI{0.1}{\percent}$ from the nominal prediction.
Additional studies were performed to investigate the large impact of the \tty parton-shower model choice.
It was found that the difference in rate between the two samples is mostly introduced by the kinematic requirements on the clustered \bjets at parton level, and by the required distances of these \bjets to other parton-level objects in the \etaphi plane, but not by different photon kinematics in the two samples.
The \tty \Pythia \emph{A14 var3c} tune variations yield an asymmetric impact of $+\SI{1.2}{\percent}$ and $-\SI{0.2}{\percent}$.
Relative differences from the \tty \NNPDF variations, obtained by taking the standard deviation of the one hundred replicas, amount to approximately $\pm \SI{0.14}{\percent}$, that from the \tWy variations to one order of magnitude less, which justifies them being dropped for the fit setup in \cref{sec:systematics-modelling}.

A large relative difference is found for the uncertainty on the \tWy parton definition.
As described in the previous section, this uncertainty assumes a doubling of the \tWy events in the fiducial volume, while the efficiency~$\epsilon$ is kept constant.
The resulting relative difference for the correction factor~$C$ with respect to the nominal predictions amounts to $-\SI{2.75}{\percent}$.
As this uncertainty is purely based on the definition of the fiducial volume, it is added directly to the uncertainty of the final result in quadrature.
According to \cref{eq:strategy-fidxsec-final}, the dependence of the fiducial \xsec on~$C$ is inversely proportional, thus, the observed relative difference with a negative sign is added in quadrature to the upper systematic uncertainty of the fiducial \xsec in the following chapter.



\chapter{Results}
\label{chap:results}

Before proceeding to the results of the fiducial inclusive and differential \xsec measurements, the predicted event yields are re-evaluated with the systematic uncertainties included that were introduced in the previous chapter.
The pruning criteria are applied to reduce the number of nuisance parameters of the model.
However, correlations of the remaining nuisance parameters are only calculated \emph{during} the fit to data and, thus, the pre-fit predictions assume no correlation among them.%
\footnote{%
  In technical terms, this means that the Hessian matrix of second-order partial derivatives of the log-likelihood function with respect to the nuisance parameters is diagonal.
  Off-diagonal elements with mixed partial derivatives, \ie $\partial^2 \mathcal{L}/\partial \theta_i \partial\theta_j$ with $i \neq j$, are only estimated in the fitting.
}
The resulting predictions for each simulation category are listed in \cref{tab:results-prefit-yields}.
As done for the table in \cref{cha:selection} with \MC-statistical uncertainties only, the predictions of the \tty and \tWy categories marked with (*) were scaled in such a way that the total \MC prediction matches the numbers of reconstructed events in data in each column of the table.
Compared to the table without systematic uncertainties, the uncertainties on the predictions are enlarged significantly.
This is visible in particular for the \cathfake, \catefake and \catprompt categories, the uncertainties of which are dominated by the conservative \SI{50}{\percent} normalisation uncertainties introduced in \cref{sec:systematics-modelling}.

\begin{table}
  \centering
  \caption[Predicted event yields with all uncertainties included]{%
    Predicted event yields for all \MC categories and numbers of reconstructed events in \ATLAS data in the \emu signal region.
    The quoted uncertainties are combined statistical and systematic uncertainties.
    As done in \cref{tab:selection-statonly-yields}, the predictions of the \tty and \tWy categories marked with (*) were scaled to match the numbers of reconstructed events in data in each column.
    The scaling factors are \num{1.358}, \num{1.445}, \num{1.414} and \num{1.408}, respectively.
    Rounding to significant digits is applied to the quoted uncertainties, but the central values are given in integer values to have exact correspondence with the yields in data.
  }
  \label{tab:results-prefit-yields}
  \footnotesize
  \begin{tabular}{%
    l
    S[table-format=3.0] @{${}\pm{}$} S[table-format=2.0]
    S[table-format=3.0] @{${}\pm{}$} S[table-format=2.0]
    S[table-format=4.0] @{${}\pm{}$} S[table-format=2.0]
    S[table-format=4.0] @{${}\pm{}$} S[table-format=3.0]
    }
    \toprule
    & \multicolumn{2}{c}{2015/16} & \multicolumn{2}{c}{2017} & \multicolumn{2}{c}{2018} & \multicolumn{2}{c}{full dataset} \\
    \midrule
    \catttyemu{}* & 643 & 34 & 760 & 40 & 989  & 50 & 2391 & 130 \\
    \cattWyemu{}* & 43  & 3  & 50  & 3  & 63   & 9  & 156  & 15  \\
    \catother{}*  & 76  & 4  & 88  & 5  & 115  & 7  & 279  & 15  \\
    \cathfake     & 23  & 12 & 23  & 12 & 31   & 16 & 78   & 40  \\
    \catefake     & 6   & 3  & 7   & 4  & 10   & 5  & 23   & 12  \\
    \catprompt    & 19  & 10 & 30  & 15 & 38   & 19 & 88   & 40  \\
    \midrule
    Total \MC     & 809 & 40 & 958 & 50 & 1247 & 70 & 3014 & 160 \\
    \midrule
    Data                 & \multicolumn{2}{l}{809} & \multicolumn{2}{l}{958} & \multicolumn{2}{l}{1247} & \multicolumn{2}{l}{3014} \\
    \bottomrule
  \end{tabular}
\end{table}

\begin{figure}[p]
  \centering
  \includegraphics[width=0.48\textwidth]{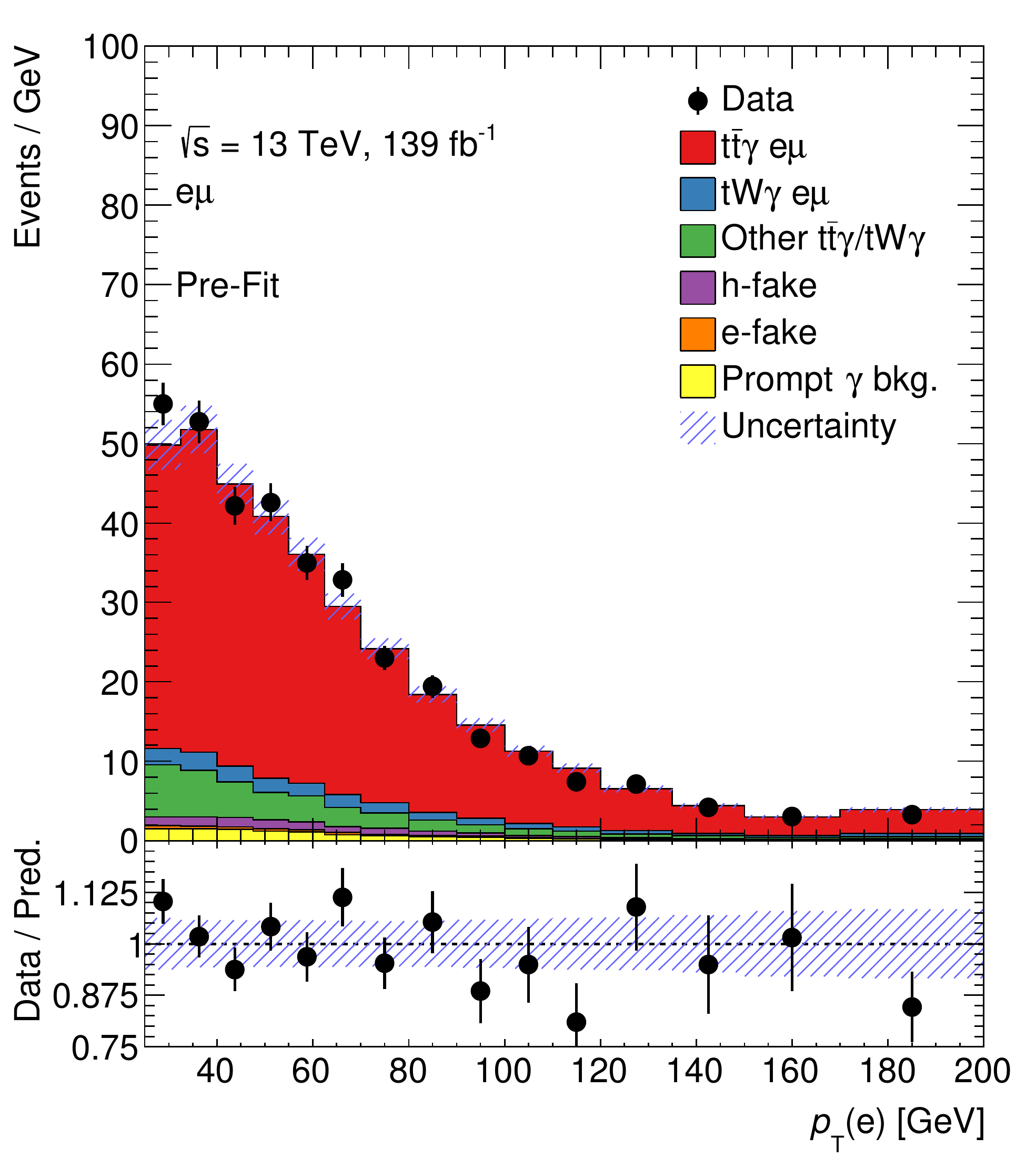}%
  \includegraphics[width=0.48\textwidth]{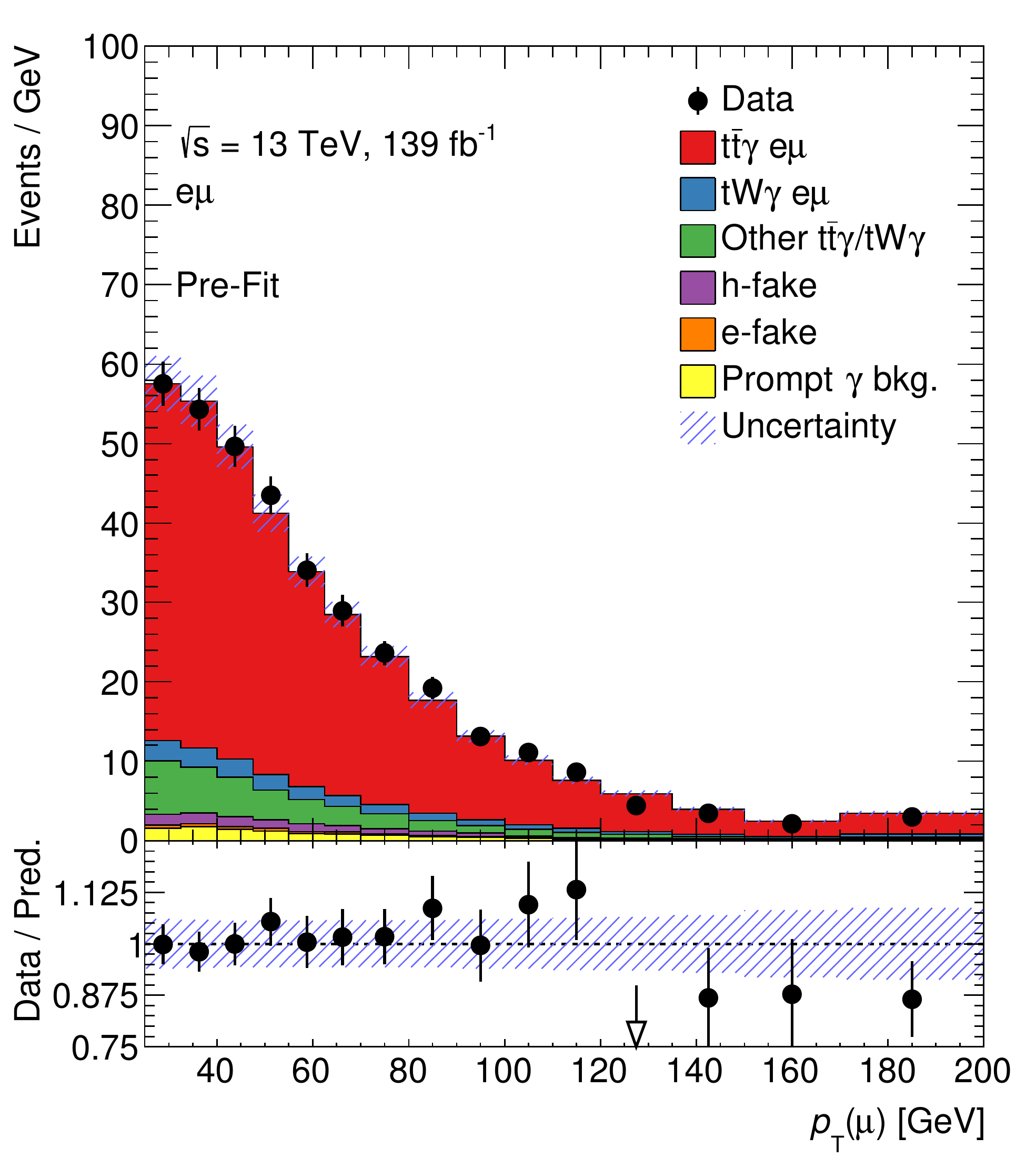}%
  \\
  \includegraphics[width=0.48\textwidth]{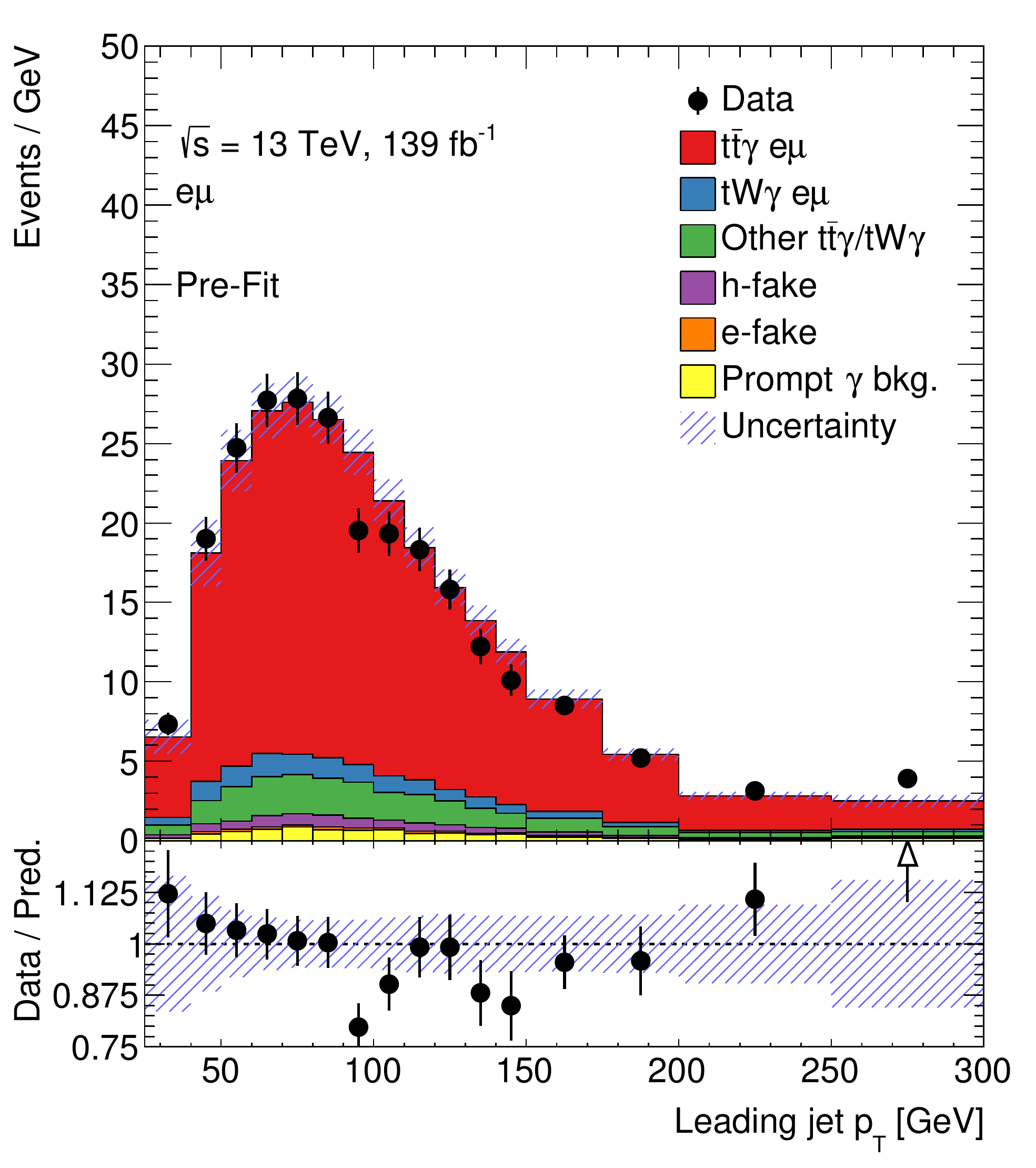}%
  \includegraphics[width=0.48\textwidth]{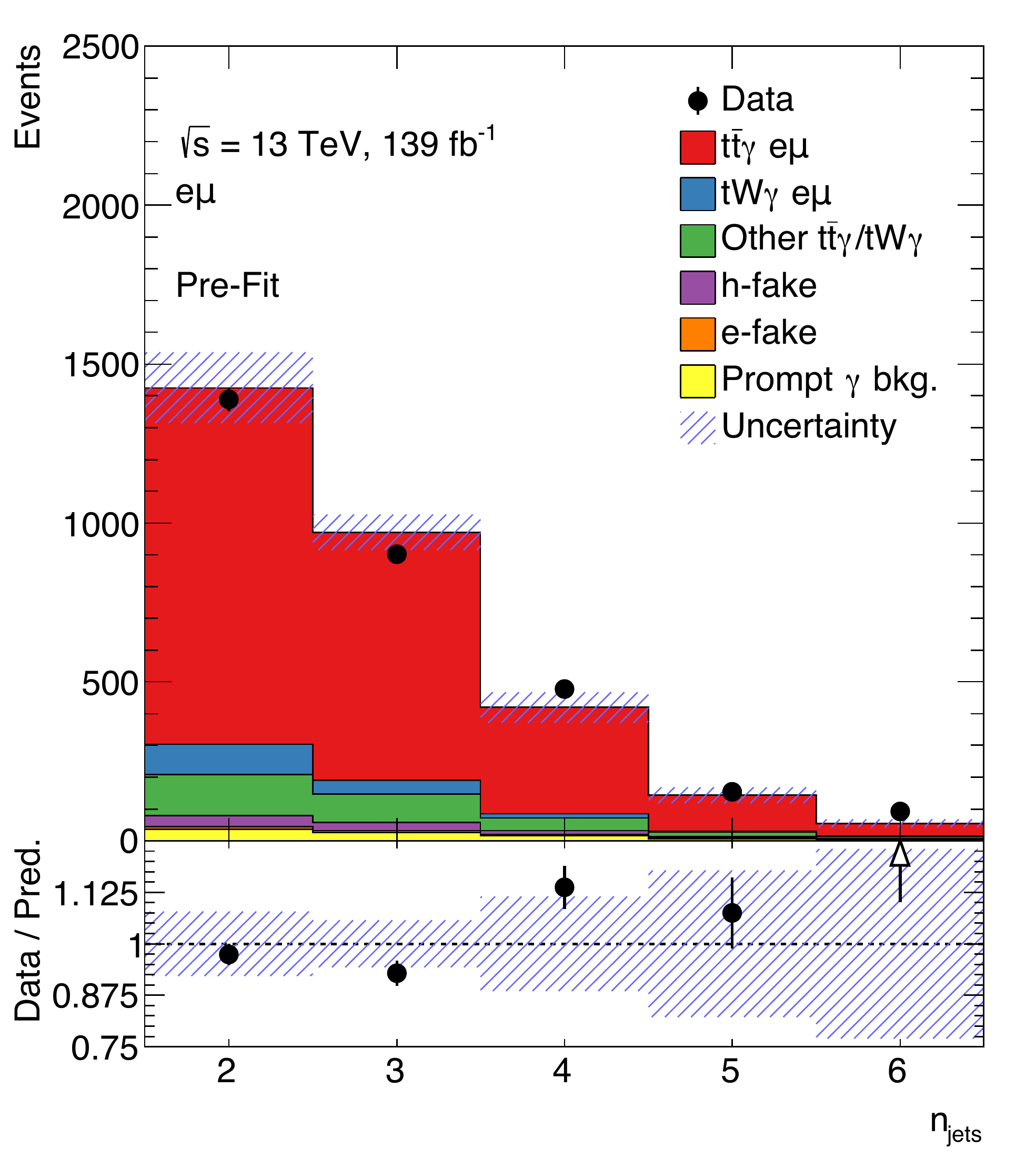}%
  \caption[Control plots with all uncertainties included (1)]{%
    Control plots for a data/\MC comparison in the \emu signal region.
    The shaded error bands of the prediction are combined statistical and systematic uncertainties.
    As in \cref{tab:results-prefit-yields}, the predictions of the \tty and \tWy categories were scaled to match the numbers of reconstructed events in data.
    The shown observables are the transverse momenta of the electron, of the muon and of the leading jet as well as the jet multiplicity.
  }
  \label{fig:results-controlplots-1}
\end{figure}

\begin{figure}[p]
  \centering
  \includegraphics[width=0.48\textwidth]{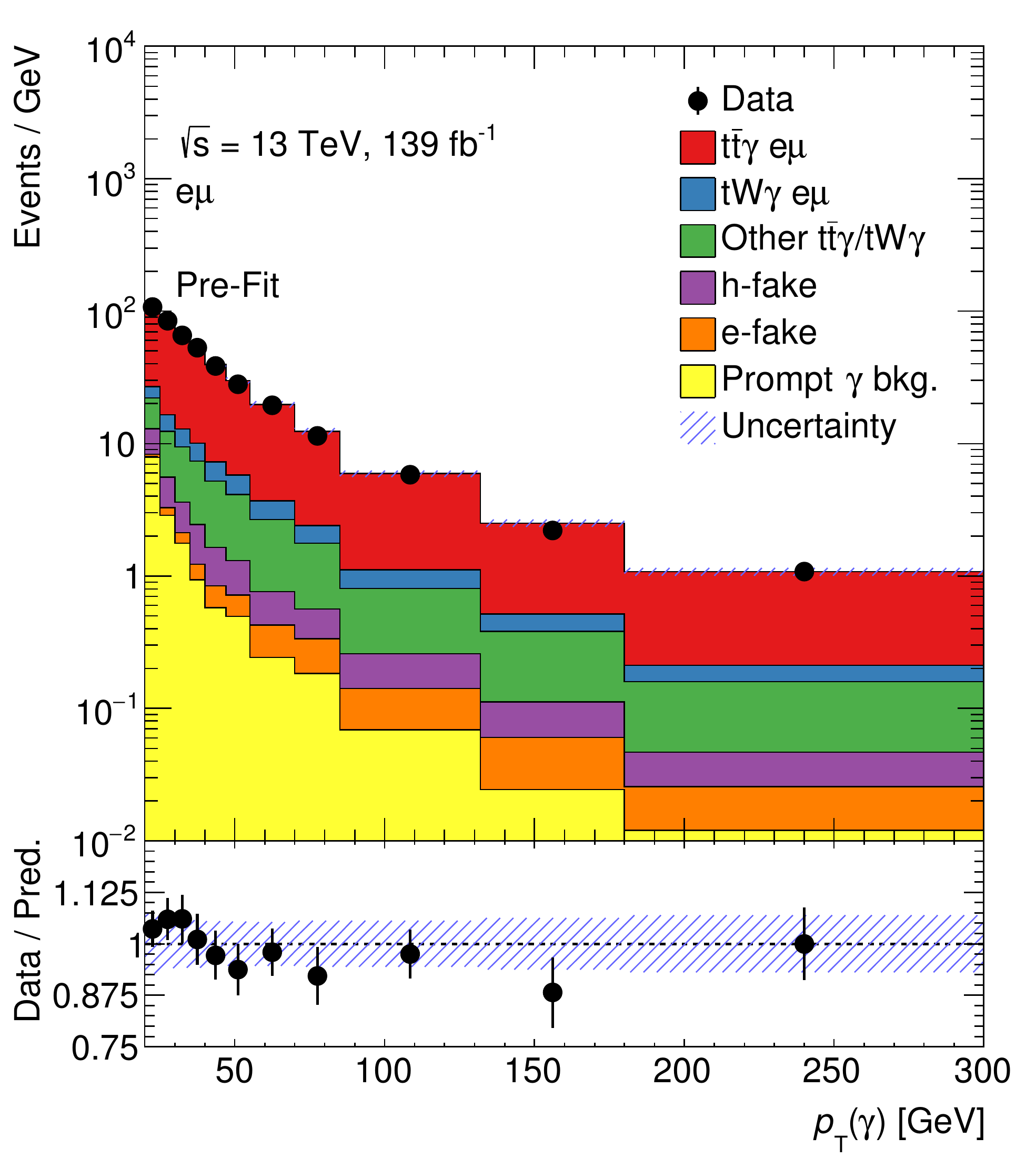}%
  \includegraphics[width=0.48\textwidth]{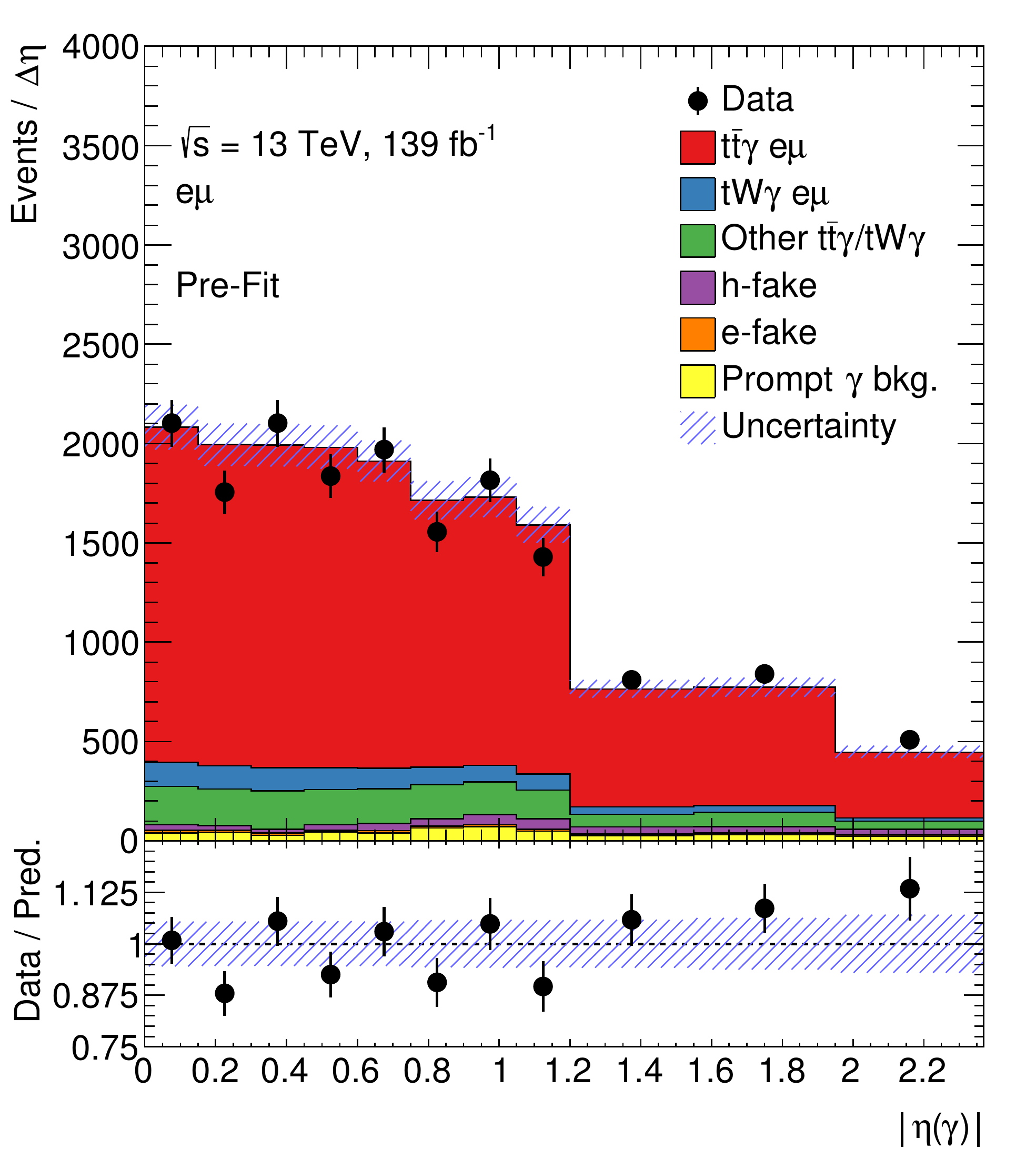}%
  \\
  \includegraphics[width=0.48\textwidth]{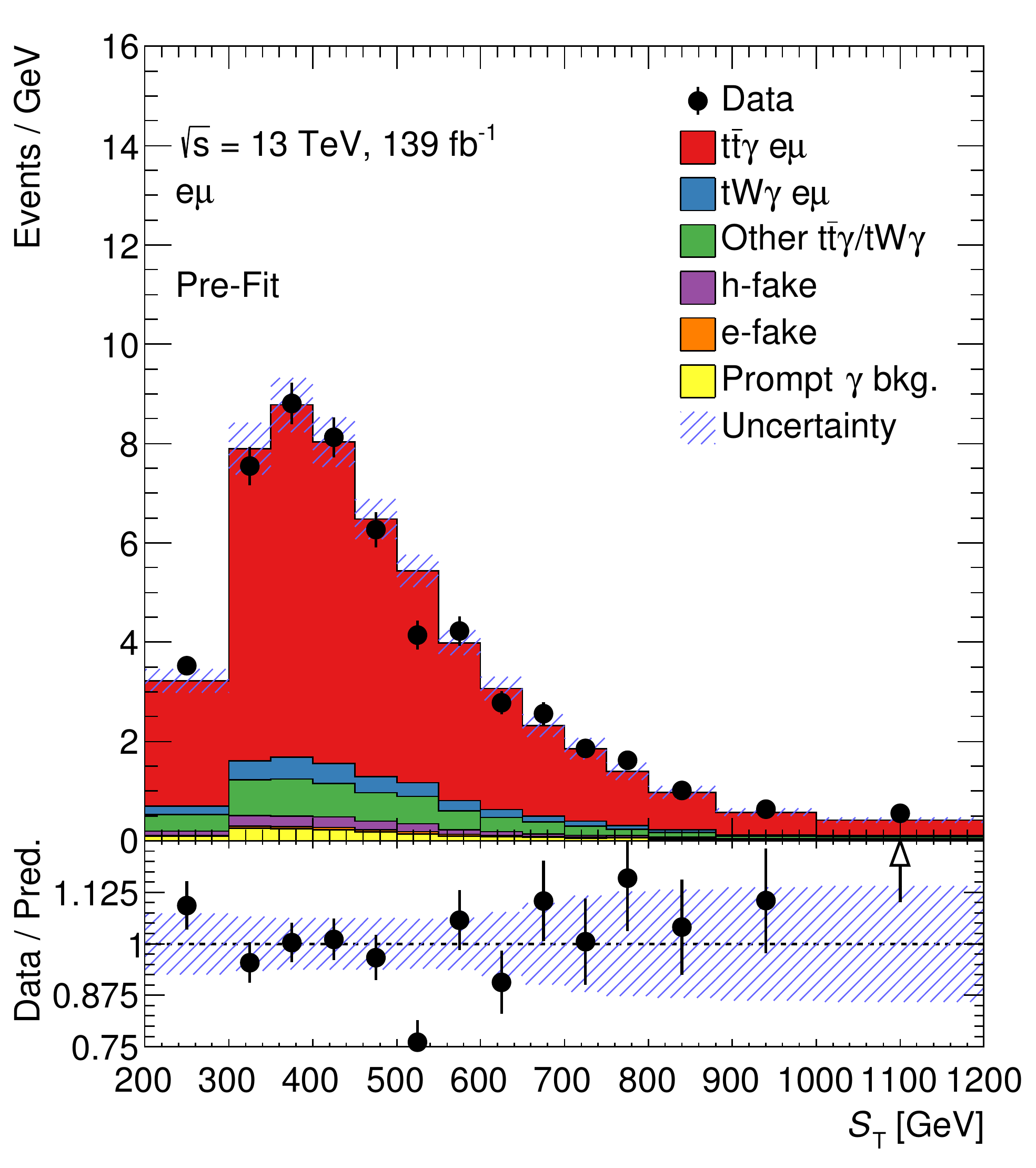}%
  \includegraphics[width=0.48\textwidth]{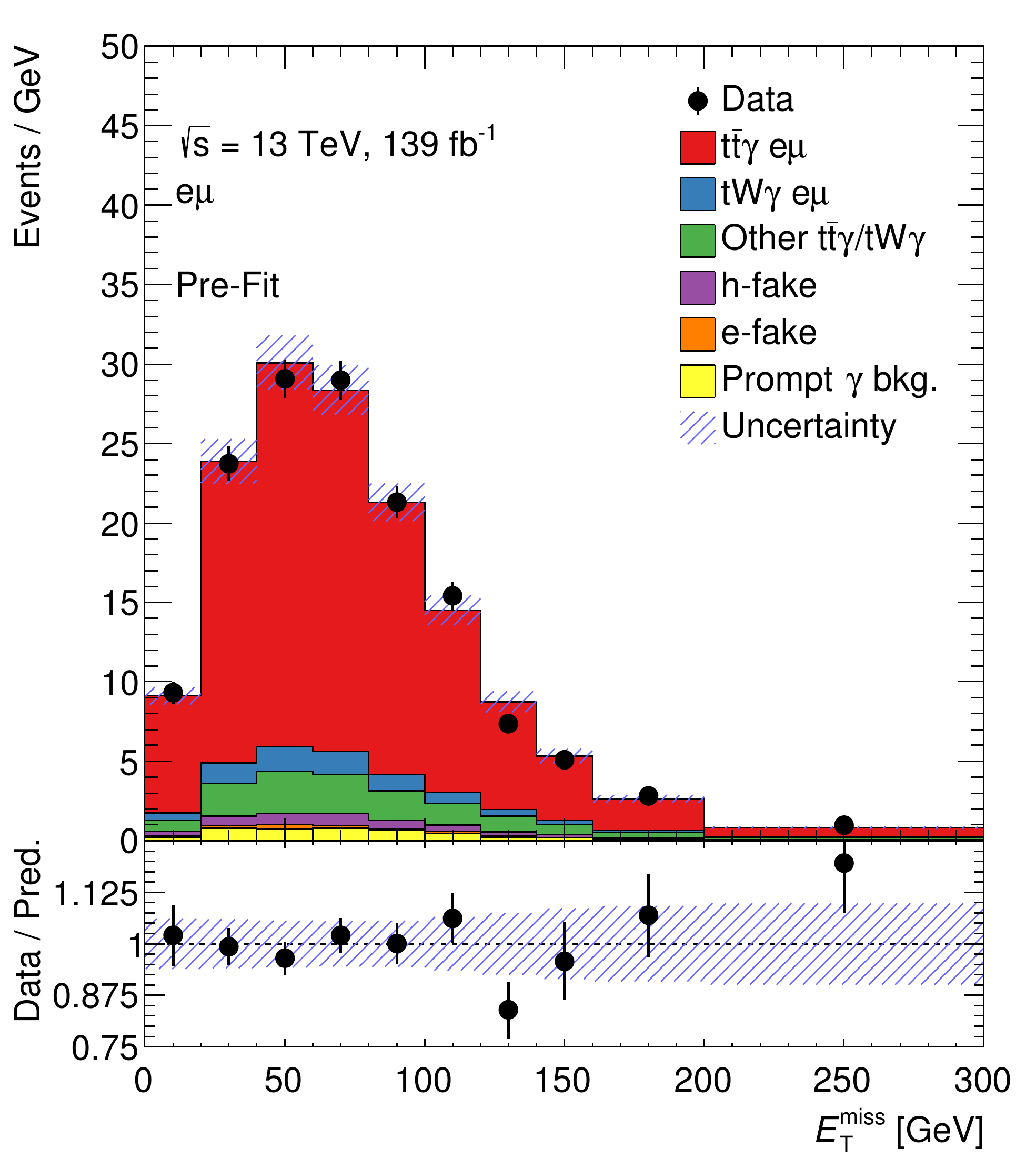}%
  \caption[Control plots with all uncertainties included (2)]{%
    Control plots for a data/\MC comparison in the \emu signal region.
    The shaded error bands of the prediction are combined statistical and systematic uncertainties.
    As in \cref{tab:results-prefit-yields}, the predictions of the \tty and \tWy categories were scaled to match the numbers of reconstructed events in data.
    The shown observables are the transverse momentum and absolute pseudorapidity of the photon, the missing transverse momentum \MET, and the scalar sum \ST of all transverse momenta of the event, including \MET.
  }
  \label{fig:results-controlplots-2}
\end{figure}

In addition, \cref{fig:results-controlplots-1,fig:results-controlplots-2} present data/\MC control plots in the \emu signal region of the same observables as shown in \cref{fig:selection-controlplots-1,fig:selection-controlplots-2}.
The shaded uncertainty bands represent combined statistical and systematic uncertainties.
\Cref*{fig:results-controlplots-1} shows the transverse momenta of the electron, of the muon and of the leading jet as well as the jet multiplicity.
\Cref*{fig:results-controlplots-2} shows the transverse momentum and absolute pseudorapidity of the photon, the missing transverse momentum \MET, and the scalar sum \ST of all transverse momenta of the event.
As done for the table, the combined integrals of the \tty and \tWy categories were scaled in such a way that the total integral of the \MC prediction matches the data yields for each plot.
The broadened pre-fit uncertainty bands now cover almost all discrepancies between \MC prediction and data that were pointed out previously:
the data points in the 3-jet and 4-jet bins of the jet multiplicity distribution lie within the shaded uncertainty bands, and also shape discrepancies at the tails of the \ST and \MET observables are covered by the uncertainty bands.
Control plots of additional observables, including those used for the differential \xsec measurements, are shown in \cref{chap:app-add-controlplots}.

The following sections of this chapter summarise the results of the analysis.
Initially, the fitting framework of the fiducial inclusive \xsec measurement was set up exactly as described in the previous chapters of this thesis.
Studies were then performed to test the stability of the fit, the behaviour of the systematic uncertainties and the expected sensitivity of the measurement.
The latter was estimated using \emph{Asimov} pseudo-data.
\Cref{sec:results-configuration} discusses these studies of the fit configuration and the chosen final fit setup.
\Cref{sec:results-Asimov} summarises the fit results obtained with the Asimov pseudo-data in this configured setup.
\Cref{sec:results-data} details the results of the profile likelihood fit, when the model is fit to \ATLAS data, and presents the central results of this work.
\Cref{sec:results-differential} shows the differential \xsec distributions obtained from unfolding.

\section{Studies of the fit configuration}
\label{sec:results-configuration}

With the strategy for extracting a fiducial inclusive \xsec value from \ATLAS data laid out in \cref{sec:strategy-fit}, and with the uncertainty model introduced in \cref{chap:systematics}, the fit framework can be set up using the \HistFactory package.
Before performing the profile likelihood fit on the \ATLAS dataset, various studies of the fit configuration were done.
This includes the studies on the pruning thresholds for systematic uncertainties and on the shape amplification of the \ttbar modelling uncertainties discussed in the previous chapter.
In addition, the stability of the fit procedure, the behaviour of the systematic uncertainties and the expected sensitivity of the measurement were evaluated using an \emph{Asimov} dataset.
As explained before, this set of pseudo-data is created based on the predicted number of total events in \MC simulation in each bin of the \ST distribution and, hence, constitutes a dataset that matches the \MC predictions perfectly.
Results obtained with Asimov pseudo-data in the final configured fit setup are summarised in \cref{sec:results-Asimov}.

\begin{figure}
  \centering
  \includegraphics[scale=0.7,clip,trim= -80pt 0 170pt 0]{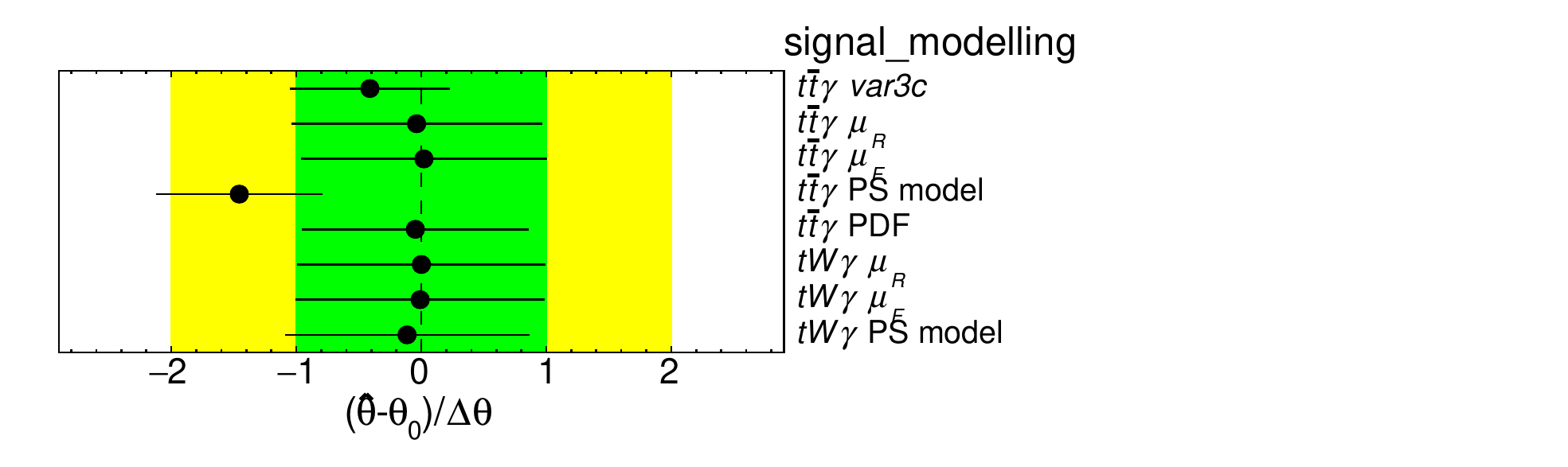}
  \caption[Nuisance-parameter pulls/constraints in early data fits]{%
    Pull distributions of those nuisance parameters associated with the signal modelling in early fits to \ATLAS data.
    The nuisance parameters of the \tty \emph{var3c} and the \tty \PS model uncertainties are strongly pulled and constrained.
  }
  \label{fig:results-correlated-scenario-modelling-NPs}
\end{figure}

This section addresses a few caveats of the initial fit setup that were spotted in early fitting tests to Asimov pseudo-data and to \ATLAS data.
More specifically, it was noticed that two of the nuisance parameters associated with the modelling of the \tty signal showed critical behaviour and needed a revision of the strategy:
those of the \tty \Pythia \emph{A14 var3c} and of the \tty \PS model uncertainties.
Fits to Asimov pseudo-data showed strong constraints of their post-fit uncertainties, that is, their post-fit uncertainty estimates $\Delta \estimate{\theta}$ were reduced significantly compared to their prior uncertainties $\Delta \theta$.
This hints towards a possible overestimation of the associated systematic uncertainties prior to the fit.
The same behaviour was seen when fitting to \ATLAS data.
In addition, the data fit revealed strong nuisance-parameter pulls, \ie their post-fit estimates $\estimate{\theta}$ were far away from their prior central values $\theta_0$.
This indicates that the data \enquote{favours} the alternative predictions over the nominal models, the latter of which are associated with the prior central values of the nuisance parameters.
\Cref{fig:results-correlated-scenario-modelling-NPs} gives a visual overview of the post-fit estimates $\estimate{\theta}$ for all modelling uncertainties in early tests with \ATLAS data.
To allow an easier comparison, the post-fit values $\estimate{\theta}$ are shifted by $\theta_0$ and displayed as a fraction of $\Delta\theta$.
While all other modelling uncertainties show post-fit values around zero and error bars that approximately match the width of the green prior-uncertainty band, the \tty \emph{var3c} and \tty \PS model parameters are not centred and their error bars are visibly shorter than those of the others.

\begin{figure}
\centering
\includegraphics[width=0.48\textwidth]{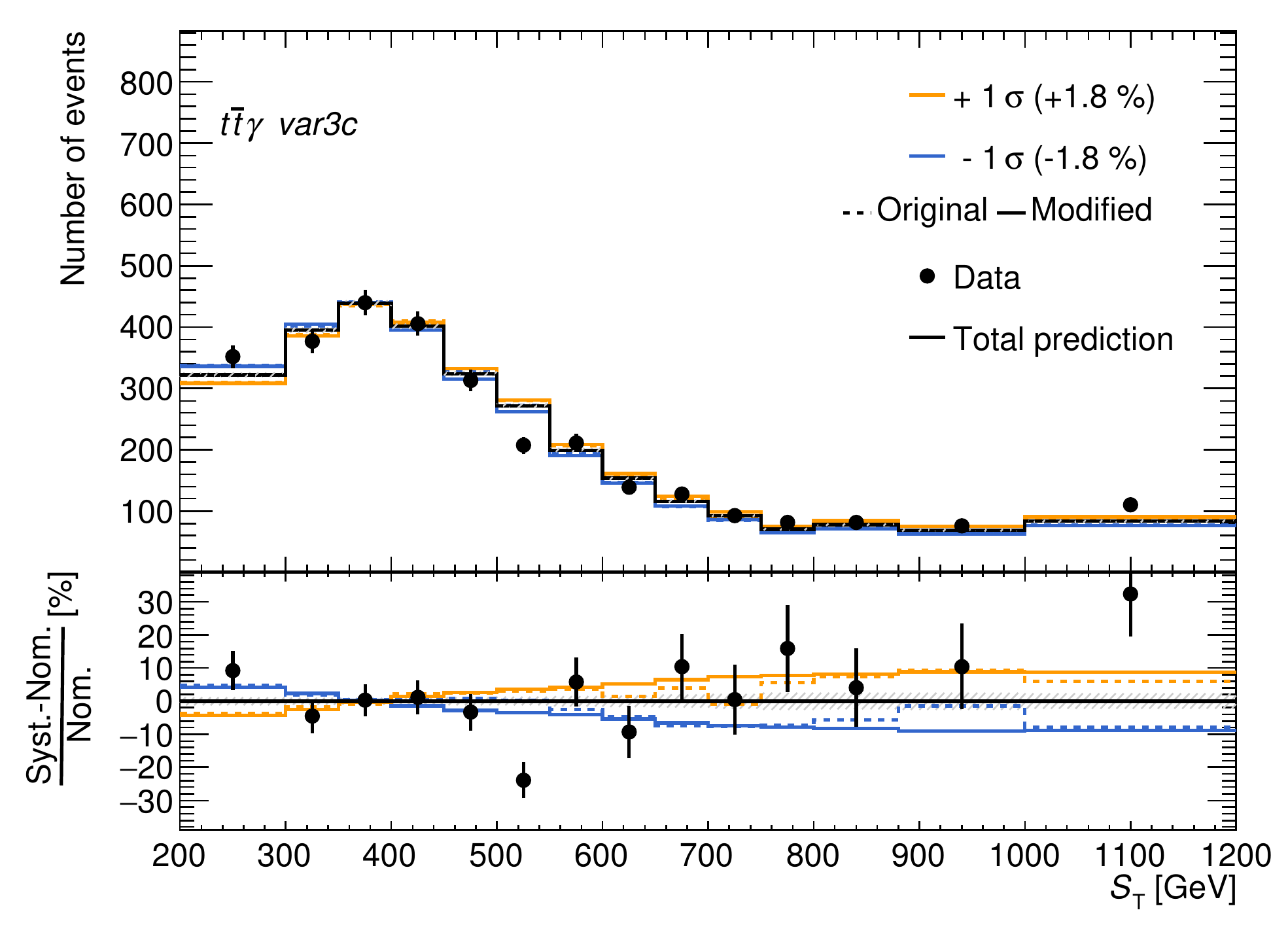}
\includegraphics[width=0.48\textwidth]{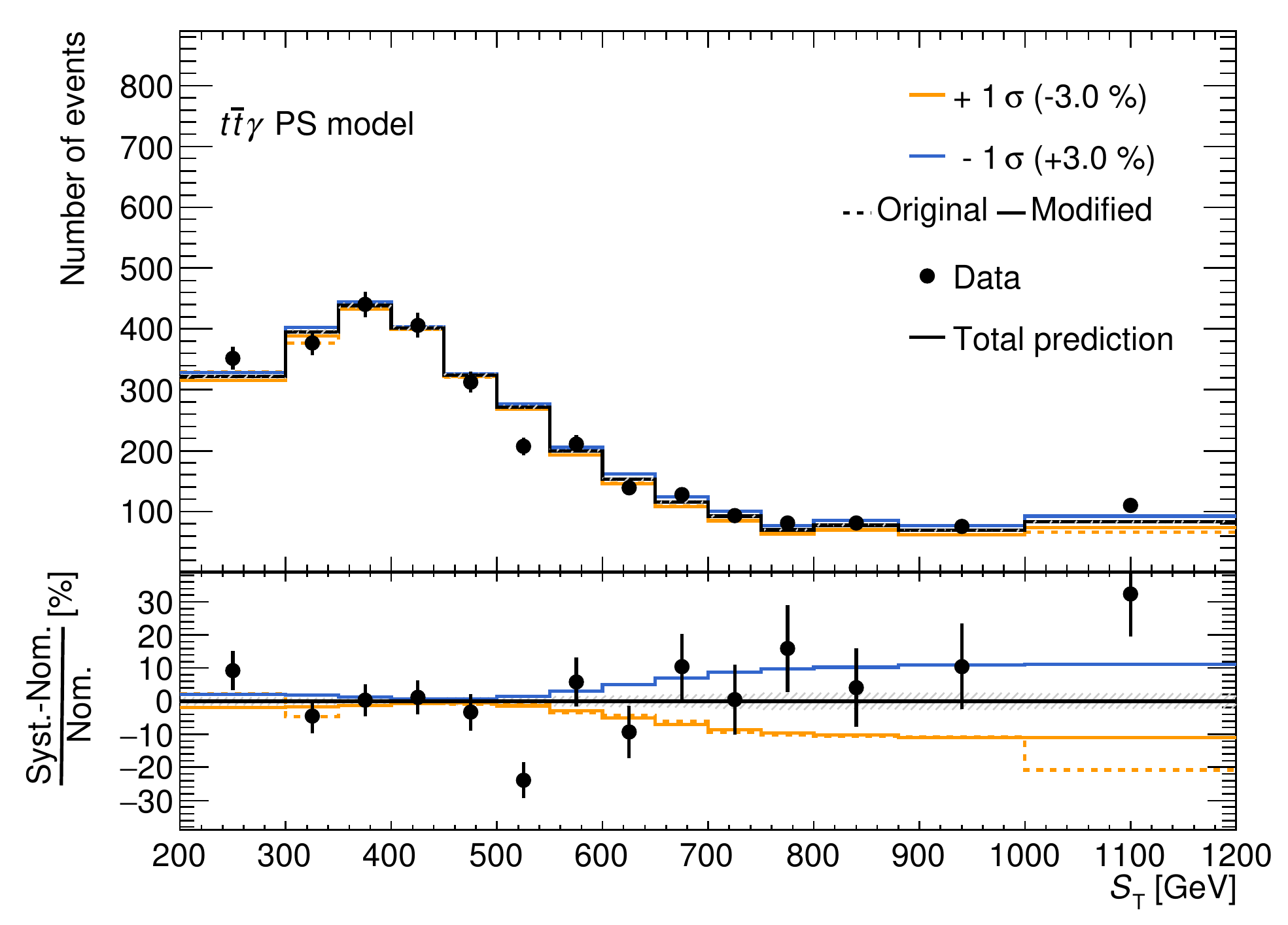}
\caption[Comparison to data for problematic \tty modelling templates]{%
  Combined systematic templates for the \tty \emph{var3c} and the \tty \PS model uncertainties in comparison with data.
  The combined templates of the \tty and \tWy categories are scaled in such a way that the total integral of the \MC prediction matches the event yields in data.
  The systematic variations were scaled with the same factors to highlight how they relate to the nominal prediction and to data.
}
\vspace*{3pt} 
\label{fig:results-problematic-NPs}
\end{figure}

The reason for the pulls and constraints can be understood when comparing the nominal and alternative models with the \ATLAS data they are fitted against.
\Cref{fig:results-problematic-NPs} shows modified versions of the template control plots presented in the previous chapter:
here, the plots highlight the combined impact of the templates on the total \MC prediction, and they are compared to data directly.
Again, the integrals of the nominal \tty and \tWy predictions are scaled in such a way that the total integral of the \MC prediction matches the event yields in data.
The systematic templates were scaled with the same factors to allow a direct comparison with the nominal prediction and with data.
Nominal prediction and data deviate visibly in three areas of the \ST distribution:
the prediction undershoots and overshoots data in the bins centred at \SI{250}{\GeV} and \SI{525}{\GeV}, respectively, and the prediction underestimates data for $\ST > \SI{750}{\GeV}$.
The strong pull of the nuisance parameter associated with the \tty \PS model uncertainty is most likely caused by the latter where data agrees more with the down variation shown in blue over the entire range than with the nominal prediction.
In addition, the bin at \SI{250}{\GeV} is described better by this variation than by the nominal prediction.
The nominal prediction disagrees with data in the bin at \SI{525}{\GeV}, but neither the up nor the down variation show remarkably better agreement with data than the nominal prediction.
Similarly, the pull of the \tty \emph{var3c} nuisance parameter towards the down variation can be explained by the bins at \SIlist{250;525}{\GeV}:
in both bins, the down variation in solid blue describes the data spectrum better than the nominal prediction.

As pulls and constraints are observed for both of these uncertainties, and as both are expected to have a large impact on the sensitivity of the result, it was decided to de-correlate rate and shape effects of their templates.
That is, the parameters $\eta(\theta)$ and $\sigma_b(\theta)$, reflecting overall normalisation differences and bin-by-bin shape differences, respectively, were included in the fit through two separate nuisance parameters.
This allows both parameters to be treated independently.
In particular, in a de-correlated scenario the overall rate uncertainty cannot be affected by any pulls and constraints due to shape information from the templates.
And while this is a conservative estimate of the impact of these two modelling uncertainties, the observed pulls and constraints can be attributed separately to the rate and shape components.
The expectation here is that both components may develop correlations with other nuisance parameters and the parameter of interest and may be pulled, but only the shape can be constrained through a bin-by-bin comparison with data.

\Cref{tab:fit_data_correlated_pulls_constraints} shows the effect of removing the rate components from the \tty \emph{var3c} and \tty \PS model uncertainty templates (which are then included into the fit model with their own nuisance parameters).
The table lists the constraints $\Delta \estimate{\theta}/\Delta\theta$ observed in a fit to Asimov pseudo-data and the constraints and pulls observed for a fit to \ATLAS data.
The arrows point out how pulls and constraints change when removing the rate uncertainty from the templates -- hence, effectively comparing combined rate and shape templates to shape-only templates:
in both fits to Asimov and \ATLAS data, the constraints are relaxed by \SIrange{6}{11}{\percent} for the two nuisance parameters.
In addition, the pull $(\estimate{\theta} - \theta_0)/\Delta\theta$ observed for the \tty \emph{var3c} is reduced by almost \SI{50}{\percent}, and that of the \tty \PS model uncertainty by about \SI{4}{\percent}.
The strong pull reduction for the \tty \emph{var3c} variations can be explained with \cref{fig:results-problematic-NPs}:
as the rate uncertainties are removed, the curves of the down and up variation move up and down by \SI{1.8}{\percent}, respectively.
This yields little difference between the templates in the bin centred at \SI{525}{\GeV}, where prediction and data deviate.
Remaining deviations in the bin at \SI{250}{\GeV} and for $\ST > \SI{750}{\GeV}$ do not favour the same, but opposing variations, thus, resulting in an overall much reduced pull.

\begin{table}
  \centering
  \caption[Effect of de-correlating shape and rate for \tty modelling]{%
    Constraints and pulls when removing the rate components from the \tty \emph{var3c} and \tty \PS model uncertainty templates.
    In a fit to Asimov pseudo-data, the post-fit constraints, $\Delta \estimate{\theta}/\Delta\theta$, are relaxed by about \SIlist{8;6}{\percent}, respectively.
    The same effect is observed when fitting \ATLAS data with reductions of \SIlist{11;7}{\percent}, respectively.
    The fit to \ATLAS data also shows reduced pulls when the rate components are removed: the value $(\estimate{\theta} - \theta_0)/\Delta\theta$ is reduced by almost \SI{50}{\percent} for the \tty \emph{var3c} uncertainty, and by about \SI{4}{\percent} for the \tty \PS model variation.
  }
  \label{tab:fit_data_correlated_pulls_constraints}
  \sisetup{round-precision=1,round-mode=places}
  \begin{tabular}{
    l
    S[table-format=2.1] @{${\,}\to{\,}$}
    S[table-format=2.2]
    S[table-format=2.1] @{${\,}\to{\,}$}
    S[table-format=2.2]
    S[round-precision=2, table-format=1.2, table-sign-mantissa] @{${\,}\to{\,}$}
    S[round-precision=2, table-format=1.3, table-sign-mantissa]
    }
    \toprule
    \multirow{2}{*}{Parameter}
    & \multicolumn{2}{c}{Fit to Asimov}
    & \multicolumn{4}{c}{Fit to \ATLAS data} \\
    & \multicolumn{2}{c}{$\Delta \estimate{\theta} / \Delta \theta$ [\si{\percent}]}
    & \multicolumn{2}{c}{$\Delta \estimate{\theta} / \Delta \theta$ [\si{\percent}]}
    & \multicolumn{2}{c}{$( \estimate{\theta} - \theta_0 ) / \Delta \theta$} \\
    \midrule
    \tty \emph{var3c} & +68.1677 & +73.6353 & +64.0305 & +71.2051 & -0.41089 & -0.211017 \\
    \tty \PS model    & +70.8728 & +75.1225 & +66.4444 & +70.9973 & -1.45585 & -1.40324 \\
    \bottomrule
  \end{tabular}
\end{table}

In total, the sensitivity of the result is expected to decrease with uncorrelated rate and shape components of two dominant modelling uncertainties as the uncorrelated rate components cannot be constrained with shape information.
More side-by-side comparisons of the impact of these changes are shown in \cref{chap:app-fit-decorrelation}, where this question is also addressed.
The following sections now summarise the fit results in the configured setup with de-correlated rate and shape components.
Results on Asimov pseudo-data including calculations of the expected sensitivity are given in \cref{sec:results-Asimov}, and results on \ATLAS data including the resulting fiducial inclusive \xsec are shown in \cref{sec:results-data}.

\section{Fit results with Asimov pseudo-data}
\label{sec:results-Asimov}

Before fitting the configured fit model against \ATLAS data, test scenarios with Asimov pseudo-data provide valuable information about the model itself.
The pseudo-data, created directly from the nominal prediction, represents a \enquote{perfect} dataset where none of the nuisance parameters is expected to be pulled from its nominal value $\theta_0$, \ie the post-fit estimate obeys $\estimate{\theta} = \theta_0$.
The same goes for the parameter of interest: as the spectra of the pseudo-data and of the nominal prediction are identical, the post-fit estimate of the signal strength remains at \num{1.0}.
However, the log-likelihood is profiled with respect to the nuisance parameters and the impact of the nuisance-parameter variations on the final result is evaluated.
Although no pulls of the nuisance parameters are expected, their post-fit uncertainty estimate $\Delta\estimate{\theta}$ can still be reduced compared to the prior uncertainty $\Delta\theta$, \ie the parameter can be constrained.
\Cref{tab:results_fit_Asimov_constraints} lists all nuisance parameters constrained during a fit to Asimov pseudo-data by more than \SI{10}{\percent}.
The only uncertainties with significant constraints are the shape components of the \tty \emph{var3c} and of the \tty \PS model variations that were de-correlated from their rate components, \cf \cref{sec:results-configuration}.

\begin{table}
  \centering
  \caption[List of constrained nuisance parameters in Asimov fit]{%
    Constrained nuisance parameters in a fit to Asimov pseudo-data.
    Only those with post-fit uncertainty estimates $\Delta \estimate{\theta}$ smaller than \SI{90}{\percent} of the prior uncertainties $\Delta\theta$ are listed.
    Large constraints are only observed for the shape components of the \tty \emph{var3c} and of the \tty \PS model variations.
    For these two uncertainties, the shape and rate components of the templates were de-correlated as detailed in \cref{sec:results-configuration}.
  }
  \label{tab:results_fit_Asimov_constraints}
  \sisetup{round-precision=1,round-mode=places}
  \begin{tabular}{l S[table-format=2.1]}
    \toprule
    \multirow{2}{*}{Nuisance parameter} & {constraint} \\
     & {$\Delta \hat{\theta} / \Delta \theta$ [\si{\percent}]} \\
    \midrule
    \tty \emph{var3c} (shape) & +73.6353 \\
    \tty \PS model (shape)    & +75.1225 \\
    \bottomrule
  \end{tabular}
  \vspace*{6pt}  
\end{table}

\begin{figure}[p]
  \vspace*{0.1\textheight}  
  \centering
  \includegraphics[width=0.90\textwidth,trim=20pt 0 -20pt 0]{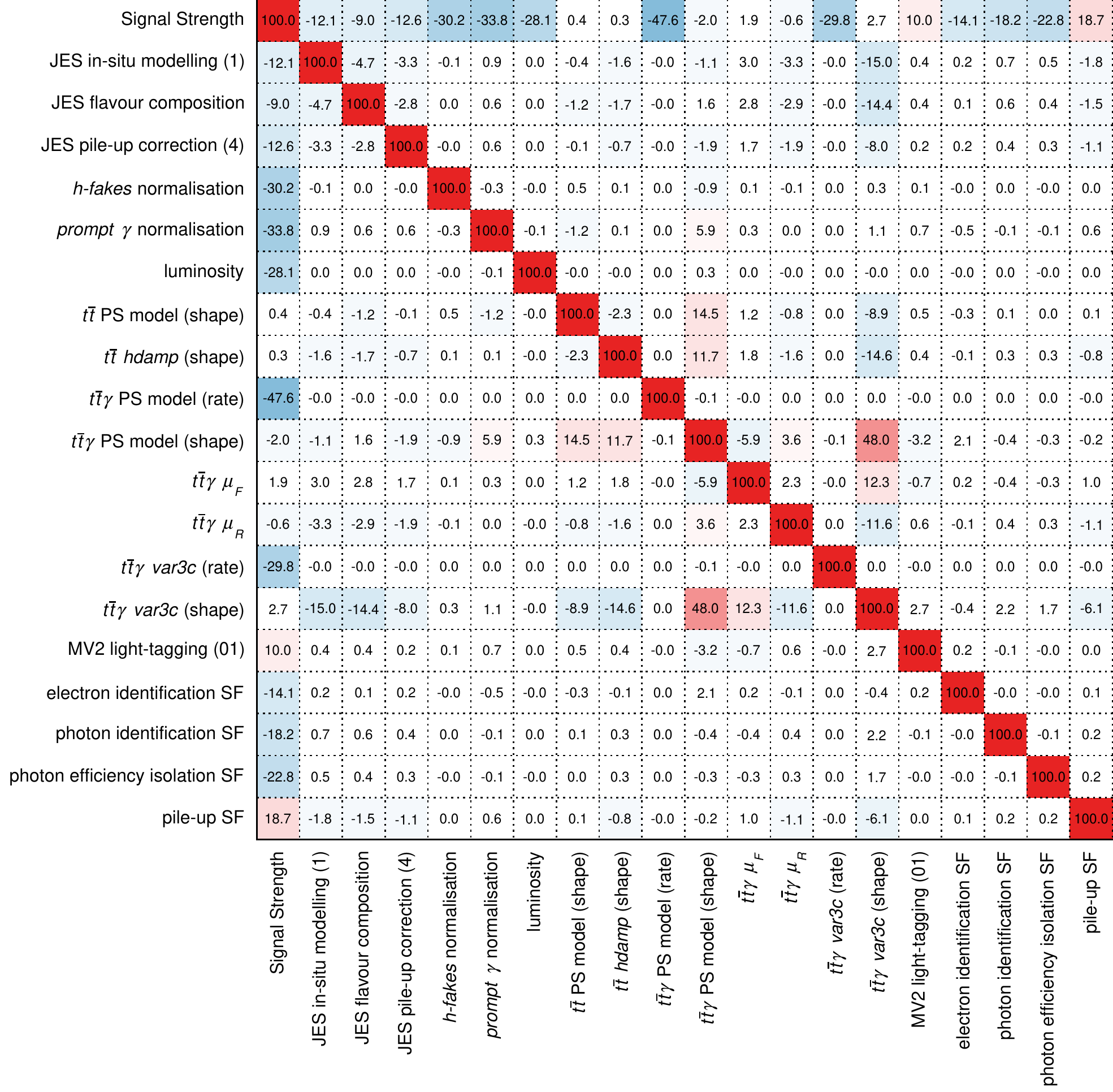}
  \caption[Correlation matrix of nuisance parameters in Asimov fit]{%
    Correlation matrix for the fit to Asimov pseudo-data.
    All nuisance parameters with \SI{10}{\percent} or higher absolute correlation value to other parameters or to the signal strength are shown.
  }
  \label{fig:results-Asimov-corrmatrix}
  \vspace*{0.1\textheight}  
\end{figure}

In addition to possible constraints, an Asimov fit scenario also gives an estimate of the correlation values between the model parameters.
A correlation matrix of all nuisance parameters with \SI{10}{\percent} or higher absolute correlation value to others or to the signal strength is presented in \cref{fig:results-Asimov-corrmatrix}.
Many of the listed nuisance parameters show large (positive or negative) correlation values with the signal strength:
among those are the rate components of the aforementioned signal modelling uncertainties, \ie the \tty \emph{var3c} and \tty \PS model rate variations.
Overall normalisation uncertainties of \SI{\pm 50}{\percent} assigned to the \cathfake and \catprompt categories show large negative correlation values with the signal strength as their rates influence the value of $\estimate{\mu}$ directly.
The same is observed for the luminosity uncertainty: as the luminosity increases, the signal strength decreases, hence a large negative correlation value.
Otherwise noteworthy are the shape components of the \tty \emph{var3c} and of the \tty \PS model uncertainties that are highly correlated with many of the other listed nuisance parameters.
They are correlated with each other with \SI{48}{\percent} which is the largest overall absolute value of correlation observed.

\begin{figure}
  \centering
  \includegraphics[width=0.71\textwidth]{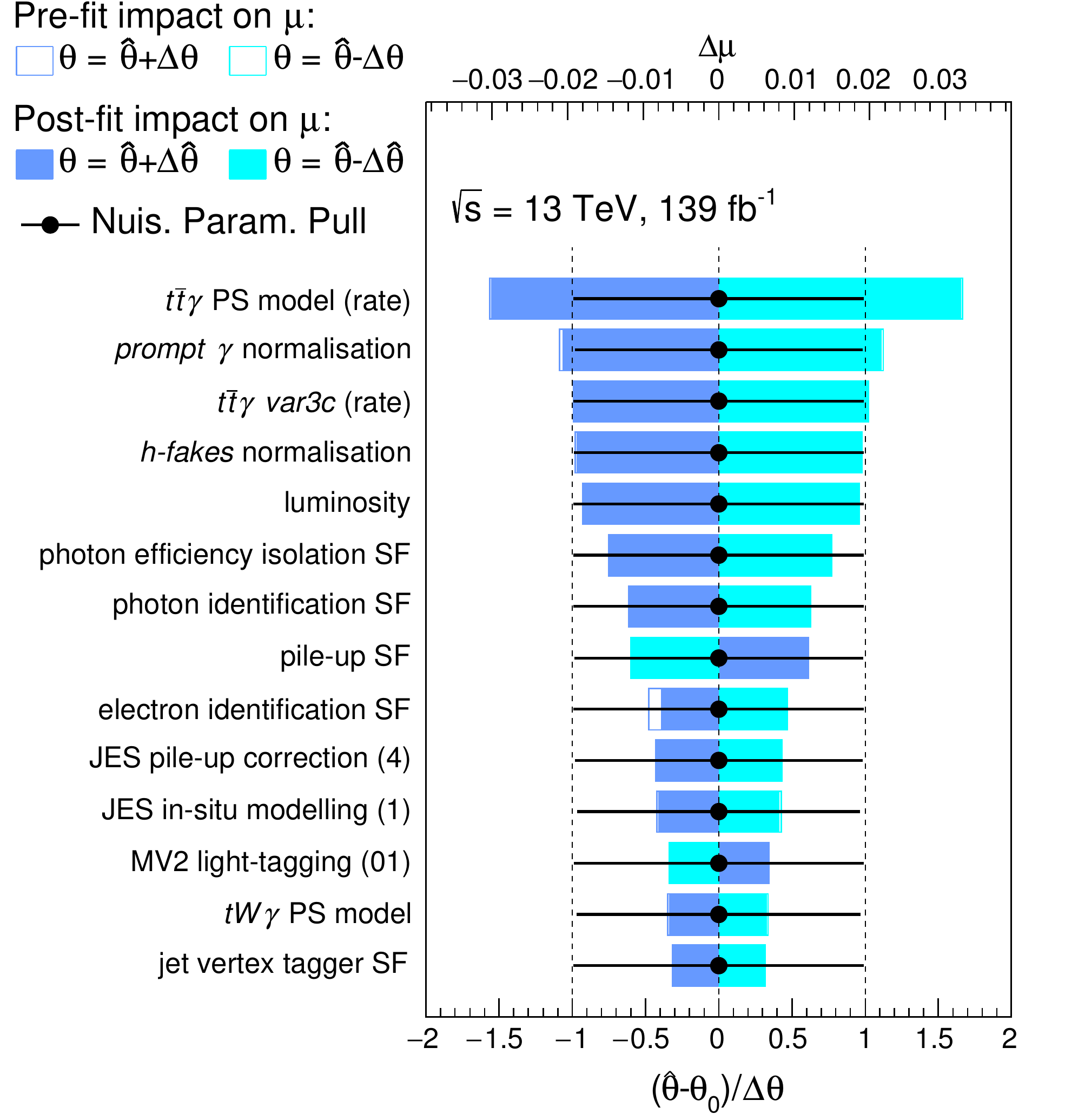}
  \caption[Ranking of nuisance parameters in Asimov fit]{%
    Nuisance parameters ranked according to their impact on the parameter of interest in a fit to Asimov pseudo-data.
    The framed blue and turquoise rectangles indicate the pre-fit impact $\Delta\mu$ of each nuisance parameter on the signal strength $\mu$, whereas the filled blue and turquoise areas show the post-fit impact.
    Possible differences between the two reflect a reduction of the impact due to nuisance-parameter constraints or correlations determined in the fit.
    The impact is overlaid with the nuisance-parameter pulls $(\estimate{\theta} - \theta_0)/\Delta \theta$, all of which are centred at zero for the Asimov fit.
  }
  \label{fig:results-Asimov-ranking}
  \vspace*{3pt}  
\end{figure}

Those nuisance parameters with large correlation values to the signal strength are also expected to have a large impact on the sensitivity of the result.
The post-fit impact of each nuisance parameter can be evaluated by performing a fit, where the tested parameter is fixed to its upper and lower post-fit estimates $\estimate{\theta} \pm \Delta \estimate{\theta}$, while the others remain free.
The same can be done for the pre-fit impact by fixing the parameter to its upper and lower prior values $\theta_0 \pm \Delta \theta$.
The resulting difference in the signal strength $\Delta\mu$ with respect to that obtained in the nominal fit scenario is an estimate of the impact of the tested parameter on the result.
The impact calculation was performed for all nuisance parameters after pruning, the results of which are shown in \cref{fig:results-Asimov-ranking}.
The plot ranks those nuisance parameters with the highest expected post-fit impact on the signal strength.
The framed blue and turquoise rectangles indicate the pre-fit impact, the solid boxes the post-fit impact.
They are overlaid with the nuisance-parameter values $(\estimate{\theta} - \theta_0)/\Delta\theta$ post fit.
The highest ranked nuisance parameters are identical with those that show large correlation values to the signal strength in \cref{fig:results-Asimov-corrmatrix}:
among the top five are the rate components of the \tty \emph{var3c} and \tty \PS model uncertainties, the normalisation uncertainties of the \cathfake and \catprompt categories and the luminosity uncertainty.
Others affecting the result significantly are the uncertainties on the photon identification and isolation efficiencies, on the electron identification efficiency and on the pile-up reweighting, followed by uncertainties on the jet energy-scale calibration.

The Asimov fit also gives an expected uncertainty for the parameter of interest, \ie the \tty signal strength.
The calculated value of the signal strength is
\begin{align}
  \label{eq:results-Asimov-mu}
  \mu
  = 1.000 \pm 0.023 \stat ^{+0.065}_{-0.059} \syst
  = 1.000 \, ^{+0.069}_{-0.063} \, ,
\end{align}
that is, the measurement is expected to have a sensitivity corresponding to less than \SI{7}{\percent} relative uncertainty.
The first and second quoted uncertainties correspond to statistical and systematic uncertainties, respectively.
The first are estimated in a \emph{stat-only} fit where all nuisance parameters are fixed to their best-fit values and only the statistical uncertainties are evaluated.
Systematic uncertainties are calculated as the quadratic difference between the total uncertainties in the nominal fit scenario and those obtained from the stat-only fit.

\section{Fit results with ATLAS data}
\label{sec:results-data}

%
%

This section summarises the results of the fit to the \runii \ATLAS dataset as introduced in \cref{cha:selection}.
The fitting model uses de-correlated rate and shape components of the \tty \emph{var3c} and \tty \PS model variations as discussed in \cref{sec:results-configuration}, \ie it is identical to that used in the previous section to fit Asimov pseudo-data.
This section largely provides the same plots and information to allow a direct comparison with the Asimov fit.

\begin{figure}[p]
  \centering
  \begin{adjustbox}{clip,trim= 0 {0.50\height} 0 0,margin*=0 {0.04\height} 0 0}%
    \includegraphics[width=0.48\textwidth, clip,%
    trim= 0 0 40pt 0]{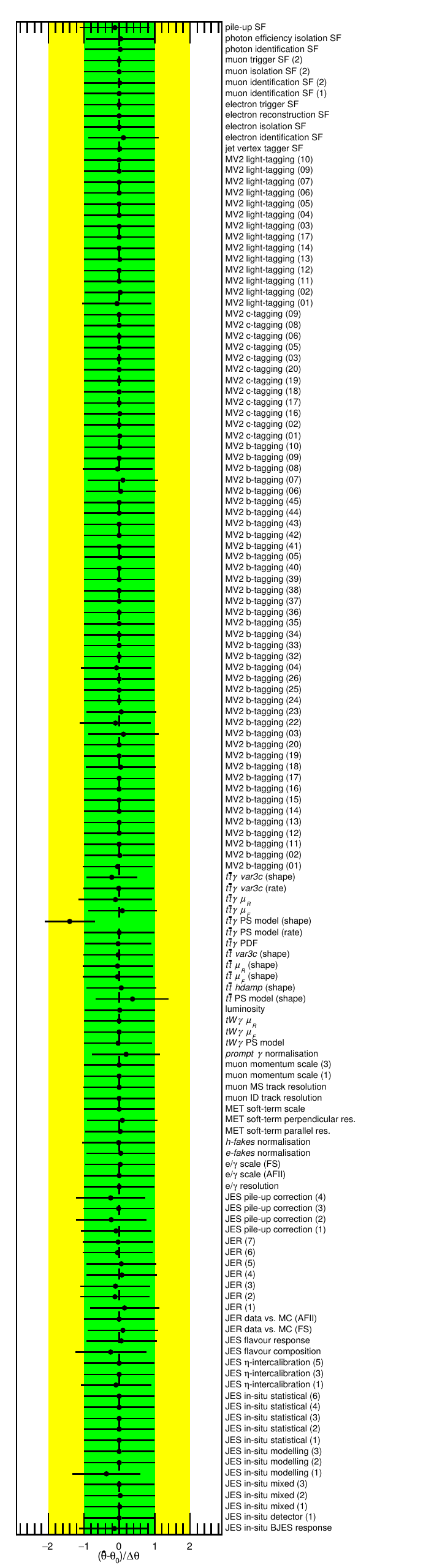}%
  \end{adjustbox}%
  \begin{adjustbox}{clip,trim=0 0 0 {.49\height},margin=0 0 0 {0.02\height}}%
    \includegraphics[width=0.48\textwidth, clip,%
    trim= 0 0 40pt 0]{figures/results/data/NuisPar}%
  \end{adjustbox}
  \caption[Nuisance-parameter pulls in the fit to data]{%
    Nuisance-parameter pulls in the fit to data.
    The plot is cut in half at the bottom of the left-hand column and continues on the top right.
    The post-fit estimates $\estimate{\theta}$ are shifted by the central prior value $\theta_0$ and shown as a fraction of $\Delta\theta$.
  }
  \label{fig:results-data-pulls}
\end{figure}

The post-fit estimates $\estimate{\theta}$ of the nuisance parameters are calculated during the fit as done for Asimov pseudo-data.
In the data fit, in addition to possible constraints, they may also be pulled from their nominal prior values, \ie $\estimate{\theta} \neq \theta_0$.
A visual overview of all nuisance parameters and their pulls observed in the fit to \ATLAS data is given in \cref{fig:results-data-pulls}.
As this overview contains some one hundred fifty nuisance parameters, the most strongly pulled and constrained parameters are listed in \cref{tab:fit_data_pulls_constraints}.
Similarly to the Asimov fit in \cref{tab:results_fit_Asimov_constraints}, the shape components of the two problematic modelling uncertainties, \ie the \tty \emph{var3c} and \tty \PS model uncertainties, show large constraints -- albeit relaxed with respect to a scenario, where the rate and shape components of these uncertainties remain correlated, \cf \cref{sec:results-configuration}.
Compared to the Asimov fit, the constraints in data are slightly tightened, from values of \SIlist{73.6;75.1}{\percent} to \SIlist{71.2;71.0}{\percent} for the \emph{var3c} and \PS model variations, respectively.
A large pull, where the post-fit estimate $\estimate{\theta}$ deviates from the prior nominal value $\theta_0$, is observed for the \tty \PS model shape variation:
the nuisance parameter is pulled beyond the prior interval of one standard deviation to $(\estimate{\theta} - \theta_0)/\Delta\theta = \num{-1.40}$.
This may be explained with the distribution of the corresponding templates shown in \cref{sec:results-configuration}, where this pull was also discussed.
Smaller pulls are observed for the shape-only uncertainty associated to the \PS model choice for \ttbar, and for one of the modelling uncertainties of the jet energy-scale in-situ calibration.

\begin{table}
  \centering
  \caption[List of pulled and constrained nuisance parameters in the fit to data]{%
    Pulled and constrained nuisance parameters in the fit to data.
    Only those with post-fit uncertainty estimates $\Delta \estimate{\theta}$ smaller than \SI{90}{\percent} of the prior uncertainties $\Delta \theta$, and those with pull values larger than \num{\pm 0.3} are listed.
    As in the Asimov fit, \cf \cref{tab:results_fit_Asimov_constraints}, large constraints are only observed for the shape components of the \tty \emph{var3c} and the \tty \PS model variations.
    The latter is also pulled strongly, whereas the \tty \emph{var3c} shape component remains well below the chosen pull threshold.
  }
  \label{tab:fit_data_pulls_constraints}
  \sisetup{round-precision=2,round-mode=places}
  \begin{tabular}{
    l
    S[table-format=1.2, table-sign-mantissa]
    S[table-format=2.1, round-precision=1]
    }
    \toprule
    \multirow{2}{*}{Nuisance parameter} & {pull value} & {constraint} \\[0.3ex]
     & {$( \hat{\theta} - \theta_0 ) / \Delta \theta$} & {$\Delta \hat{\theta} / \Delta \theta$ [\si{\percent}]} \\
    \midrule
    \tty \emph{var3c} (shape)  & {---}    & 71.2051 \\
    \tty \PS model (shape)     & -1.40324 & 70.9973 \\
    \ttbar \PS model (shape)   & 0.366817 & {---} \\
    JES in-situ modelling (1)  & -0.36722 & {---} \\
    \bottomrule
  \end{tabular}
\end{table}

\begin{figure}[p]
  \vspace*{0.1\textheight}  
  \centering
  \includegraphics[width=0.90\textwidth,trim=20pt 0 -20pt 0]{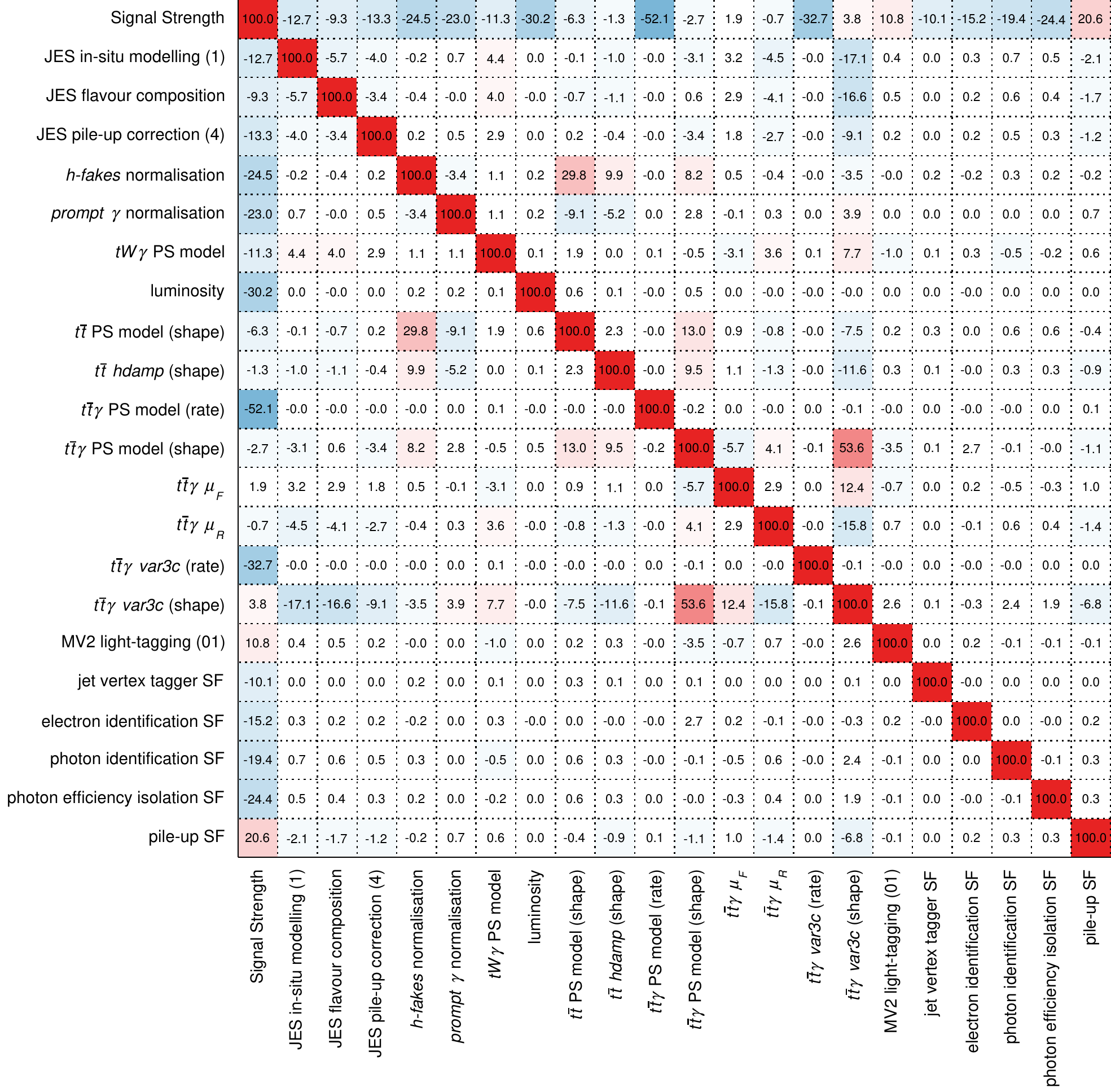}
  \caption[Correlation matrix of nuisance parameters in the fit to data]{%
    Correlation matrix for the fit to data.
    All nuisance parameters with \SI{10}{\percent} or higher absolute correlation value to other parameters or to the signal strength are shown.
  }
  \label{fig:results-data-corrmatrix}
  \vspace*{0.1\textheight}  
\end{figure}

\Cref{fig:results-data-corrmatrix} presents a correlation matrix of those nuisance parameters that show \SI{10}{\percent} or larger absolute correlation value to other nuisance parameters or to the parameter of interest.
The correlations observed are largely similar to those observed in the Asimov fit, \cf \cref{fig:results-Asimov-corrmatrix}:
the rate components of the two aforementioned signal modelling uncertainties, \ie the \tty \emph{var3c} and \tty \PS model variations, are strongly correlated to the signal strength.
With respect to the Asimov scenario, the calculated correlation values are slightly increased from \SIlist{-29.8;-47.8}{\percent} to \SIlist{-32.7;-52.1}{\percent}, respectively.
The normalisation uncertainties of the \cathfake and \catprompt categories, and the luminosity uncertainty are among those with strong negative correlation values with the signal strength, with all magnitudes increased with respect to those in the Asimov fit.
As observed for the Asimov fit, the shape components of the \tty \emph{var3c} and the \tty \PS model variations are correlated with many of the other nuisance parameters, including a high correlation among each other.
Their correlation value increased from \SI{48}{\percent} to over \SI{50}{\percent} in the fit to \ATLAS data.
Previously unobserved was a correlation between the shape-only \ttbar \PS model variation with the normalisation uncertainty of the \cathfake category, the value of which increased from \SI{0.5}{\percent} to \SI{29.8}{\percent}.

\begin{figure}
  \centering
  \includegraphics[width=0.70\textwidth]{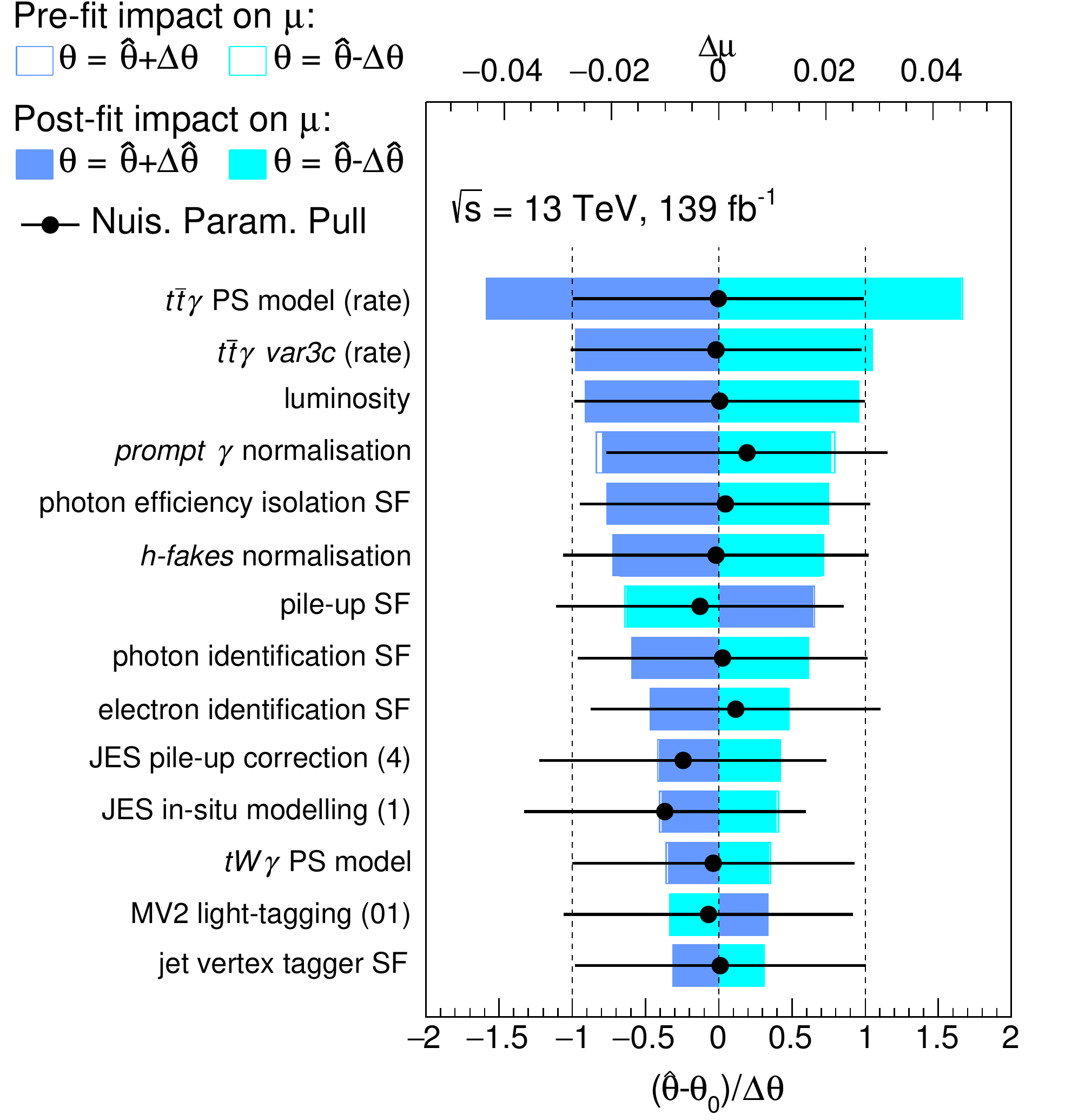}
  \caption[Ranking of nuisance parameters in the fit to data]{%
    Nuisance parameters ranked according to their impact on the parameter of interest in the fit to data.
    The framed blue and turquoise rectangles indicate the pre-fit impact $\Delta\mu$ of each nuisance parameter on the signal strength $\mu$, whereas the filled blue and turquoise areas show the post-fit impact.
    Possible differences between the two reflect a reduction of the impact due to nuisance-parameter constraints or correlations determined in the fit.
    The impact is overlaid with the nuisance-parameter pulls $(\estimate{\theta} - \theta_0)/\Delta \theta$.
  }
  \label{fig:results-data-ranking}
\end{figure}

The nuisance parameters largely correlated with the parameter of interest also show a large impact on the uncertainty associated with that parameter.
Pre-fit and post-fit impact values of each nuisance parameter are evaluated the same way as described in \cref{sec:results-Asimov}:
the nuisance parameter under test is fixed to its pre-fit and post-fit one-standard deviation values, \ie $\theta_0 \pm \Delta\theta$ and $\estimate{\theta} \pm \Delta \estimate{\theta}$, and then the fit is redone for all these scenarios.
The resulting impact, given as difference in the obtained signal strength with respect to the nominal fit, $\Delta\mu$, is shown in \cref{fig:results-data-ranking}.
The framed blue and turquoise rectangles indicate the pre-fit impact, the solid boxes the post-fit impact.
They are overlaid with the nuisance-parameter values $(\estimate{\theta} - \theta_0)/\Delta\theta$ post fit.
The highest-ranked nuisance parameters are identical with those of the Asimov fit scenario:
the rate components of the \tty \emph{var3c} and \tty \PS model uncertainties, the normalisation uncertainties of the \cathfake and \catprompt background categories and the luminosity uncertainty are among the top six.
All of them show large correlations with $\mu$ in the correlation matrix, \cf \cref{fig:results-data-corrmatrix}.
As in the fit to Asimov pseudo-data, others affecting the result are the uncertainties on the photon identification and isolation efficiencies, on the electron identification efficiency and on the pile-up reweighting, all measured in data and corrected in \MC simulation through scale factors.
None of the highly ranked uncertainties show large pulls or constraints, which underlines the stability of the fit.

\begin{figure}
  \centering
  \includegraphics[width=0.6\textwidth,clip,trim=0 12pt 0 14pt]{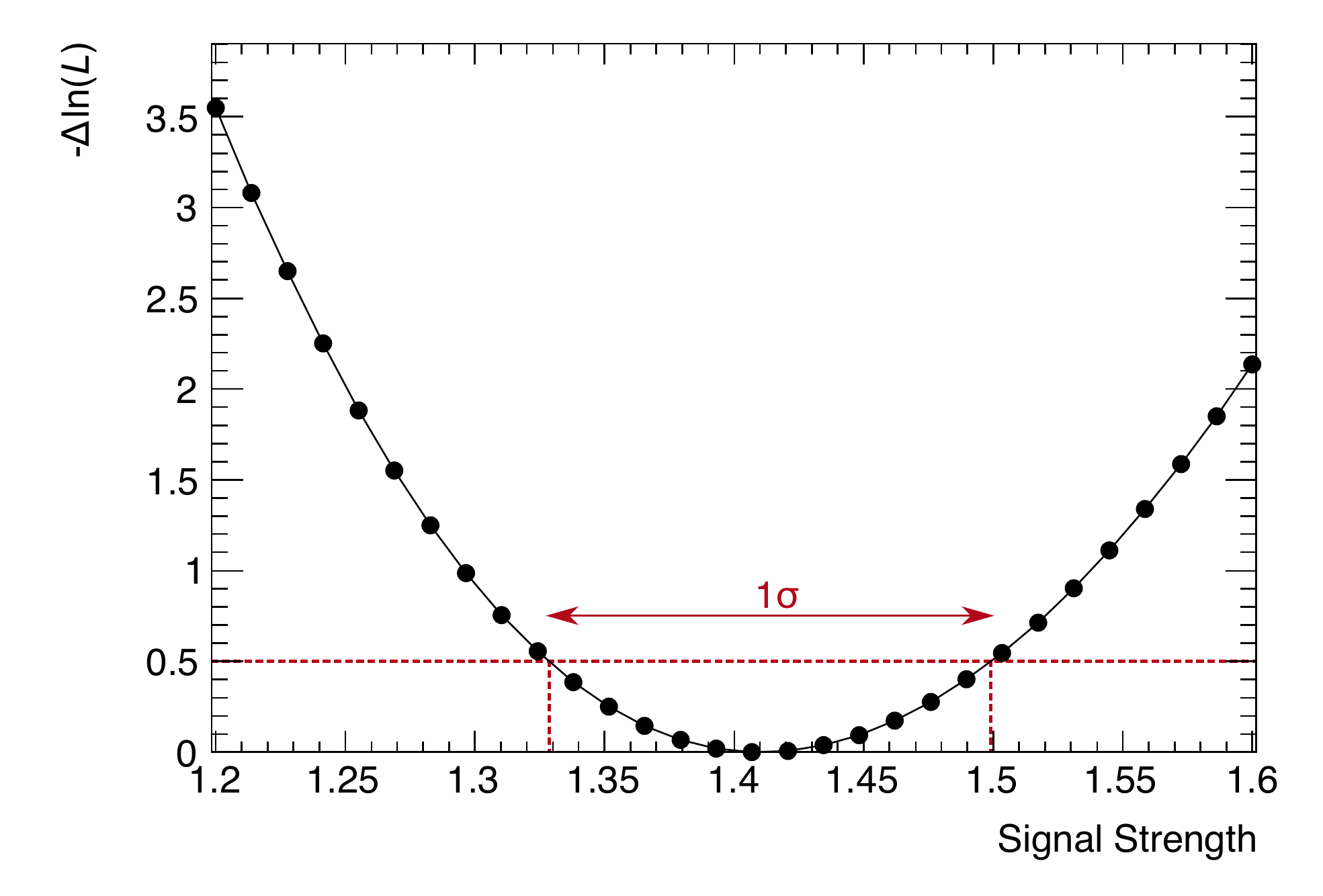}
  \caption[Profile likelihood scan with respect to $\mu$ in the fit to data]{%
    Scan of the profile likelihood in the fit to data for the parameter of interest.
    The $1 \sigma$ interval, defined by $-\Delta \ln(\mathcal{L}) = 0.5$ and quoted as an uncertainty on the result, is marked with dashed lines.
    The fit shows high stability over the entire scanned range.
  }
  \label{fig:results-data-LHscan}
\end{figure}

This can also be demonstrated by a scan of the (profiled) log-likelihood.
In an interval of $\estimate{\mu} \pm 0.2$ around the post-fit estimate of the parameter of interest, the fit is redone with the signal strength $\mu$ fixed to \num{30} equidistant sampling points in that interval.
The resulting value of the log-likelihood is then compared against the value obtained in the nominal fit scenario and their difference $\Delta \ln(\mathcal{L})$ is calculated for each sampling point.
The result of this log-likelihood scan is shown in \cref{fig:results-data-LHscan}.
Over the entire scanned spectrum of the signal strength, the fit shows high stability and there are no hints towards local minima or other artefacts that could point towards problems in the likelihood function.

The post-fit value of the signal strength measured with the \ATLAS dataset amounts to
\vspace*{3pt}  
\begin{align}
  \label{eq:results-data-mu}
  \mu
  = 1.411 \, ^{+0.028}_{-0.027} \stat ^{+0.084}_{-0.077} \syst
  = 1.411 \, ^{+0.089}_{-0.082} \, ,
\end{align}
corresponding to total relative uncertainties of $^{+\SI{6.3}{\percent}}_{-\SI{5.8}{\percent}}$.
The first and second quoted uncertainties correspond to statistical and systematic uncertainties, respectively.
The first are estimated in a \emph{stat-only} fit where all nuisance parameters are fixed to their best-fit values and only the post-fit statistical uncertainties are evaluated.
Systematic uncertainties are calculated as the quadratic difference between the total uncertainties in the nominal fit scenario and those obtained from the stat-only fit.
As the obtained central value deviates from \num{1.0}, the na{\"i}ve conclusion would be an incompatibility with the \SM prediction.
However, the predictions of the signal processes only use \LO \xsecs in \QCD, while large \NLO corrections are expected, \cf \cref{chap:simulation,cha:selection}.
Thus, deviating values are anticipated when fitting the model against \ATLAS data and are not problematic.

\begin{figure}
  \centering
  \includegraphics[width=0.48\textwidth]{figures/controlplots-with-syst/event_ST}%
  \includegraphics[width=0.48\textwidth]{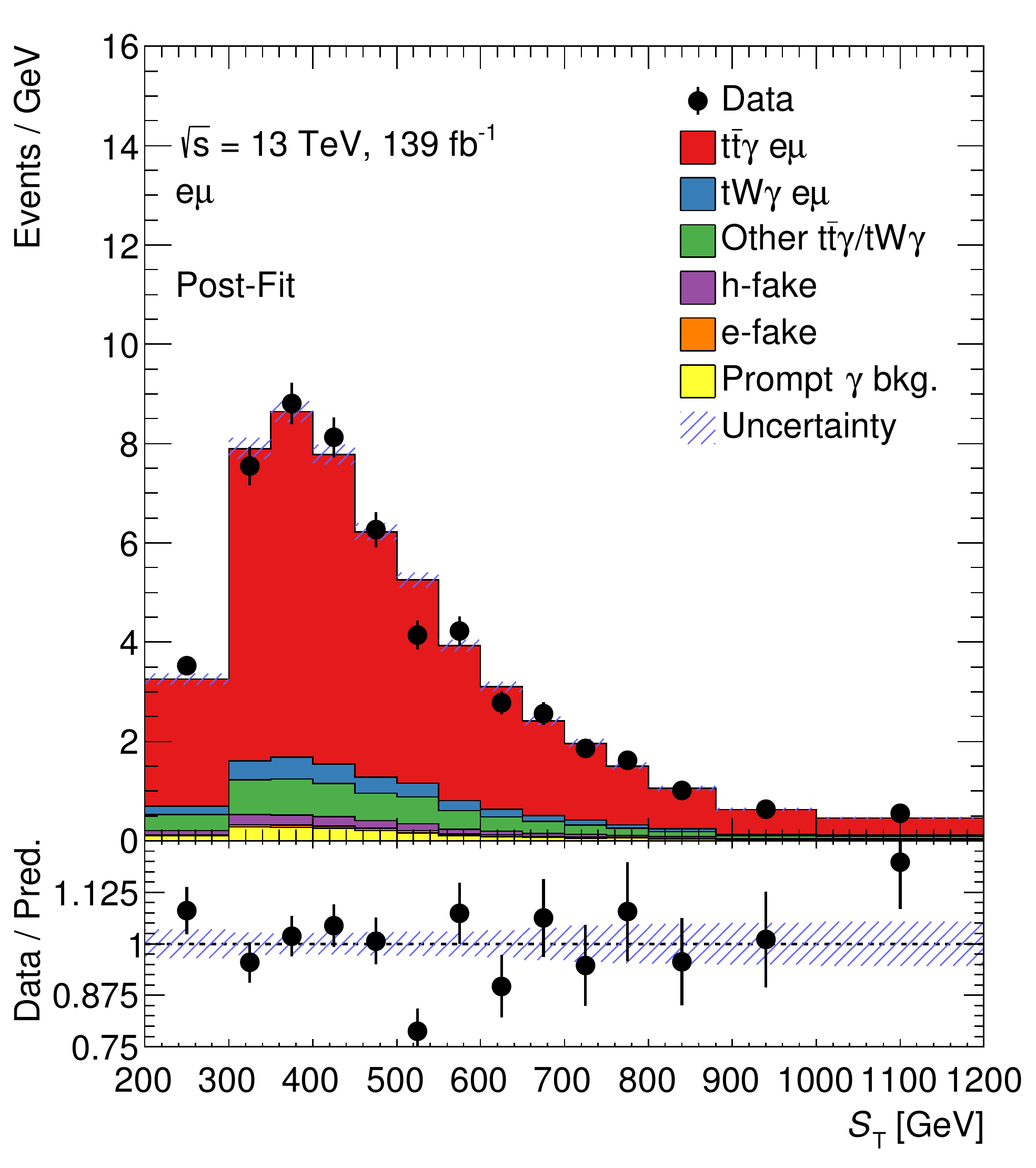}
  \caption[Pre-fit and post-fit distributions of \ST in comparison]{%
    Comparison of pre-fit and post-fit distributions of the \ST observable.
    The agreement between post-fit prediction and data was evaluated with a Pearson $\chi^2$ test and amounts to $\chi^2/\text{\NDF} = 21.2/14$.
    This corresponds to a $p$-value of \num{0.10}.
  }
  \label{fig:results-data-postfit-ST}
  \vspace*{5pt} 
\end{figure}

The post-fit distribution of \ST is shown in \cref{fig:results-data-postfit-ST} on the right-hand side, where it can be compared directly to its pre-fit distribution.
In comparison, the post-fit distribution shows a much narrower hatched uncertainty band for the total prediction.
On the one hand, this is due to the obtained constraints of some of the nuisance parameters.
On the other hand, the log-likelihood is now profiled with respect to all nuisance parameters and, hence, correlations between the parameters are taken into account.
In technical terms, the Hessian matrix of second-order derivatives of the log-likelihood has non-zero off-diagonal elements post-fit.
Strongly pronounced deviations in the pre-fit distributions, \eg those in the bins centred at \SIlist{525;1100}{\GeV}, are compensated for by the fit and the overall agreement between the total prediction and data is improved.

Using the best-fit values of all nuisance parameters and the post-fit estimate of the signal strength, the predicted event yields can be recalculated.
The obtained values are shown in \cref{tab:results-data-yield-table} and are compared to the pre-fit yields as listed in \cref{tab:results-prefit-yields} for the total \runii dataset.
The values show little to no change in the central values from pre-fit to post-fit, thus, none of the background predictions is adjusted drastically in the profile likelihood fit.
Instead, the scaling applied to the predictions of the \tty and \tWy categories marked with (*) for the pre-fit values has approximately the same effect as the fitting of the free-floating signal-strength parameter that controls these categories.
The uncertainties associated with the background categories show no large changes from pre-fit to post-fit, but the pre-fit uncertainties of the \tty-related categories are reduced considerably.
Due to cross-correlations between the uncertainties of the individual event categories, the total \MC prediction also shows strongly reduced post-fit uncertainties.

\begin{table}
  \centering
  \caption[Predicted pre-fit and post-fit event yields]{%
    Predicted pre-fit and post-fit event yields for all \MC categories and numbers of reconstructed events in \ATLAS data in the \emu signal region.
    The quoted uncertainties are combined statistical and systematic uncertainties.
    For the pre-fit yields, the predictions of the \tty and \tWy categories marked with (*) were scaled in such a way that the total \MC prediction matches the number of reconstructed events in data.
    As opposed to \cref{tab:results-prefit-yields}, the pre-fit yields are rounded to significant digits.
  }
  \label{tab:results-data-yield-table}
  \begin{tabular}{%
    l
    S[table-format=4.0] @{${}\pm{}$} S[table-format=3.0]
    S[table-format=4.0] @{${}\pm{}$} S[table-format=2.0]
    }
    \toprule
    & \multicolumn{2}{c}{pre-fit} & \multicolumn{2}{c}{post-fit} \\
    \midrule
    \catttyemu{}* & 2390 & 130 & 2390 & 70 \\
    \cattWyemu{}* & 156  & 15  & 154  & 15 \\
    \catother{}*  & 279  & 15  & 278  & 9  \\
    \cathfake     & 80   & 40  & 80   & 40 \\
    \catefake     & 23   & 12  & 23   & 11 \\
    \catprompt    & 90   & 40  & 100  & 40 \\
    \midrule
    Total \MC     & 3010 & 160 & 3010 & 60 \\
    \midrule
    Data         & \multicolumn{2}{l}{3014} & \multicolumn{2}{l}{3014} \\
    \bottomrule
  \end{tabular}
  \vspace*{3pt} 
\end{table}

With the post-fit predictions of the event yields, the fiducial inclusive \xsec value can be calculated.
This is done using \cref{eq:strategy-fidxsec-final} which avoids a direct dependence of the \xsec on the signal strength by considering the ratio of signal-like \emu-channel events to all signal-like events.
With this equation, the fiducial \xsec is
\begin{align}
  \label{eq:results-data-xsec}
  \sigma^{\text{fid}} (\tty \to \emu)
  = 39.6 \pm 0.8 \stat ^{+2.6}_{-2.2} \syst \si{\fb}
  = 39.6 \, ^{+2.7}_{-2.3} \; \si{\fb} \, ,
\end{align}
corresponding to total relative uncertainties of $^{+\SI{6.9}{\percent}}_{-\SI{5.8}{\percent}}$.
The statistical uncertainties are propagated directly from the statistical uncertainties on the signal strength in \cref{eq:results-data-mu}.
For the systematic uncertainties, as detailed in \cref{sec:systematics-acceptance}, the uncertainties associated with the \tWy parton definition were added in quadrature to the upper uncertainties.
This explains the different relative uncertainties of the signal strength in \cref{eq:results-data-mu} and of the \xsec in \cref{eq:results-data-xsec}.
The measured \xsec is in good agreement with the value predicted by the fixed-order theory computation~\cite{Bevilacqua:2018woc,Bevilacqua:2018dny} in \cref{eq:strategy-prediction-xsec}, which expects a value of $  \sigma^{\text{fid}}_{\NLO} = 38.50 \, ^{+0.56}_{-2.18} \, (\text{scale}) \, ^{+1.04}_{-1.18} \, (\text{\PDF}) \, \si{\fb}$.

As systematics are the dominant source of uncertainties for this measurement, their composition is of particular interest.
While the ranked nuisance parameters in \cref{fig:results-data-ranking} give an idea about which of the individual parameters have the largest impact on the result, classifying the parameters into groups of uncertainty sources gives a more qualitative picture of the uncertainty composition.
\Cref{tab:results-data-impact} lists the impact on the signal strength $\Delta\mu$ with respect to the signal-strength uncertainty in the nominal fit scenario for a total of nine groups of uncertainties.
As done for the total systematic uncertainties, the impact values are obtained by fixing the group of nuisance parameters to their best-fit values.
The computed uncertainties on $\mu$ are then subtracted in quadrature from those of the nominal fit scenario, which yields an estimate of the group's impact.
The quoted values in \cref{tab:results-data-impact} are averaged for upper and lower uncertainties obtained with this method.

\begin{table}
  \centering
  \caption[Impact of systematic uncertainties on the result]{%
    Groups of systematic uncertainties and their relative impact on the sensitivity of the result.
    The values are determined by fixing the nuisance parameters of the tested group to their best-fit values.
    The obtained uncertainties on $\mu$ are subtracted in quadrature from those of the nominal fit scenario.
    In addition to the systematic uncertainties evaluated in the profile likelihood fit, the uncertainty on the \tWy parton definition is added in quadrature to the final result and shows a large impact.
  }
  \label{tab:results-data-impact}
  \renewcommand{\arraystretch}{1.06}  
  \begin{tabular}{l S}
    \toprule
    Group                & {Uncertainty} \\
    \midrule
    Signal modelling        & \SI{3.8}{\percent} \\  
    Background modelling    & \SI{2.1}{\percent} \\  
    Photons                 & \SI{1.9}{\percent} \\  
    Luminosity              & \SI{1.8}{\percent} \\  
    Jets                    & \SI{1.7}{\percent} \\  
    Pile-up                 & \SI{1.3}{\percent} \\  
    Leptons                 & \SI{1.1}{\percent} \\  
    Flavour-tagging         & \SI{1.1}{\percent} \\  
    MC statistics           & \SI{0.4}{\percent} \\  
    \MET soft-term          & \SI{0.2}{\percent} \\  
    \midrule
    Total systematic impact & \SI{5.7}{\percent} \\  
    \midrule
    $\oplus$ \tWy parton definition & \SI{-2.75}{\percent} \\
    \bottomrule
  \end{tabular}
\end{table}

The nuisance parameters for modelling uncertainties are grouped into signal and background modelling to disentangle these two.
Uncertainties on the signal-process modelling have the largest impact on the result with \SI{3.8}{\percent}, as was already seen when the individual nuisance parameters were ranked in \cref{fig:results-data-ranking}.
Background modelling, which also includes the \SI{50}{\percent} normalisation uncertainties assigned to the \cathfake, \catefake and \catprompt categories, is the second-most dominant class of uncertainties with an impact of \SI{2.1}{\percent}.
Uncertainties associated with the reconstruction, identification and calibration of physics objects, \cf \cref{sec:exp_objects}, are grouped into separate categories for each type of object, \eg photon uncertainties.
Those associated with photons and with jets show larger impacts of \SIlist{1.9;1.7}{\percent}, respectively.
In addition, the luminosity uncertainty limits the sensitivity of the measurement significantly, as also seen in the ranking plot.
This single-nuisance-parameter group has a total impact of \SI{1.8}{\percent}.
Other uncertainties include those on the pile-up and on flavour-tagging, whereas statistical limitations of the used \MC samples and uncertainties on the missing transverse momentum have little impact.
In addition, the \tWy parton-definition uncertainty, added in quadrature to the final result, limits the sensitivity.

\section{Differential \xsec measurements}
\label{sec:results-differential}

\vspace*{2pt plus 3pt}  

Fiducial differential \xsec measurements and their associated methodology were not the main focus of the author's work, but they provide valuable complementary results to the fiducial inclusive \xsec measurement.
Thus, this section briefly summarises the results obtained in the fiducial differential \xsec measurements in Ref.~\cite{TOPQ-2020-03}%
\footnote{%
  After this thesis had been finalised and submitted in its first version, minor adjustments to the smoothing and symmetrisation of systematics were made by the \ATLAS Collaboration for the publication of the results in Ref.~\cite{TOPQ-2020-03}.
  This section was now updated to summarise the published results of Ref.~\cite{TOPQ-2020-03}.
  Thus, the results of this section are not fully consistent with the treatment described in \cref{chap:systematics}, although no visible difference in the results or in the associated tests is observed.
}.
The five observables,
\ie the transverse momentum and absolute pseudorapidity of the photon,
the distance \DRlph in the \etaphi plane between the photon and the closer of the two charged leptons,
and the absolute differences in pseudorapidities \Detall and in azimuthal angles \Dphill between the two charged leptons,
are first optimised in their binning.
Non-equidistant binning is used to overcome large statistical uncertainties in less populated areas of the observable distributions.
Two criteria are considered for the binning of these unfolding observables:
firstly, the statistical uncertainty in each bin of the distributions ought to be less than \SI{10}{\percent}.
This ensures a minimum number of predicted events per bin.
Secondly, the bin width has to be larger than twice the expected resolution of the variable to minimise migration effects during unfolding.

As introduced in \cref{sec:strategy-differential}, the binning is then stress-tested using the bootstrap method:
one thousand sets of bootstrap replicas are generated from \MC simulation using Poisson bin-by-bin fluctuations.
Each of these sets, generated with the expected statistical uncertainty in data, is unfolded with the same unfolding setup.
Then, for the ensemble of one thousand sets, the pulls in each bin with respect to the truth distributions are calculated.
These pulls should be unbiased, \ie centred around zero, and their average spread should follow that of a Gaussian distribution with a width equal to the expected statistical uncertainty in that bin.
All five observables show high stability in these tests with the chosen binning.
The pull tests are followed by two stress tests related to the shape stability of the observables:
the nominal unfolding setups are used to unfold observable spectra that are reweighted in their shape to assess whether the setups are biased towards the shapes they were created with.
The first reweighting introduces linear slopes to the pseudo-data, the other enhances bin-by-bin differences between the reconstructed spectra in \MC simulation and \ATLAS data.
In both stress tests, the nominal unfolding setups recover the used truth distributions for all five observables, thus, certifying high stability of the setups.

\begin{figure}
  \centering
  \includegraphics[width=0.48\textwidth, clip, trim=5pt 0 5pt 0]{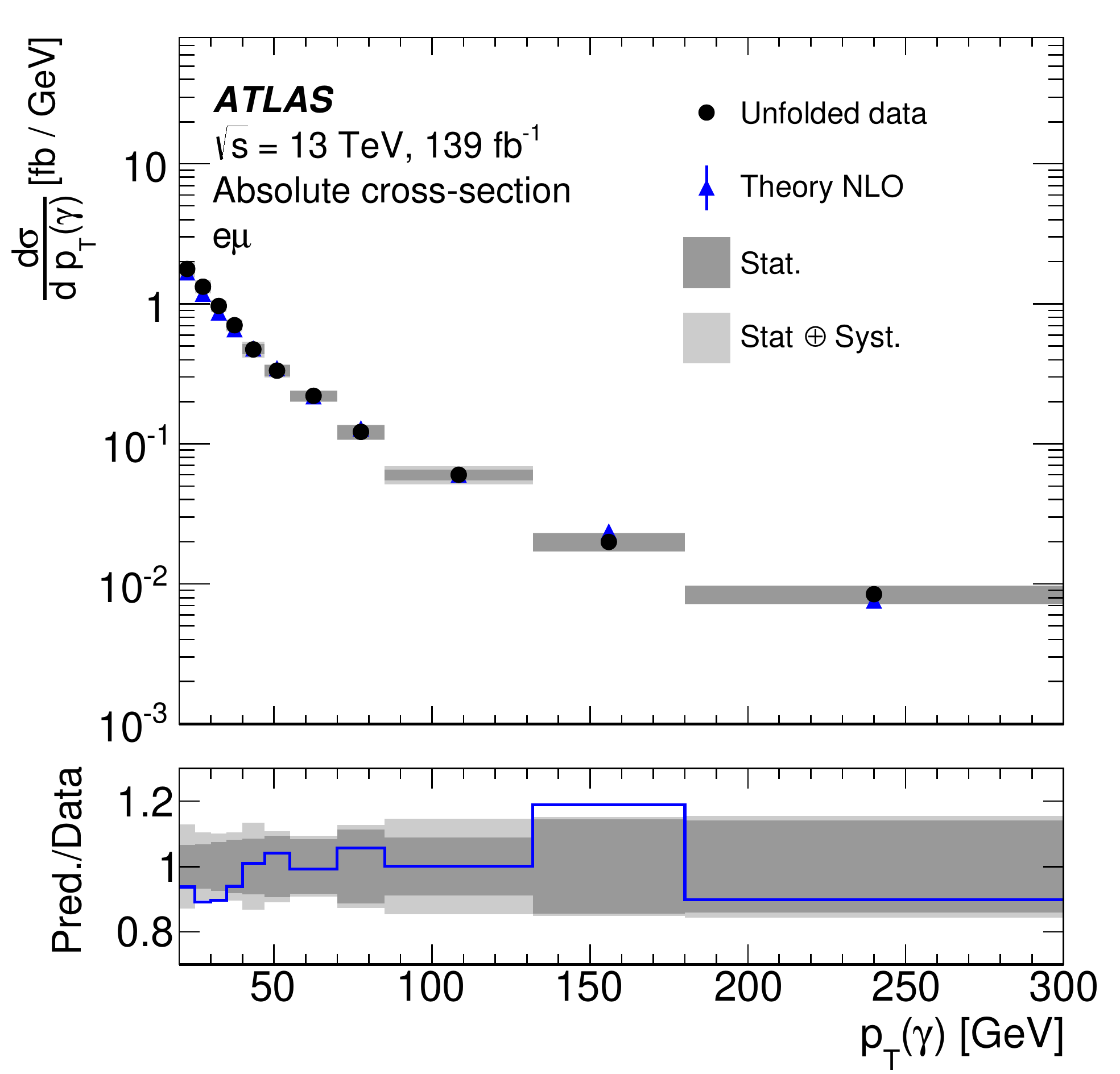}
  \includegraphics[width=0.48\textwidth, clip, trim=5pt 0 5pt 0]{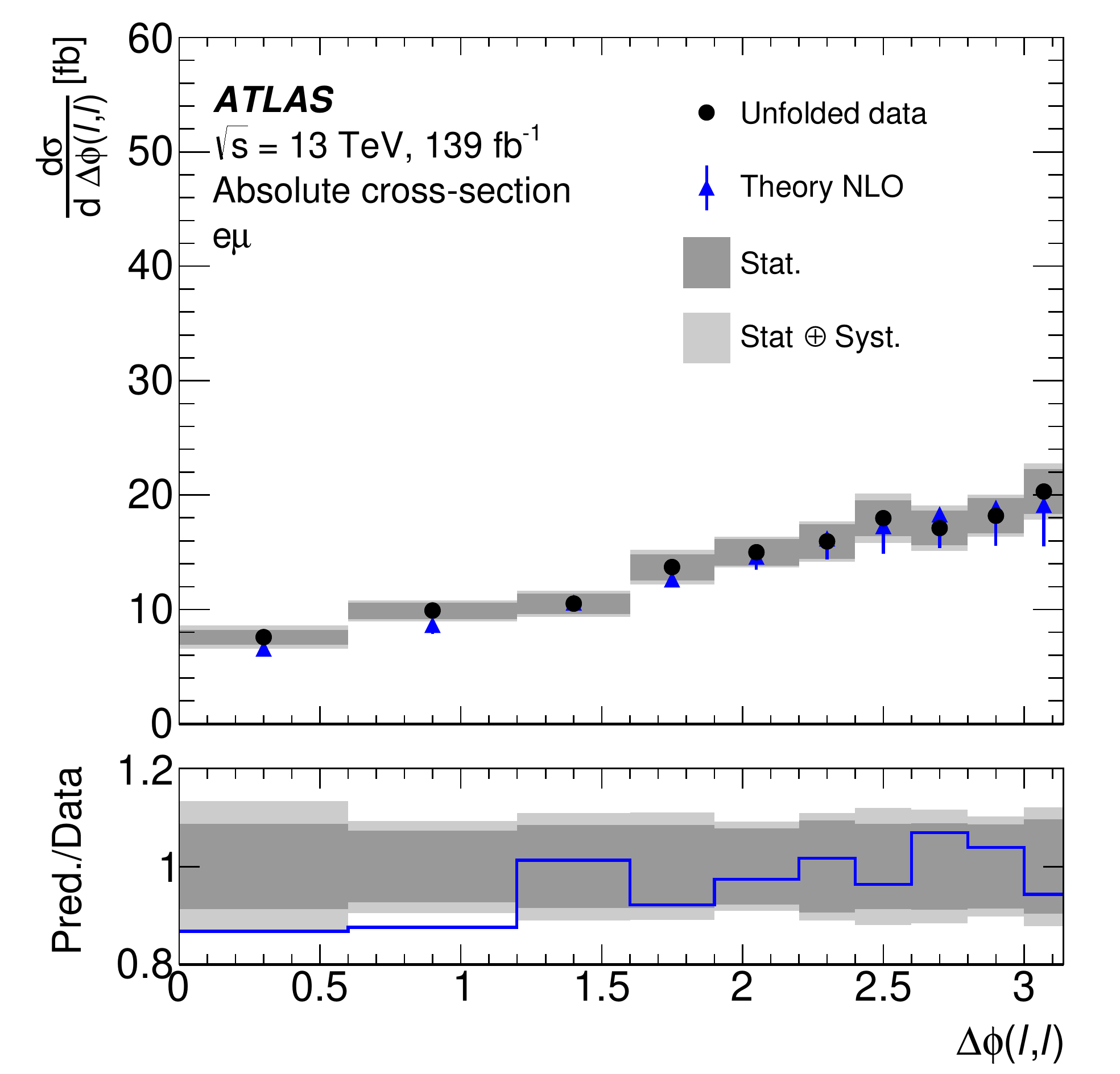}
  \caption[Differential \xsecs in $\pT(\gamma)$ and \Dphill (absolute)]{%
    Differential \xsecs as functions of the photon transverse momentum and of the difference in azimuthal angles \Dphill of the two charged leptons.
    The plots show the absolute bin-by-bin \xsec values obtained from unfolding \ATLAS data and from the fixed-order theory computation~\cite{Bevilacqua:2018woc,Bevilacqua:2018dny}.
    Figures taken from Ref.~\cite{TOPQ-2020-03}.
  }
  \label{fig:results-diff-abs}
\end{figure}

\Cref{fig:results-diff-abs} presents measured differential \xsecs as a function of the photon transverse momentum and of the difference in azimuthal angles \Dphill of the two charged leptons.
The plots allow a direct comparison of the unfolded \ATLAS data with the fixed-order theory computation~\cite{Bevilacqua:2018woc,Bevilacqua:2018dny}, which show little difference in both spectra.
Statistical as well as combined statistical and systematic uncertainties on the unfolded spectra are represented by the differently shaded uncertainty bands, respectively.
The agreement can also be calculated with a Pearson $\chi^2$ test.
\Cref{tab:results-diff-abs-chi2} shows the obtained $\chi^2$ values and the corresponding numbers of degrees of freedom.
A comparison with the respective $\chi^2$ distributions yields the $p$-values listed in the table.
For both distributions, the unfolded \ATLAS data and \NLO theory computation show good agreement, with $p$-values of \numlist{0.87;0.83} for the photon transverse momentum and \Dphill, respectively.

\begin{table}
  \centering
  \caption[$\chi^2$/\NDF and $p$-values for the $\pT(\gamma)$ and \Dphill \xsecs (absolute)]{%
    Agreement between the unfolded \ATLAS data in \cref{fig:results-diff-abs} and the fixed-order \NLO theory computation~\cite{Bevilacqua:2018woc,Bevilacqua:2018dny}.
    Pearson $\chi^2$ tests are performed and the results are given in combination with the numbers of degrees of freedom (\NDF).
    The $p$-value is taken from a comparison with the corresponding $\chi^2$ distribution.
    Values taken from Ref.~\cite{TOPQ-2020-03}.
  }
  \label{tab:results-diff-abs-chi2}
  \begin{tabular}{l S[table-format=1.1] @{${\,}/{\,}$} S[table-format=2.2] S[table-format=1.2]}
    \toprule
    Observable & {$\chi^2$} & {\NDF} & {$p$-value} \\
    \midrule
    $\pT(\gamma)$ & 6.1 & 11 & 0.87 \\
    \Dphill       & 5.8 & 10 & 0.83 \\
    \bottomrule
  \end{tabular}
\end{table}

\begin{figure}
  \centering
  \includegraphics[width=0.48\textwidth, clip, trim=5pt 0 5pt 0]{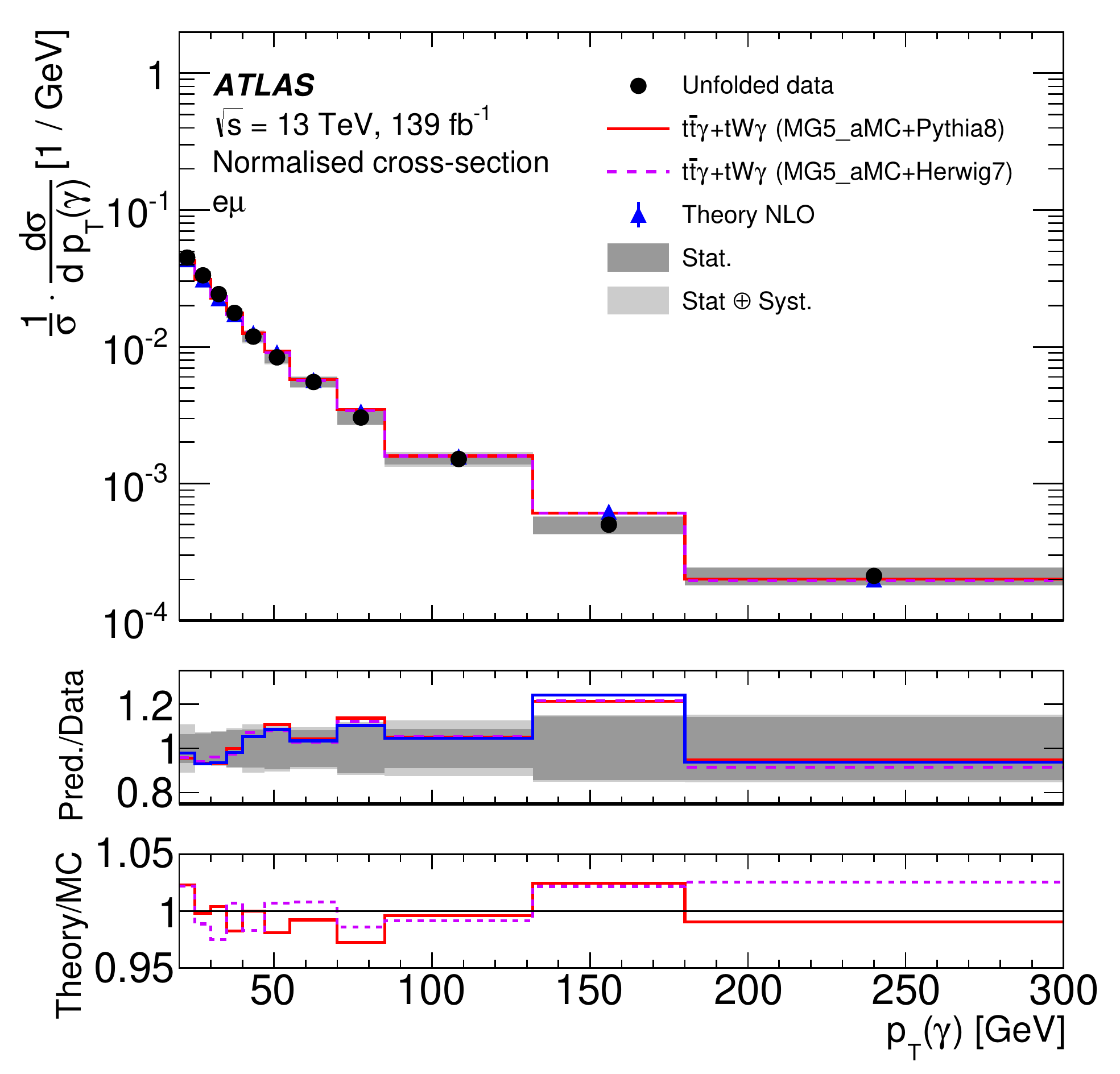}
  \includegraphics[width=0.48\textwidth, clip, trim=5pt 0 5pt 0]{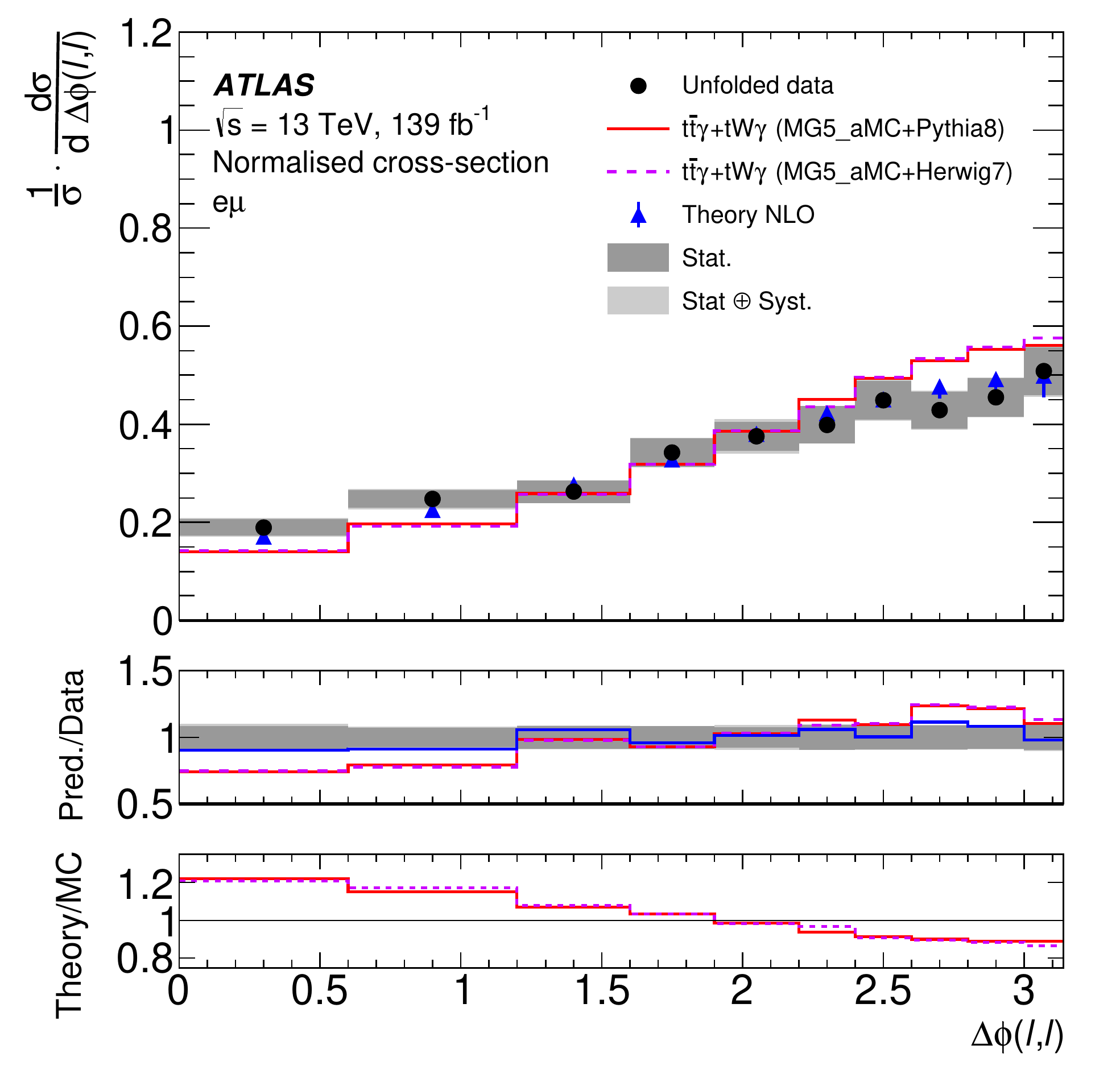}
  \caption[Differential \xsecs in $\pT(\gamma)$ and \Dphill (normalised)]{%
    Differential \xsecs of the same observables as in \cref{fig:results-diff-abs}, but with the bin-by-bin \xsec values normalised to the measured integrals of the distributions.
    The plots show unfolded \ATLAS data, predictions by the fixed-order theory computation~\cite{Bevilacqua:2018woc,Bevilacqua:2018dny} as well as the combined \tty and \tWy \LOPS simulations used in this analysis: \Madgraph interfaced to \Pythia and to \Herwig.
    Figures taken from Ref.~\cite{TOPQ-2020-03}.
  }
  \vspace*{-3pt}  
  \label{fig:results-diff-norm}
\end{figure}

Many of the modelling and experimental systematics show effects that impact the normalisation of the entire distribution of an observable, such as global rate uncertainties of the \xsec.
The effects of these can be reduced for differential distributions if the values are normalised to the measured integral of the distribution.
These normalised spectra are shown in \cref{fig:results-diff-norm}, again for the photon transverse momentum and \Dphill.
In addition to the fixed-order \NLO theory prediction, the plots also show the \LOPS predictions for the \tty and \tWy signal processes (\Madgraph interfaced with \Pythia and with \Herwig).%
\footnote{%
  For the absolute \xsecs in \cref{fig:results-diff-abs}, these would deviate from \ATLAS data significantly given the measured signal strength of about $1.41$, as shown in \cref{eq:results-data-mu}.
  Therefore, they were not shown in those plots to focus on the comparison to the more accurate \NLO theory computation.
}
The uncertainty bands are visibly reduced compared to those shown in \cref{fig:results-diff-abs}.
The \LOPS predictions and the unfolded \ATLAS data agree for the spectrum of $\pT(\gamma)$, but deviate significantly for \Dphill.
Both \LOPS simulations show a consistent slope when compared to data.
The \NLO theory computation, on the other hand, agrees well with unfolded \ATLAS data across both spectra.
The agreement is quantified with Pearson $\chi^2$ tests as listed in \cref{tab:results-diff-norm-chi2}.
The table supports good agreement for the $\pT(\gamma)$ spectrum for all three theory predictions, whereas the obtained values in the $\chi^2$ test in \Dphill yield $p$-values below \num{0.01} for the \LOPS predictions.
This hints towards systematic issues in predicting the \Dphill spectrum in \LOPS simulations as it is not specific to one of the two parton-shower algorithms.
The observable is sensitive to spin correlations of the top-quark pair, \cf \eg Ref.~\cite{TOPQ-2016-10}, which could be modelled insufficiently in the used \LOPS simulations.
Good agreement of the \NLO theory computation with unfolded \ATLAS data is observed.

\begin{table}
  \centering
  \caption[$\chi^2$/\NDF and $p$-values for the $\pT(\gamma)$ and \Dphill \xsecs (normalised)]{%
    Agreement between the unfolded \ATLAS data in \cref{fig:results-diff-norm}, the combined \tty and \tWy \LOPS simulations and the fixed-order theory computation~\cite{Bevilacqua:2018woc,Bevilacqua:2018dny}.
    The results of Pearson $\chi^2$ tests and the numbers of degrees of freedom are given.
    In addition, the $p$-value is taken from a comparison with the corresponding $\chi^2$ distribution.
    Values taken from Ref.~\cite{TOPQ-2020-03}.
  }
  \label{tab:results-diff-norm-chi2}
  \begin{tabular}{l l S[table-format=2.1] @{${\,}/{\,}$} S[table-format=2.2] S[table-format=1.2,table-comparator=true]}
    \toprule
    Observable && {$~~\chi^2$} & {\NDF} & {$p$-value} \\
    \midrule
    \multirow{3}{*}{$\pT(\gamma)$}
    & \textsc{mg5}+\Pythia & 6.3 & 10 & 0.79 \\
    & \textsc{mg5}+\Herwig & 5.3 & 10 & 0.87 \\
    & \NLO theory          & 6.0 & 10 & 0.82 \\
    \midrule
    \multirow{3}{*}{\Dphill}
    & \textsc{mg5}+\Pythia & 30.8 & 9 & <0.01 \\
    & \textsc{mg5}+\Herwig & 31.6 & 9 & <0.01 \\
    & \NLO theory          &  5.8 & 9 & 0.76 \\
    \bottomrule
  \end{tabular}
\end{table}

\begin{figure}
  \centering
  \includegraphics[width=0.48\textwidth, clip, trim=5pt 25pt 5pt 10pt]{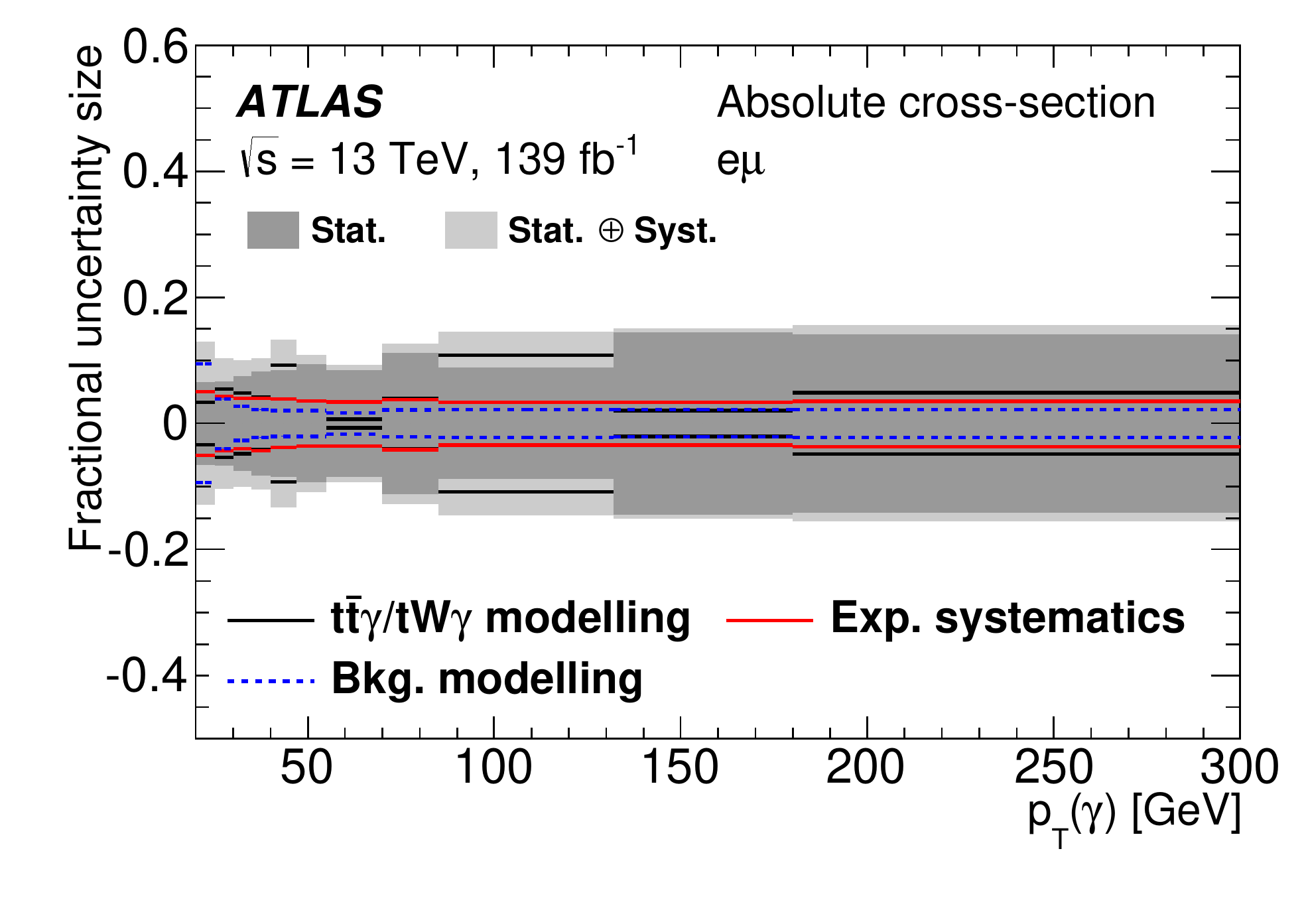}
  \includegraphics[width=0.48\textwidth, clip, trim=5pt 25pt 5pt 10pt]{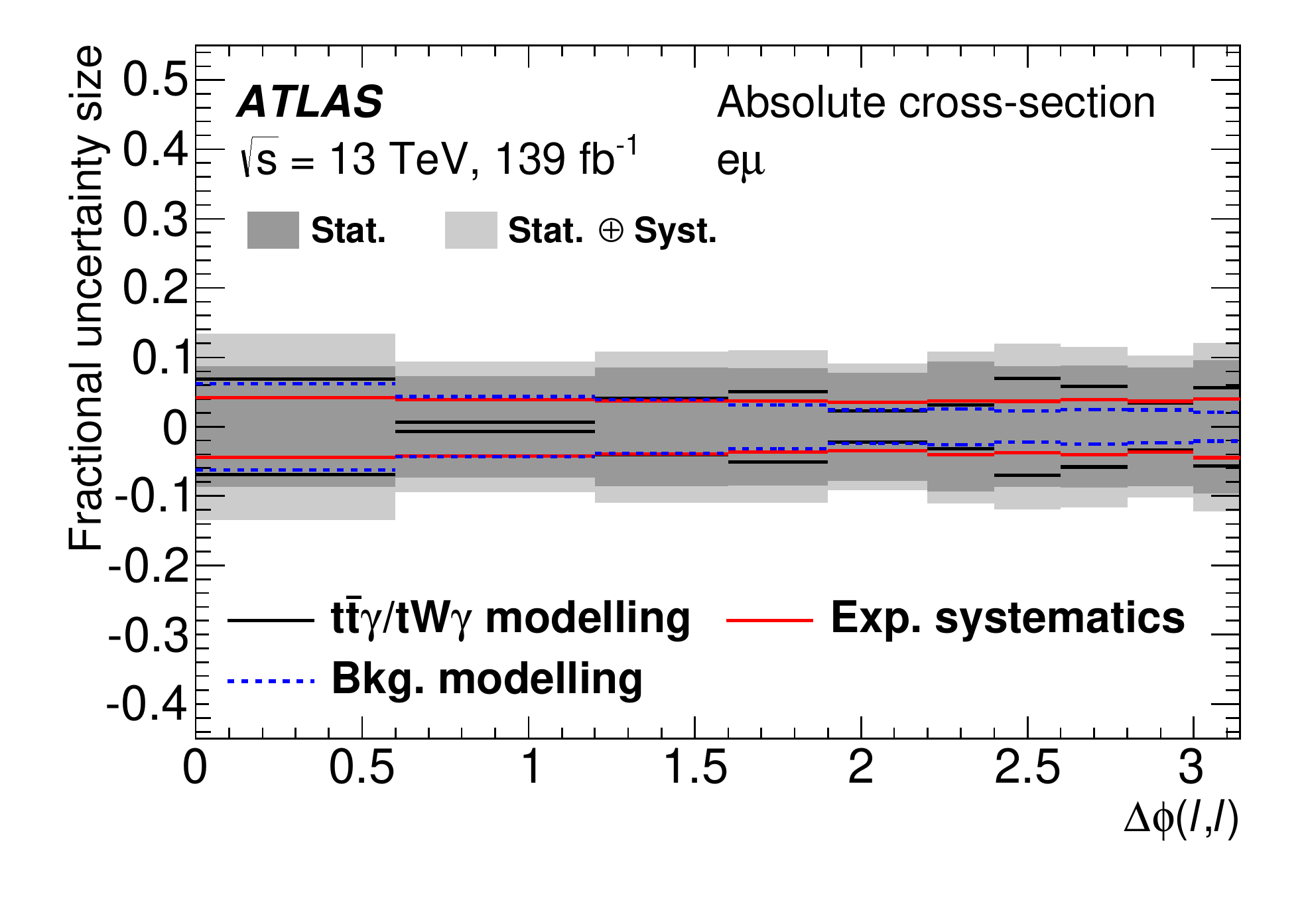}
  \\
  \includegraphics[width=0.48\textwidth, clip, trim=5pt 25pt 5pt 10pt]{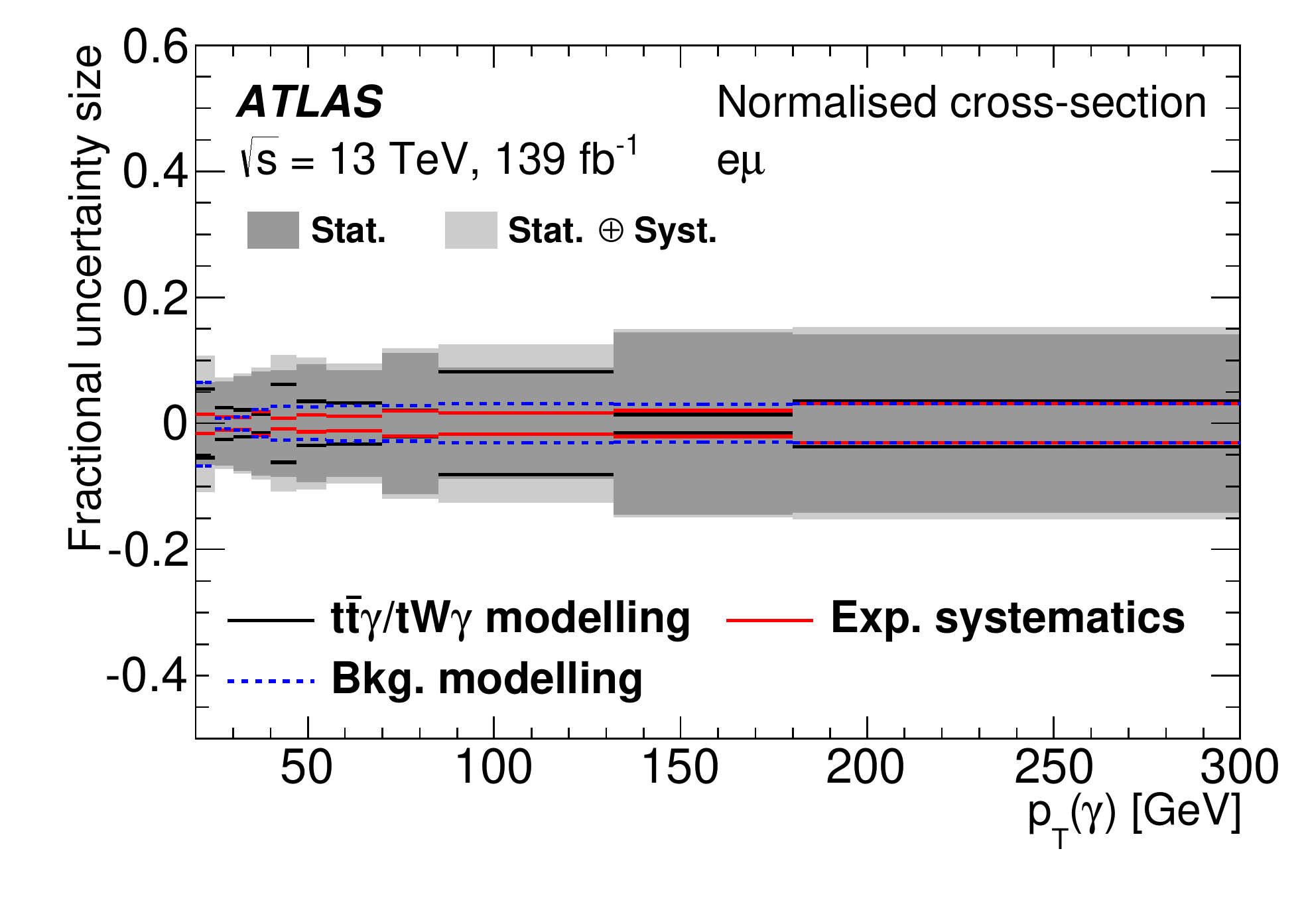}
  \includegraphics[width=0.48\textwidth, clip, trim=5pt 25pt 5pt 10pt]{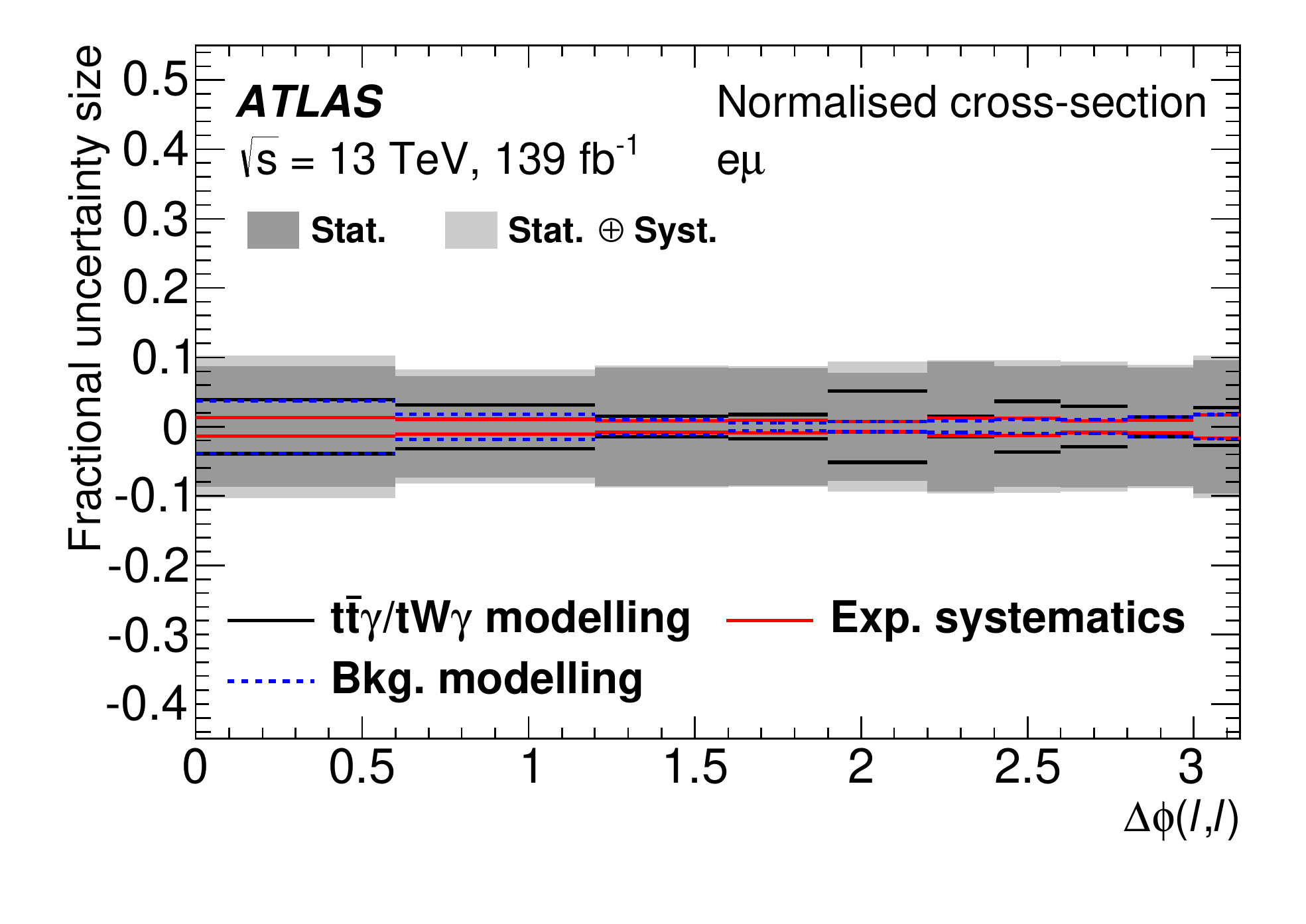}
  \caption[Uncertainty composition for differential \xsecs]{%
    Composition of the bin-by-bin uncertainties for the differential \xsec distributions as a function of the photon transverse momentum and \Dphill.
    The top row shows the uncertainties for the absolute bin-by-bin spectra in \cref{fig:results-diff-abs}, the bottom row those of the normalised spectra in \cref{fig:results-diff-norm}.
    The systematic uncertainties are reduced visibly in the normalised spectra.
    Figures taken from Ref.~\cite{TOPQ-2020-03}.
  }
  \label{fig:results-diff-uncertainties}
\end{figure}

\Cref{fig:results-diff-uncertainties} shows the composition of the uncertainties for each of the presented differential distributions.
The top row of the figure gives details about the systematics of the absolute bin-by-bin spectra shown in \cref{fig:results-diff-abs}, the bottom row about those of the normalised \xsec distributions in \cref{fig:results-diff-norm}.
When comparing the two rows, the overall uncertainties are reduced visibly when normalising the \xsec distributions to their integrals.
The systematic uncertainties, displayed in combination with statistical uncertainties in the outer uncertainty band, are split further in three categories: signal modelling, background modelling and experimental systematics.
The categories contain the same sources of uncertainties as those listed in \cref{tab:results-data-impact} for the fiducial inclusive \xsec.

This section only presented unfolded \ATLAS data for the photon transverse momentum and \Dphill.
Results for the other three observables unfolded in Ref.~\cite{TOPQ-2020-03},
\ie the absolute pseudorapidity of the photon,
the distance \DRlph in the \etaphi plane between the photon and the closer of the two charged leptons,
and the absolute difference in pseudorapidities \Detall between the two charged leptons,
are shown in \cref{chap:app-diff-xsec}.

\chapter{Summary and conclusions}
\label{chap:conclusions}

The electromagnetic coupling between the top quark and the photon is accessible through the measurement of \tty production, where a top-quark pair is produced in association with a hard photon.
Both strength and structure of the coupling are sensitive to physics beyond the Standard Model.
This thesis presents results of \tty \xsec measurements using data recorded with the \ATLAS detector at the \LHC at \CERN.
The examined dataset was taken during \runii of the \LHC in the years 2015 to 2018 in proton-proton collisions at a \com energy of \SI{13}{TeV} and corresponds to an integrated luminosity of \SI{139}{\ifb}.
The analysis focuses on the \emu final state due to its high signal purity and precise available theory predictions~\cite{Bevilacqua:2018woc,Bevilacqua:2018dny}.
In order to be consistent with these predictions that also include non-doubly-resonant diagrams, a combined measurement of $\tty + \tWy$ is performed.
The signal region of the measurement is defined by selecting events with exactly one photon, one electron and one muon of opposite electric charge, and at least two jets, of which one or more must be \btagged.

With a signal-to-background ratio of almost $6:1$, only very few predicted background events contaminate the chosen signal region.
Other decay channels of the signal processes that migrated into the \emu selection constitute the largest fraction.
This includes indirect \emu final states through the decay of one or two \tauleptons and a small fraction of \ljets events with an additional fake lepton.
Other background processes are categorised into those containing \hfakephotons, \efakephotons and prompt photons.
Due to their low contributions, the rate of fake photons is estimated from \MC simulation and no data-driven corrections are applied to the estimate.
Instead, all three background categories are assigned conservative normalisation uncertainties.

\begin{figure}
  \centering
  \includegraphics[width=\textwidth,clip=true,trim=0 8pt 0 8pt]{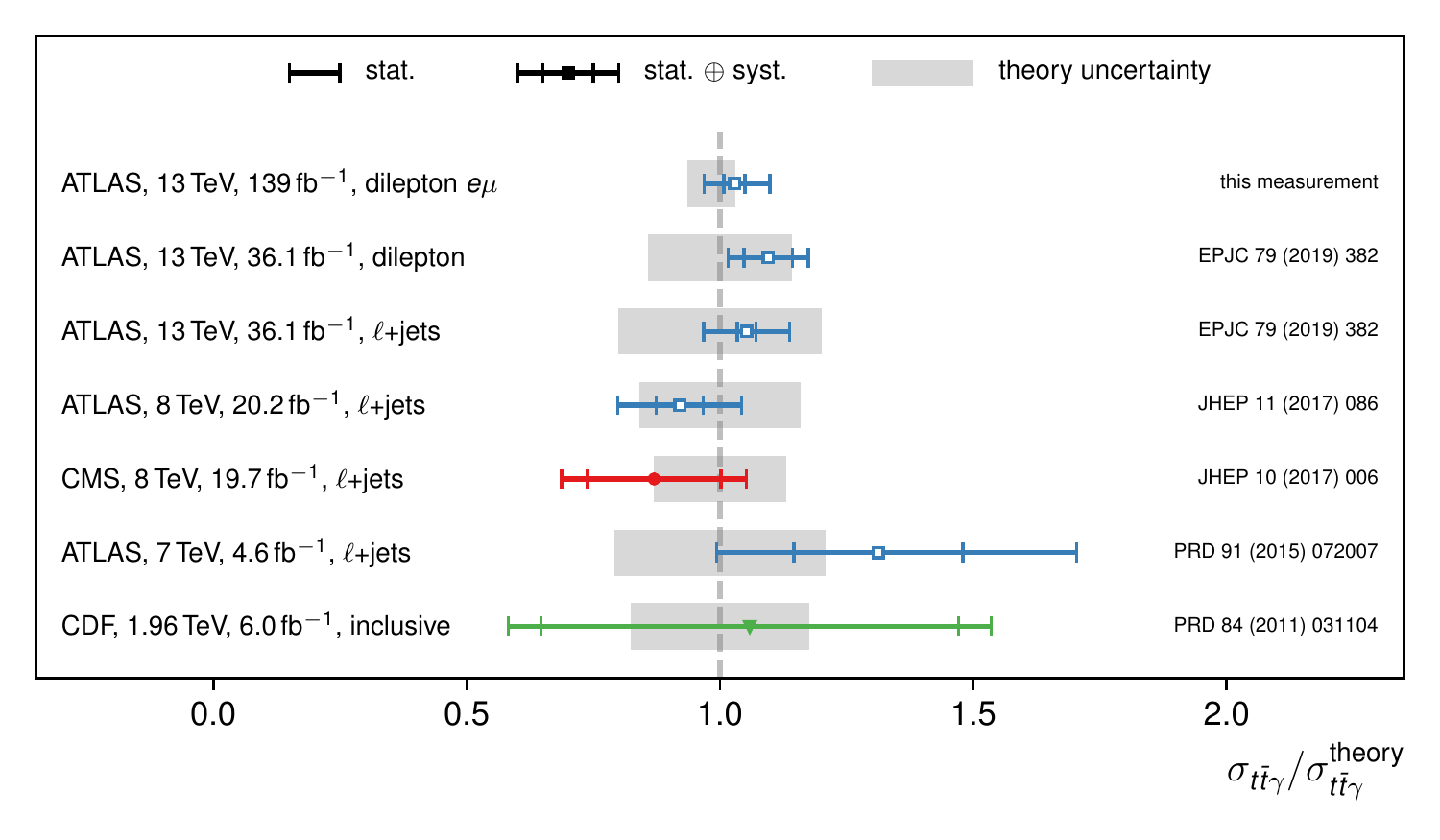}
  \caption[Presented measurement in comparison with previous results]{%
    This measurement in comparison with previous results of \tty production at hadron colliders~%
    \cite{Aaltonen:2011sp,TOPQ-2012-07,CMS-TOP-14-008,TOPQ-2015-21,TOPQ-2017-14}.
    As a figure of merit, the measured \xsec over \SM prediction is given.
    The inner error bars are statistical uncertainties only, the outer bars are combined statistical and systematic uncertainties.
    The grey blocks represent theory uncertainties.
  }
  \label{fig:conclusions-tty-measurements}
\vspace{4pt}  
\end{figure}

The focus of this thesis lies on the measurement of the fiducial inclusive \xsec.
\ATLAS data is compared with \MC simulation in a binned distribution of the observable \ST, defined as the sum of transverse momenta of all objects of the event, including \MET.
The signal strength of the combined $\tty + \tWy$ production is estimated through a maximum likelihood fit to data.
Systematic uncertainties on the predictions are included into the fitting procedure using the template method and the profiling technique.
The results are then corrected for efficiency and migration effects with respect to a fiducial phase-space volume at parton level that is defined in such a way that it mimics the experimental cuts of the signal region.
The fit yields a fiducial \xsec of
\vspace{0pt plus 2pt}  
\begin{align}
  \label{eq:conclusions-xsec}
  \sigma^{\text{fid}} (\tty \to \emu)
  = 39.6 \pm 0.8 \stat ^{+2.6}_{-2.2} \syst \si{\fb}
  = 39.6 \, ^{+2.7}_{-2.3} \; \si{\fb} \, ,
\end{align}
corresponding to total relative uncertainties of $^{+\SI{6.9}{\percent}}_{-\SI{5.8}{\percent}}$.
The authors of Ref.~\cite{Bevilacqua:2018woc,Bevilacqua:2018dny} provide a dedicated computation of the \tty \xsec at \NLO in \QCD in the identical fiducial phase space.
The measured value is in good agreement with the computed fiducial \xsec of $\sigma^{\text{fid}}_{\NLO} = 38.50 \, ^{+0.56}_{-2.18} \, (\text{scale}) \, ^{+1.04}_{-1.18} \, (\text{\PDF}) \, \si{\fb}$.
Compared to previous results, this constitutes the most precise measurement of the \tty production \xsec to date.
\Cref{fig:conclusions-tty-measurements} picks up the previous results shown in \cref{fig:theory-tty-measurements} and adds the result of this measurement to the diagram.
While the previous \ATLAS measurement in the dilepton channels performed with \SI{36}{\ifb} at \SI{13}{\TeV}~\cite{TOPQ-2017-14} was limited by statistical and systematic uncertainties equally, the result presented here shows significantly lower statistical limitations.
The statistical uncertainties decreased from \SI{3.8}{\percent} to \SI{1.9}{\percent}, whereas relative systematic uncertainties remain approximately identical despite a conservative treatment of modelling uncertainties in this measurement.
Statistical and systematic uncertainties combined result in an overall uncertainty reduction.
Exploiting the full \runii dataset corresponding to an integrated luminosity of \SI{139}{\ifb}, this measurement in the \emu channel has reached the same statistical precision as the \ljets channels at \SI{36}{\ifb}, while the total uncertainties are some \SI{20}{\percent} lower.
Apart from the improvements on the measurement side, the dedicated fixed-order computation~\cite{Bevilacqua:2018woc,Bevilacqua:2018dny} is a milestone in terms of precision of \tty theory predictions.

Besides the fiducial inclusive measurement, the ATLAS data is unfolded to parton level using an iterative technique based on Bayes' theorem.
Measurements of differential cross-sections as functions of the transverse momentum of the photon and of the difference in azimuthal angles \Dphill between the two leptons are presented.
Three additional differential \xsecs as functions of $\left| \eta(\gamma) \right|$, \DRlph  and \Detall are summarised in \cref{chap:app-diff-xsec}.
All differential distributions show agreement with the fixed-order theory computation in the examined fiducial phase-space volume.
However, \MC simulations using matrix-element generators at leading order in \QCD interfaced to parton-shower algorithms deviate significantly from \ATLAS data in some of the spectra.
This is particularly pronounced in the \Dphill distribution, which is sensitive to spin-correlation effects.
The disagreement between \MC simulation and \ATLAS data in \Dphill shows trends that are similar to those in unfolded \Dphill distributions of \ttbar production in the \emu final state~\cite{TOPQ-2016-10}.

An additional focus of this thesis is placed on studies concerning machine-learning techniques to identify photons in the \SI{36}{\ifb} analysis performed by \ATLAS.
The developed tool, the \PPT, provides binary classification of photon candidates into prompt photons and \hfakephotons based on the lateral and longitudinal evolution of photon showers in the \ATLAS calorimeters.
In the \tty analysis it is used as an input to an event-level neural network in the \ljets channels and provides large separation power between the \tty signal and background processes with \hfakephotons.

\paragraph{outlook.}
Rapid developments in the area of machine learning over the past fifteen years have not spared the field of experimental particle physics.
The \PPT was one of the first tools on analysis level within \ATLAS to use novel open-source machine-learning libraries, such as \textsc{keras}~\cite{keras} and \textsc{tensorflow}~\cite{tensorflow}.
Since then, these have become increasingly popular and are used in various places of the \ATLAS software infrastructure.
For example, while the results presented here still rely on the \MVtwo high-level algorithm for \btag, the \ATLAS Collaboration is slowly moving towards a new deep-learning-based tagger named \textsc{dl1} that was trained with \textsc{keras} and the \textsc{theano} library as backend~\cite{theano}.
With a plethora of new techniques and tools at hand, these will change the way statistical data analysis is carried out in experimental particle physics fundamentally over the next years.

For the presented \tty measurement in the \emu final state, the \PPT or a similar tool was of little interest for this particular channel due to the low background contamination.
However, with the full \runii dataset available, precise measurements of \tty final states in the \ljets channels would profit from such a tool and make it a tool worth re-investing into.
Various improvements could be considered to boost the overall performance of the \PPT.
Corrections implemented through simulation-to-data scale factors were needed in \cref{chap:PPT} to apply the \PPT to \MC samples with very different photon kinematics than those of the training data.
These problems could be circumvented by retraining the \PPT using results from the field of domain adaption:
two neural networks could compete in an \emph{adversarial} structure~\cite{Schmidhuber1992:aa,Ganin2016:aa,Edwards2015:aa,Goodfellow2014:aa}, where the discriminating model is advised by another in order to remain uncorrelated with kinematic variables of the photon, \ie with $\pT(\gamma)$ and $\left| \eta(\gamma) \right|$.
This could help to adapt to unseen data with different photon kinematics as shown in studies performed within \ATLAS~\cite{Nagel2019}.
On the other hand, a machine-learning tool can only be as good as its input data.
One of the limiting factors of deep-learning structures based on photon shower-shape variables remains the poor modelling of such in \MC simulation.
As discussed in \cref{sec:exp_objects}, fudge factors are used to mitigate discrepancies in the peak positions of these variables.
While this is sufficient for particle identification based on one-dimensional cuts on these observables, deep neural networks exploit correlations and non-linear relations between these observables.
With a limited understanding of the photon shower development from \MC simulation, tools such as the \PPT will remain constrained by systematic uncertainty estimates needed to close the residual gaps between simulation and data.

The measurement in the \emu channel presented here set a new benchmark for fiducial inclusive \xsec measurements and will probably not be outperformed by any of the other channels for mainly two reasons:
firstly, it provides the best signal-to-background ratio, while other channels have larger background contamination, thus, relying on more sophisticated estimation techniques and their associated systematic uncertainties to control them.
Secondly, despite the small branching ratio of the \emu channel, the \runii dataset has already provided enough statistics to render a systematically limited measurement.
However, measurements in the \emu channel itself could yet be improved, \eg by understanding and reducing its dominant systematic uncertainties.
The ranking of nuisance parameters in \cref{fig:results-data-ranking} and the impact of groups of systematic uncertainties listed in \cref{tab:results-data-impact} both point the finger at signal modelling as the most dominant uncertainty source.
The limitations are manifold:
on the one hand, the \tWy final state was measured in combination with \tty for the first time, but only \LO simulations in \QCD in the five-flavour scheme were available for \tWy when this analysis was performed.
This required the introduction of an additional \tWy \emph{parton definition} uncertainty detailed in \cref{sec:systematics-acceptance} due to the expected, but often missing second \bquark at parton level.
The assigned uncertainty has considerable impact on the measurement precision.
A \LOPS simulation of the \tWby final state would resolve the observed problems with the parton definition as the second \bquark is generated explicitly in the matrix-element simulation.%
\footnote{%
  As discussed earlier, the leading-order diagram of \tWby is equivalent to the lowest-order diagram for \tWy production in the four-flavour scheme, where the \bquark is not described by a \PDF.
}

On the other hand, the \LOPS simulation of the doubly-resonant \tty production shows large associated uncertainties, when varying radiation parameters of the parton shower (\Pythia \emph{A14 var3c} eigentune) or varying the overall parton-shower model (\Pythia vs. \Herwig).
Significant differences are observed in both rate and shape of the \ST distribution.
Including \NLO effects in \QCD into the matrix-element calculation could potentially alleviate these differences.
The parton-shower algorithm would then have to be matched to the \NLO matrix element, but would only have to take care of radiation beyond the \NLO real-emission corrections -- providing an overall more precise simulation of the doubly-resonant \tty final state.
Variations of parton-shower radiation parameters or of the overall parton-shower model would be expected to have a smaller impact on the prediction.
Compared to the fixed-order theory calculation, the ultimate \MC simulation would be a matrix-element calculation of the \WbWby final state at \LO or even \NLO in \QCD, which includes all resonant and non-resonant diagrams and their interference effects into one single simulated \MC sample.
A similar computation was implemented into \Powheg for combined $\ttbar + \tW b$ simulation~\cite{Jezo:2015aia,Jezo:2016ujg} and has shown good agreement with unfolded \ATLAS data in regions that were designed to be sensitive to interference effects~\cite{TOPQ-2017-05}.

All in all, the \ATLAS \runii dataset provides more opportunities for precision measurements of the \tty process.
Although not expected to surpass the \emu channel in precision, the other channels can provide valuable complementary measurements.
With much larger branching ratios, the \ljets channels have enough data available to enable high-resolution differential measurements, possibly even doubly-differential, \eg differential \xsecs as functions of kinematic variables, such as photon or lepton transverse momenta, and of the jet or \bjet multiplicity.
With many \ATLAS calibrations and efficiency measurements improved over the course of \runii, systematic uncertainties are expected to be reduced considerably with respect to the results obtained with \SI{36}{\ifb}.
Furthermore, studies within a MSc thesis project in G{\"o}ttingen have revealed the potential of machine-learning techniques using multiclass classification at event level in the \ljets channels~\cite{Korn2019}.
With the possibility to discriminate not only signal and background, but signal and different classes of background processes, such tools could outperform the event-level classifier used in the analysis of 2015 and 2016 data and provide additional separation power through multiple output nodes.
Another MSc thesis project investigated possibilities to enhance the fraction of photons radiated directly by one of the top quarks in \tty production with machine-learning techniques, and has shown promising results~\cite{Kirchhoff2018}.
Suppressing photons from the \bquarks, the \Wbosons and the decay products of the \Wbosons would increase the sensitivity of a \tty measurement to the top-photon coupling, which in turn would allow more thorough tests of its strength and structure.

Eventually, the focus of properties measurements in the top-quark sector will shift from \SM precision tests to interpreting results in the view of \BSM theories.
One powerful tool at hand is the \EFT framework, in which multiple theories may be tested simultaneously through their interpretations as effective theories that modify coupling behaviours.
As introduced in \cref{sec:theory-tty}, \tty production is sensitive to some of the lowest-order, dimension-6 \EFT operators that involve the top quark.
Fiducial inclusive \ATLAS measurements, but even more so differential distributions enable constraints of the Wilson coefficients that describe the magnitude of these operators in an \EFT-extended \SM Lagrangian.
Such measurements require dedicated \MC simulations with modified coupling behaviours that reflect changes in individual \EFT operators.
Then, templates of observable distributions can be generated for these \EFT-extended predictions and can be compared to (unfolded) \ATLAS data as done in this measurement with \SM predictions.
With the full \runii dataset recorded with the \ATLAS detector at hand, the \ljets channels already provide enough statistics to allow interpreting \tty measurements in the context of \EFT.
This will become even more relevant with the scheduled \runiii of the \LHC, which expects proton-proton collisions from 2021 and an estimated integrated luminosity of \SI{300}{\ifb}.

Particular attention should also be given to \xsec ratios of \tty and \ttbar production, where $\mathcal{R} = \sigma(\tty)/\sigma(\ttbar)$.
The authors of Ref.~\cite{Bevilacqua:2018dny} argue that this ratio shows higher stability against radiative corrections and reduced dependence on the choice of renormalisation and factorisation scales and on \PDFs.
In addition, measurements of $\mathcal R$ instead of direct \tty \xsec measurements would yield reduced experimental uncertainties.
The luminosity uncertainty, one of the limitations of the measurement presented in this thesis, would be cancelled in the ratio.
Uncertainties associated with physics objects common to the \tty and \ttbar final states, \eg uncertainties on the energy and \pT calibrations of leptons and jets, would be reduced considerably.
Such ratio measurements could be performed differentially as well:
distributions such as
\begin{align}
  \label{eq:conclusions-ratio}
  \mathcal{R}_{\Dphill} = \left( \frac{\mathrm{d}\sigma(\tty)}{\mathrm{d}\left( \Dphill \right)} \right) \left( \frac{\mathrm{d}\sigma(\ttbar)}{\mathrm{d}\left( \Dphill \right)} \right)^{-1}
\end{align}
could give further insight into the observed shape discrepancies in the \Dphill observable in both \tty and \ttbar production, while reducing theoretical and experimental uncertainties to a minimum with respect to direct \xsec measurements.
The magnitude of these ratios would be sensitive to the top-photon coupling strength directly, and differential distributions of $\mathcal R$ could be used to probe \ATLAS data for modified coupling structures.


\printbibliography

\appendix

\chapter{Templates of systematic variations}
\label{cha:app-red-blue-plots}

\begin{figure}
  \centering
  \includegraphics[width=0.48\textwidth]{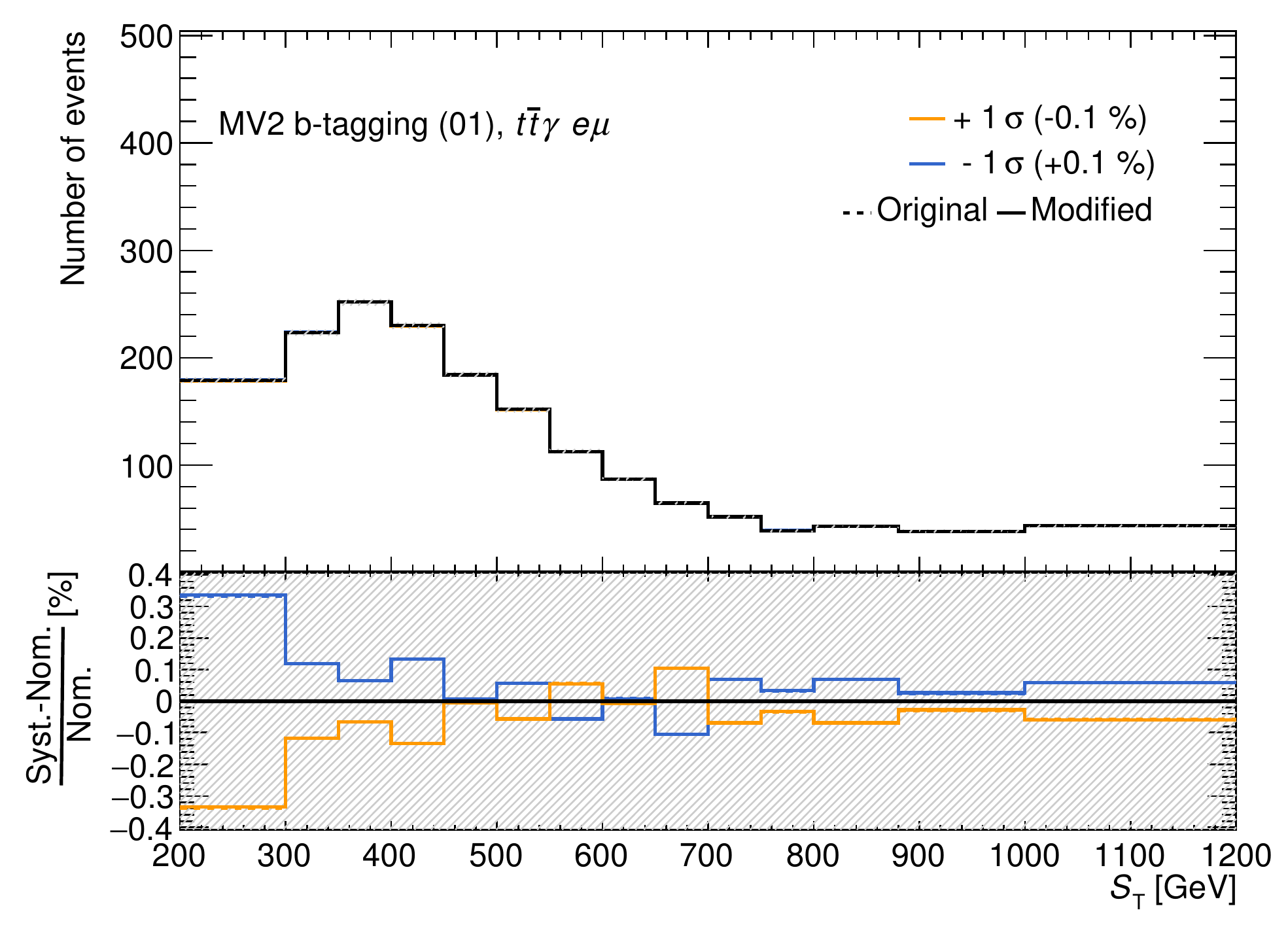}
  \includegraphics[width=0.48\textwidth]{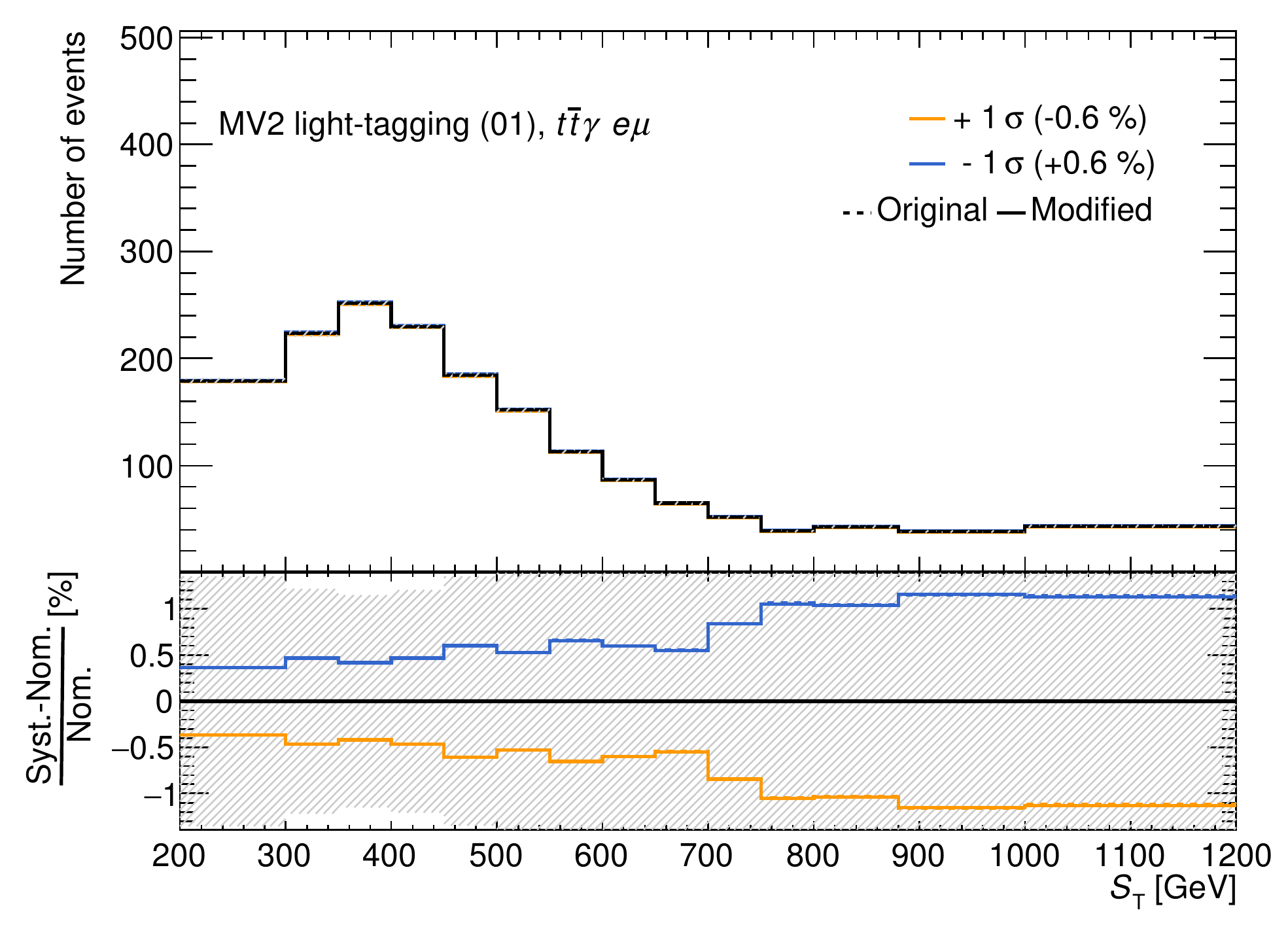}
  \caption[Templates for two \tty flavour-tagging systematics]{%
    Two examples of systematic templates for flavour-tagging uncertainties in the \catttyemu event category:
    one of the variations associated to the \btag efficiency and
    one of the variations of the light-mis-tagging rate.
    The templates are symmetric by construction.
    The shaded uncertainty bands represent \MC-statistical uncertainties on the nominal prediction.
  }
  \label{fig:syst_btagging}
\end{figure}

\begin{figure}
  \centering
  \includegraphics[width=0.48\textwidth]{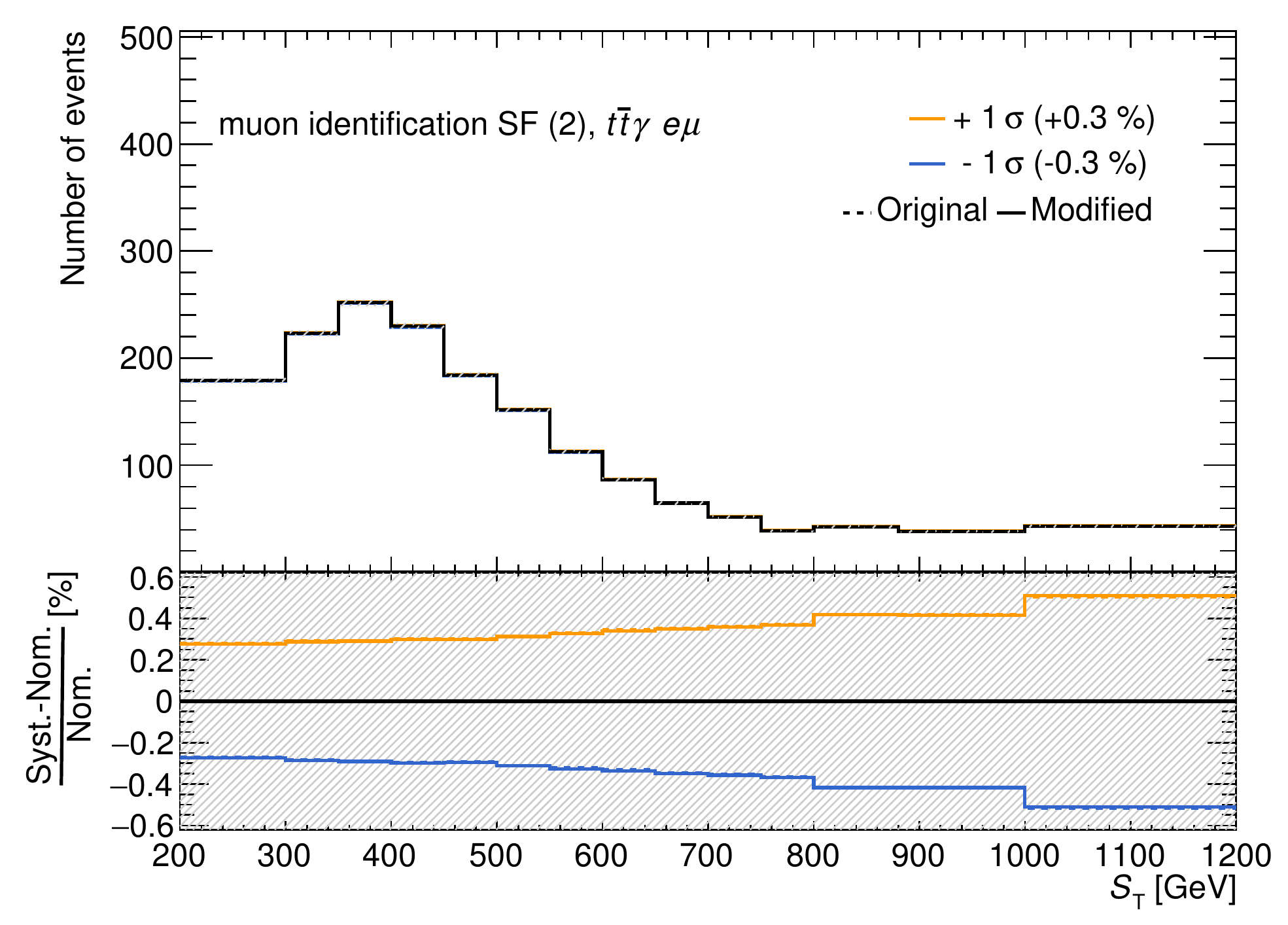}
  \includegraphics[width=0.48\textwidth]{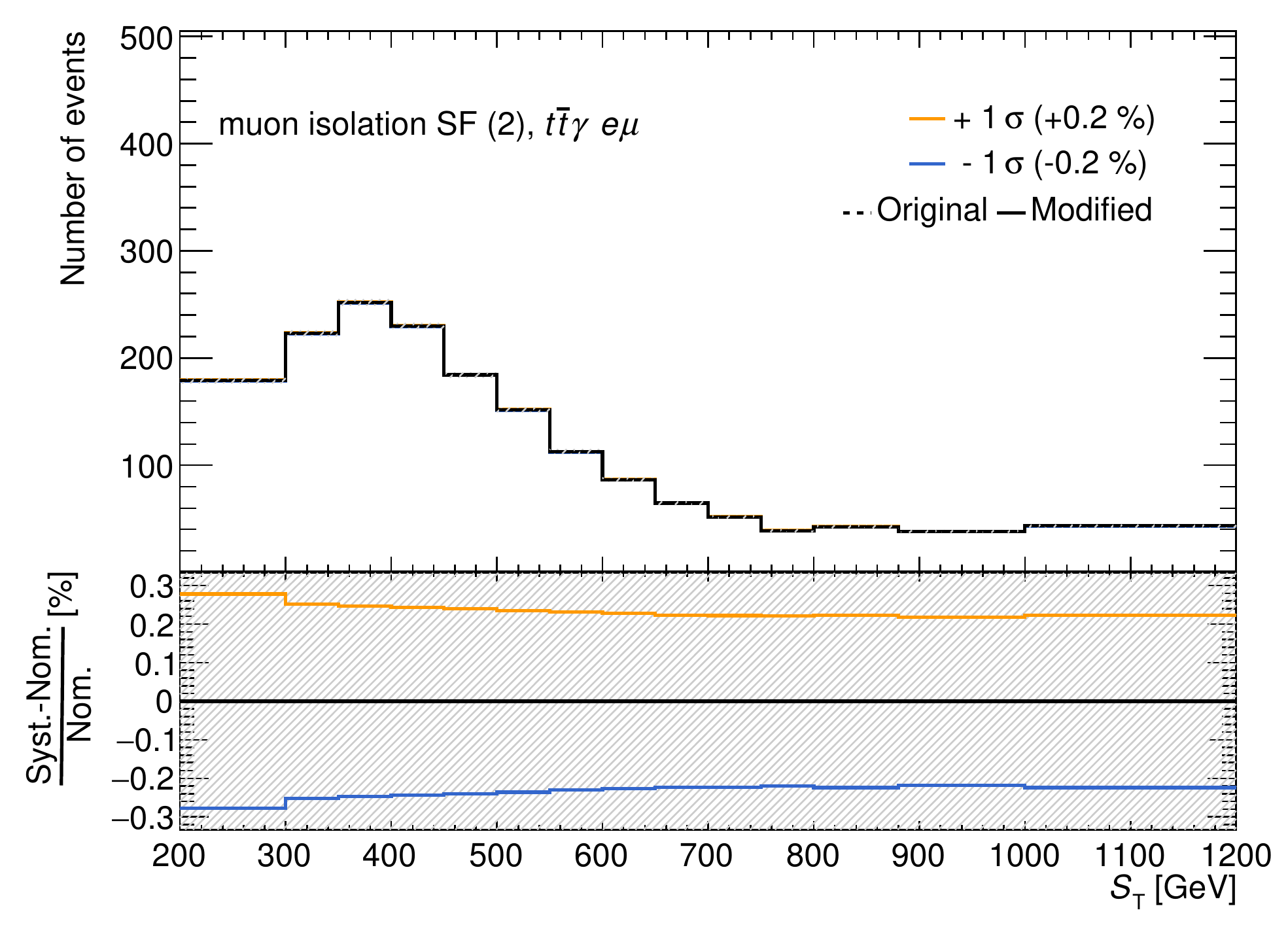}
  \caption[Templates for muon ID and muon isolation systematics]{%
    Systematic templates for two simulation-to-data scale-factor uncertainties in the \catttyemu event category.
    Both uncertainties shown are related to muons:
    the systematic uncertainty on the muon identification scale factors and
    the systematic uncertainty on the muon isolation-efficiency scale factors.
    The final templates are shown in solid orange and blue to be compared against the nominal prediction in black.
    The shaded uncertainty bands represent \MC-statistical uncertainties on the nominal prediction.
  }
  \label{fig:syst_muons}
\end{figure}

\begin{figure}
  \centering
  \includegraphics[width=0.48\textwidth]{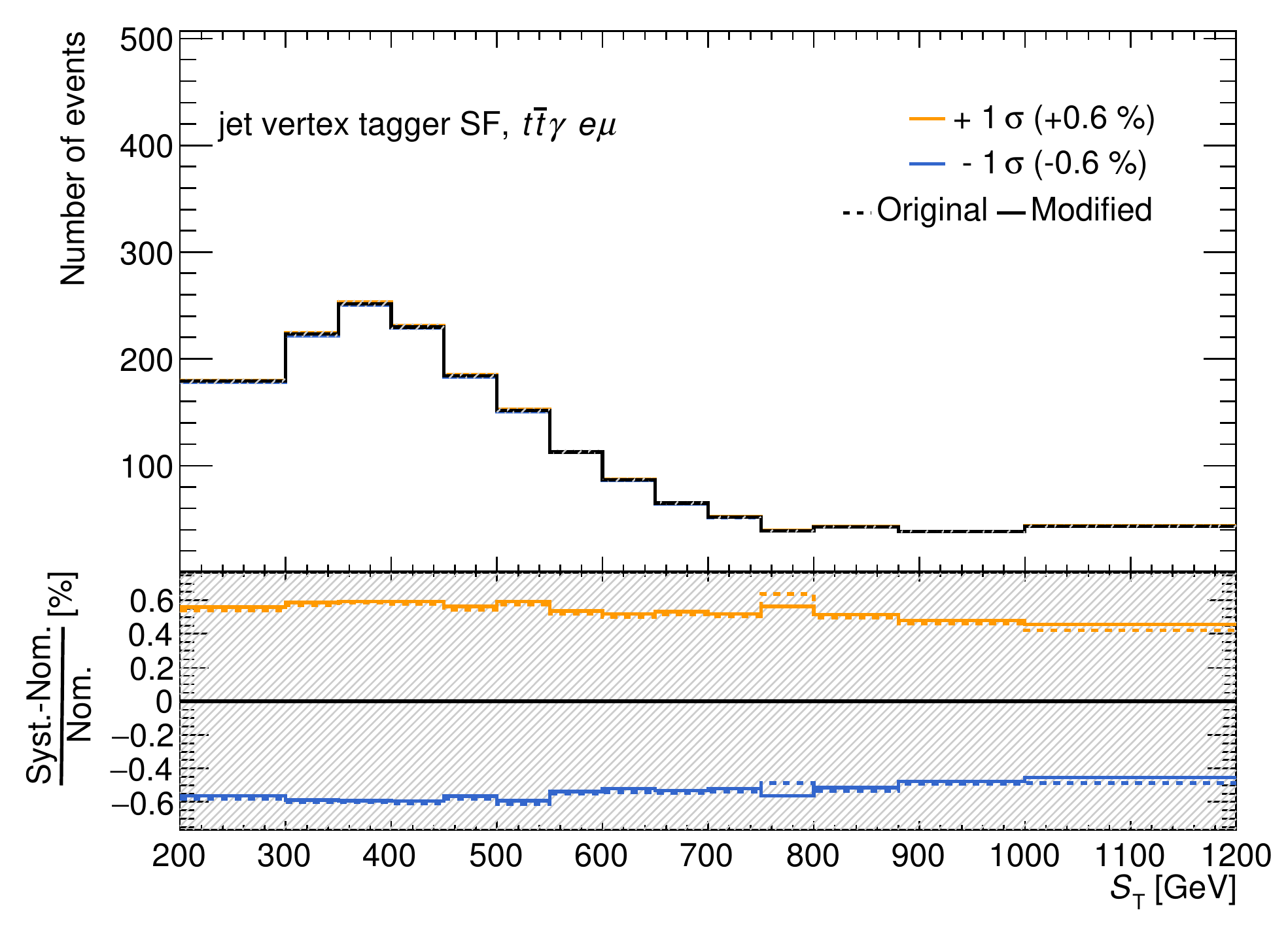}
  \includegraphics[width=0.48\textwidth]{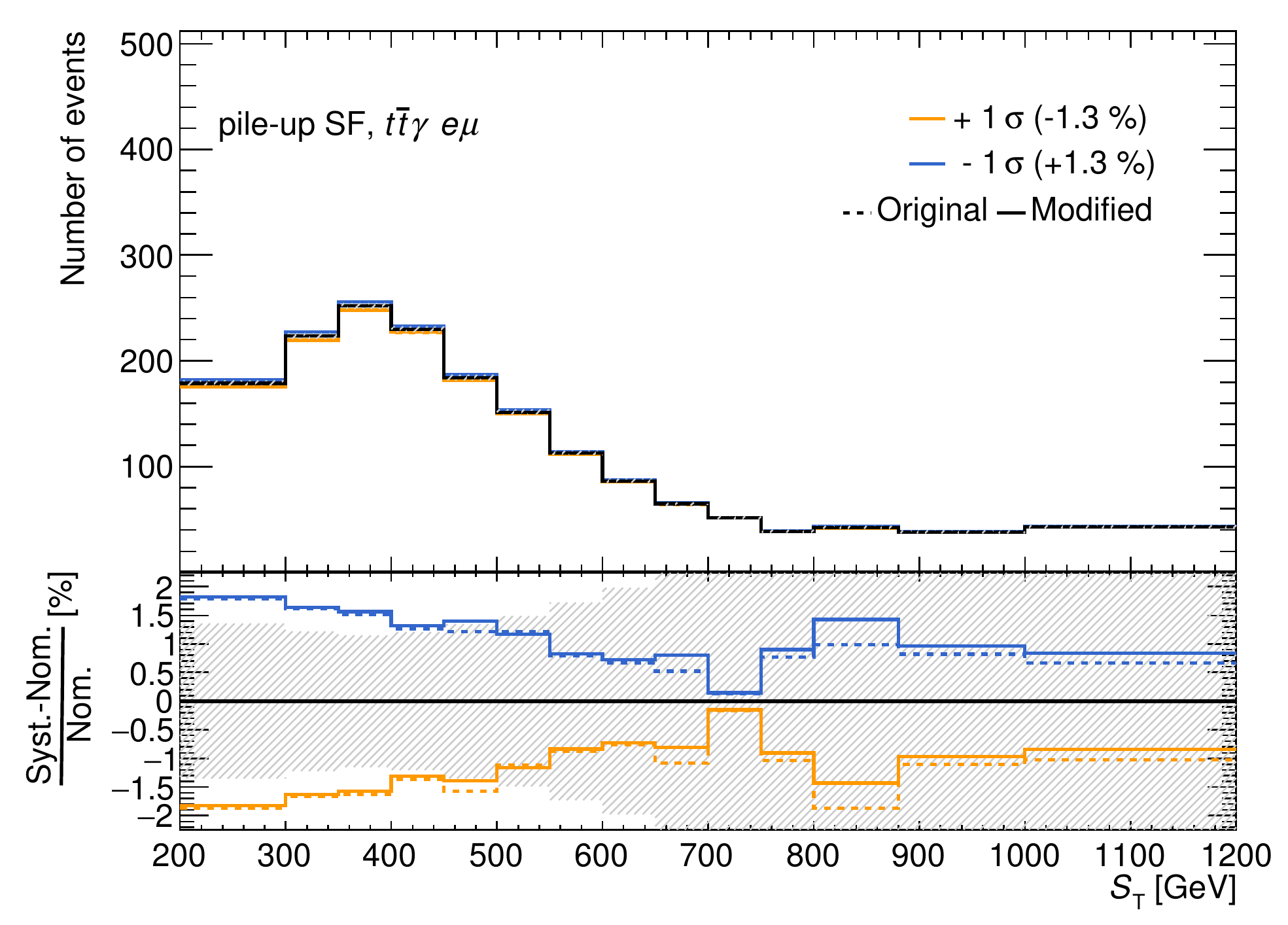}
  \caption[Templates for JVT and pile-up systematics]{%
    Systematic templates of two simulation-to-data scale-factor uncertainties in the \catttyemu event category:
    (1) for the scale factors associated with the jet vertex fraction and
    (2) for the pile-up scale factors.
    Both templates use two-sided symmetrisation.
    The dashed lines are the non-smoothed templates; the final templates after smoothing and symmetrisation are shown in solid orange and blue.
    The shaded uncertainty bands represent \MC-statistical uncertainties on the nominal prediction.
  }
  \label{fig:syst_JVT_pileup}
\end{figure}

\begin{figure}
  \centering
  \includegraphics[width=0.48\textwidth]{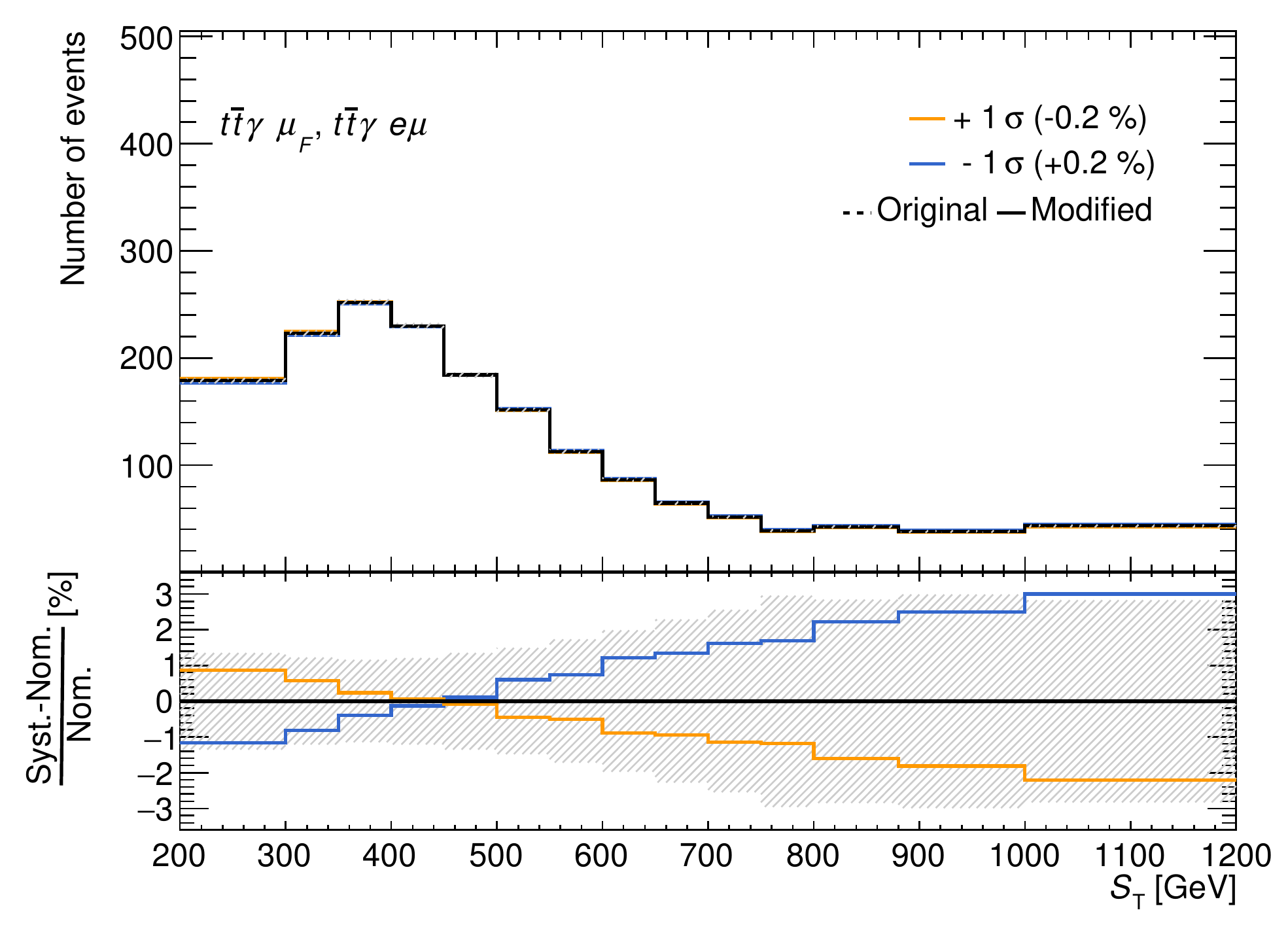}
  \includegraphics[width=0.48\textwidth]{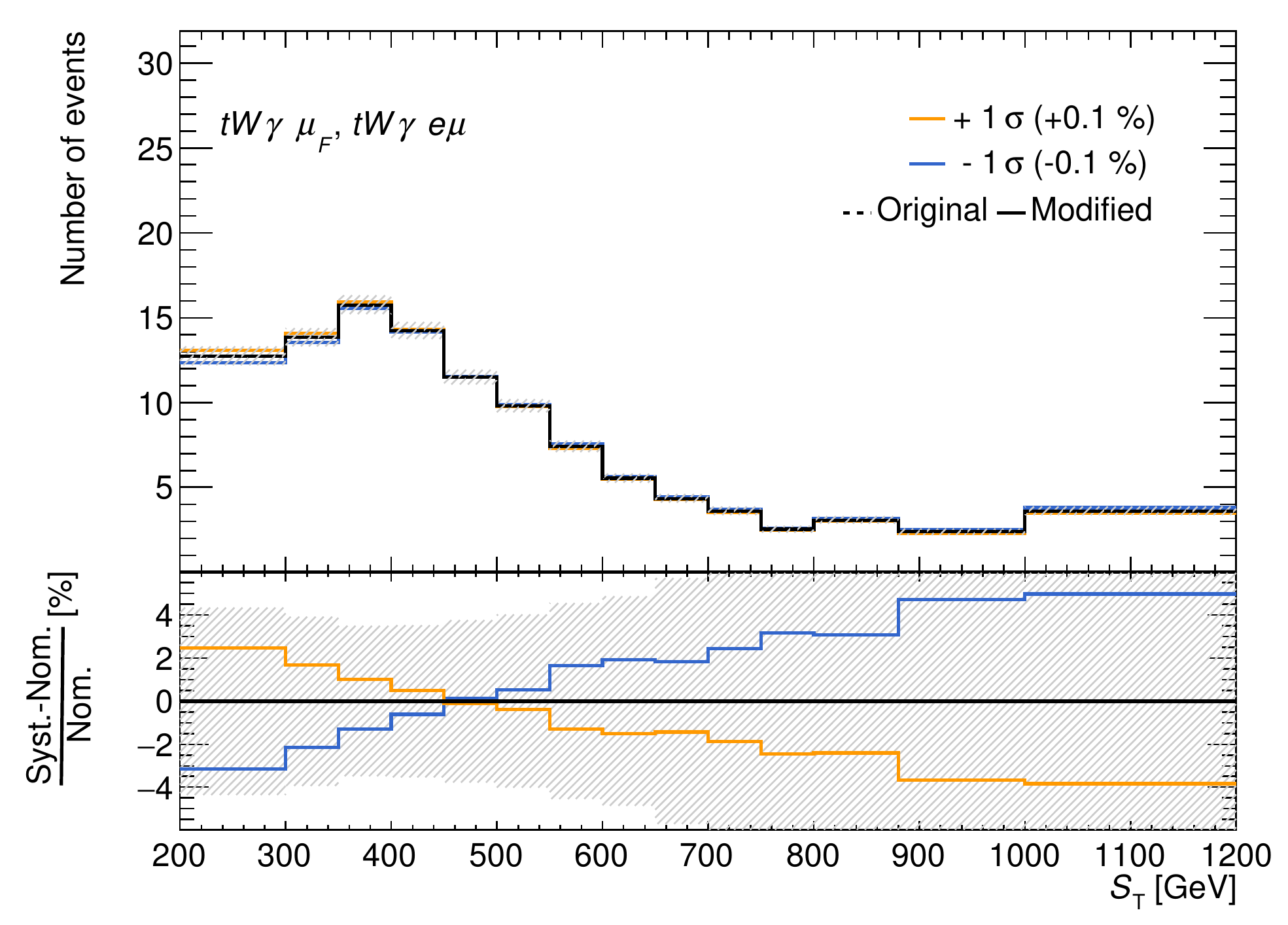}
  \caption[Templates for \tty and \tWy factorisation scale systematics]{%
    Systematic templates of the uncertainty on the factorisation scales for
    the \tty signal, shown in the \catttyemu signal category, and
    the \tWy signal, shown in the \cattWyemu signal category.
    The templates are neither smoothed nor symmetrised and are shown in solid orange and blue.
    The shaded uncertainty bands represent \MC-statistical uncertainties on the nominal prediction.
  }
  \label{fig:syst_muF}
\end{figure}

\begin{figure}
  \centering
  \includegraphics[width=0.48\textwidth]{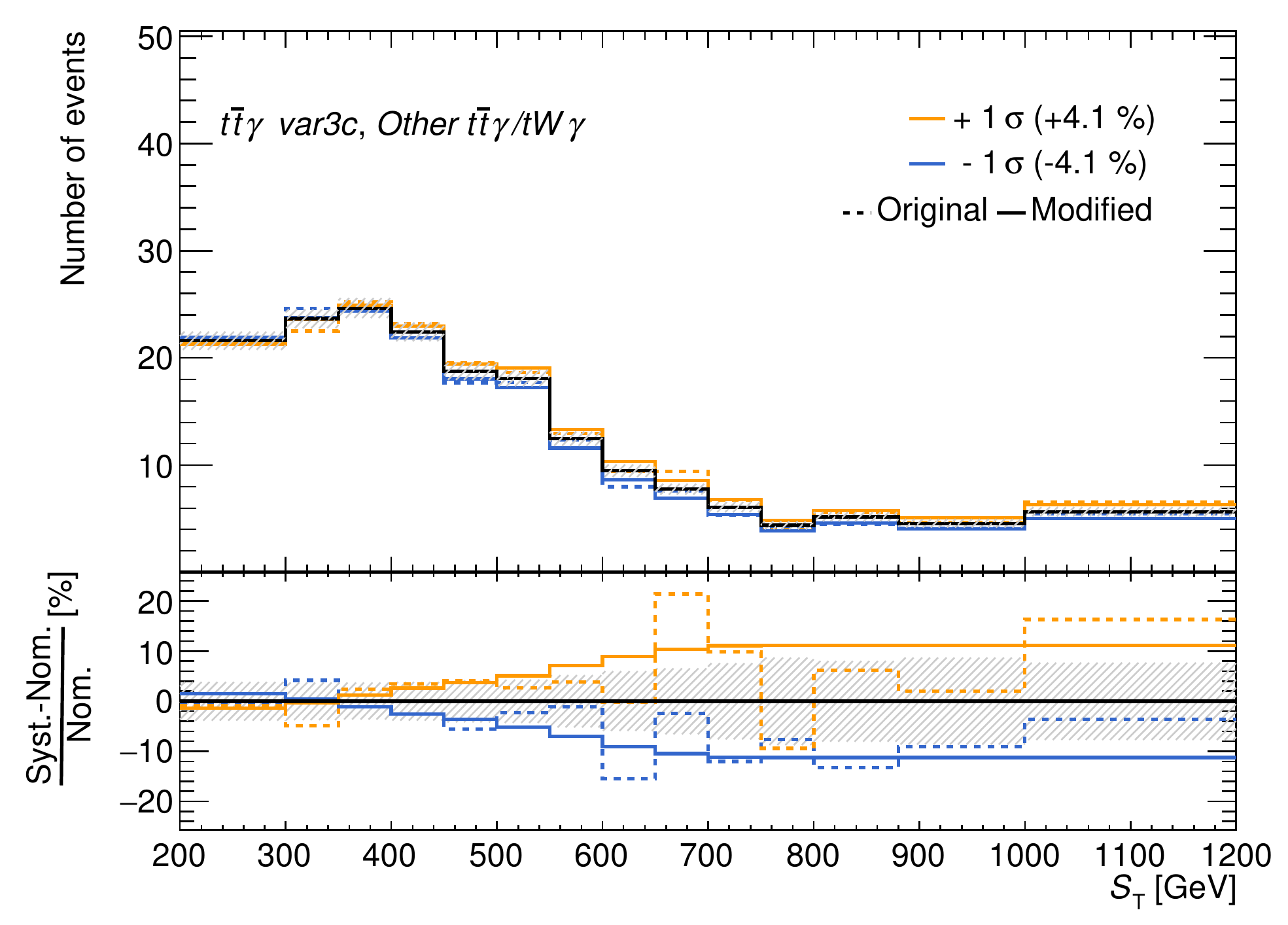}
  \includegraphics[width=0.48\textwidth]{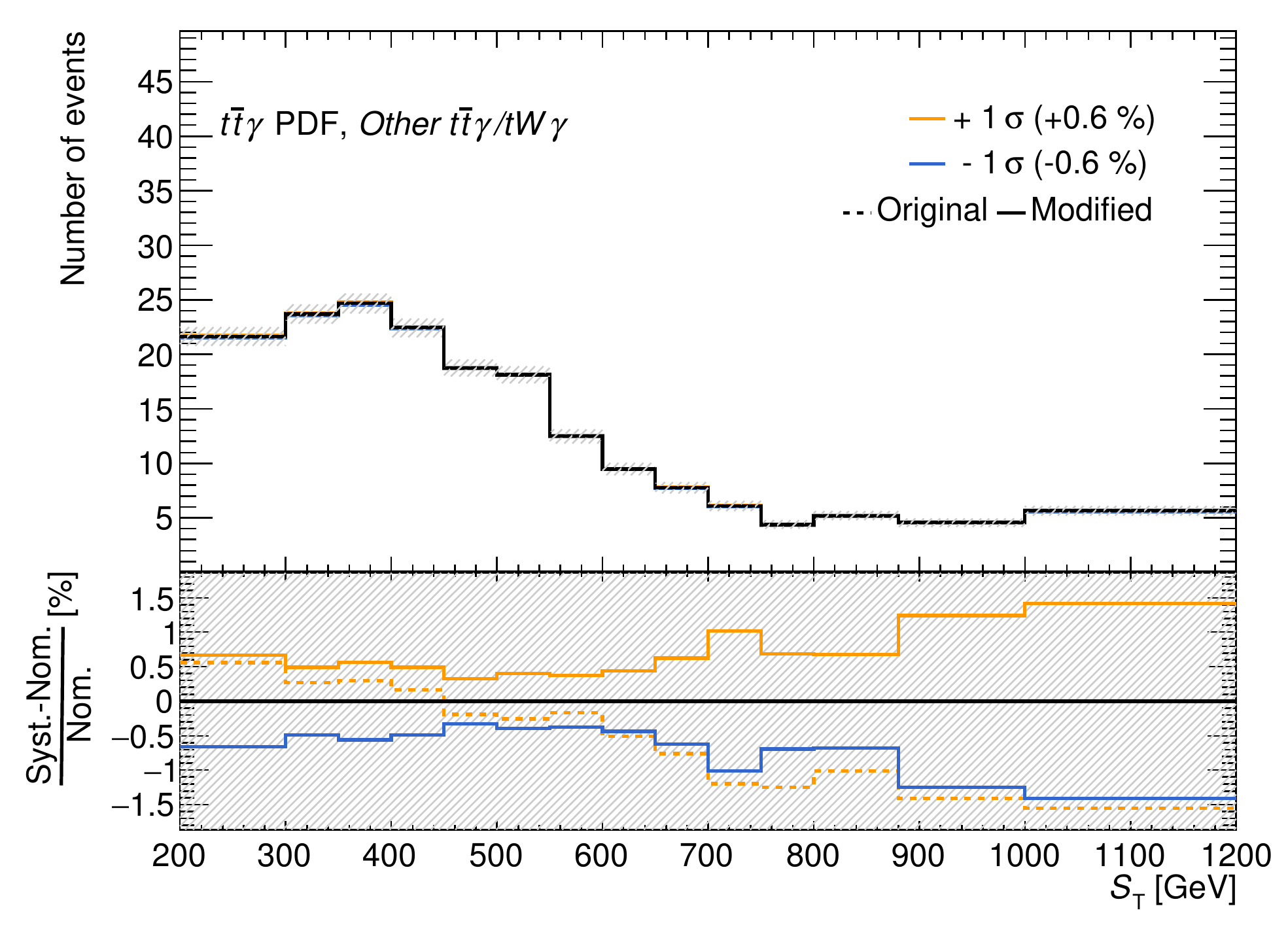}
  \caption[Templates for \tty \Pythia eigentune and \PDF systematics]{%
    On the left:
    systematic templates for the \tty radiation uncertainty (\Pythia eigentune) in the \catother event category.
    The templates are smoothed and use \emph{maximum} symmetrisation.
    The dashed lines are the non-smoothed templates; the final templates are shown in solid orange and blue.
    On the right:
    combined systematic template for the \tty \PDF uncertainty in the \catother event category.
    The dashed line in the background shows one of the \NNPDF replicas.
    The shaded uncertainty bands represent \MC-statistical uncertainties on the nominal prediction.
  }
  \label{fig:syst_other_tty_var3c_PDF}
\end{figure}

\begin{figure}
  \centering
  \includegraphics[width=0.48\textwidth]{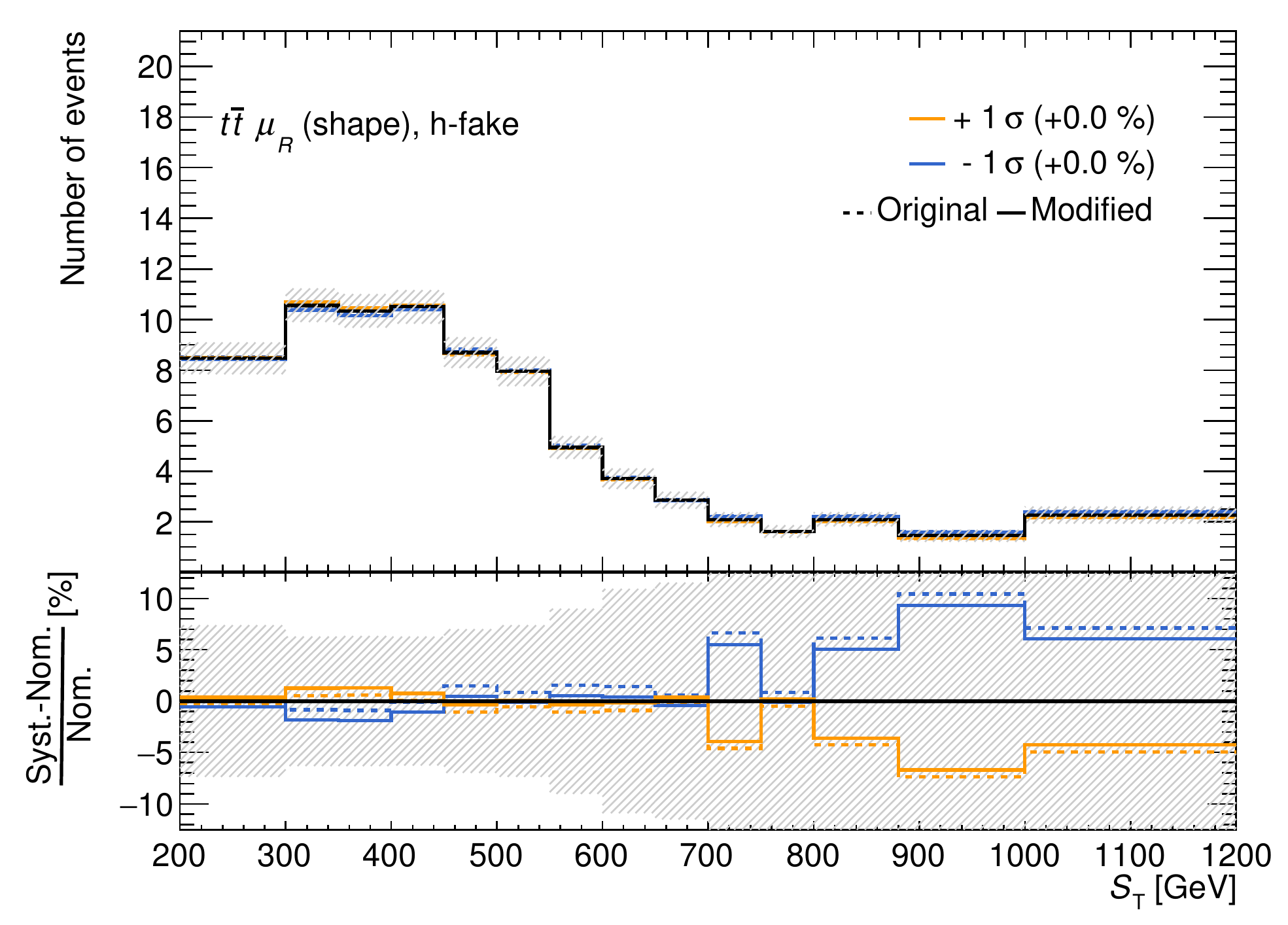}
  \includegraphics[width=0.48\textwidth]{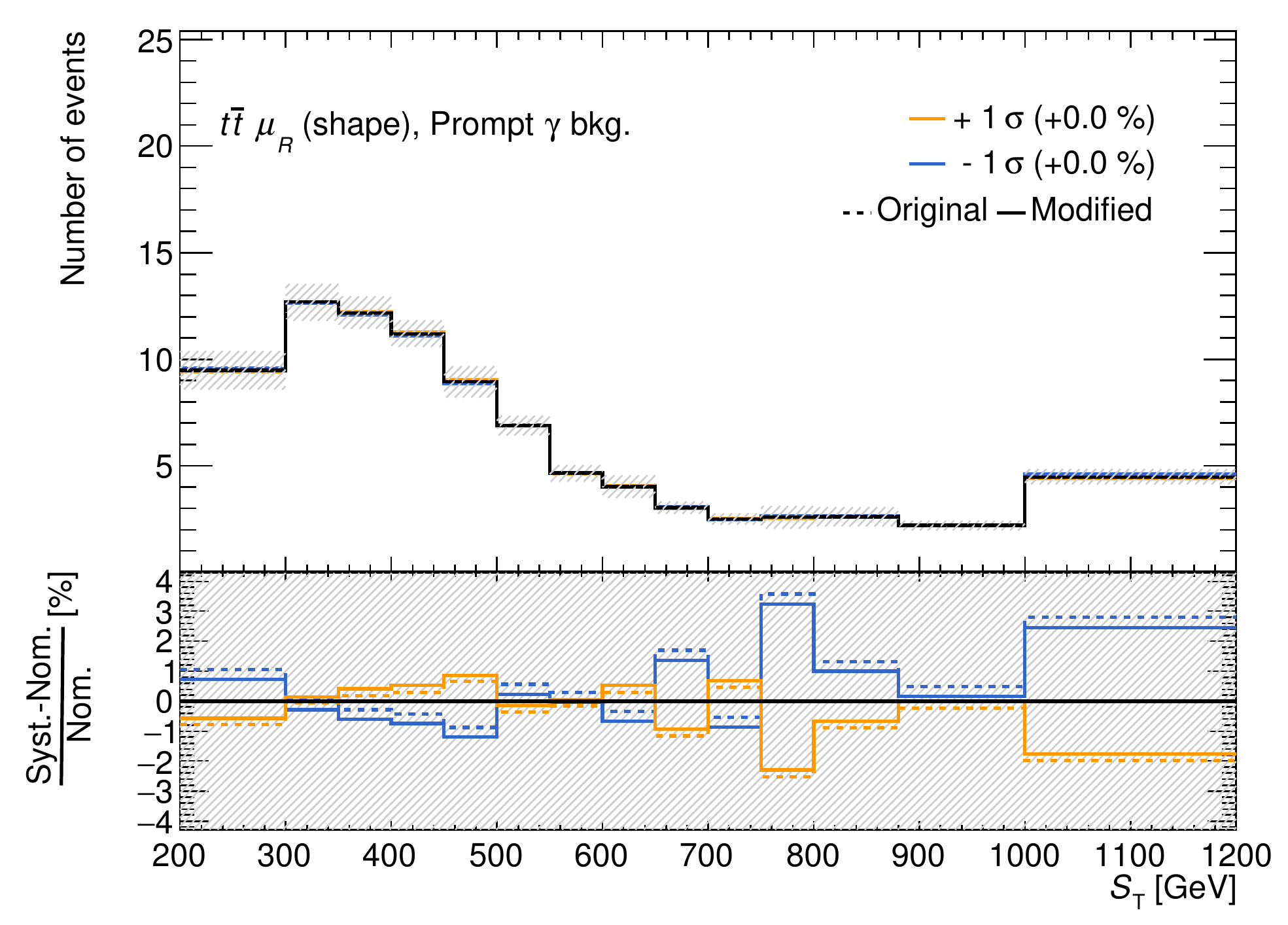}
  \caption[Templates for \ttbar renormalisation scale systematics]{%
    Systematic templates for the \ttbar renormalisation scale uncertainty in
    the \cathfake and the \catprompt categories.
    The dashed lines are the non-smoothed templates; the final templates after smoothing and symmetrisation are shown in solid orange and blue.
    The shaded uncertainty bands represent \MC-statistical uncertainties on the nominal prediction.
  }
  \label{fig:syst_ttbar_muR}
\end{figure}

\begin{figure}
  \centering
  \includegraphics[width=0.48\textwidth]{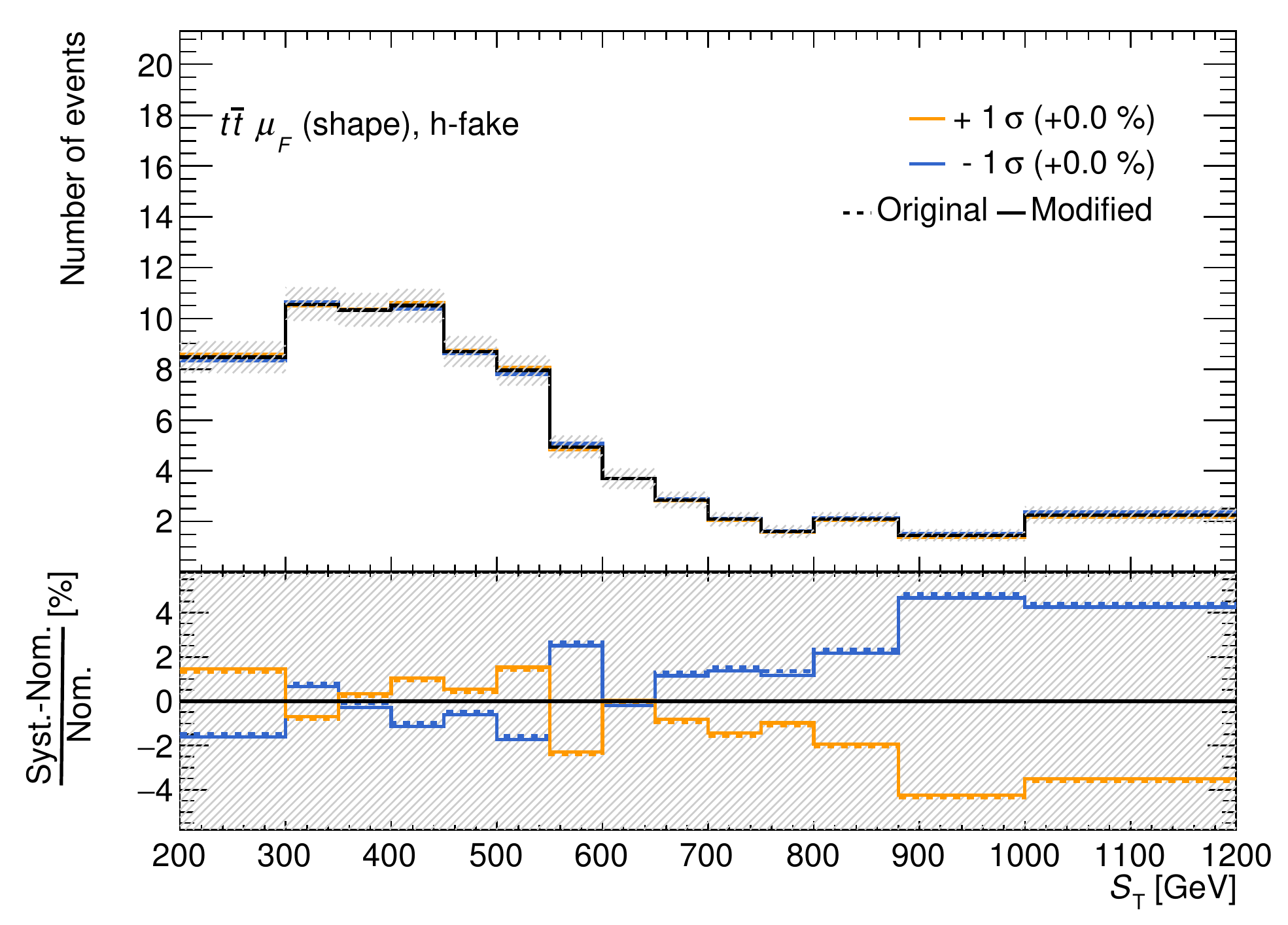}
  \includegraphics[width=0.48\textwidth]{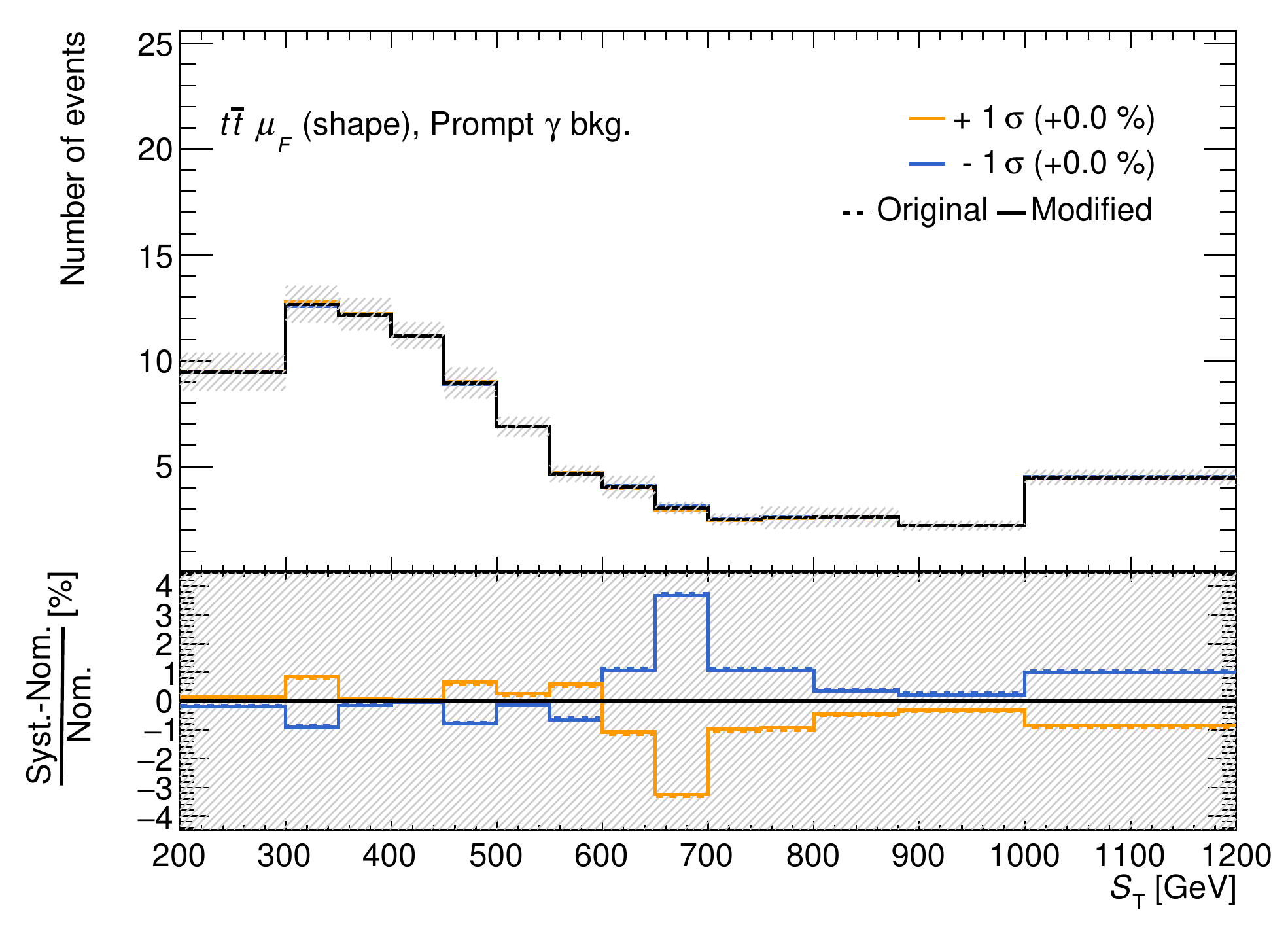}
  \caption[Templates for \ttbar factorisation scale systematics]{
    Systematic templates for the \ttbar factorisation scale uncertainty in
    the \cathfake and
    the \catprompt category.
    The dashed lines are the non-smoothed templates; the final templates after smoothing and symmetrisation are shown in solid orange and blue.
    The shaded uncertainty bands represent \MC-statistical uncertainties on the nominal prediction.
  }
  \label{fig:syst_ttbar_muF}
\end{figure}

This appendix shows additional plots of systematic templates that enter the profile-likelihood fit, but were not displayed in \cref{chap:systematics}.
\Cref{fig:syst_btagging} shows systematic templates of some flavour-tagging uncertainties.
\Cref{fig:syst_muons} shows systematic templates of two simulation-to-data scale-factor uncertainties related to muons.
\Cref{fig:syst_JVT_pileup} shows systematic templates of the jet vertex fraction and pile-up simulation-to-data scale-factor uncertainties.
\Cref{fig:syst_muF} shows templates for the uncertainties on the \tty and \tWy factorisation scales.
\Cref{fig:syst_other_tty_var3c_PDF} shows the \tty radiation uncertainty and the \tty \PDF uncertainty as displayed in \cref{fig:syst_tty_var3c_PDF} in \cref{chap:systematics}, but instead the templates for the \catother category are plotted.
\Cref{fig:syst_ttbar_muR,fig:syst_ttbar_muF} show additional \ttbar modelling uncertainties.

\chapter{Additional control plots}
\label{chap:app-add-controlplots}

\begin{figure}[b]
  \centering
  \includegraphics[width=0.48\textwidth]{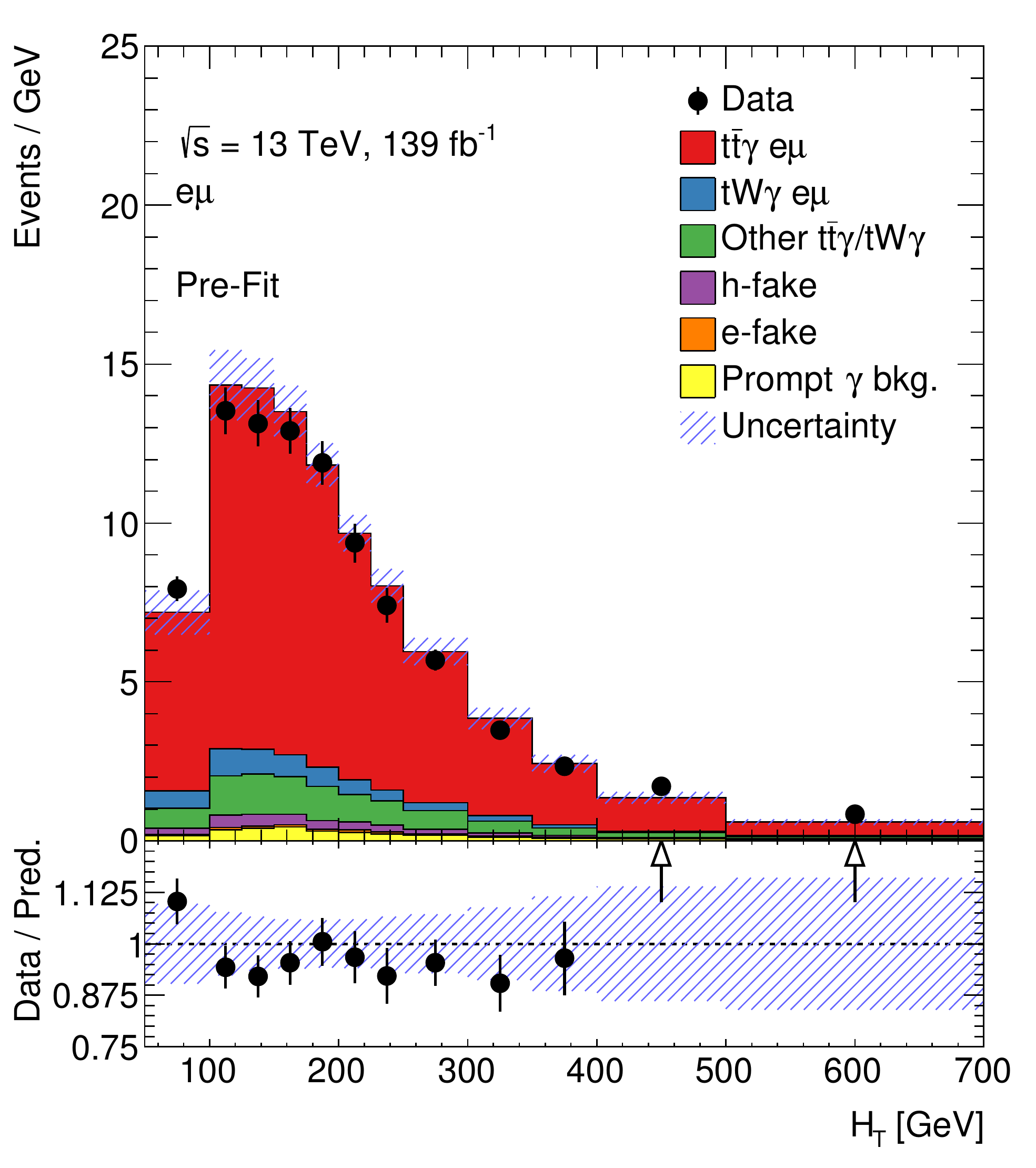}%
  \includegraphics[width=0.48\textwidth]{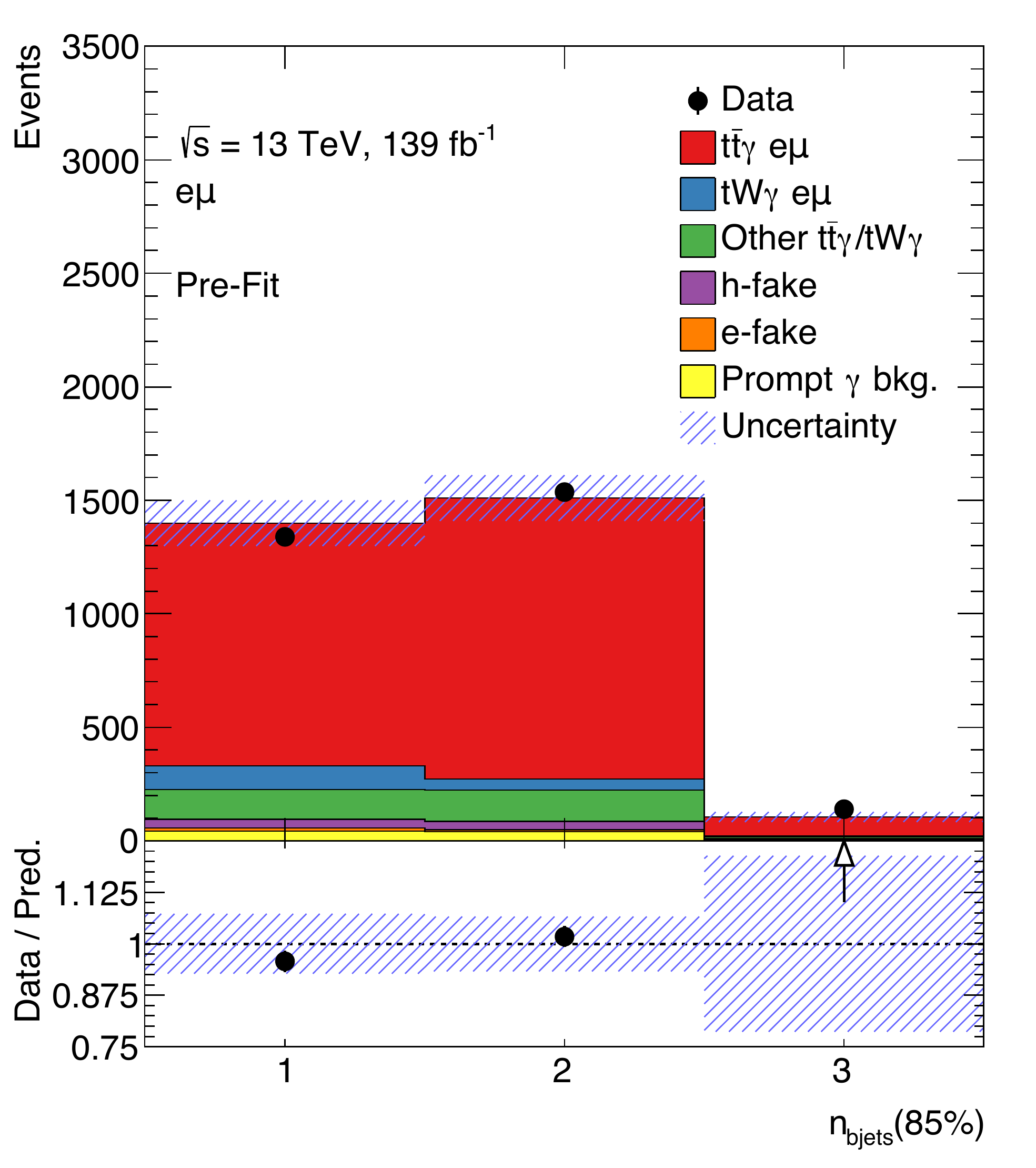}%
  \caption[Additional control plots with all uncertainties included (1)]{%
    Control plots for a data/\MC comparison in the \emu signal region with all statistical and systematic uncertainties included.
    Note that the predictions of the \tty and \tWy categories were scaled to match the numbers of reconstructed events in data.
    The shown observables are the scalar sum \HT of all jet \pT and the \bjet multiplicity.
  }
  \label{fig:app-add-controlplots-1}
\end{figure}

\begin{figure}[p]
  \centering
  \includegraphics[width=0.48\textwidth]{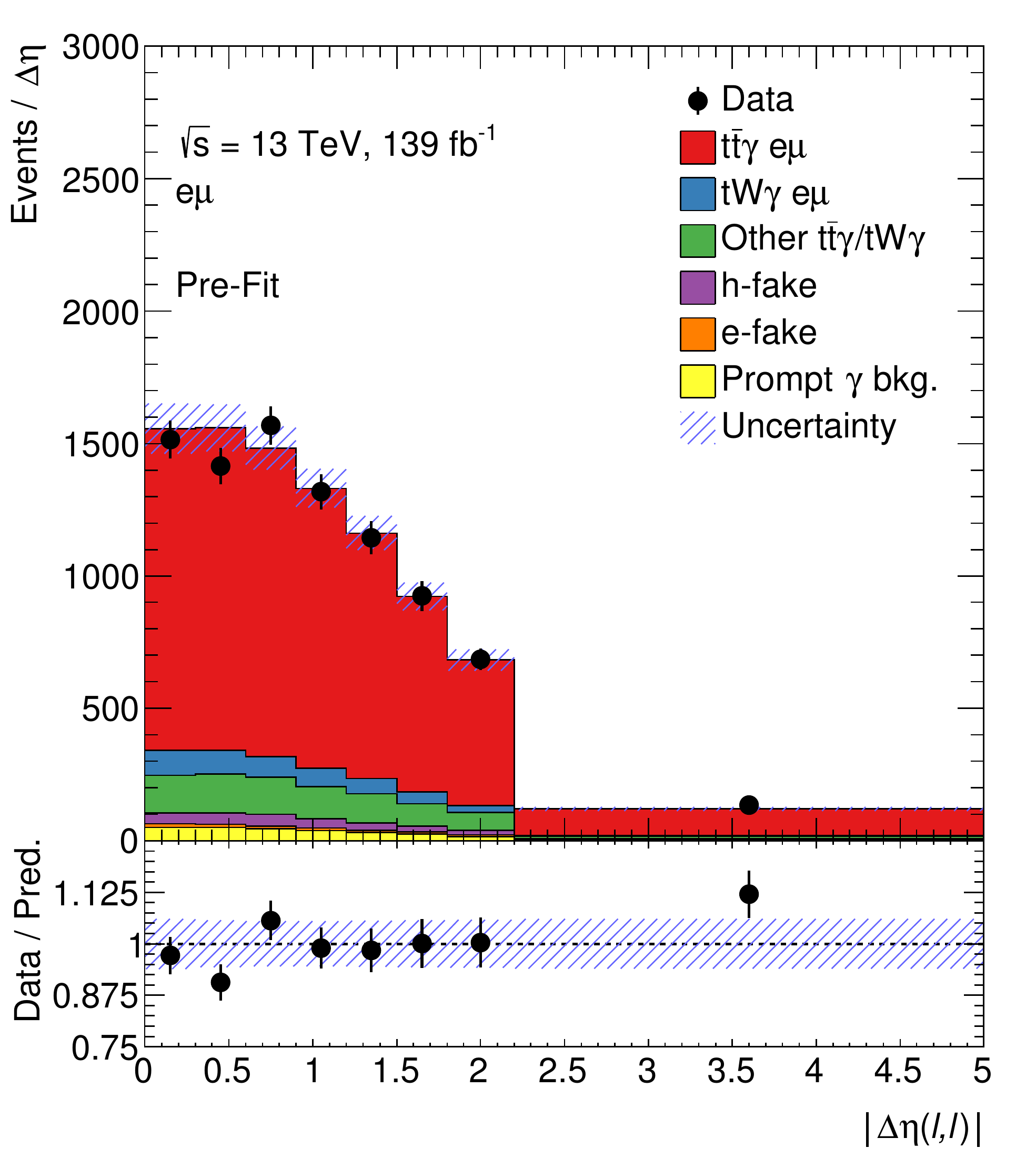}%
  \includegraphics[width=0.48\textwidth]{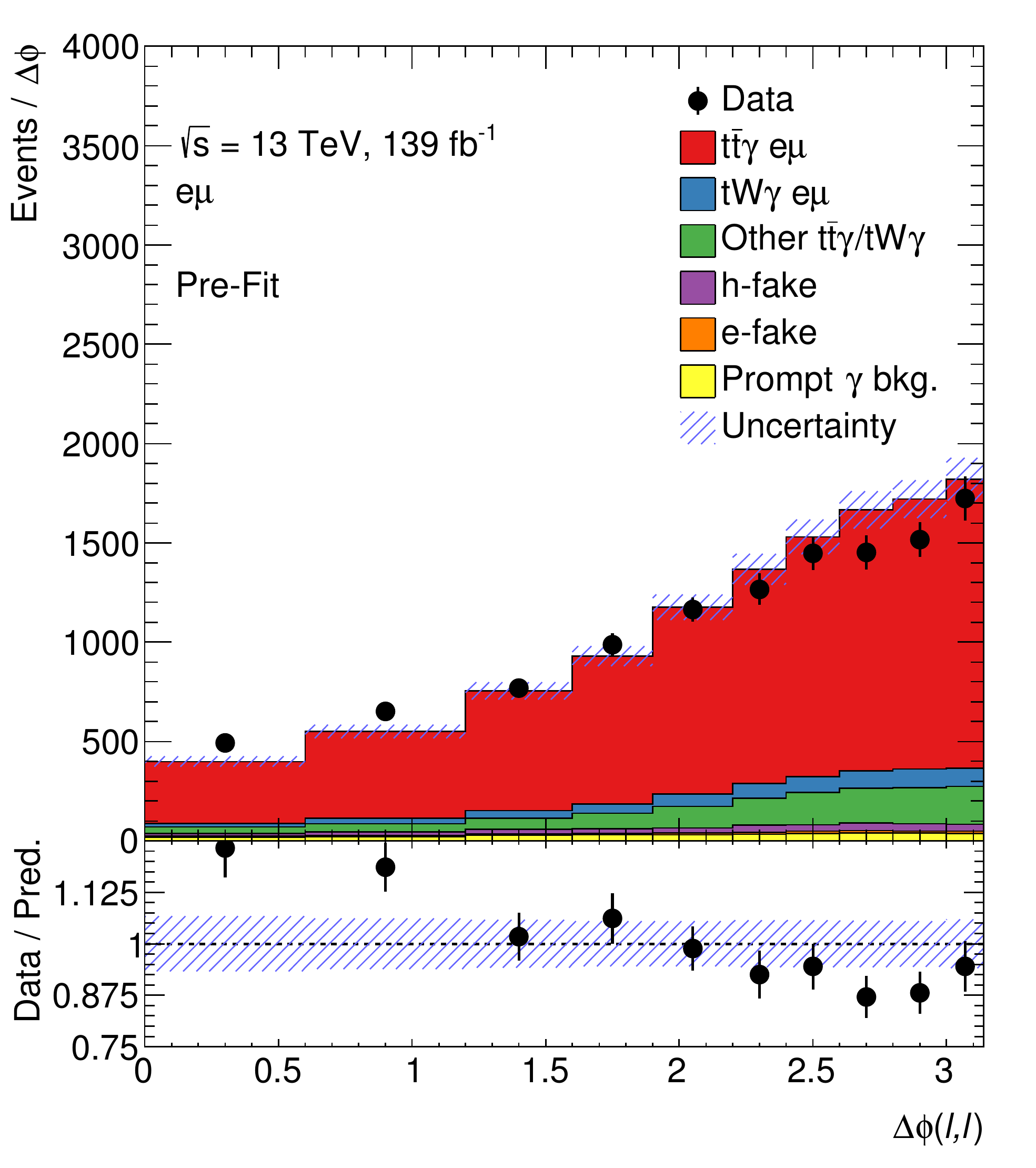}%
  \\
  \includegraphics[width=0.48\textwidth]{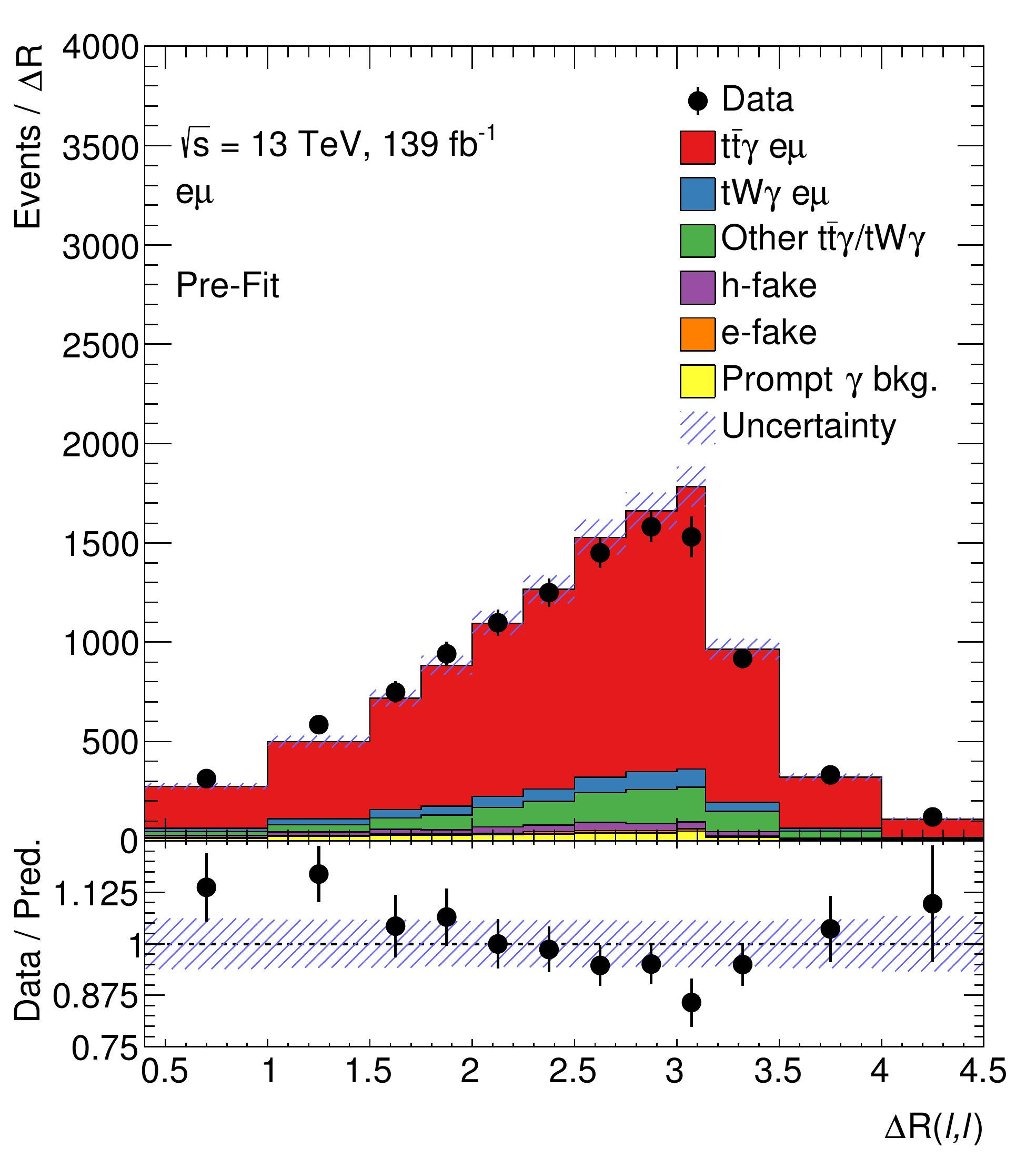}%
  \includegraphics[width=0.48\textwidth]{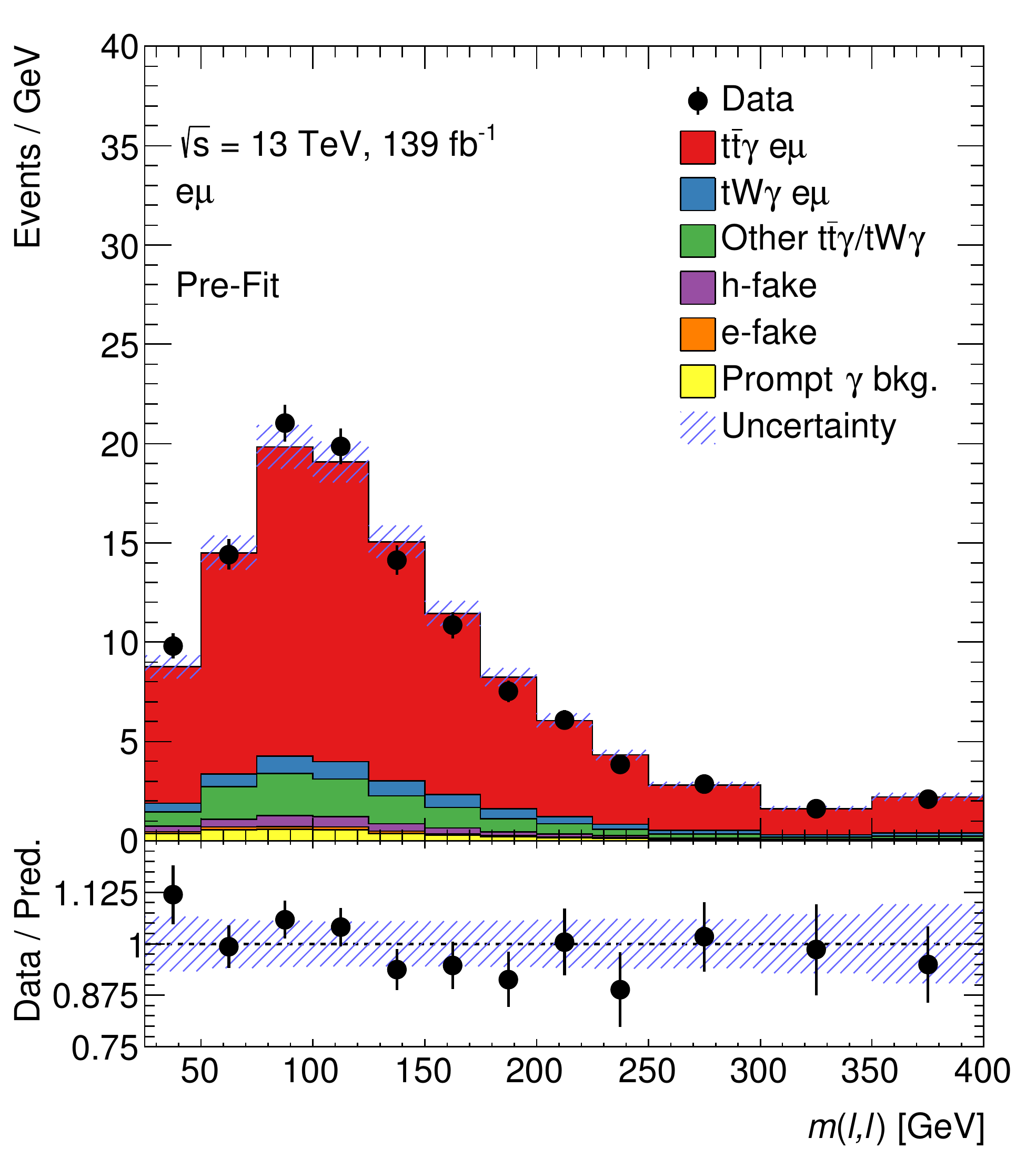}%
  \caption[Additional control plots with all uncertainties included (2)]{%
    Control plots for a data/\MC comparison in the \emu signal region with all statistical and systematic uncertainties included.
    Note that the predictions of the \tty and \tWy categories were scaled to match the numbers of reconstructed events in data.
    The shown observables are related to the two charged leptons: their absolute differences in $\eta$ and $\phi$ and their combined distance $\Delta R$ in the \etaphi plane, and their invariant mass.
  }
  \label{fig:app-add-controlplots-2}
\end{figure}

\begin{figure}[p]
  \centering
  \includegraphics[width=0.48\textwidth]{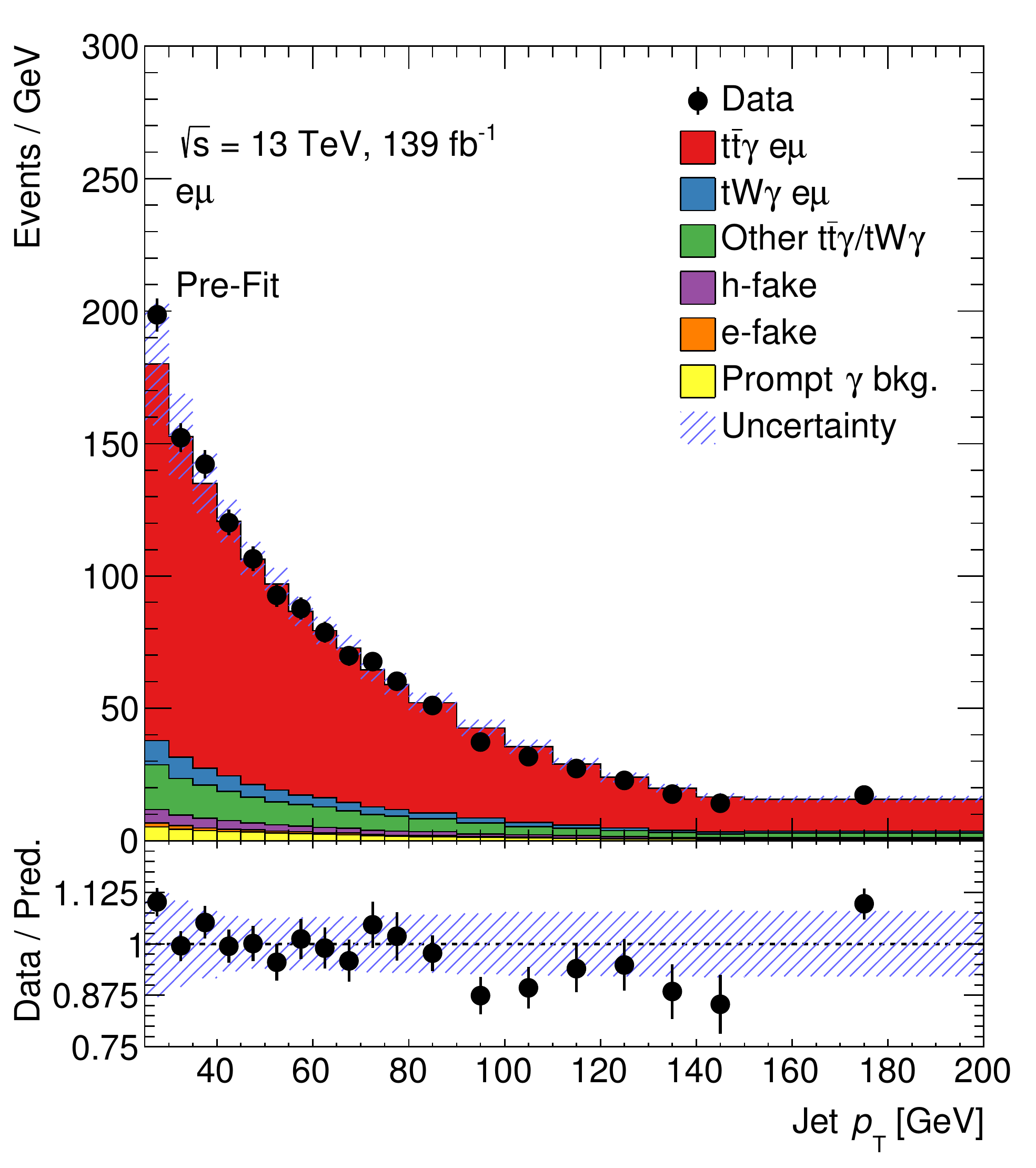}%
  \includegraphics[width=0.48\textwidth]{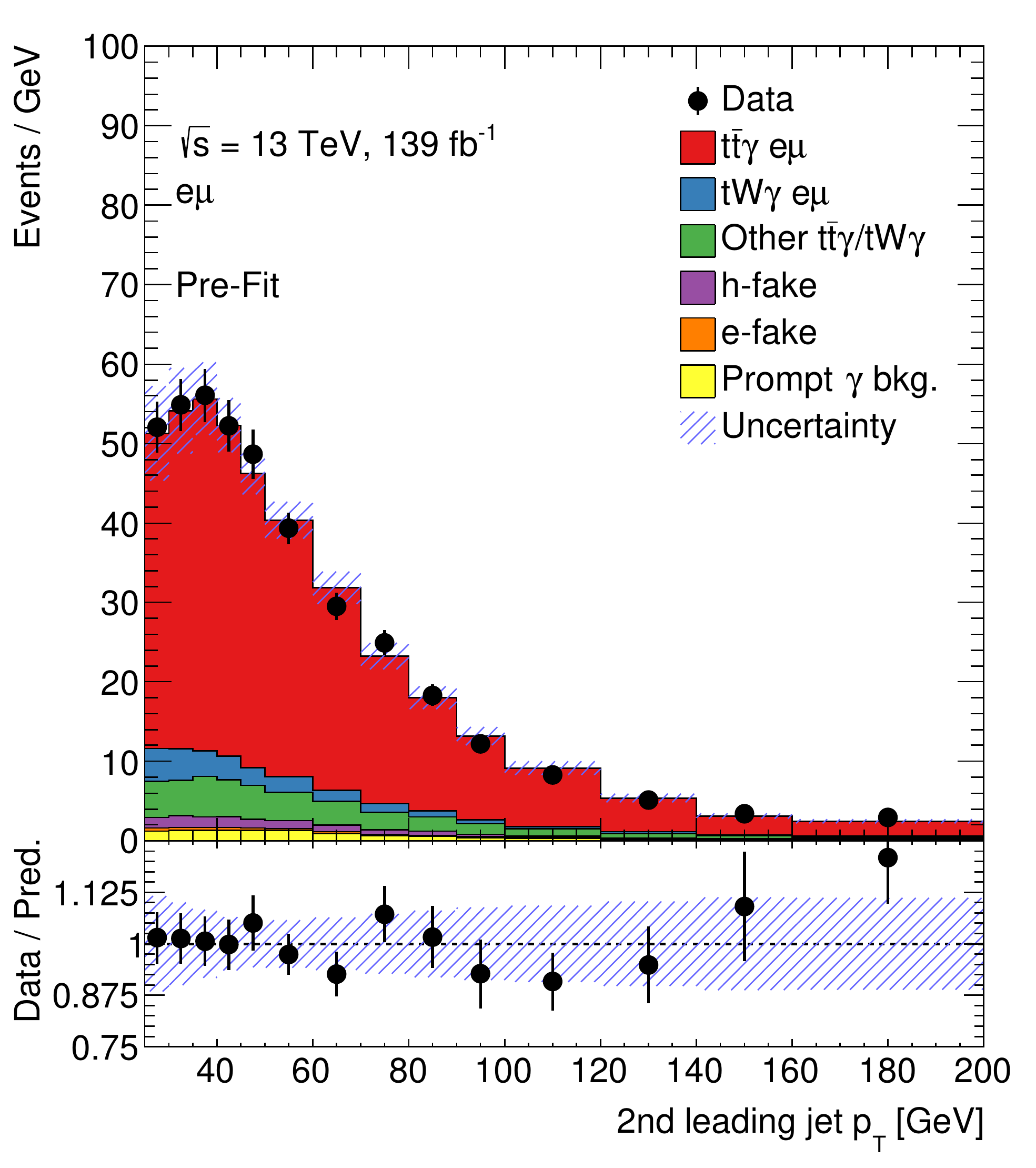}%
  \\
  \includegraphics[width=0.48\textwidth]{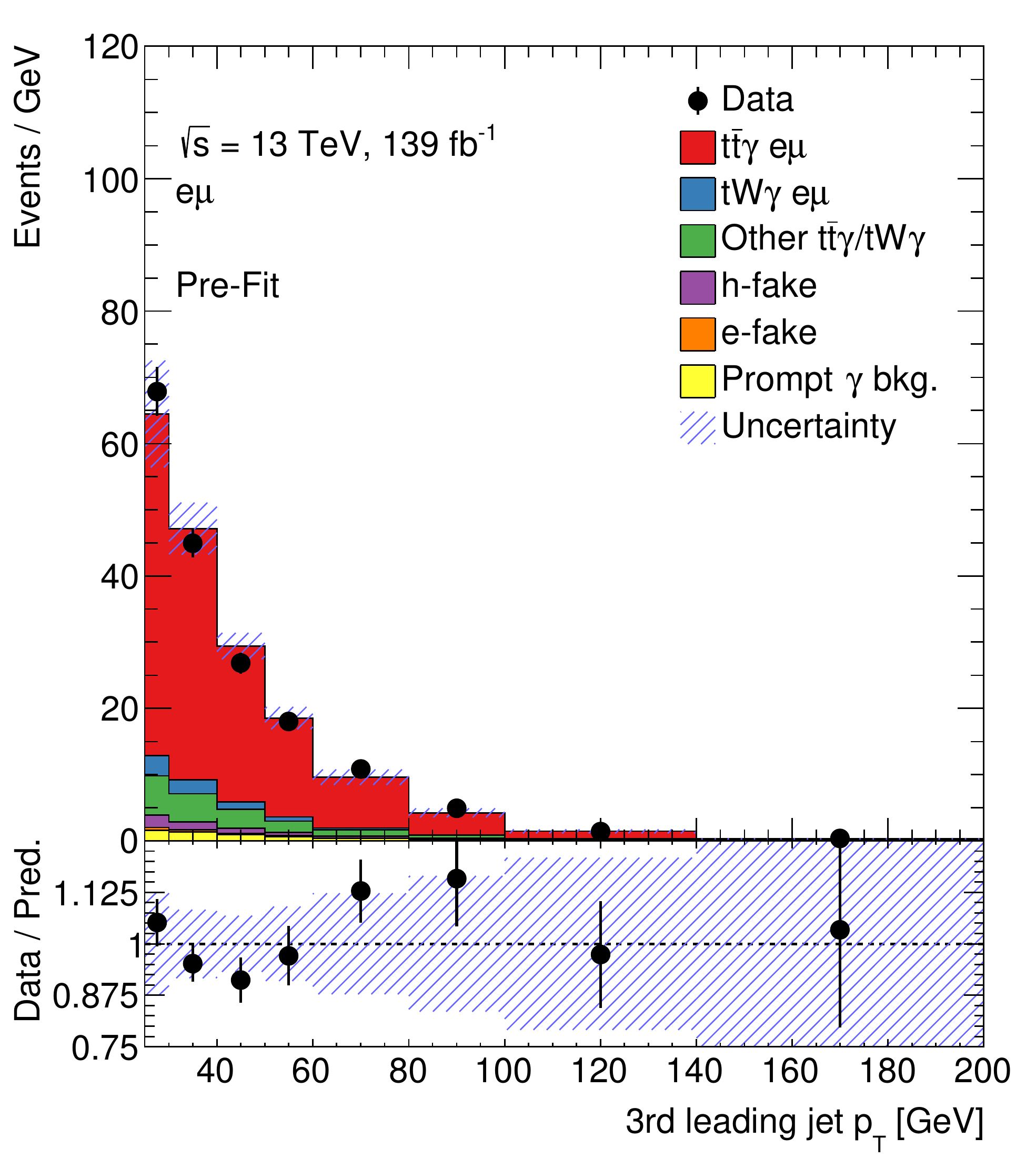}%
  \includegraphics[width=0.48\textwidth]{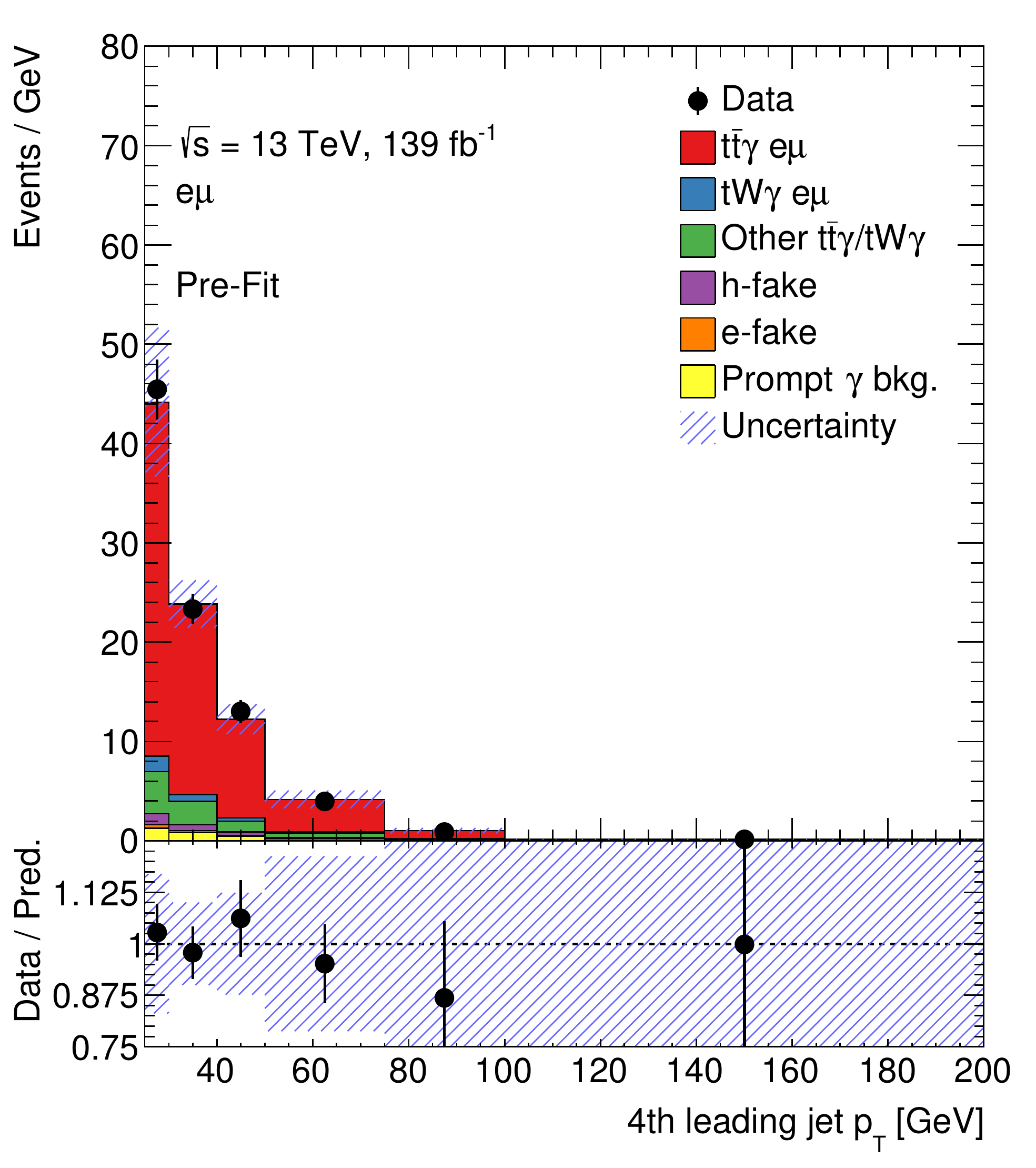}%
  \caption[Additional control plots with all uncertainties included (3)]{%
    Control plots for a data/\MC comparison in the \emu signal region with all statistical and systematic uncertainties included.
    Note that the predictions of the \tty and \tWy categories were scaled to match the numbers of reconstructed events in data.
    The shown observables are the transverse momenta of all jets of the event, and the transverse momenta of the second, third and fourth leading jets.
  }
  \label{fig:app-add-controlplots-3}
\end{figure}

\begin{figure}[p]
  \centering
  \includegraphics[width=0.48\textwidth]{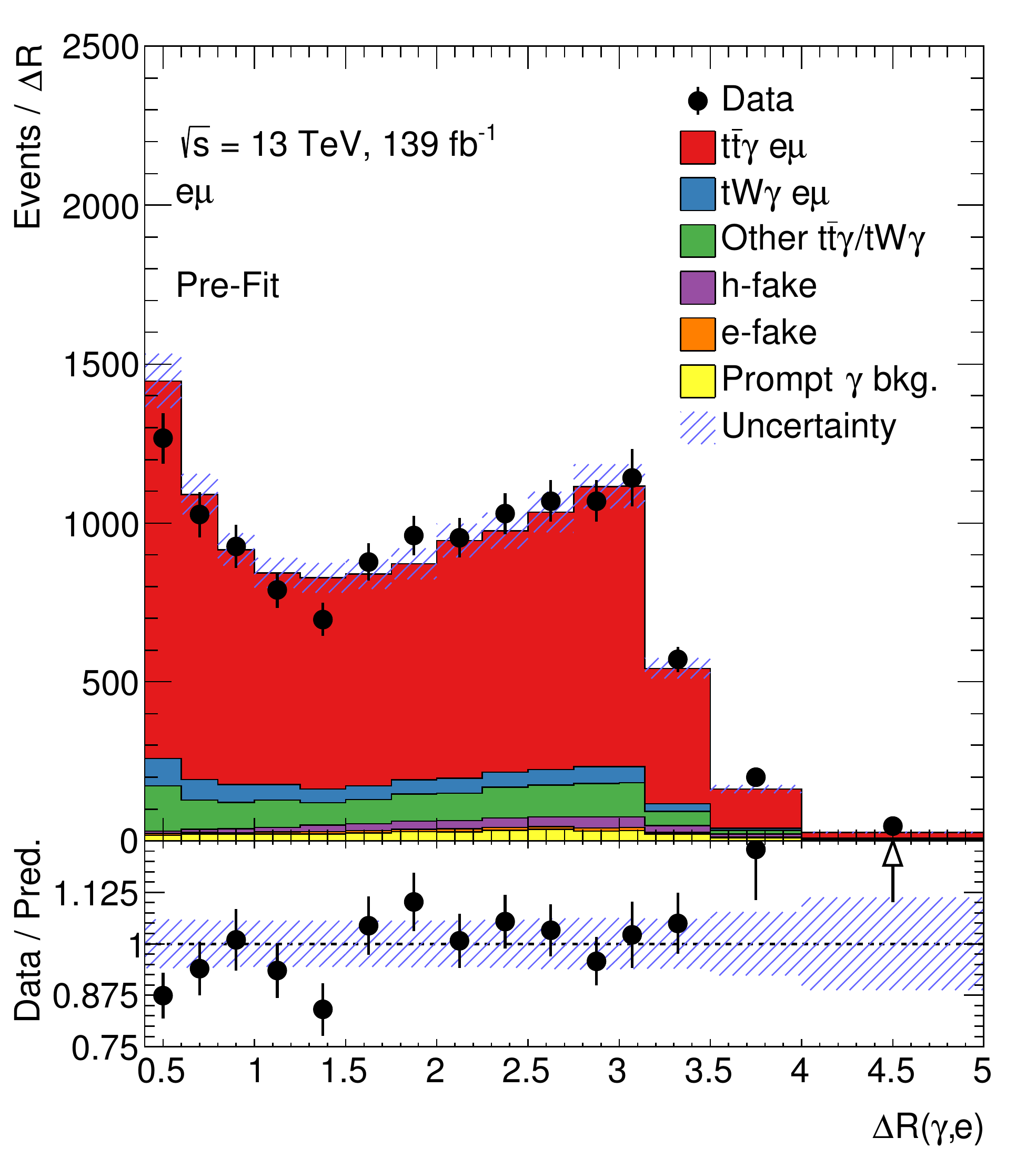}%
  \includegraphics[width=0.48\textwidth]{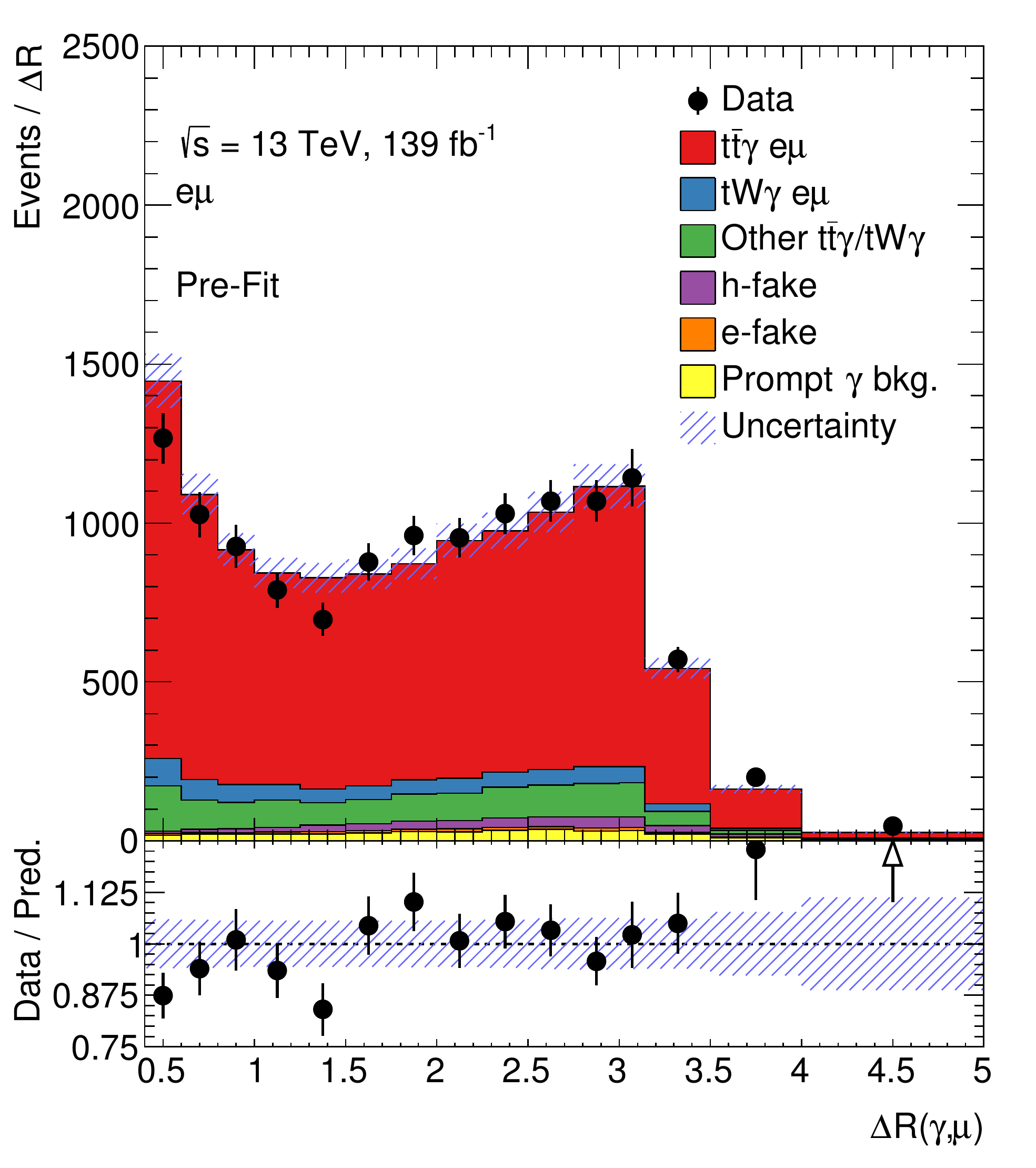}%
  \\
  \includegraphics[width=0.48\textwidth]{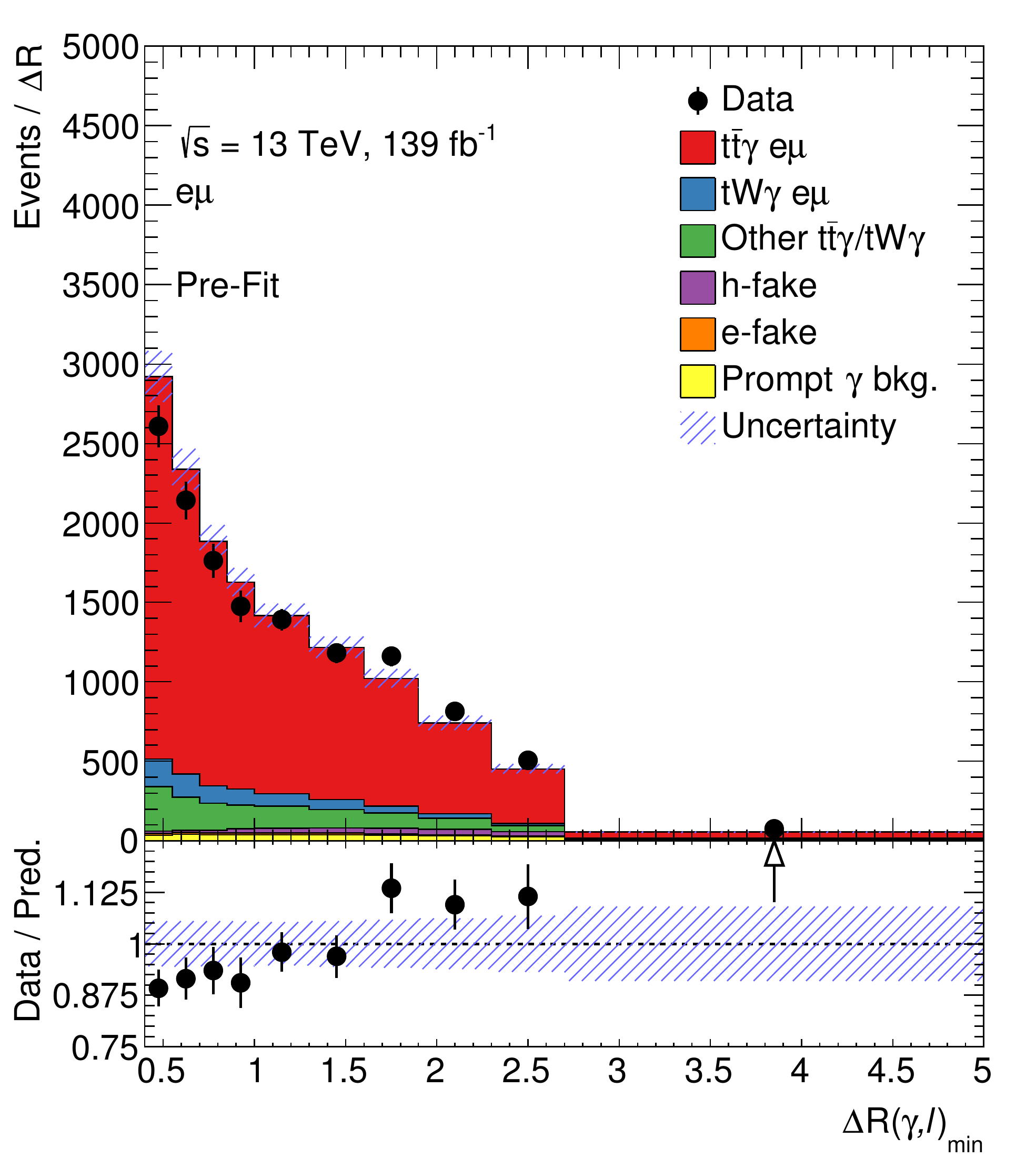}%
  \includegraphics[width=0.48\textwidth]{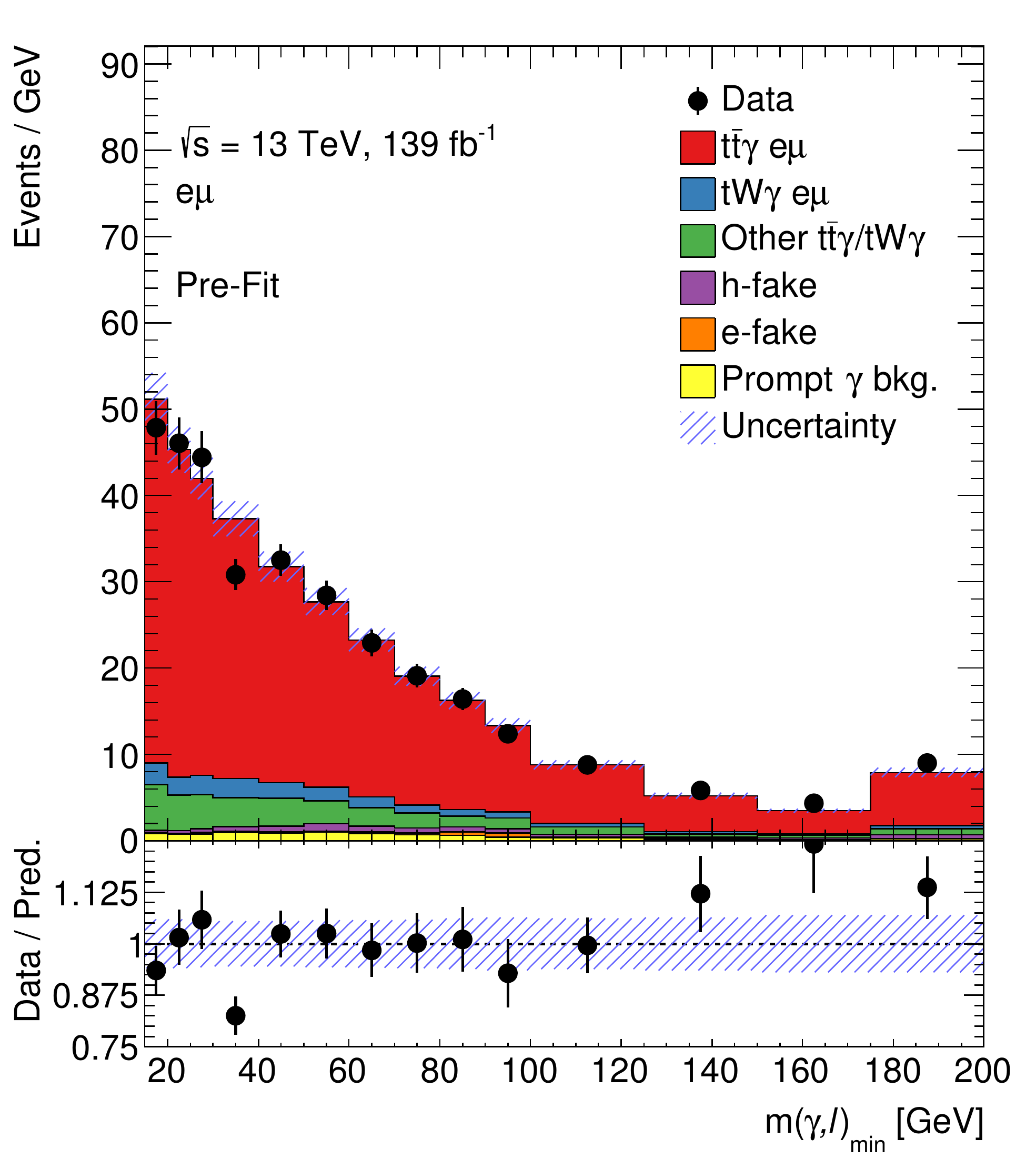}%
  \caption[Additional control plots with all uncertainties included (4)]{%
    Control plots for a data/\MC comparison in the \emu signal region with all statistical and systematic uncertainties included.
    Note that the predictions of the \tty and \tWy categories were scaled to match the numbers of reconstructed events in data.
    The shown observables are the distance $\Delta R$ between photon and electron/muon in the \etaphi plane, as well as the distance $\Delta R$ and invariant mass between the photon and its closest charged lepton.
  }
  \label{fig:app-add-controlplots-4}
\end{figure}

\begin{figure}[p]
  \centering
  \includegraphics[width=0.48\textwidth]{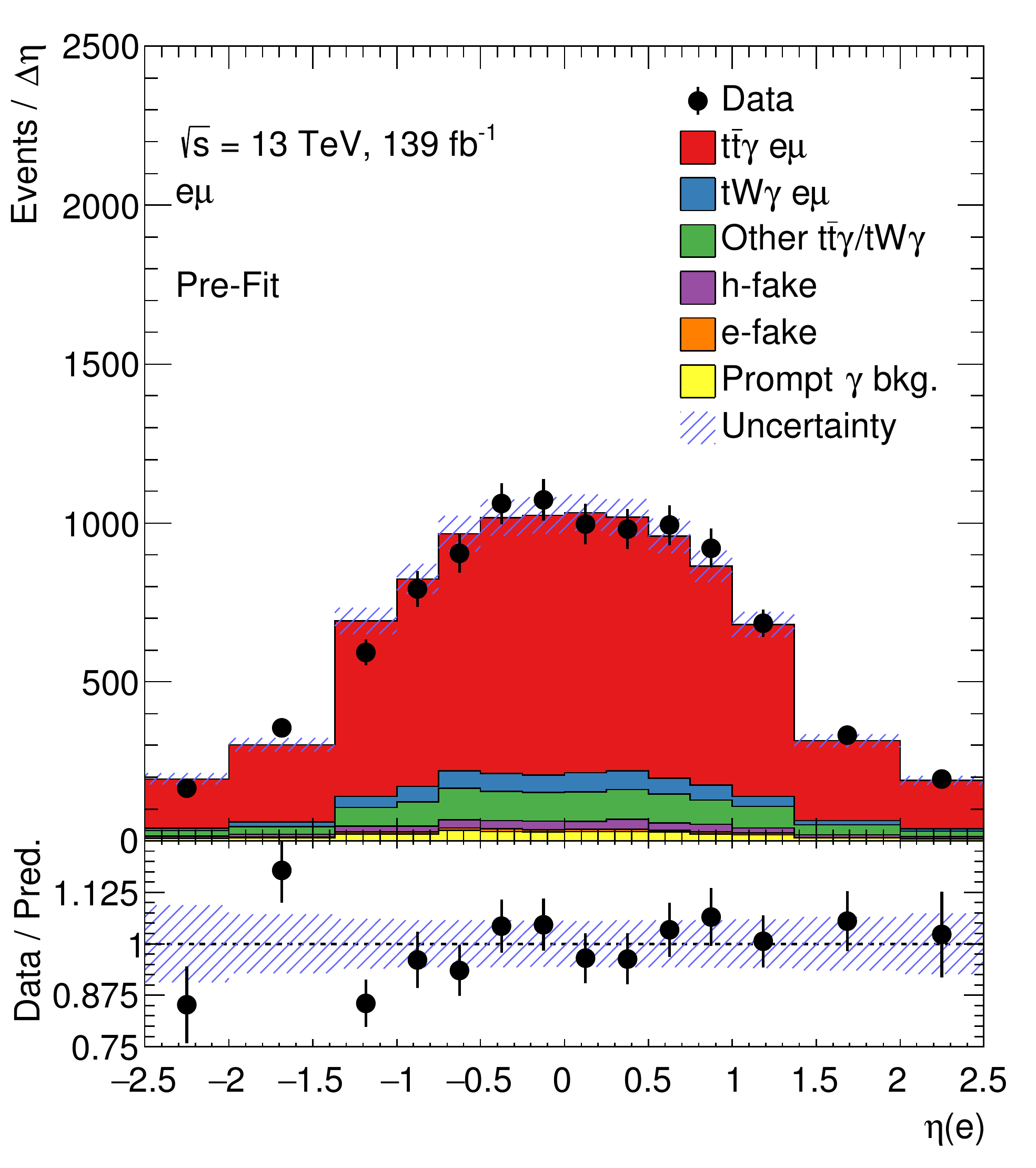}%
  \includegraphics[width=0.48\textwidth]{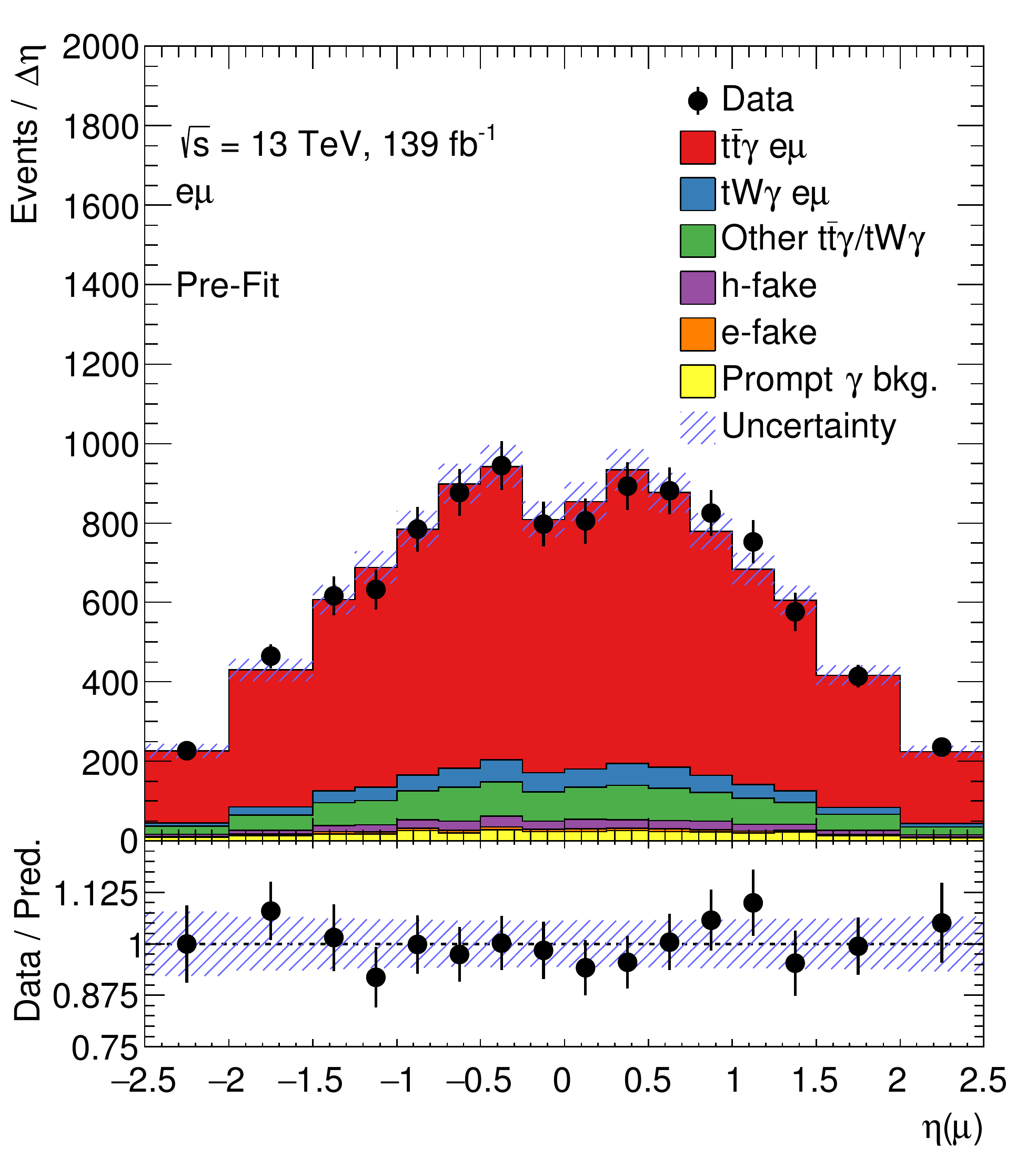}%
  \\
  \includegraphics[width=0.48\textwidth]{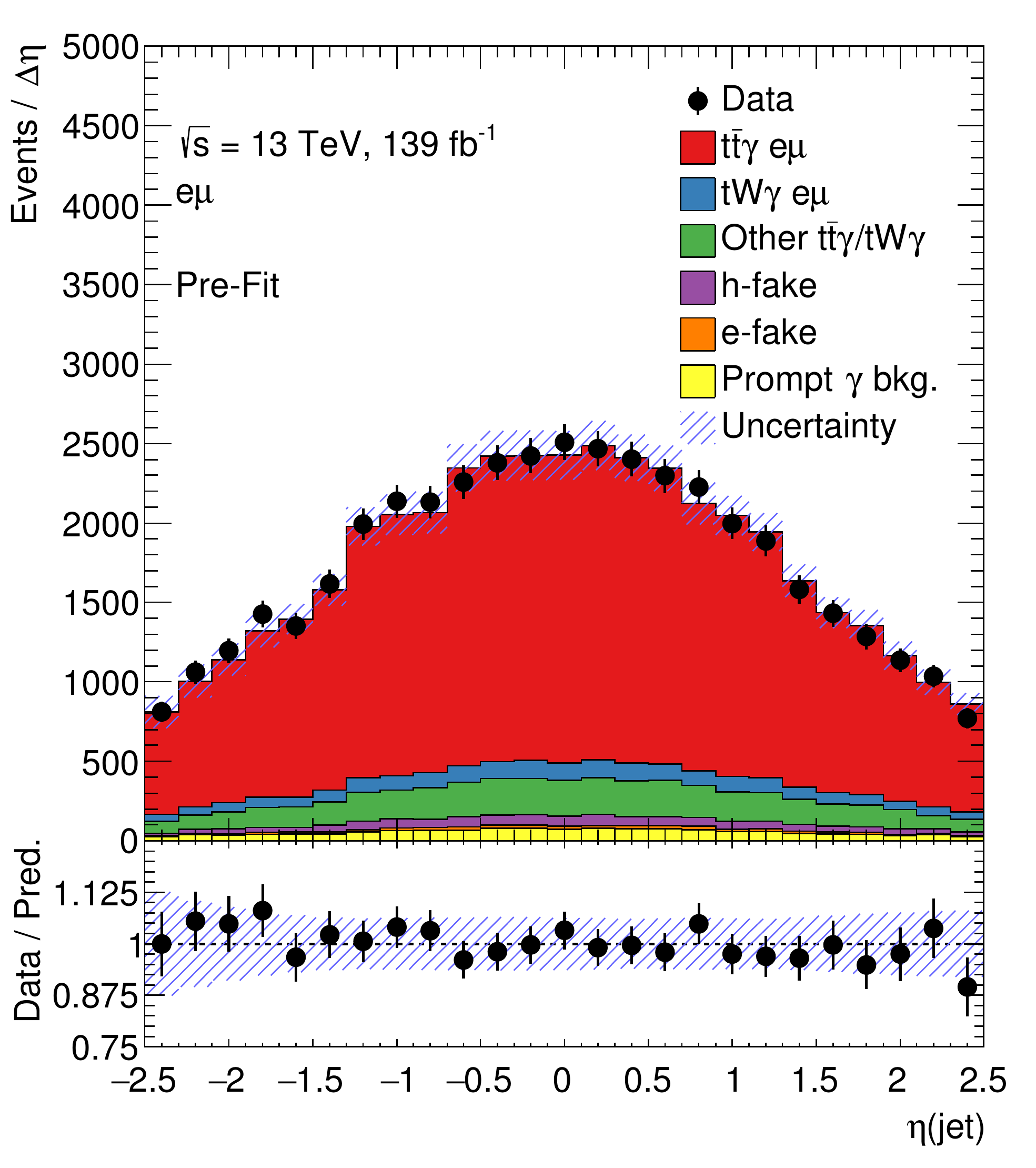}%
  \caption[Additional control plots with all uncertainties included (5)]{%
    Control plots for a data/\MC comparison in the \emu signal region with all statistical and systematic uncertainties included.
    Note that the predictions of the \tty and \tWy categories were scaled to match the numbers of reconstructed events in data.
    The shown observables are the pseudorapidities of the electron, of the muon, and of all jets of the event.
  }
  \label{fig:app-add-controlplots-5}
\end{figure}

\Cref{fig:app-add-controlplots-1,fig:app-add-controlplots-2,fig:app-add-controlplots-3,fig:app-add-controlplots-4,fig:app-add-controlplots-5} of this appendix show control plots for data/\MC comparison in addition to those presented in \cref{fig:results-controlplots-1,fig:results-controlplots-2} in the main body.
\Cref{fig:app-add-controlplots-1} shows the scalar sum \HT of all jet \pT and the \bjet multiplicity,
\cref{fig:app-add-controlplots-2} observables relating the charged leptons,
\cref{fig:app-add-controlplots-3} additional jet transverse momenta distributions,
\cref{fig:app-add-controlplots-4} observables relating the photon and the charged leptons,
\cref{fig:app-add-controlplots-5} additional pseudorapidity distributions.

\chapter{Fit results in the non-configured setup}
\label{chap:app-fit-decorrelation}

%
%

\Cref{sec:results-configuration} discusses issues with the \tty \emph{var3c} and \tty \PS model variations, and for the final results presented in \cref{sec:results-Asimov,sec:results-data}, the fit was configured to use the following setup:
the rate and shape components of the two modelling uncertainties were de-correlated due to strong pulls and constraints.
This appendix now provides some additional plots for the correlated fit scenario that was not chosen for the final fit setup.
Similarly to what is done in \cref{sec:results-Asimov}, the expected uncertainty for the parameter of interest can be calculated in an Asimov fit scenario. The obtained value is
\begin{align}
  \label{eq:app-fit-Asimov-mu}
  \mu
  = 1.000 \pm 0.023 \stat ^{+0.055}_{-0.051} \syst
  = 1.000 \, ^{+0.060}_{-0.056} \, ,
\end{align}
thus, the expected statistical uncertainties are identical to those predicted in \cref{eq:results-Asimov-mu}, but the expected systematic uncertainties are reduced by approximately \SI{15}{\percent} compared to the de-correlated scenario.
A fit to data yields a signal strength of
\begin{align}
  \label{eq:app-fit-data-mu}
  \mu
  = 1.357 \, ^{+0.027}_{-0.026} \stat ^{+0.064}_{-0.062} \syst
  = 1.357 \, ^{+0.069}_{-0.067} \, ,
\end{align}
corresponding to total relative uncertainties of $^{+\SI{5.2}{\percent}}_{-\SI{4.9}{\percent}}$.
Thus, removing the de-correlation leads to a down shift of the signal strength by \num{0.054} and a reduction of the total relative uncertainties by approximately \SI{15}{\percent}.
The resulting post-fit yields are listed in \cref{tab:app-fit-data-yield-table}, equivalent to \cref{tab:results-data-yield-table} in the main body for the de-correlated scenario.
Despite the down shift of the signal strength, which is reflected in the reduced post-fit \tWy yields, the post-fit \tty yields remain unchanged with respect to the scenario in the main body.
This is because the reduced signal strength is compensated by the strongly pulled \tty \PS model variation.
Translated into a fiducial inclusive \xsec value using \cref{eq:strategy-fidxsec-final}, this yields
\begin{align}
  \label{eq:app-fit-data-xsec}
  \mu^{\text{fid}} (\tty \to \emu)
  = 39.6 \pm 0.8 \stat ^{+2.2}_{-1.8} \syst \si{\fb}
  = 39.6 \, ^{+2.3}_{-2.0} \; \si{\fb} \, ,
\end{align}
corresponding to total relative uncertainties of $^{+\SI{5.8}{\percent}}_{-\SI{4.9}{\percent}}$.
Compared to the final result in the main body in \cref{eq:results-data-xsec}, the central value of the measured fiducial inclusive \xsec is \emph{unchanged}, but the systematic and total uncertainties are decreased by approximately \SI{15}{\percent}.
This is reassuring as the decisions made about the fit setup have no impact on the central values of the result, but only affect the uncertainty estimate.

\begin{table}
  \centering
  \caption[Predicted pre-fit and post-fit event yields (correlated)]{%
    Pre-fit and post-fit event yields for all \MC categories and numbers of reconstructed data events when the rate and shape components of the modelling uncertainties remain correlated -- \cf \cref{tab:results-data-yield-table} for the de-correlated scenario.
  }
  \label{tab:app-fit-data-yield-table}
  \begin{tabular}{%
    l
    S[table-format=4.0] @{${}\pm{}$} S[table-format=3.0]
    S[table-format=4.0] @{${}\pm{}$} S[table-format=2.0]
    }
    \toprule
    & \multicolumn{2}{c}{pre-fit} & \multicolumn{2}{c}{post-fit} \\
    \midrule
    \catttyemu{}* & 2390 & 130 & 2390 & 70 \\
    \cattWyemu{}* & 156  & 15  & 147  & 13 \\
    \catother{}*  & 279  & 15  & 275  & 10 \\
    \cathfake     & 80   & 40  & 80   & 40 \\
    \catefake     & 23   & 12  & 23   & 11 \\
    \catprompt    & 90   & 40  & 100  & 40 \\
    \midrule
    Total \MC     & 3010 & 160 & 3010 & 60 \\
    \midrule
    Data         & \multicolumn{2}{l}{3014} & \multicolumn{2}{l}{3014} \\
    \bottomrule
  \end{tabular}
\end{table}

\begin{figure}
  \centering
  \includegraphics[width=0.70\textwidth]{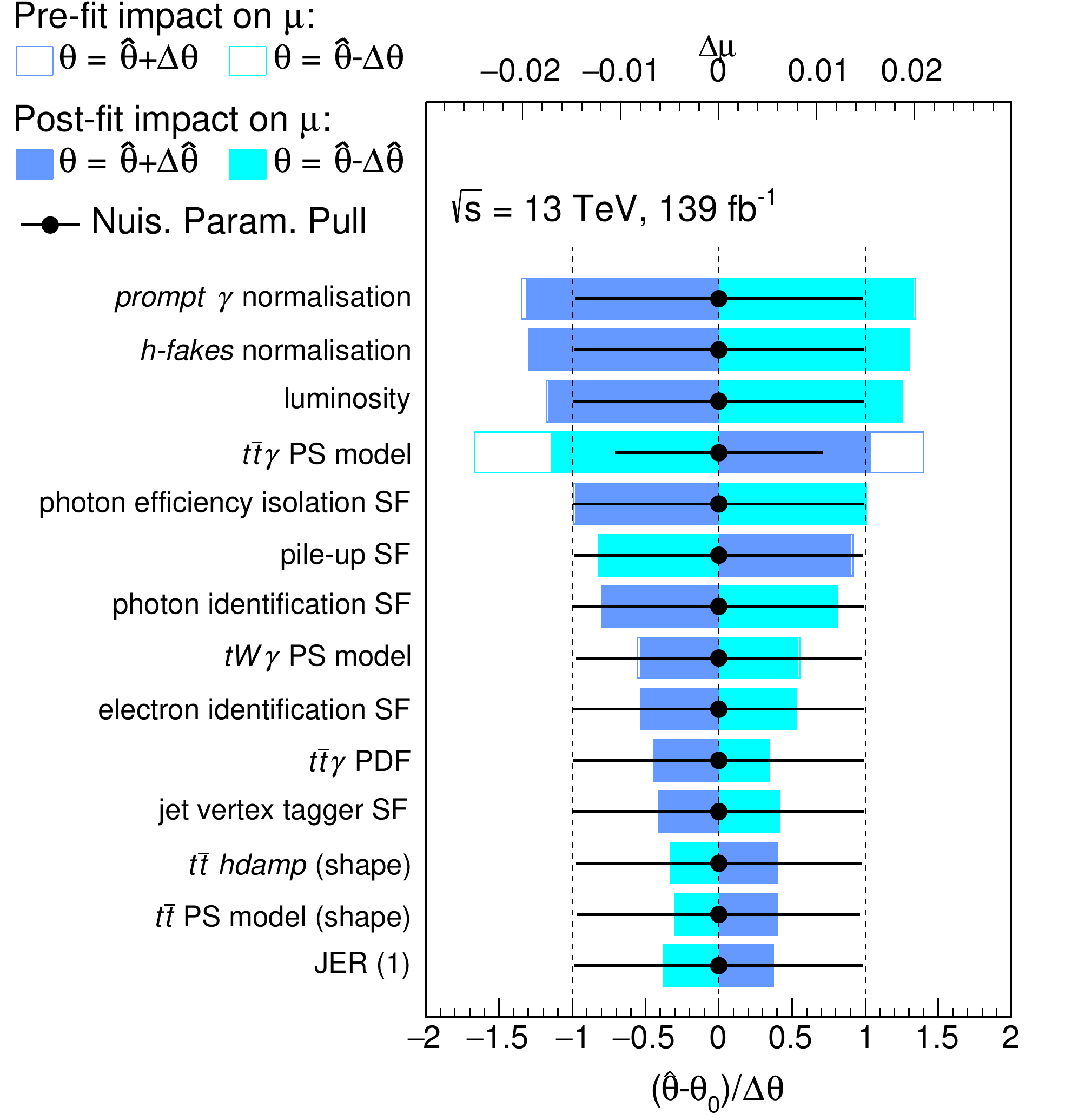}
  \caption[Ranking of nuisance parameters in Asimov fit (correlated)]{%
    Nuisance parameters ranked according to their impact on the parameter of interest in the fit to Asimov pseudo-data when the rate and shape components of the modelling uncertainties remain correlated.
  }
  \label{fig:app-fit-Asimov-ranking}
\end{figure}

\begin{figure}
  \centering
  \includegraphics[width=0.70\textwidth]{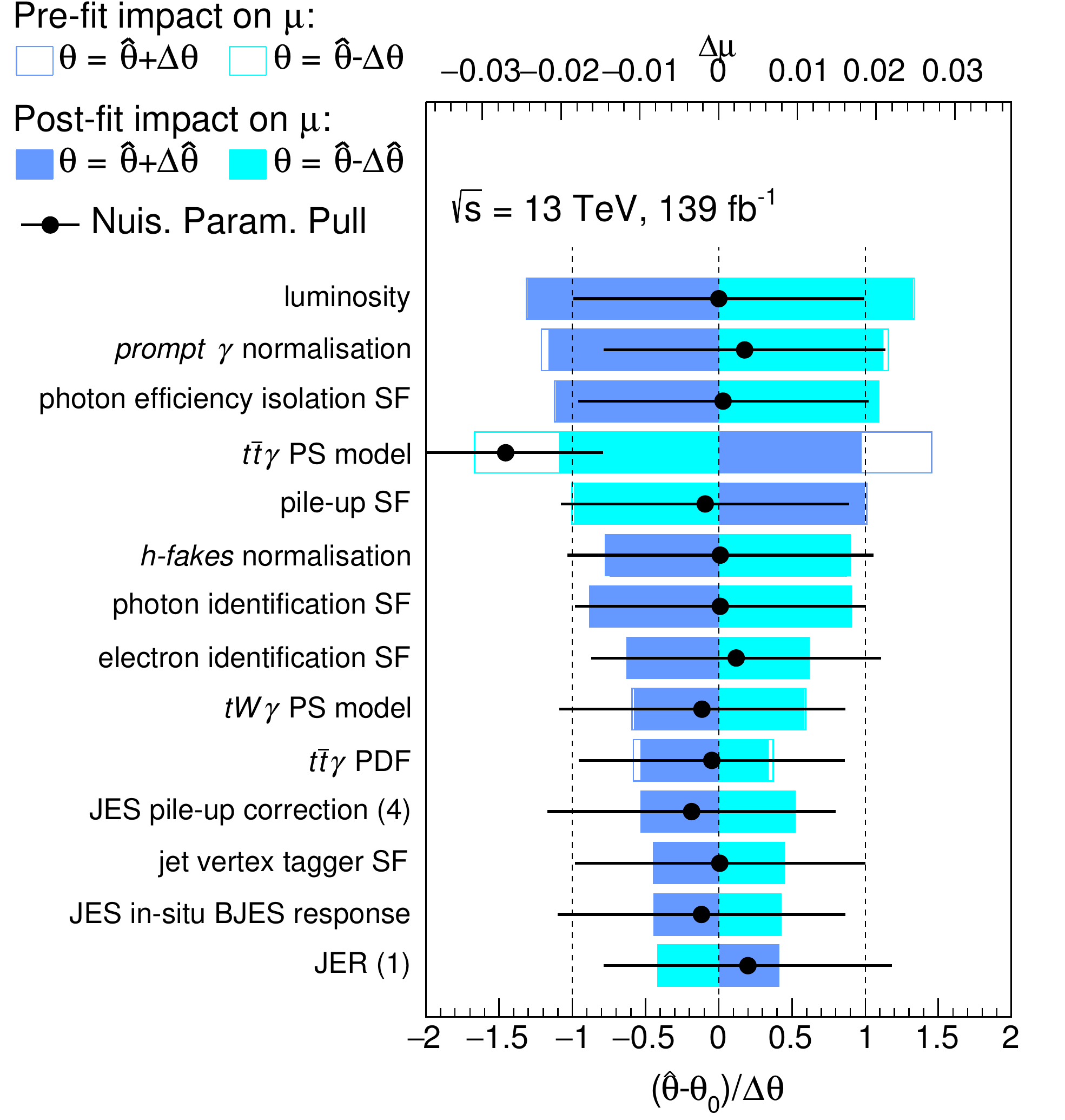}
  \caption[Ranking of nuisance parameters in the fit to data (correlated)]{%
    Nuisance parameters ranked according to their impact on the parameter of interest in the fit to data when the rate and shape components of the modelling uncertainties remain correlated.
  }
  \label{fig:app-fit-data-ranking}
\end{figure}

\Cref{tab:app-fit-Asimov-constraints} lists the nuisance parameters strongly constrained in a fit to Asimov pseudo-data.
The corresponding values in the main body are in \cref{tab:results_fit_Asimov_constraints}.
\Cref{fig:app-fit-Asimov-ranking} shows those nuisance parameters with the highest expected impact on the fit result in an Asimov fit scenario.
They are ranked according to their post-fit impact, \cf \cref{fig:results-Asimov-ranking} in the main body.
\Cref{tab:app-fit-data-constraints} lists the nuisance parameters strongly pulled and constrained in the fit to \ATLAS data, \cf \cref{tab:fit_data_pulls_constraints} in the main body.
\Cref{fig:app-fit-data-ranking} shows those nuisance parameters with the highest impact on the parameter of interest in the fit to \ATLAS data.
They are ranked according to their post-fit impact, \cf \cref{fig:results-data-ranking} in the main body for the identical plot in the de-correlated scenario.

\begin{table}
  \centering
  \caption[List of constrained nuisance parameters in Asimov fit (correlated)]{%
    Constrained nuisance parameters in a fit to Asimov pseudo-data, when the rate and shape components of the modelling uncertainties remain correlated -- \cf \cref{tab:results_fit_Asimov_constraints} for the de-correlated scenario.
    }
  \label{tab:app-fit-Asimov-constraints}
  \sisetup{round-precision=1,round-mode=places}
  \begin{tabular}{l S[table-format=2.1]}
    \toprule
    \multirow{2}{*}{Nuisance parameter} & {constraint} \\
     & {$\Delta \hat{\theta} / \Delta \theta$ [\si{\percent}]} \\
    \midrule
    \tty \emph{var3c} & 68.1677 \\
    \tty \PS model    & 70.8728 \\
    \bottomrule
  \end{tabular}
\end{table}

The ranking plots show that the rate component of the \PS model, \ie the highest-ranked nuisance parameter in the fit results in the main body, has a weaker impact on the result when correlated with its shape information.
It is only ranked the fourth-highest nuisance parameter in this section.
In values, the post-fit impact of the rate-only component in the main body was calculated to be $\Delta\mu = ^{+0.045}_{-0.043}$ for the fit to data.
The post-fit impact of the combined rate and shape uncertainty in this section is $\Delta\mu = ^{+0.018}_{-0.020}$ in the data fit.
This is the reason for the decreased relative uncertainties of the result presented here.

However, this scenario puts a lot more importance on the shape information of the \tty \PS~model variation as it remains correlated with the rate.
Through comparison with the de-correlated scenario it becomes clear that the shape component is what causes the large pull and constraint of the combined nuisance parameter.
Looking at the template distributions in \cref{fig:results-problematic-NPs}, it was discussed that the large pull of the \tty \PS model is due to discrepancies between data and nominal \MC prediction.
In the de-correlated scenario of the main body, this is compensated by the shape-only nuisance parameter -- a parameter that has negligible impact on the parameter of interest and is not ranked among the top twenty high-impact parameters.
Its post-fit impact is calculated to be $|\Delta\mu| < 0.004$.
In the correlated scenario, however, the discrepancy is compensated for by the combined rate-and-shape nuisance parameter, which has a major impact on the parameter of interest.
The compensation of this discrepancy through the \tty \PS model shape component is expected to only be an artefact of the model, \ie it is simply that template in the fit model which compensates these discrepancies \enquote{best}.
Hence, it is much preferable to have this compensation done by a lowly ranked nuisance parameter that has little influence on the final result, although this increases the overall uncertainties of the final result.

\begin{table}
  \centering
  \caption[List of pulled and constrained nuisance parameters in the fit to data (correlated)]{%
    Pulled and constrained nuisance parameters in the fit to data, when the rate and shape components of the modelling uncertainties remain correlated -- \cf \cref{tab:fit_data_pulls_constraints} for the de-correlated scenario.
  }
  \label{tab:app-fit-data-constraints}
  \sisetup{round-precision=2,round-mode=places}
  \begin{tabular}{
    l
    S[table-format=1.2, table-sign-mantissa]
    S[table-format=2.1, round-precision=1]
    }
    \toprule
    \multirow{2}{*}{Nuisance parameter} & {pull value} & {constraint} \\
     & {$( \hat{\theta} - \theta_0 ) / \Delta \theta$} & {$\Delta \hat{\theta} / \Delta \theta$ [\si{\percent}]} \\
    \midrule
    \tty \emph{var3c}        & -0.41089 & 64.0305 \\
    \tty \PS model           & -1.45585 & 66.4444 \\
    \tty \PDF                & {---}    & 90.6724 \\
    \ttbar \PS model (shape) & 0.415503 & {---}   \\
    \bottomrule
  \end{tabular}
\end{table}

\chapter{Additional unfolded results}
\label{chap:app-diff-xsec}

\begin{figure}[p]
  \centering
  \includegraphics[width=0.48\textwidth, clip, trim=5pt 0 5pt 0]{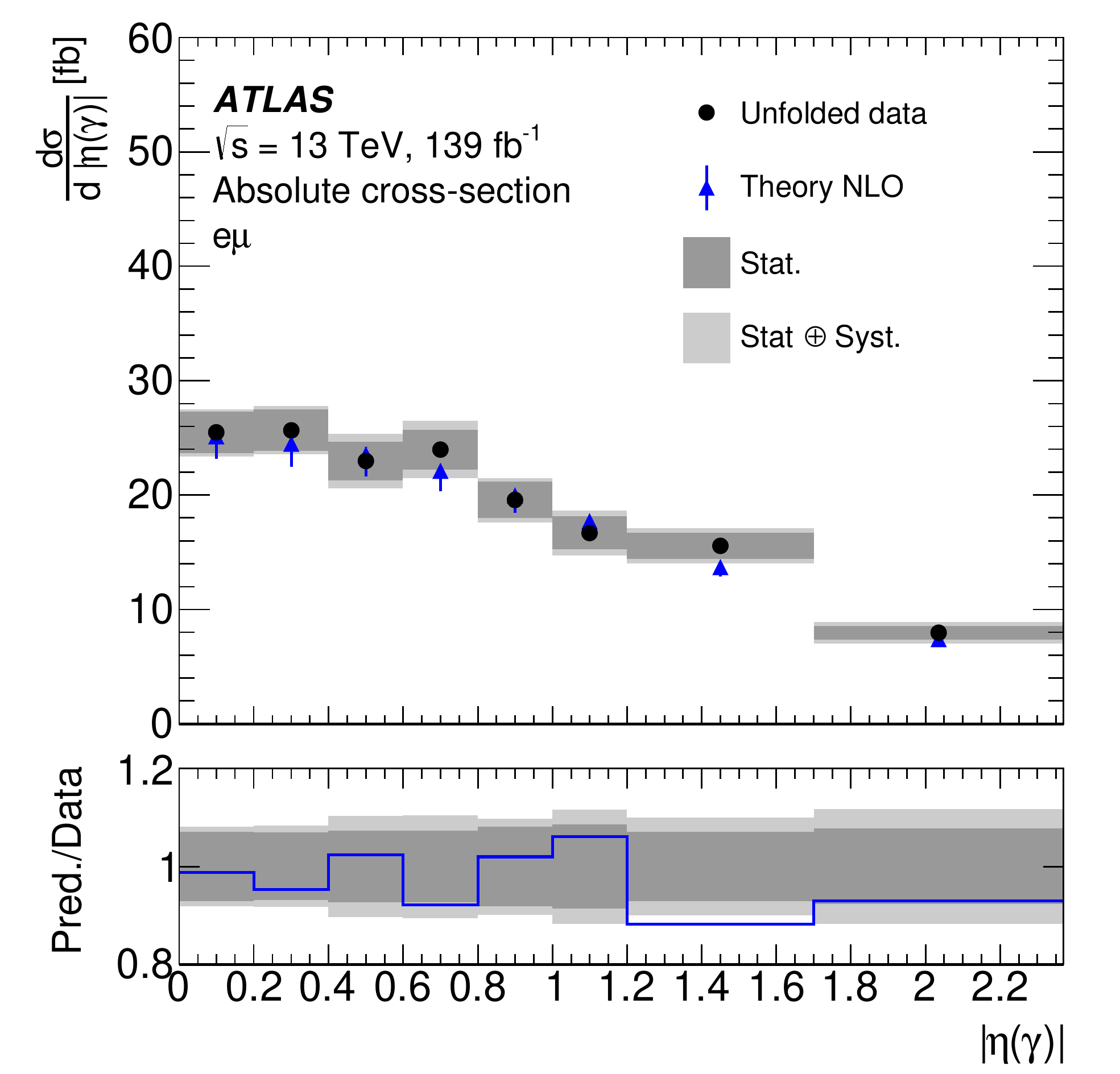}
  \\
  \includegraphics[width=0.48\textwidth, clip, trim=5pt 0 5pt 0]{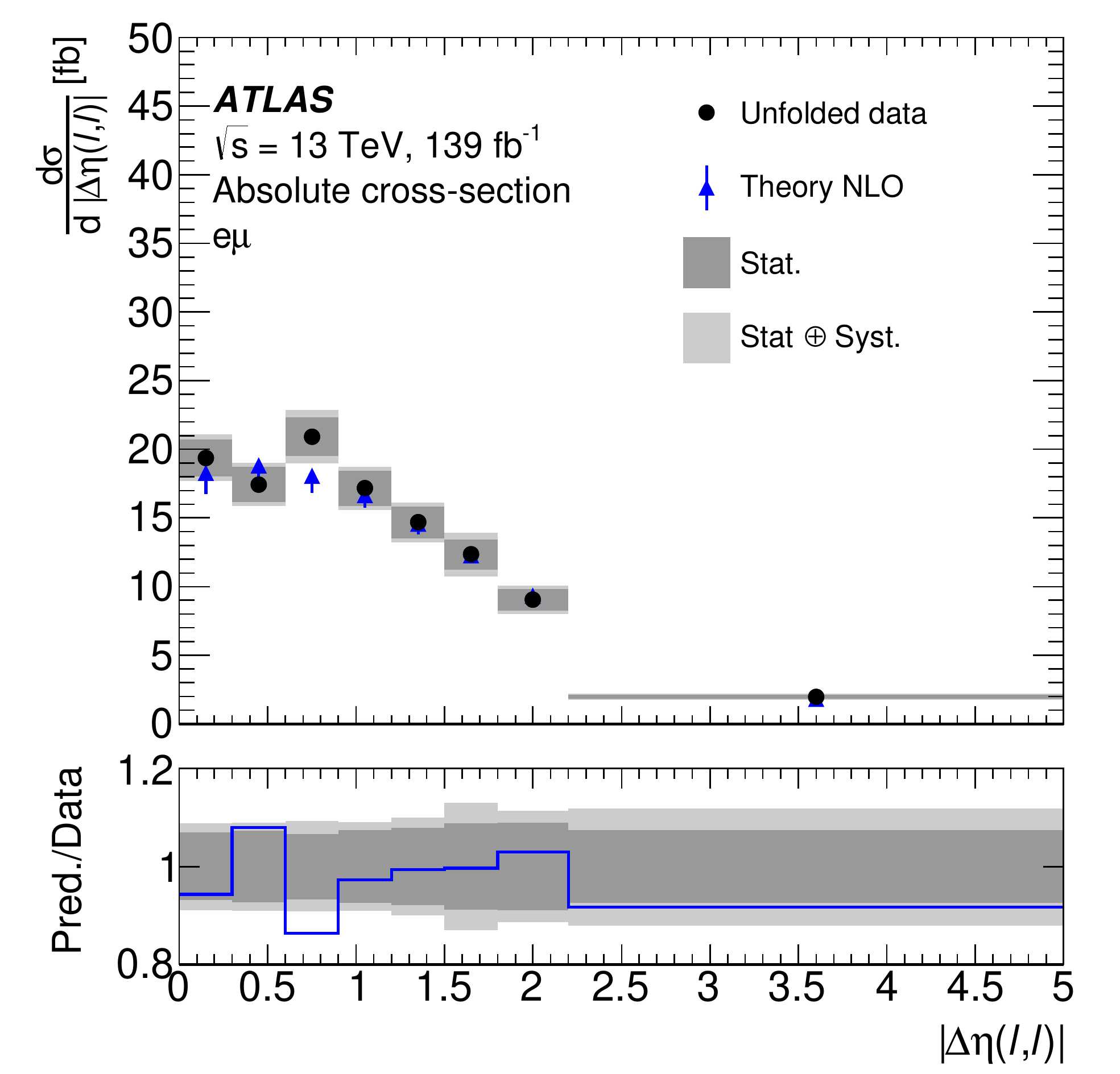}
  \includegraphics[width=0.48\textwidth, clip, trim=5pt 0 5pt 0]{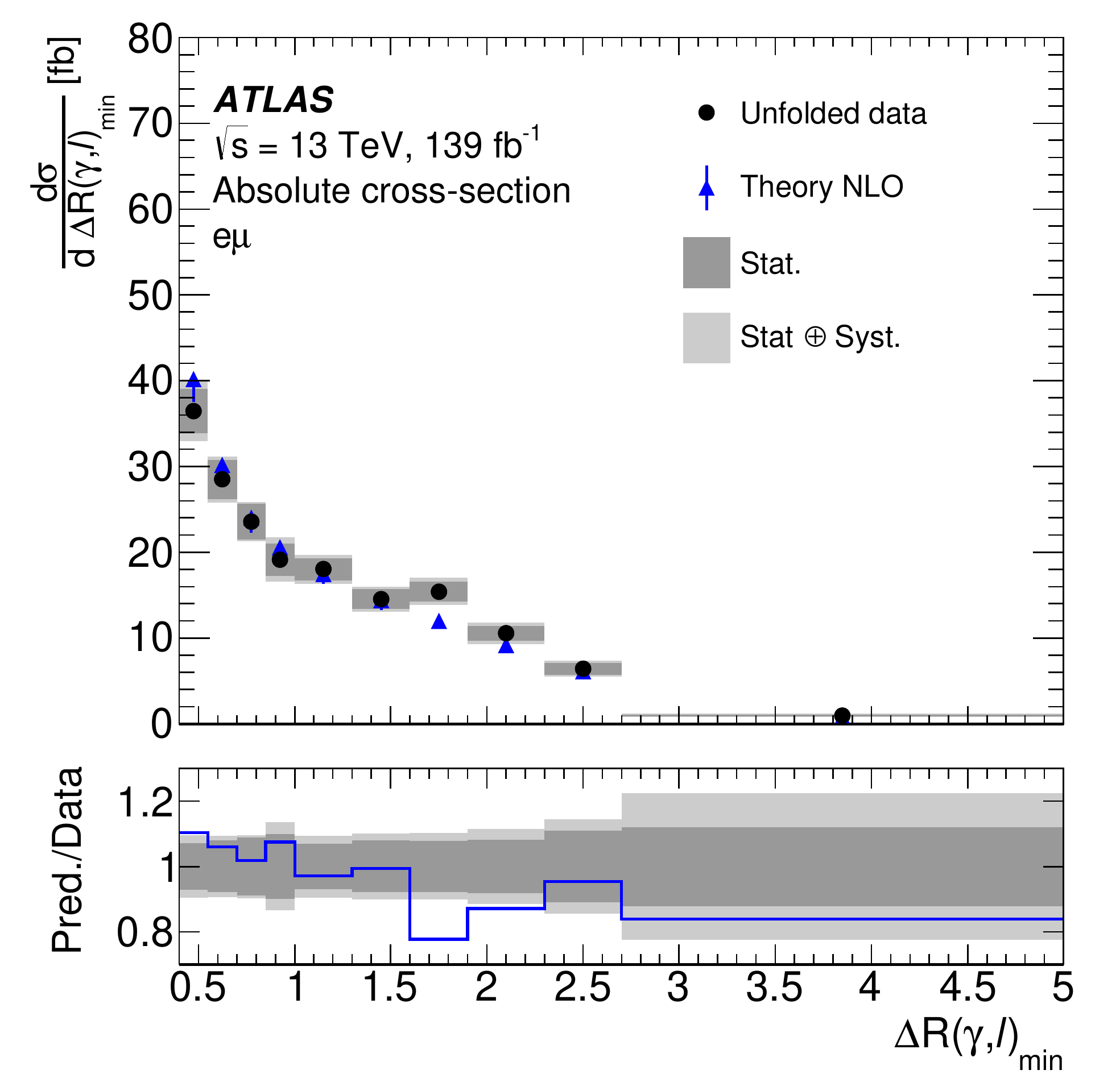}
  \caption[Absolute differential \xsecs for unshown observables]{%
    Differential \xsecs as functions of the photon absolute pseudorapidity, $\left| \eta(\gamma) \right|$, the absolute difference in pseudorapidities of the two charged leptons, \Detall, and the distance between the photon and the closer of the two charged leptons in the \etaphi plane, \DRlph.
    The plots show the absolute bin-by-bin \xsec values obtained from unfolding \ATLAS data and from the fixed-order theory computation~\cite{Bevilacqua:2018woc,Bevilacqua:2018dny}.
    Figures taken from Ref.~\cite{TOPQ-2020-03}.
  }
  \label{fig:app-diff-xsec-abs}
\end{figure}

\begin{figure}[p]
  \centering
  \includegraphics[width=0.48\textwidth, clip, trim=5pt 0 5pt 0]{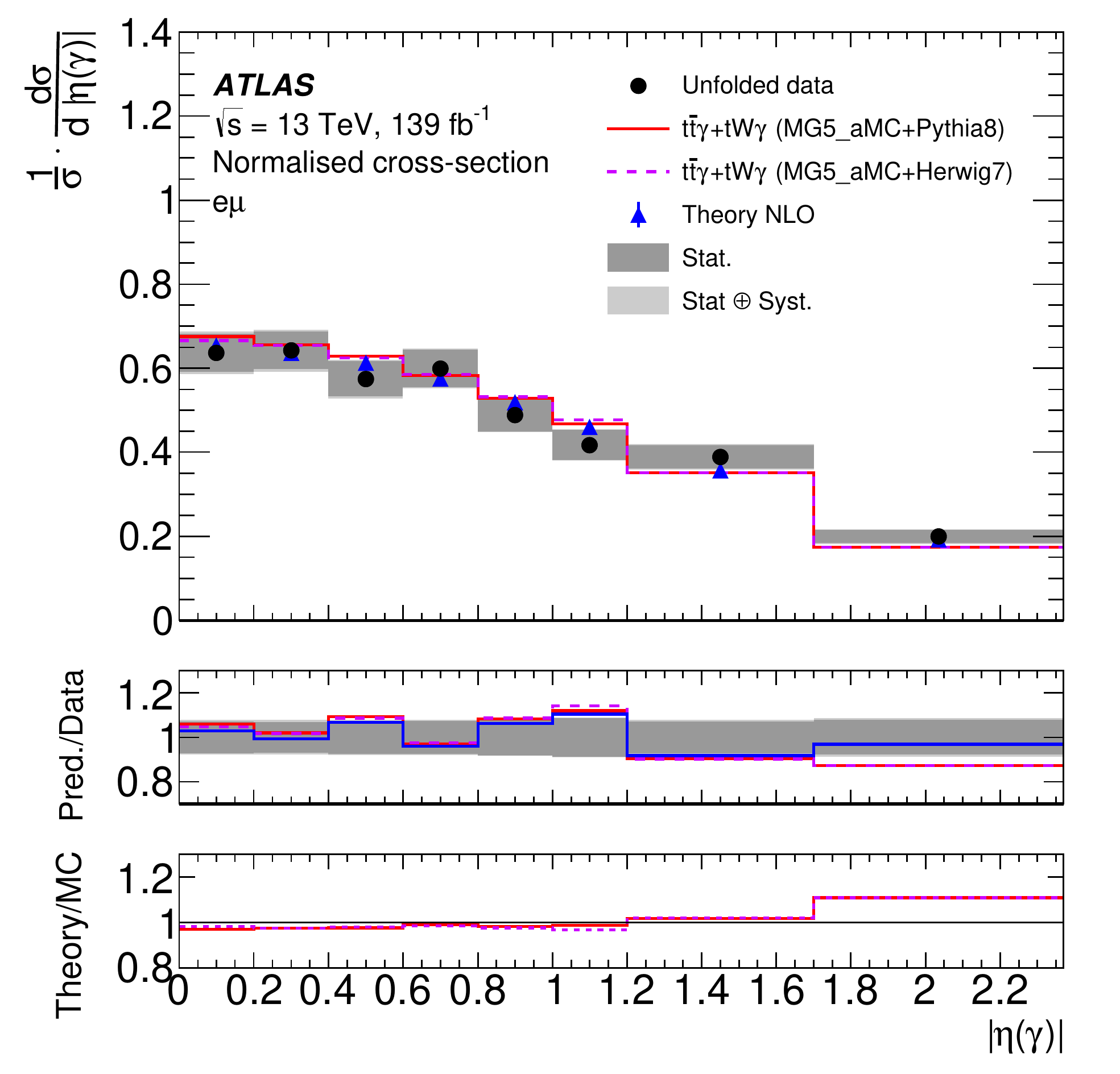}
  \\
  \includegraphics[width=0.48\textwidth, clip, trim=5pt 0 5pt 0]{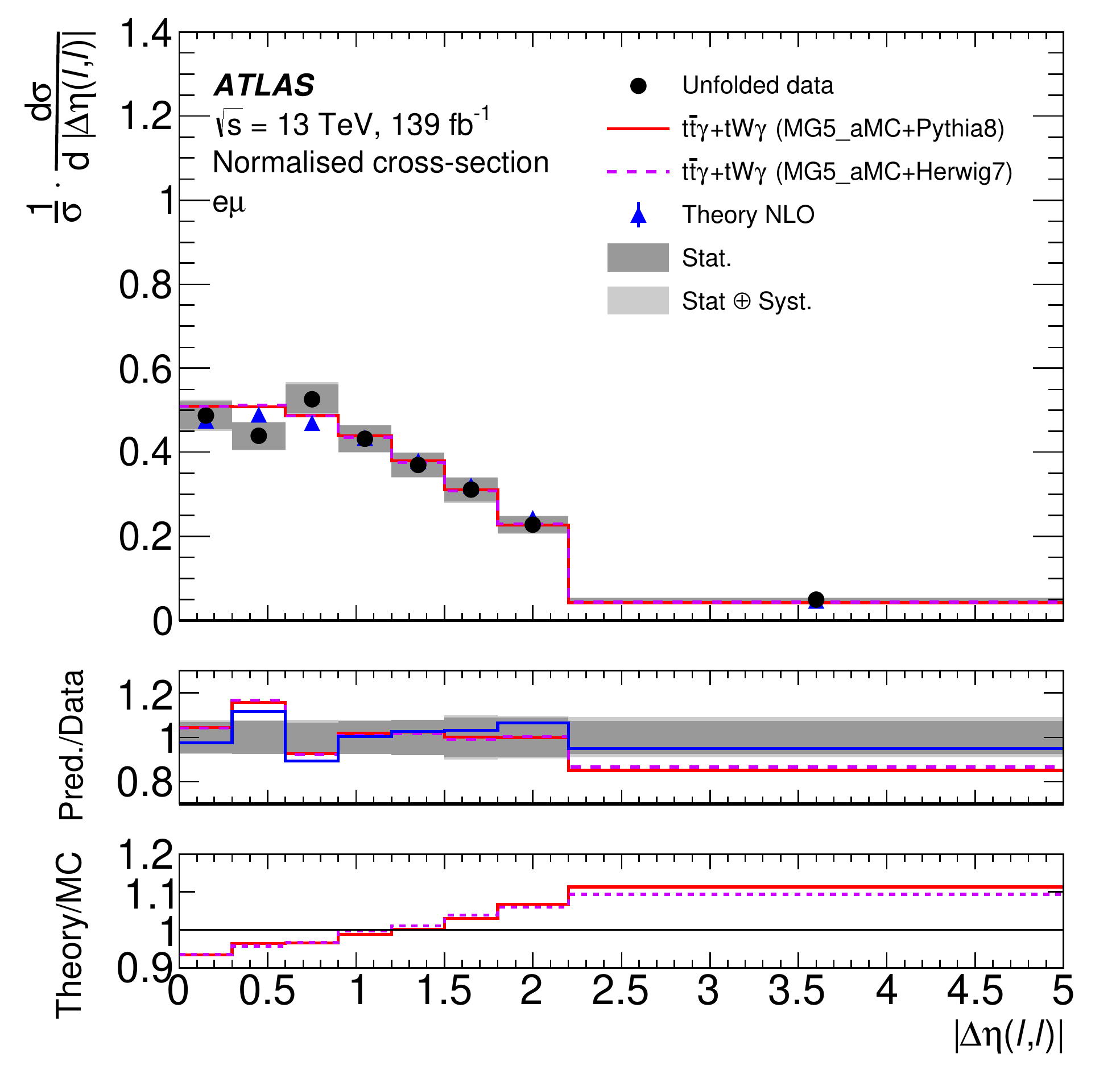}
  \includegraphics[width=0.48\textwidth, clip, trim=5pt 0 5pt 0]{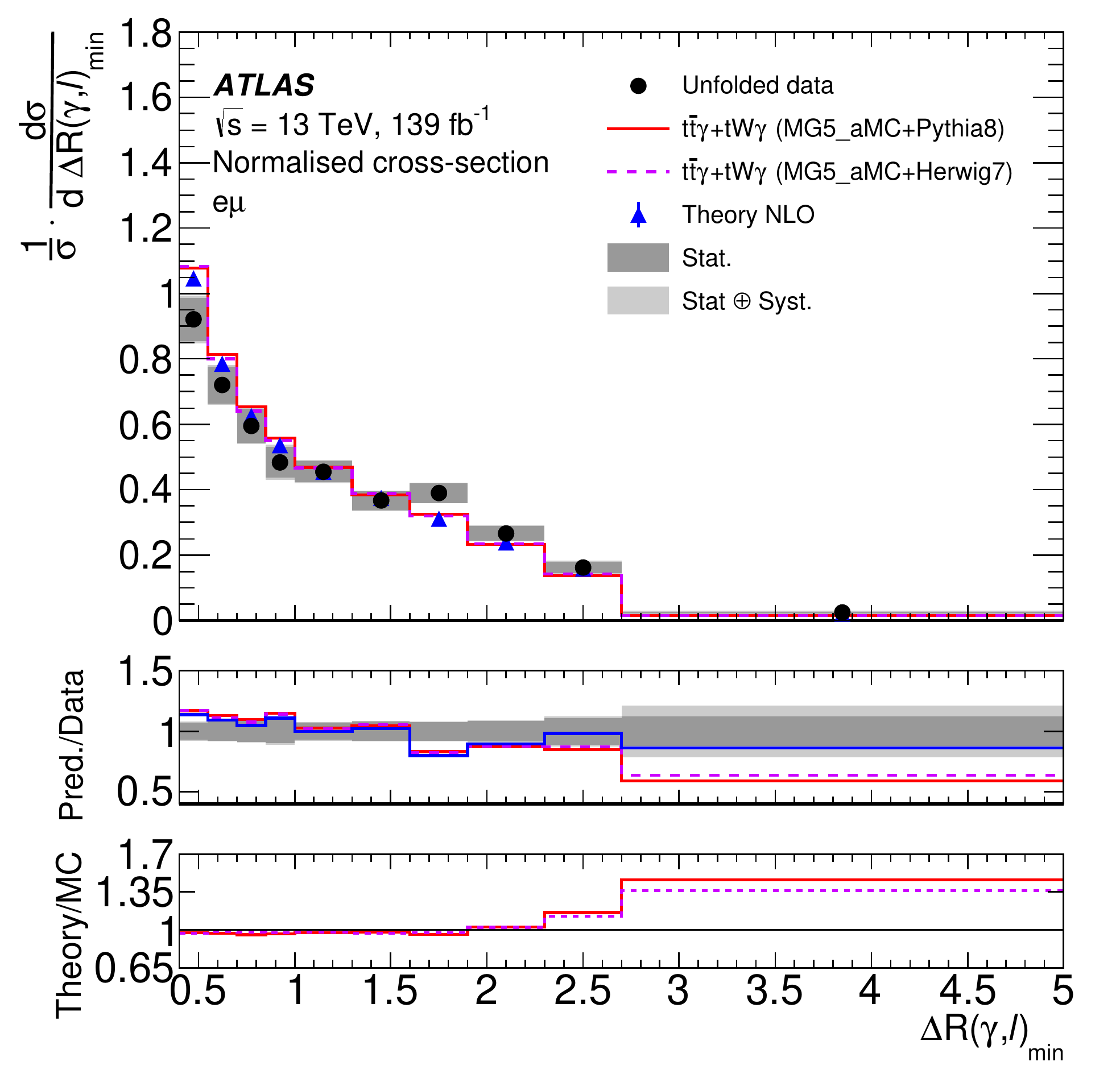}
  \caption[Normalised differential \xsecs for unshown observables]{%
    Differential \xsecs as functions of the photon absolute pseudorapidity, $\left| \eta(\gamma) \right|$, the absolute difference in pseudorapidities of the two charged leptons, \Detall, and the distance between the photon and the closer of the two charged leptons in the \etaphi plane, \DRlph.
    The plots show the bin-by-bin \xsec values normalised to the measured integrals of the distributions.
    Apart from unfolded \ATLAS data, the plots contain the values predicted by the fixed-order theory computation~\cite{Bevilacqua:2018woc,Bevilacqua:2018dny} as well as those of the \LOPS simulations of the \tty and \tWy processes used in this analysis: \Madgraph interfaced to \Pythia and to \Herwig.
    Figures taken from Ref.~\cite{TOPQ-2020-03}.
  }
  \label{fig:app-diff-xsec-norm}
\end{figure}

\begin{figure}[p]
  \centering
  \includegraphics[width=0.48\textwidth, clip, trim=5pt 25pt 5pt 10pt]{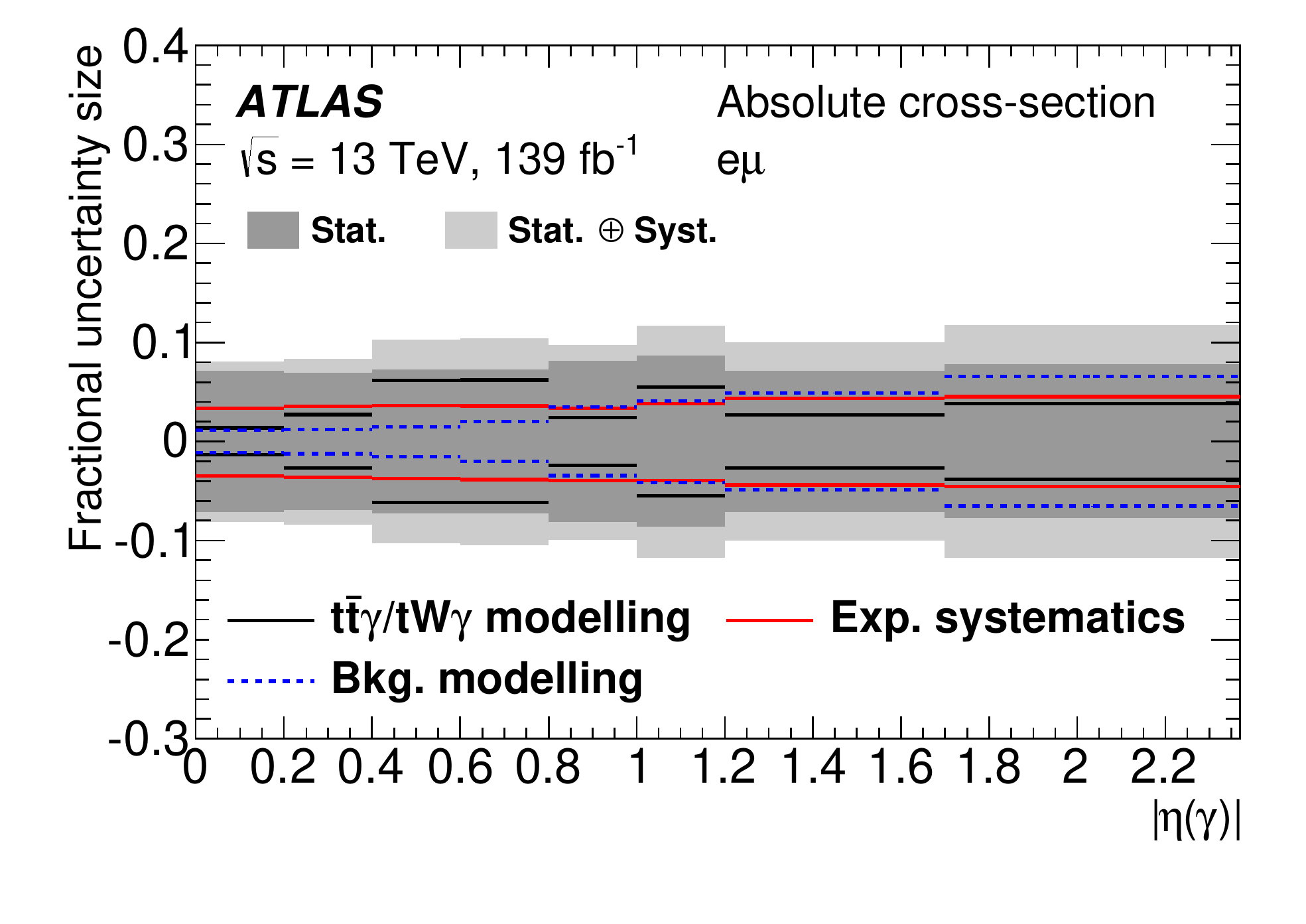}
  \includegraphics[width=0.48\textwidth, clip, trim=5pt 25pt 5pt 10pt]{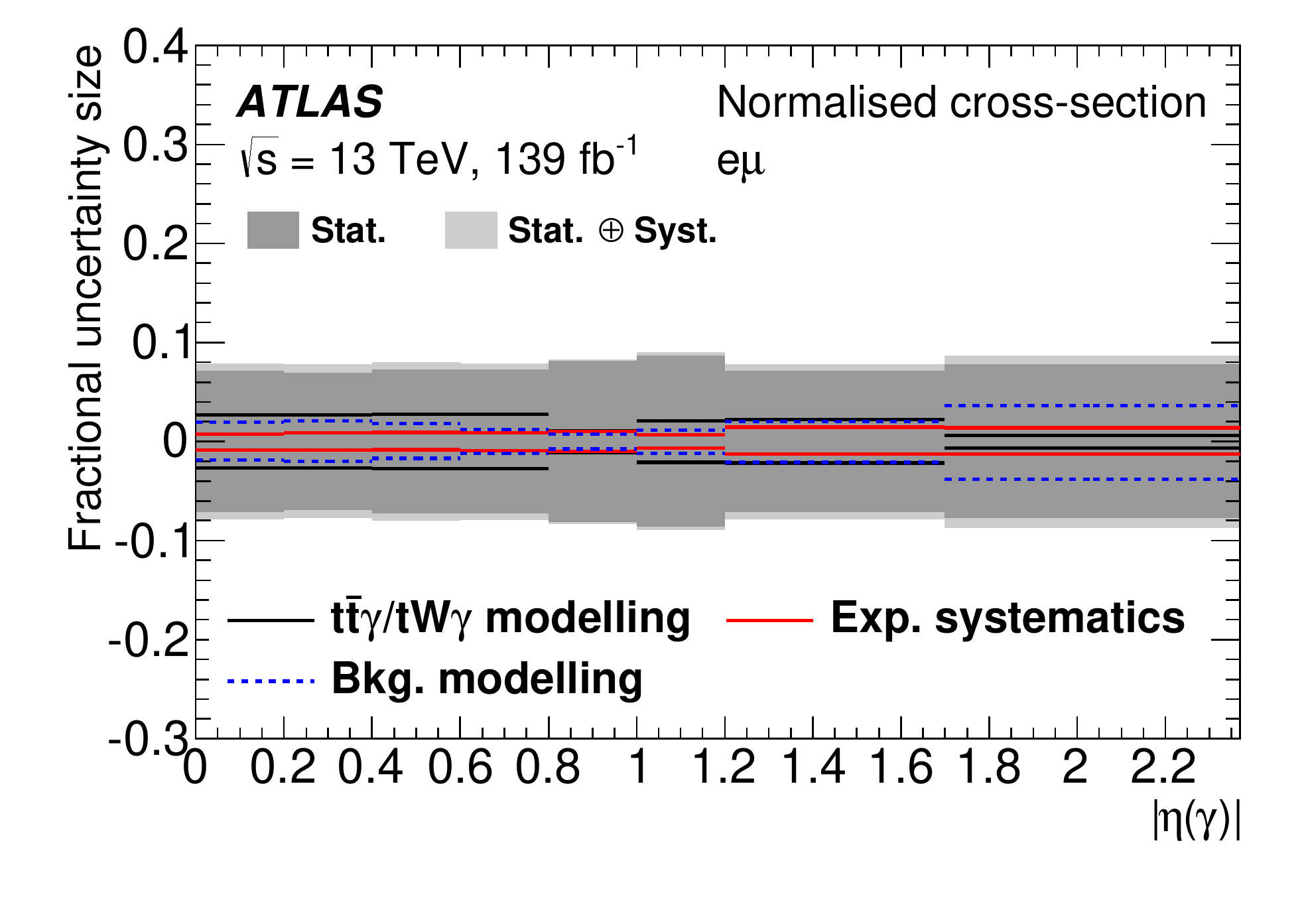}
  \\
  \includegraphics[width=0.48\textwidth, clip, trim=5pt 25pt 5pt 10pt]{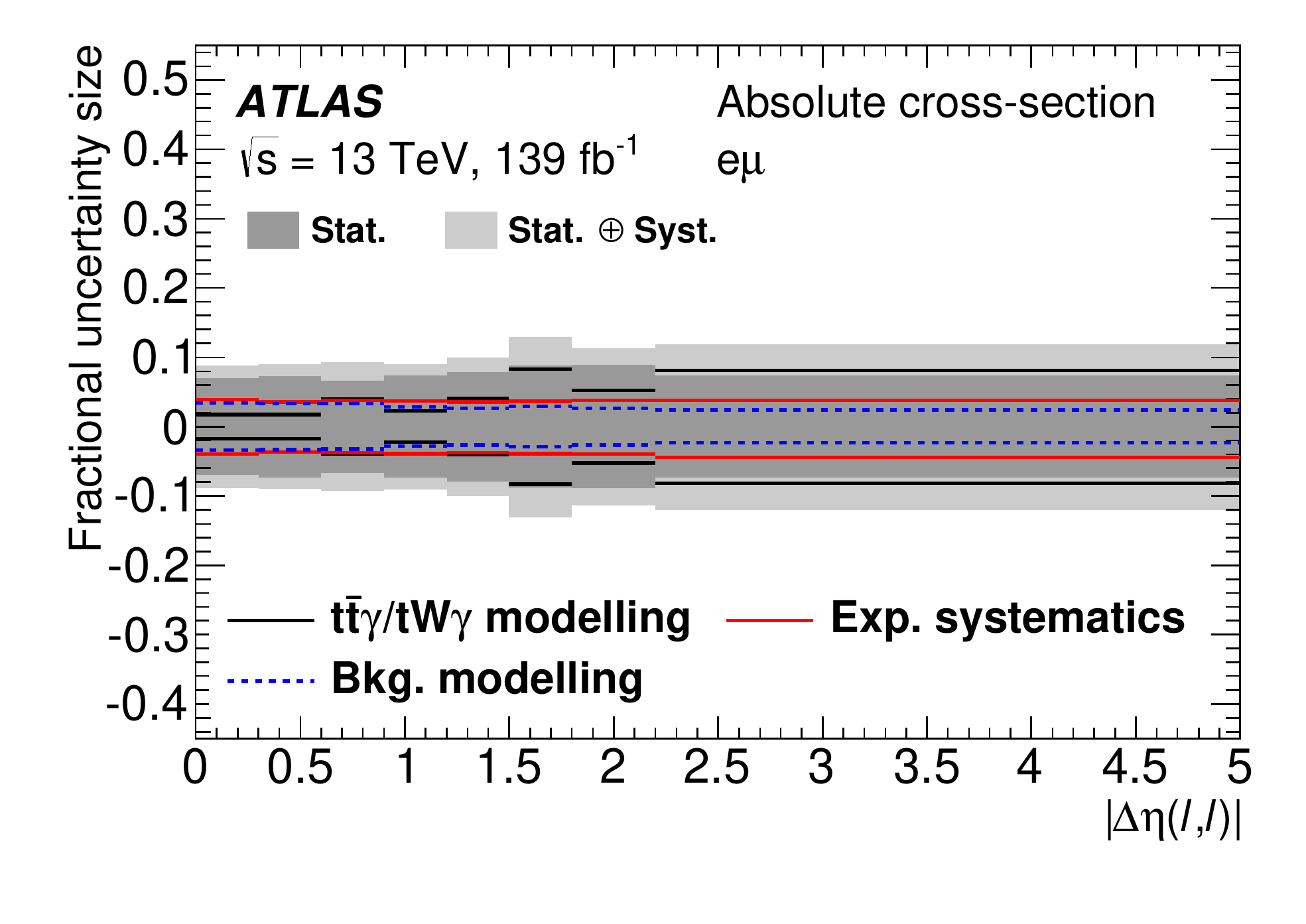}
  \includegraphics[width=0.48\textwidth, clip, trim=5pt 25pt 5pt 10pt]{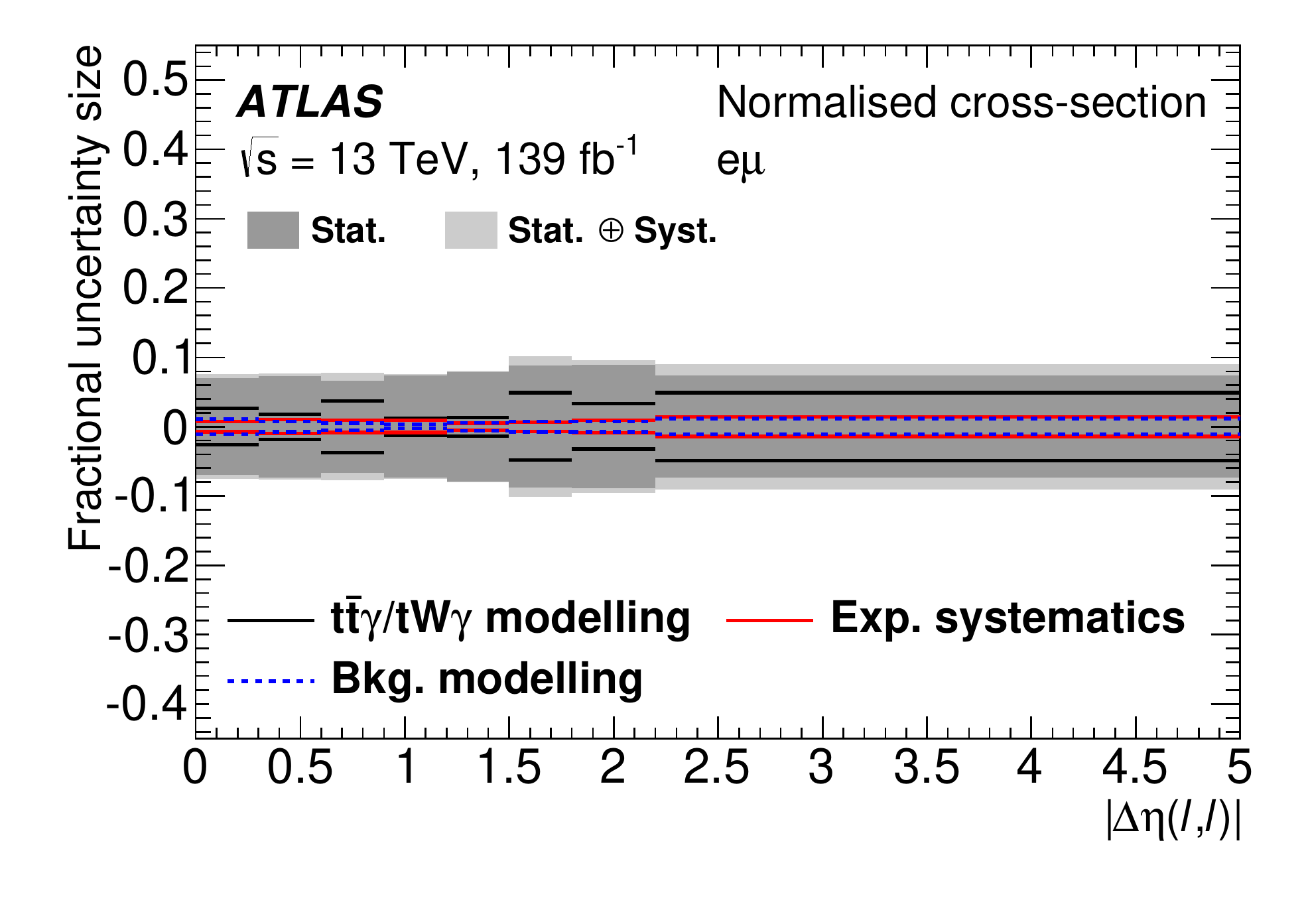}
  \\
  \includegraphics[width=0.48\textwidth, clip, trim=5pt 25pt 5pt 10pt]{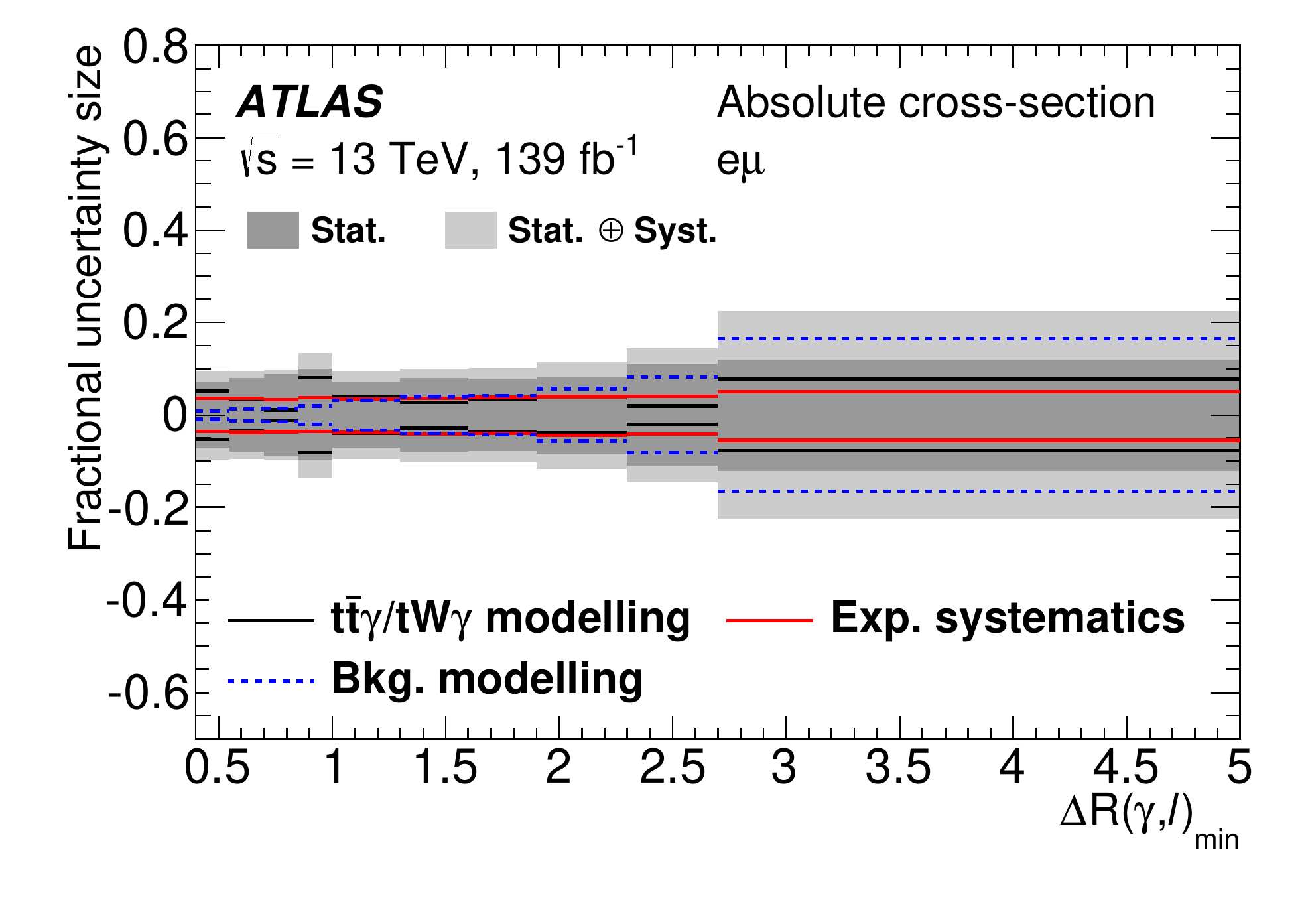}
  \includegraphics[width=0.48\textwidth, clip, trim=5pt 25pt 5pt 10pt]{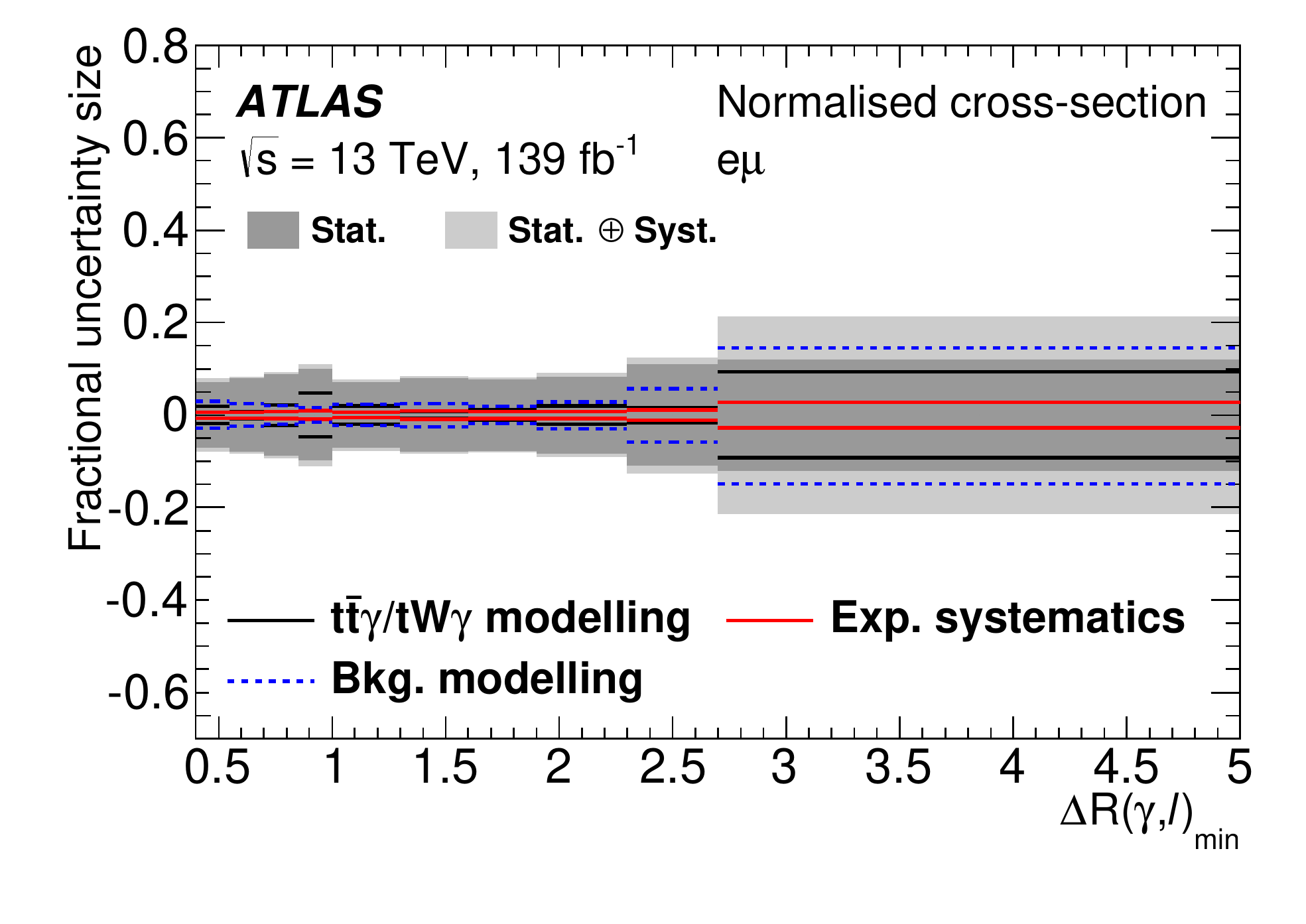}
  \caption[Uncertainty composition for \cref{fig:app-diff-xsec-abs,fig:app-diff-xsec-norm}]{%
    Composition of the bin-by-bin uncertainties for the differential \xsecs as functions of the photon absolute pseudorapidity, $\left| \eta(\gamma) \right|$, the absolute difference in pseudorapidities of the two charged leptons, \Detall, and the distance between the photon and the closer of the two charged leptons in the \etaphi plane, \DRlph.
    The left column shows the uncertainties for the absolute bin-by-bin spectra in \cref{fig:app-diff-xsec-abs}, the right column those of the normalised spectra in \cref{fig:app-diff-xsec-norm}.
    The systematic uncertainties are reduced visibly for the normalised spectra.
    Figures taken from Ref.~\cite{TOPQ-2020-03}.
  }
  \label{fig:app-diff-xsec-uncertainties}
\end{figure}

This appendix summarises additional results with unfolded \ATLAS data published in Ref.~\cite{TOPQ-2020-03}.
None of these results were the main focus of the author's work.
The following observables are shown:
the photon absolute pseudorapidity, $\left| \eta(\gamma) \right|$, the absolute difference in pseudorapidities of the two charged leptons, \Detall, and the distance between the photon and the closer of the two charged leptons in the \etaphi plane, \DRlph.
\Cref{fig:app-diff-xsec-abs} shows the distributions with absolute bin-by-bin values compared to the \NLO theory prediction.
Agreements are calculated with Pearson $\chi^2$ tests and summarised in \cref{tab:app-diff-xsec-abs-chi2}.
\Cref{fig:app-diff-xsec-norm} shows the normalised differential \xsecs compared with the \NLO theory prediction and the \LOPS predictions for the \tty and \tWy signal (\Madgraph interfaced with \Pythia and \Herwig).
Pearson $\chi^2$ test results are summarised in \cref{tab:app-diff-xsec-norm-chi2}.
The composition of uncertainties of all shown spectra is given in \cref{fig:app-diff-xsec-uncertainties}.

\begin{table}
  \centering
  \caption[$\chi^2$/\NDF and $p$-values for absolute \xsecs shown in \cref{fig:app-diff-xsec-abs}]{%
    Agreement between the unfolded \ATLAS data in \cref{fig:app-diff-xsec-abs} and the fixed-order \NLO theory computation~\cite{Bevilacqua:2018woc,Bevilacqua:2018dny}.
    Pearson $\chi^2$ tests are performed and the results are given in combination with the numbers of degrees of freedom (\NDF).
    The $p$-value is taken from a comparison with the corresponding $\chi^2$ distribution.
    Values taken from Ref.~\cite{TOPQ-2020-03}.
  }
  \label{tab:app-diff-xsec-abs-chi2}
  \begin{tabular}{l S[table-format=2.1] @{${\,}/{\,}$} S[table-format=2.2] S[table-format=1.2]}
    \toprule
    Observable & {$~~\chi^2$} & {\NDF} & {$p$-value} \\
    \midrule
    $\left| \eta(\gamma) \right|$ &  4.5 &  8 & 0.81 \\
    \Detall                       & 6.2  & 8  & 0.62 \\
    \DRlph                        & 11.7 & 10 & 0.31 \\
    \bottomrule
  \end{tabular}
\end{table}

\begin{table}
  \centering
  \caption[$\chi^2$/\NDF and $p$-values for absolute \xsecs shown in \cref{fig:app-diff-xsec-norm}]{%
    Agreement between the unfolded \ATLAS data in \cref{fig:app-diff-xsec-norm} and the \LOPS simulations of the \tty and \tWy processes as well as with the fixed-order \NLO theory computation~\cite{Bevilacqua:2018woc,Bevilacqua:2018dny}.
    Pearson $\chi^2$ tests are performed and the results are given in combination with the numbers of degrees of freedom (\NDF).
    The $p$-value is taken from a comparison with the corresponding $\chi^2$ distribution.
    Values taken from Ref.~\cite{TOPQ-2020-03}.
  }
  \label{tab:app-diff-xsec-norm-chi2}
  \begin{tabular}{l l S[table-format=2.1] @{${\,}/{\,}$} S[table-format=2.2] S[table-format=1.2]}
    \toprule
    Observable && {$~~\chi^2$} & {\NDF} & {$p$-value} \\
    \midrule
    \multirow{3}{*}{$\left| \eta(\gamma) \right|$}
    & \textsc{mg5}+\Pythia & 7.3 & 7 & 0.40 \\
    & \textsc{mg5}+\Herwig & 7.7 & 7 & 0.36 \\
    & \NLO theory          & 4.5 & 7 & 0.72 \\
    \midrule
    \multirow{3}{*}{\Detall}
    & \textsc{mg5}+\Pythia &  6.5 & 7 & 0.48 \\
    & \textsc{mg5}+\Herwig &  6.8 & 7 & 0.45 \\
    & \NLO theory          &  5.6 & 7 & 0.59 \\
    \midrule
    \multirow{3}{*}{\DRlph}
    & \textsc{mg5}+\Pythia & 20.1 & 9 & 0.02 \\
    & \textsc{mg5}+\Herwig & 18.9 & 9 & 0.03 \\
    & \NLO theory          & 13.5 & 9 & 0.14 \\
    \bottomrule
  \end{tabular}
\end{table}

\end{document}